%% file: main.tex
\documentclass[12pt]{article}
\usepackage{appendix}
\usepackage[utf8]{inputenc}
\usepackage[numbers]{natbib}
\usepackage{authblk, changepage}
\usepackage{pdfpages}
\usepackage{multicol}
\usepackage{multirow}
\usepackage{tabularx}
\usepackage{longtable}
\usepackage{amsmath}
\usepackage{amssymb}
\usepackage{amsfonts}
\usepackage{booktabs}
\usepackage{comment}
\usepackage{color}
\usepackage{subcaption}
\usepackage{bm}
\usepackage{pdflscape}
\usepackage{pdfpages}
\usepackage{array}
\usepackage{float}
\usepackage{fancyhdr}
\usepackage{adjustbox}
\usepackage{threeparttable}
\usepackage{makecell}
\newcommand{\bbeta}{ {\boldsymbol{\beta}} }
\newcommand{\bx}{ {\boldsymbol{x}} }
\newcommand{\balpha}{ {\boldsymbol{\alpha}} }

\usepackage{lineno}
\usepackage{hyperref}
\hypersetup{colorlinks=true, pdfstartview=FitV, linkcolor=blue, citecolor=blue, urlcolor=blue}
\usepackage{xurl}
\usepackage[
  paper  = letterpaper,
  left   = 1.0in,
  right  = 1.0in,
  top    = 1.0in,
  bottom = 1.0in,
  ]{geometry}

\title{Social Network Structure, Wealth, and Wealth Inequality Across Cultures\thanks{This material is based upon work supported by the National Science Foundation under Award Nos. 1743019, 2218860, and 2218861. Authors marked with $^\ast$ contributed substantially to the conceptualization, organization, analysis, and writing of this paper.}}

\author{%
\footnotesize
Eleanor~A.~Power$^{15,24,\dagger,\ast}$,
Monique~Borgerhoff~Mulder$^{34,24,17,32,\ast}$,
Samuel~Bowles$^{24,\ast}$,
Matthew~O.~Jackson$^{25,24,\ast}$,
Jeremy~Koster$^{17,\ast}$,
Daniel~Redhead$^{40,17,\ast}$,
Thomas~Rutter$^{25,\ast}$,
Sahana~Subramanyam$^{25,\ast}$,
Justin~Weltz$^{24,\ast}$,
Nurul~Alam$^{12}$,
Sarah~Alami$^{29}$,
Alexandra~Alvergne$^{28,5,10}$,
Curtis~Atkisson$^{48}$,
Michele~Barnes$^{46}$,
Bret~Beheim$^{17}$,
Christine~M.~Beitl$^{41}$,
Madeline~Brown$^{42}$,
Mark~Caudell$^{50}$,
Wendy~Ch\'{a}vez-P\'{a}ez$^{8}$,
Komal~Chauhan$^{15}$,
Joshua~Cinner$^{46}$,
Siobh\'{a}n~Cully$^{23}$,
Augusto~Dalla~Ragione$^{17}$,
Angelina~L.~DeMarco$^{47}$,
Ivan~Deschenaux$^{15}$,
Federico~Fernandez$^{6,26}$,
Juan~Pablo~Ferreiro$^{6,26}$,
Drew~Gerkey$^{19}$,
Matthew~Gervais$^{4}$,
Christopher~Golden$^{9}$,
Gianluca~Grimalda$^{16,20,11,14}$,
Werner~Hertzog$^{49}$,
Paul~L.~Hooper$^{45}$,
Karen~Kramer$^{47}$,
Geoff~Kushnick$^{1}$,
Banrida~Langstieh$^{18}$,
Rodrigo~Lazo$^{43,22}$,
Sheina~Lew-Levy$^{7}$,
Shane~Macfarlan$^{47}$,
Emmanuel~Maliti$^{13}$,
Karl~J.~Mertens$^{2}$,
Madalena~Monteban$^{6,26}$,
Rafael~Morais~Chiaravalloti$^{30}$,
Daniel~Murphy$^{38}$,
Kathryn~Oths$^{31}$,
Alejandro~P\'{e}rez~Velilla$^{36}$,
Emily~Post$^{47}$,
Sean~Prall$^{35}$,
Cody~Ross$^{17}$,
Anirudh~Sankar$^{25}$,
Brooke~Scelza$^{35}$,
Michael~Schnegg$^{27}$,
Edmond~Seabright$^{29}$,
Mary~K.~Shenk$^{21}$,
Kathrine~E.~Starkweather$^{37}$,
Chun-Yi~Sum$^{3}$,
Bram~Tucker$^{39}$,
Bapu~Vaitla$^{34}$,
Vivek~Venkataraman$^{33}$,
John~P.~Ziker$^{2}$%
}
\date{August 2026}

\begin{document}

\maketitle

\vspace{-1cm}

%\begin{adjustwidth}{1.5cm}{1.5cm}

\begin{abstract}
Despite theory tying wealth inequality to social structure, empirical evidence has been limited to a few studies based on online social media data.
This study uses a very different type of data, expands the global coverage to very different types of societies, and investigates new questions. In particular, we collect data from $\sim 3500$ sharing units (households) in 46 communities across the globe, representing considerable human social and cultural diversity. In each, we analyze the relationship between people's material wealth and the structure of social networks: borrowing money, sharing food, working together, socializing, etc.
In almost all communities, a sharing unit's material wealth is positively associated with the number of other sharing units it both helps and is helped by.
A sharing unit's wealth is also associated with the relative wealth of the sharing units to which it is linked---a form of economic homophily.
Notably, communities with greater wealth inequality are also characterized by a network structure in which poorer sharing units are less well connected to wealthier ones.
We augment our unique cross-cultural data with other community-level environmental, institutional, and economic attributes, opening new avenues for future research into the co-determination of wealth and social networks.
\end{abstract}

\newgeometry{top = 2cm, bottom = 2cm, left = 1.5cm, right = 1.5cm}

\section{Introduction}

\begin{figure}[t]
 \includegraphics[width = \linewidth]{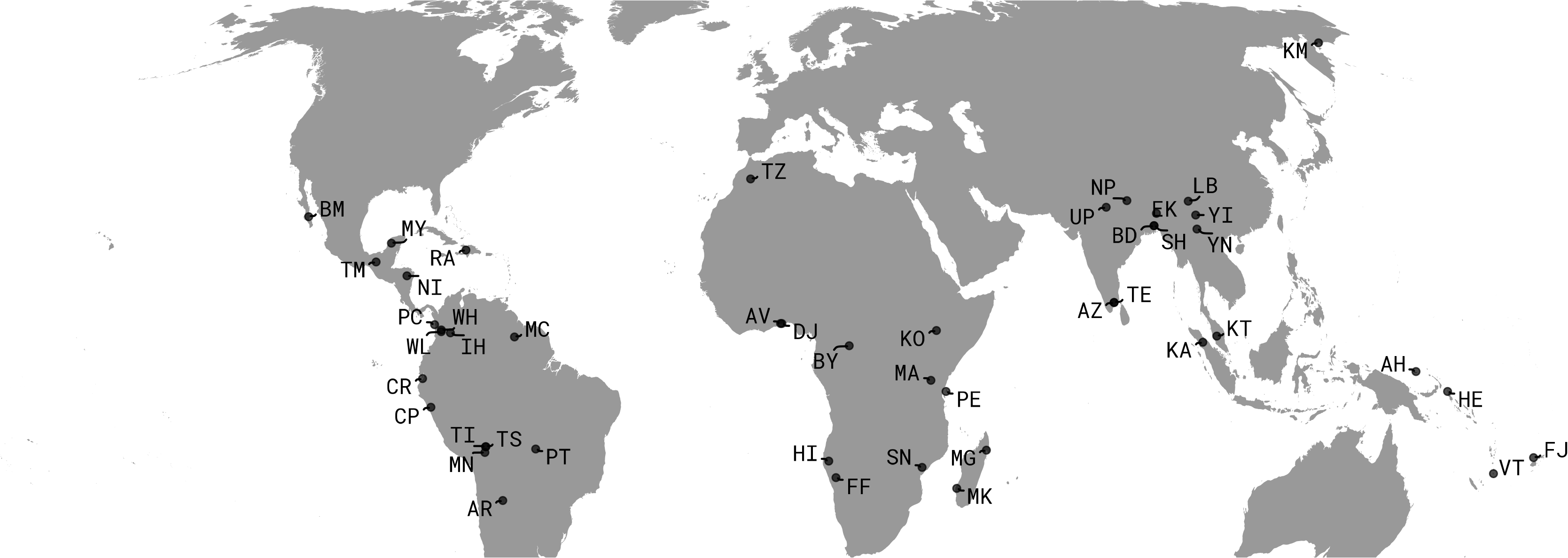}
 \caption{\textbf{Locations of the communities.} See Table~\ref{tab:su_summary} for site codes.}
 \label{fig:sites}
\end{figure}

Multiple theories have linked differential prosperity to social connections \citep{loury1977,bourdieu_forms_1986,coleman_social_1988,woolcock_social_1998,bebbington_capitals_1999, lin_social_2001-1}, and yet we know little about the extent to which this holds true across the diversity of livelihoods that characterize the human species.
In this paper, we leverage extensive economic, demographic, and social support network data that we have collected across a diverse set of 46 small, mostly rural communities that sit at various stages of market-integration, across 28 countries (see Figure~\ref{fig:sites}). The livelihoods and social structures represented in these data cover a wide range of human communities, from pastoralists in Namibia and Argentina to fishers in Benin and Russia to farmers in India and Colombia to hunter-gatherers in the Republic of the Congo. Many of these communities are ones that fall beyond the reach of government services, let alone survey enumerators or social media companies, making this a decidedly different sample than those used in prior work and one that captures more of the true breadth of human social and economic arrangements.

This unique dataset lets us investigate how social capital---here in the form of network connections---relates to material wealth.
%There is good reason to believe that network connections should be related to wealth.
Given that people rely on their social networks to buffer subsistence risks, to learn, to mobilize collective action, to access opportunities, and to secure resources, reliance on ``social capital'' has long been theorized to be a primary determinant of wealth and well-being \citep{loury1977,bourdieu_forms_1986,coleman_social_1988,woolcock_social_1998,bebbington_capitals_1999, lin_social_2001-1,dasgupta2005economics,jackson2025inequality}.
Many case studies have found that various forms of social capital correlate with welfare \citep{putnam_bowling_1995,burt_brokerage_2005,tabellini2008institutions,fafchamps2013development}.  However, social capital remains a general term that can take a variety of different forms \citep{portes_social_1998,durlauf1999case,sobel2002can, jackson_typology_2020}, which can lead to confusion both in terms of its measurement and its implications.
To clarify and distinguish different forms of social capital, research using US Facebook data examined a dozen distinct measures of social capital and found that they have different implications for economic mobility, with a specific network-based measure termed ``Economic Connectedness'' standing out as having a strong and positive correlation with a community's economic mobility \citep{chetty_social_2022,chetty_social_2022-1}.%, and causing education attainment and higher future earnings \citep{chetty2026causal}.

While a recent study has expanded the study of economic connectedess to the global Facebook user base \cite{johnston_SocialCapitalAroundTheWorld}, it remains unknown whether such patterns hold across the full breadth of human societies, where modes of production, the forms and patterns of sociality, and the dynamics of wealth inequality differ \cite{borgerhoff_mulder_intergenerational_2009}.
Prior investigations of social capital are based largely on what could be described as ``WEIRD'' settings and samples \citep{henrich_weirdest_2010}, so it is unclear how access to the global market economy, state services, and financial institutions influence patterns of day-to-day sociality and economic exchange.
One recent study in South India \cite{banerjee_changes_2024} found that the introduction of formal microfinance caused a subsequent retreat from supportive relationships even among poorer households less likely to use the services, suggesting that such integration can lead to profound changes and disproportionately impact those in greatest need.
Social support---especially household exchange and food sharing---may be particularly important for buffering risks in rural, subsistence economies \cite{winterhalder_diet_1986, baird_livelihood_2014}, and the incentives to build such social capital may vary across households based on their wealth and status \cite{boyd_coalitions_1992, gurven_does_2015, kasper_who_2015, ready_why_2018, hackman_kin_2021}.
Any effort to understand the relationship between material wealth and social capital writ large must therefore entail a much wider range of settings and include a focus on the supportive relationships that are crucial to daily life and livelihoods.

With this background in mind, we examine (\textit{i}), within each community, the relationship between each sharing unit's (i.e., household's) wealth and its network connections, as well as
(\textit{ii}), across communities, the relationship between the structure of a community's network and its wealth inequality.
We find that the relationship between a sharing unit's material wealth and its network position is generally consistent even across these highly diverse communities.  Moreover, we find that communities that have greater inequality in material wealth are those that are lacking in economic connectedness.
%We further examine how a community's level of wealth inequality relates to its other characteristics.

In terms of distinguishing forms of social capital based on networks, we examine three basic types:
($i$) the extent to which a sharing unit can solicit support from others \citep[e.g.,][]{henly_contribution_2005}, ($ii$) the extent to which a sharing unit provisions support to others \citep[e.g.,][]{ready_sharing-based_2018}, and ($iii$) a sharing unit's economic connectedness: the extent to which it has access to (relatively) \emph{wealthy} others \citep[e.g.,][]{chetty_social_2022, harris2025social}. This paper presents positive correlations between material wealth and measures of all three of these explicit network forms of social capital in the majority of the communities---significantly more than expected at random.

The fact that sharing units that have more and wealthier social connections tend to be wealthier in most communities reveals only something about relative ranks in wealth and does not indicate what to expect in terms of whether social networks predict
{\sl inequality} in the absolute levels of material wealth in a community.
Nonetheless, there are theoretical bases for expecting that specific details of social structure will be associated with material wealth inequality.
For example, differences in access to information and opportunities across a society that exhibits homophily in wealth can lead to increased (and persistent) inequality (e.g., see the discussion and references in \citep{jackson2025inequality}).
If some sharing units are disproportionately endowed with advantageous social connections, then this should be reflected in their endowments of material wealth.
Notably, we find that it is not differences in the number of support partners, but rather who is connected to whom that seems to be of consequence:
while a community's inequality in access to or provisioning of support are not associated with its material wealth inequality, its economic connectedness---the extent to which the poor within a community are well-connected to the wealthy in the community---is a strong and robust predictor. Places with relatively more segregated networks between rich and poor exhibit greater inequality.
%In brief, we expect that \emph{communities that exhibit greater inequality in these measures of ``relational wealth'' should also exhibit greater inequality in material wealth.} Further, \emph{communities where the materially poor are less well connected to the materially rich should exhibit greater inequality in material wealth.}

Differences in the strength of these associations can be partly explained by other attributes of these communities (such as their reliance on wage or salaried labor and the inclusiveness of the country's private property rights) that we additionally consider.

%Our contribution is both in the collection of this unique and compelling dataset that spans a spectrum of different societies across the world (and that are not WEIRD in the terminology of \citep{henrich_weirdest_2010}) and in uncovering patterns between key wealth and network variables that are remarkably consistent across these diverse communities.
Our contribution lies in (\textit{i}) sampling a wide spectrum of different communities across the world with varying levels of state and market incorporation (and that are generally not ``WEIRD'' in the terminology of \citep{henrich_weirdest_2010}), (\textit{ii}) amassing a unique and detailed dataset based on interviews with household members about their complete set of possessions and their multiplex social relationships (food sharing, information exchange, money lending, etc.) that play a crucial role in their daily lives and livelihoods, and (\textit{iii}) uncovering strong patterning in the relationship between key material wealth and network variables. While these patterns are quite consistent even across the highly diverse communities sampled, we refrain from claims regarding causality precisely because of the plethora of potential causal pathways likely involved in communities as distinct as mobile pastoralists, intensive farmers, and market-integrated fisherfolk.

%We make no claims regarding causality and recognize that the observed associations could result from a plethora of causal pathways; efforts at causal identification are left for future work.

\section{Fieldsites and Data}
\label{data}

\begin{figure}[t]
    \centering
    \includegraphics[width = \textwidth]{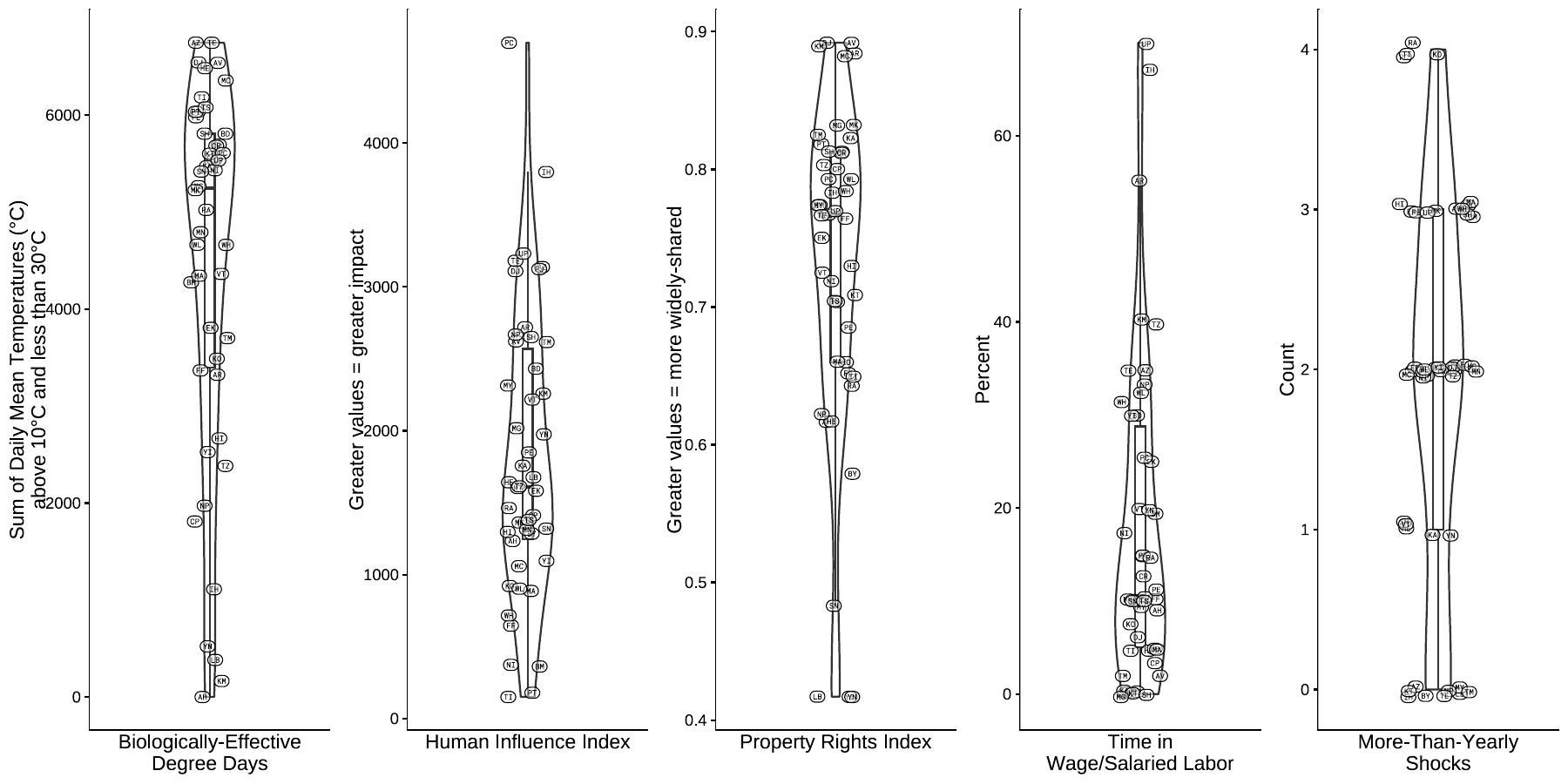}
    \caption{\textbf{Summary measures illustrating the diversity of the communities.}
    %Biologically-Effective Degree Days: \citep{nobakht_agroclimatic_2019},
    %Human Influence Index: \citep{sanderson_march_2022},
    %National Property Rights Index: \citep{vdem2025},
    %Time in wage or salaried labor and the number of shocks experienced by the community more than yearly: anthropologist-provided.
    See Table~\ref{tab:variables} for further description of these (and other) variables characterizing the communities.}
    \label{fig:site_characteristics}
\end{figure}

We gathered data from 46 communities around the world (see Figure~\ref{fig:sites}).%\footnote{Due to the opportunistic nature of this sample (willing anthropologists with active fieldsites) our inferences are primarily of a wide range of communities transitioning into market-based economies.}
The communities are diverse across multiple dimensions (see Figure~\ref{fig:site_characteristics}).
Although most are rural and remote, some are in high-population-density areas.
Some lack substantial infrastructure and reliable electricity, while others are located near major highways and have access to government services and facilities.
Although they can be roughly classified as horticulturalists, pastoralists, fishers, or similar discrete categories (see Figure~\ref{fig:subsistence}), almost all exhibit a range of livelihood strategies \citep{ellis_rural_2000}.
Most communities are not fully integrated into the global economy, but are increasingly (and to various extents) supplementing or replacing subsistence practices with market enterprises and wage labor. They are also differingly autarkic with respect to falling under national state governance and vary substantially in ecology and the extent to which they are impacted by climate and other shocks (natural or otherwise).

The communities range in size, with an average of four to five individuals grouped into 60 sharing units, on average (see Table~\ref{tab:su_summary} for descriptive statistics).
Here, we use the term ``sharing unit'' (as opposed to ``household'') as this is the unit principally within which resources are pooled, and to highlight the variable constellations of members this represents across the communities. In some communities, sharing units equate to nuclear families centered on a monogamously married couple, whereas in others they comprise larger compounds of a man and his wives and their children, or multi-generational extended families.

The breadth of this variation contrasts our work with prior studies of inequality and social capital, which commonly rely on data drawn from large global economies (because of the reliance on administrative tax data \citep[e.g.,][]{piketty_capital_2013, killewald_wealth_2017, zucman_global_2019} and/or large-scale digitally-mediated data sources \citep[e.g.,][]{eagle_network_2010, chetty_social_2022, rajkumar_causal_2022, jahani_long_2023}).
%While this literature has examined major global economies, there is reason to believe that social networks are comparably important in rural communities elsewhere in the world \citep{schweizer1997embeddedness, godoy_role_2007, baird_livelihood_2014, rockenbauch_social_2017, wiessner202244}.
Furthermore, we leverage the extensive ethnographic work of our team to collect rich, locally meaningful measures of social connections across a multitude of different forms of interactions, which are generally unavailable in other studies.
%With livelihoods that are tightly coupled to local environments, it is also possible that social connections may be more consequential in the settings under study here \citep{rockenbauch_social_2017}. %Understanding the factors and dynamics that lead to wealth accumulation and disparities in such communities is important not only for their well-being, but also because, through their diversity, they give insights into what may transpire in other societies around the world.

Data for this project, gathered by researchers working with each community, entail detailed demographic, economic, and social connections reported by virtually all sharing units in each community. Most importantly, this comprises:
detailed enumeration of each sharing unit's property and assets, from which we derive a measure of each sharing unit's overall material wealth, and
sharing unit members' nominations of who they turn to in the community for different types of support, from which we define a directed, weighted network representing the flows of support among sharing units (see the Methods section and Supplementary Section \ref{supp:databackground} for details).

With these in hand, we construct our explicit network measures of sharing units' social capital:
(\textit{i}) access to support within the community,
(\textit{ii}) provisioning of support within the community, and
(\textit{iii}) Average Alter Wealth: the average wealth of a sharing unit's connections, considering either those providing support or those to whom support is provided. (See the Methods section and Supplementary Section~\ref{supp:definitions-measures} for further details.)

\section{Within-Site Results: Sharing Units' Wealth and Social Capital}
\label{within}

First, we analyze whether there is a relationship between
a sharing unit's social capital and its material wealth \textit{within} communities.

\begin{figure}[ht!]
    \centering
    \begin{subfigure}{0.49\textwidth}
        \centering
        \includegraphics[width=\textwidth]{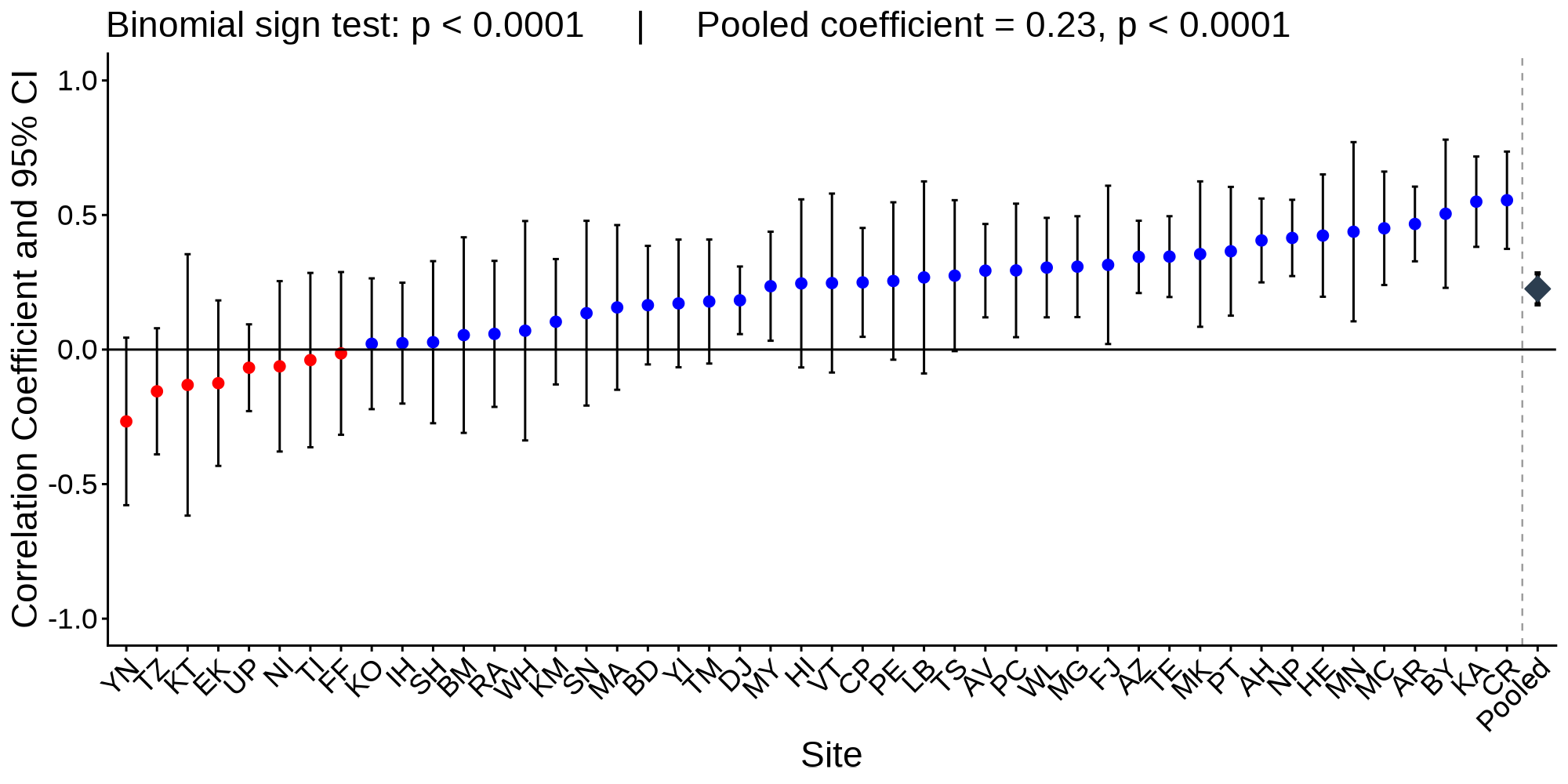}
        \caption{Support access per capita vs wealth p.c.}
    \end{subfigure}%
    \vspace{1em}
    %\hspace{2em}
    \begin{subfigure}{0.49\textwidth}
        \centering
        \includegraphics[width=\textwidth]{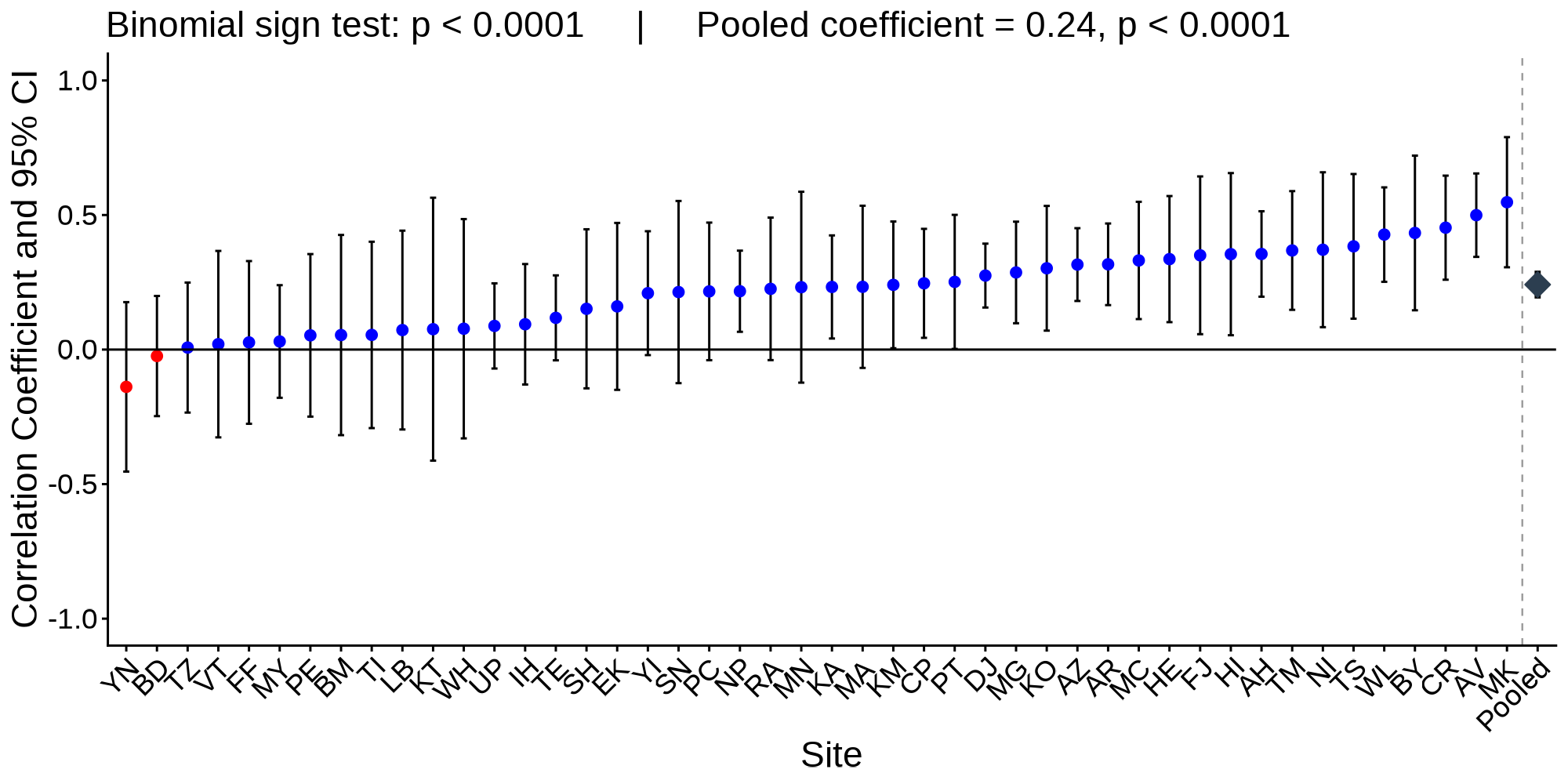}
        \caption{Support provisioning per capita vs wealth p.c.}
    \end{subfigure}
    \begin{subfigure}{0.49\textwidth}
        \centering
        \includegraphics[width=\textwidth]{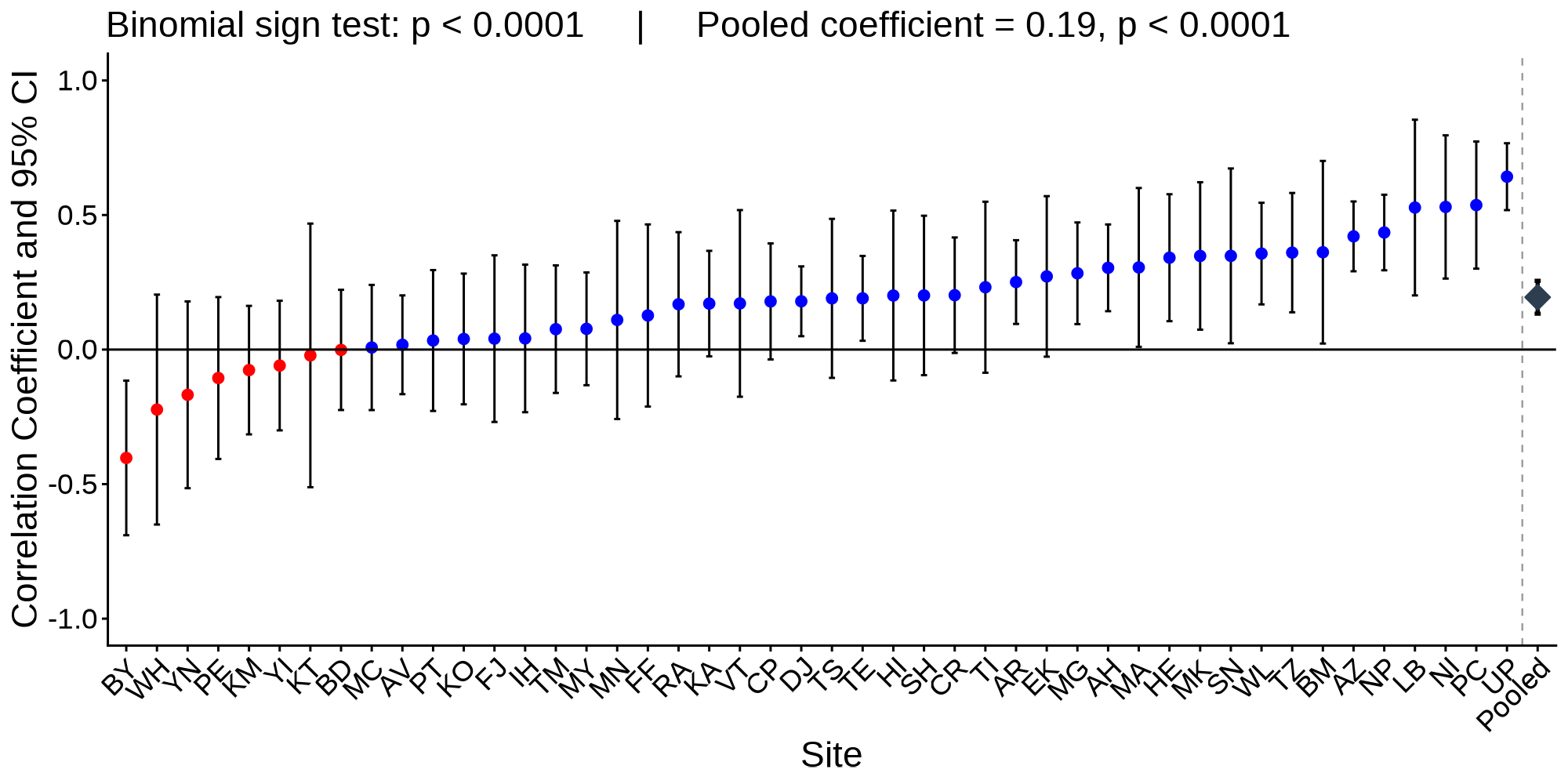}
        \caption{$\text{AAW}_i$ (of supporters) vs wealth p.c.}
    \end{subfigure}%
    %\hspace{2em}
    \begin{subfigure}{0.49\textwidth}
        \centering
        \includegraphics[width=\textwidth]{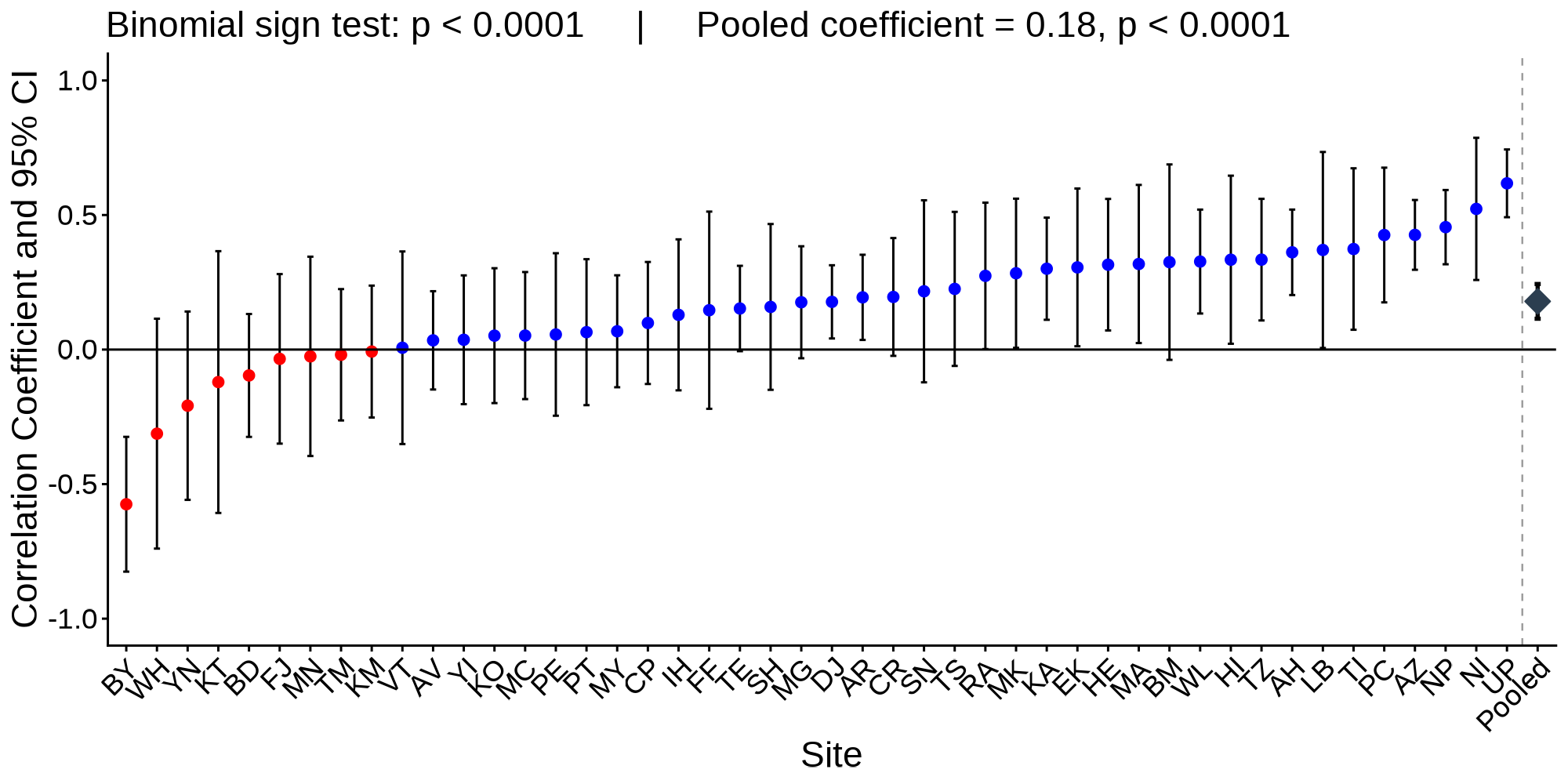}
        \caption{$\text{AAW}_i$ (of supportees) vs wealth p.c.}
    \end{subfigure}
    \caption{\textbf{Correlations between network connections and material wealth}. Each panel presents a set of correlation coefficients and 95\% confidence intervals, one for each community, for a pair of percentile ranked variables. Access to support (i.e., out-degree) per capita (a) and provisioning of support (i.e., in-degree) per capita (b) are calculated using the ``composite'' network.
    Average Alter Wealth is also calculated using the ``composite'' network, considering alters who provide support (i.e., supporters) for panel (c) and alters to whom support is provided (i.e., supportees) for panel (d). We use the percentile rank of wealth per capita (reweighting the distribution by sharing unit size). Correlation coefficients, and their standard errors, are calculated as the slope coefficient (and the corresponding standard error) of a regression of the $y$-variable, ranked across sharing units in a community, on the $x$-variable ranked across sharing units in a community. We display the two-sided $p$-value of observing so many positive correlations under the sharp null that the true correlation in each community is zero, as well as a pooled coefficient (and associated $p$-value) resulting from a random-effects meta-analysis.}
    \label{fig:within-site-multipanel}
\end{figure}

We find that, across the majority of communities, relatively wealthier sharing units have greater access to support from other sharing units (``support access'') and further have a greater propensity to be called on to provision support to others (``support provisioning''). The top panels of Figure~\ref{fig:within-site-multipanel} show the correlation in the rankings of material wealth per capita and the support access and provisioning per capita of each sharing unit respectively, in each community. (Throughout, we use percentile ranks, so that our measures are more comparable across communities and not strongly swayed by outlier wealth values.)
The correlation is positive for almost all of the communities for provisioning support and somewhat less consistently so for accessing support.
%In Figure~\ref{fig:within-site-multipanel} panels (c) and (d), we additionally present per-capita measures % to account for the fact that larger sharing units may both be wealthier and also have more support partners. Thus,
%where both material wealth and out-/in-degree are divided by the number of members of the sharing unit and then ranked within the community, so that the correlation is between the rankings of material wealth per capita and the normalized out-/in-degree per capita. %A positive correlation means that the sharing units that have relatively more wealth per capita have relatively more out-/in-degree per capita.
%This alternative accounting maintains the consistent correlation between material wealth and provisioning support and shows a strengthened correlation between material wealth and accessing support, across communities.

Beyond the number of other sharing units that a given sharing unit is connected to, we also examine the material wealth of those sharing units. To do this, we construct measures of economic connectedness following \cite{chetty_social_2022, chetty_social_2022-1}
to see if the connectedness of a sharing unit to relatively wealthy sharing units is correlated to its wealth level. These are effectively measures of wealth homophily.
The main measure of economic connectedness that we use is the average wealth of the alters of a given sharing unit, ``Average Alter Wealth'' ($\text{AAW}_i$): the average percentile rank of the wealth (per capita) of the alter sharing units to a given ego sharing unit. (See the Methods section and Supplementary Section~\ref{supp:definitions-measures} for formal definitions.)
The bottom panels of Figure~\ref{fig:within-site-multipanel} shows that, across the majority of communities, sharing units with greater material wealth are connected to sharing units that are themselves wealthier (have higher $\text{AAW}_i$), when considering connections that come either from accessing or provisioning support (i.e., considering either those who support them, or those whom they support).

For each of these four network measures of social capital, the $p$-value of observing this many positive correlations out of this many communities is close to 0, and the cross-community average effect, estimated via random-effects meta-analyses to account for variation in the strength of the relationship across communities, are positive and statistically significant.

In Supplementary Section~\ref{supp:robustness_within} we show the robustness of these associations to different accountings of sharing unit wealth, different underlying network constructions, and different versions of economic connectedness.
Section~\ref{supp:alternative_SUweightings_within} uses alternative accountings of sharing unit wealth, including using no weights (i.e., using absolute, not per capita measures), weighting by the number of adults in each sharing unit, and controlling for sharing unit size.
Section~\ref{supp:alternative_networks} constructs the network using different rules for aggregation, and Section~\ref{supp:sum_wealth_within} uses not \emph{average} alter wealth, but the median and sum of alter wealth.
Overall, we find that while the results for accessing support are not always robust and so somewhat ambiguous, our results for Average Alter Wealth and provisioning support are very consistently robust to these choices.

\section{Cross-Site Results: Communities' Wealth Inequality, Network Patterns, and Other Characteristics}
\label{across}

%Having established that material wealth of a sharing unit is correlated with network measures of centrality and economic connectedness,

We now examine the relationship between the \emph{inequality} in material wealth of a community and the community's network structure and other characteristics.

\subsection{Wealth Inequality and Network Structures}

We first consider community-level inequality in sharing units' access to and provisioning of support, and the economic connectedness of below-median wealth sharing units.

\begin{figure}[!ht]
\centering
\begin{subfigure}{0.48\textwidth}
    \centering
    \includegraphics[width=\textwidth]{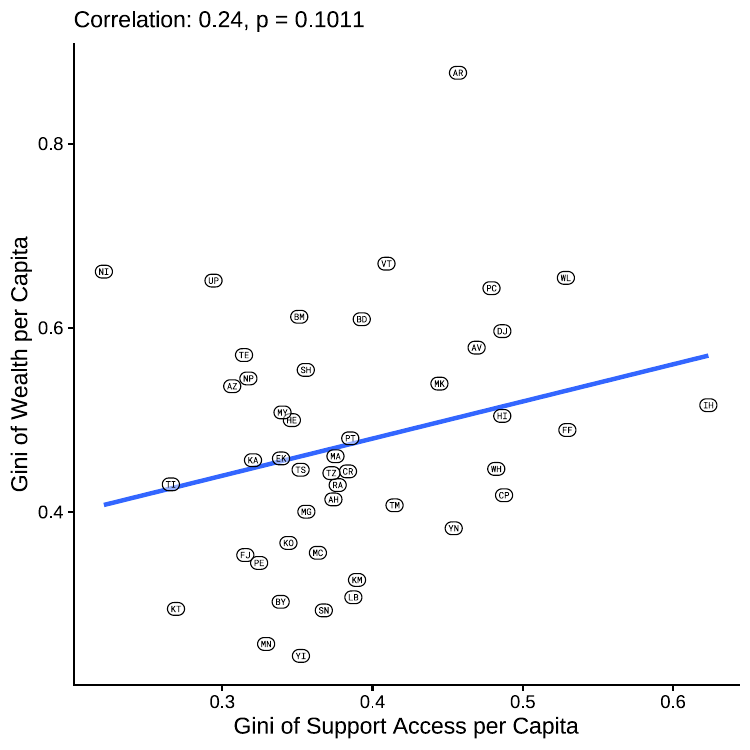}
%    \caption{In-Degree Per Capita Gini vs Wealth per Capita Gini}
\end{subfigure}
\hfill
\begin{subfigure}{0.48\textwidth}
    \centering
    \includegraphics[width=\textwidth]{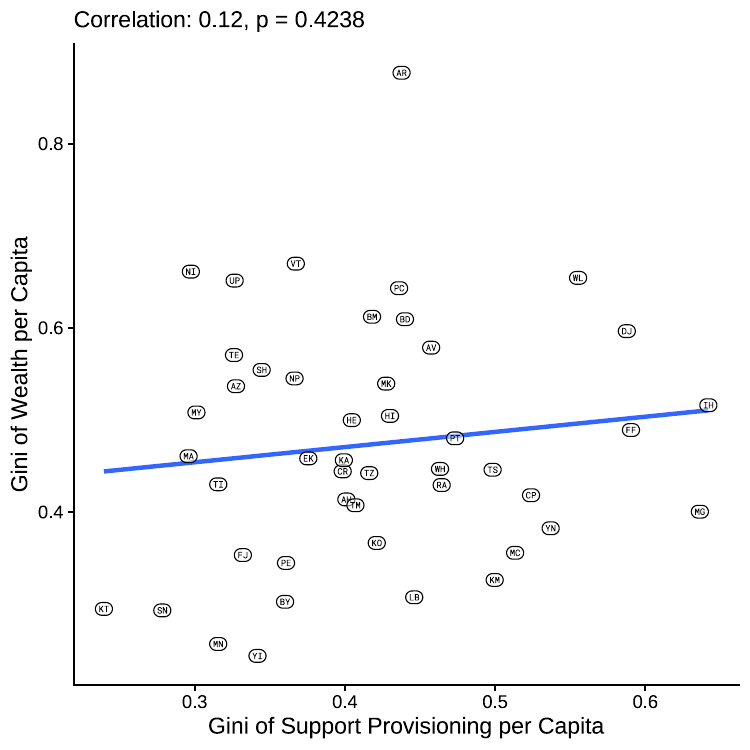}
%    \caption{In-Degree Per Capita Gini vs Wealth per Capita Gini}
\end{subfigure}
\vspace{1em}
\begin{subfigure}{0.48\textwidth}
    \centering
    \includegraphics[width=\textwidth]{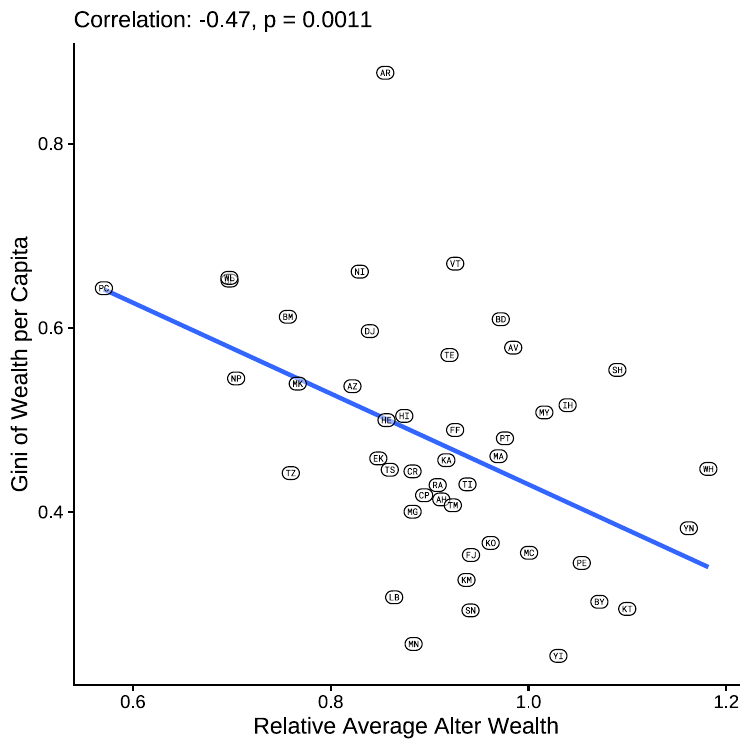}
%    \caption{AAWP vs Wealth per Capita Gini.}
\end{subfigure}
\hfill
\begin{subfigure}{0.48\textwidth}
    \centering
    \includegraphics[width=\textwidth]{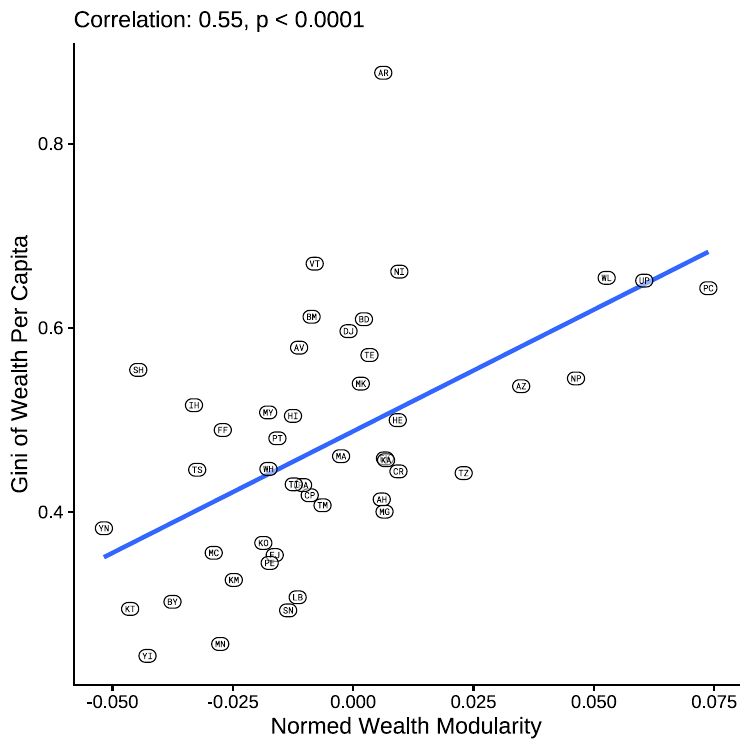}
%    \caption{In-Degree Per Capita Gini vs Wealth per Capita Gini}
\end{subfigure}
\caption{\textbf{Cross-site correlations with the wealth per capita Gini} and the Gini of support access per capita (top left), the Gini of support provisioning per capita (top right), Relative Average Alter Wealth (using per capita wealth and alters who provide support) (bottom left), and normed wealth modularity (using per capita wealth) (bottom right). For normed wealth modularity, a positive (negative) value can be interpreted as a maximum observed modularity that is larger (smaller) than the 95\% quantile of wealth modularity measures we would observe under a random distribution of wealth; larger values suggest greater assortment by wealth.}
\label{fig:wealth-ginis-vs-scatters}
\end{figure}

As shown in the top panels of
Figure~\ref{fig:wealth-ginis-vs-scatters}, there is no consistent association between a community's material wealth inequality and inequality in sharing units' accessing or provisioning of support. (The correlation between the Gini of support access and the Gini of wealth per capita is positive and at the border of significance, but as shown in Figure~\ref{fig:wealth-ginis-odeg}, this is not consistently so across different accountings of wealth).

In contrast,
as shown the bottom left panel of Figure~\ref{fig:wealth-ginis-vs-scatters},
there is a strong and significantly negative correlation between communities' material wealth Gini coefficients and the economic connectedness of the poor. The measure used here---Relative Average Alter Wealth---represents how wealthy below-median sharing units' supporters are, relative to above-median sharing units' supporters: if the poor were just as well connected as the wealthy in terms of alter wealth, then this measure would be 1. (See the Methods section and Supplementary Section~\ref{supp:wealth_modularity} for further details on this measure).
The strong negative association with Relative Average Alter Wealth means that communities that are more materially equal also have relatively more balanced connections between poorer and wealthier sharing units.
This holds across different network aggregations and accountings for sharing unit wealth (see Supplementary Section~\ref{supp:alternative_SUweightings_between}), and for different measures of the relevant connections between poorer and wealthier sharing units, as we explore in the next section.

\subsection{Wealth Inequality and Economic Connectedness}\label{sec:wealth_ineq_ec}

Given the strong associations of economic connectedness with material wealth within communities and with wealth inequality across communities, we explore economic connectedness in finer detail.

The use of a median split to divide each community into rich(er) and poor(er) sharing units allows us to standardize across settings \citep{chetty_social_2022}.
However, this split might not always capture the relevant coherent wealth classes within a community.
We therefore use network modularity \citep{newman_finding_2004} to identify the split within the wealth distribution that most cleanly divides the community into two cohesive wealth classes and then define a new measure on the basis of this split---(normed) wealth modularity---which captures how cohesive the resulting classes are (see the Methods section and Supplementary Section~\ref{supp:wealth_modularity} for further details).
The bottom right panel in Figure~\ref{fig:wealth-ginis-vs-scatters} indicates that normed wealth modularity is positively correlated with the Gini of wealth per capita, supporting the relationship between economic connectedness and wealth inequality in a more flexible context that allows for a community-specific division of sharing units into ``poor'' and ``rich.''

Supplementary Section~\ref{supp:robustness_between} contains further analysis that examines modularity and economic connectedness in more detail, as well as additional robustness checks.
In particular, Section~\ref{supp:quartiles} considers splitting communities into four wealth classes, not just two, while Section~\ref{supp:multilevel_model} presents a multilevel model (also using quartiles) that provides a close analog to our site-wise Economic Connectedness measures, while giving a different quantification of our uncertainty. Both suggest that connections between the poorest and wealthiest quartiles are most strongly correlated with a community's material wealth inequality.

\subsection{Correlations of Other Community Characteristics with Wealth Inequality}

Having established the strong association between wealth inequality and economic connectedness, we finally consider additional characteristics of the communities, to identify other potential correlates. Although causal inference is precluded with the data analyzed here, we can still descriptively explore how the relationships in the data vary by the community characteristics.

\subsubsection{Community Characteristics Beyond Social Structure}

The predictors of inequality across societies have been heavily theorized. %\citep{mattison_evolution_2016}.
Our goal here is to identify characteristics of the communities, and the countries in which they are located, that might influence wealth and its distribution among market-transitioning communities.

Specifically, we consider for the community itself:
the fraction of sharing unit's connections that are with close kin, and, based on the contributing researchers' reports,
whether there are norms that preclude certain subsets of people from doing certain productive tasks,
the percent of productive time residents spend engaged in wage or salaried labor,
the percent of goods produced in the community that are sold for cash,
and the count of climate, health, economic, and violence-related shocks that occur more than once a year.
For the immediate area of the community, we additionally consider:
a measure of local biological productivity (Biologically-Effective Degree Days, \citep{nobakht_agroclimatic_2019}) and
a measure combining population density, infrastructure, and accessibility (the Human Influence Index, \citep{sanderson_march_2022}).
Finally, at the level of the nation, we consider:
the Gross Domestic Product per capita \citep{worldbank2025wdi},
and indices on the proportion of people who enjoy private property rights,
people's freedom of movement within the country, and
political corruption \citep{vdem2025}. (See Figure~\ref{fig:site_characteristics} for distributions of many of these measures across communities and the Methods section and Supplementary Section~\ref{supp:sitevars} for further details).

%, such as: property rights,
%caste or other systems that restrict people’s relationships and actions, improved infrastructure, or
%access to wage or salaried employment, national GDP per capita, freedom of movement, corruption, a measure of price and climate shocks, environmental productivity, and others.

%The site-level characteristics that we have include key variables that fit with existing theories of wealth inequality, such as the presence of property rights, caste or other systems that restrict people's relationships and actions, etc.  Given the enormous number of theories for inequality, we do not survey that here, but include a range of site characteristics that could have an influence on wealth and its distribution.

\subsubsection{Correlations of Community Characteristics with Inequality}

\begin{figure}[t]
    \centering
    \includegraphics[width=0.8\linewidth]{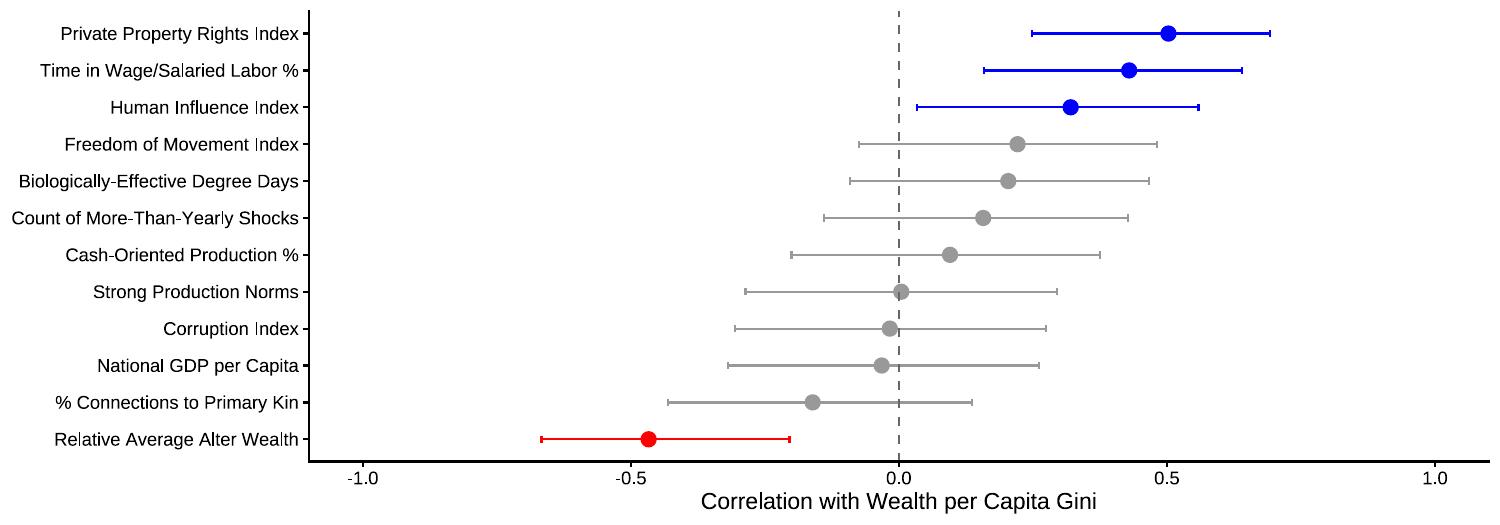}
    \caption{\textbf{Bivariate Correlations with the Gini of Wealth per Capita.} Red denotes a significant negative correlation, blue a significant positive correlation, and grey no significant correlation. Bars denote 95\% confidence intervals.}
    \label{fig:corrs-with-wpc-gini}
\end{figure}

The correlations of these variables with the Gini of wealth per capita across sites are shown in Figure \ref{fig:corrs-with-wpc-gini}.

We find a strong negative correlation of Relative Average Alter Wealth with the Gini of wealth per capita, echoing what we found in Section~\ref{across}.
Three variables show a significant positive association:
the percent of time spent in wage/salaried labor,
a national private property rights index,
and, marginally, the Human Influence Index.
Communities where residents spend more time in wage and salaried work, communities in countries where private property rights are more widely available, and communities closer to population centers with infrastructure and resources show greater material wealth inequality.

Given the large number of variables and limited correlations, it is difficult to identify relationships between these other variables with inequality that we are confident in either identifying or ruling out. To be thorough, for the interested reader we present further analyses (Lasso and multivariate regressions) in Section~\ref{supp:othervariables} of the Supplementary Materials. We leave a deeper dive for future research, when we anticipate longitudinal data that will permit cleaner identification of the causal contributions of hypothesized predictors.

\section{Discussion}\label{discussion}

Across a diverse set of communities, we find that relatively wealthier sharing units are significantly better connected within the community than their poorer counterparts.
They report greater access to support and also provide more support themselves. Further, they tend to be connected to sharing units that are themselves wealthier.
Thus, within communities, material wealth and social capital are closely tied.
Across communities, we find that the extent to which supportive relationships are unequally distributed has no consistent relationship with material wealth inequality.  Rather, a robust (negative) correlate of wealth inequality is the extent to which relatively poorer sharing units interact with relatively wealthier ones. Thus, an important element of our findings is that it is not the number of social connections that matters for the extent of wealth inequality, but instead the substance of those connections, namely who is connected to whom in terms of relative wealth.

Although our cross-sectional data do not allow for identification of the causal mechanisms underlying our results, prior literature suggests some plausible channels.
As an example, a relatively poor household's connection to relatively wealthier households could facilitate informal consumption smoothing and mitigate risks to which the poor are exposed. Less risk-exposed households will be more likely to engage in riskier, higher returns forms of investment \citep{bardhan_chapter_2000}. As a concrete example, Rosenzweig and co-authors found that poor Indian farmers disproportionately held assets they could sell in times of need (such as bullocks) rather than more profitable equipment (such as irrigation pumps) that are riskier because they have little resale value \citep{rosenzweig_wealth_1993, rosenzweig_credit_1993}. Thus, poorer farmers pursued safer strategies with lower expected returns. The risk-buffering effects of economic connectedness of poorer sharing units can thus reduce inequality by enabling riskier, higher expected-return investment by the poor and reducing differences in the rate of returns between relatively wealthier and poorer sharing units.

Other parallel channels are the information, aspirations, and norms that are dependent upon networks.  In societies with few connections between the relatively poor and wealthy, the poor can be cut off from information and unaware of opportunities or investments that the wealthy pursue, and if the poor emulate those around them they may choose not take advantage of such opportunities even when they are aware (e.g., see the discussion and references in \cite{bourdieu_forms_1986,jackson_human_2019,jackson2025inequality}).

Taken together, the above reasoning and our findings are consistent with evidence from Facebook users that economic connectedness is strongly associated with upward income mobility \cite{chetty_social_2022,harris2025social}, and comparable work in villages in Honduras \cite{shridhar_network-cycle_2025}.
As such, this study substantially broadens the geographic and social contexts within which economic connectedness exhibits such salience.
We also deliberately collected a range of relationships capturing the breadth of social and economic exchange that are at the heart of sharing units' daily lives and livelihoods, expanding the study of economic connectedness beyond the nominal realm of ``friendship.''

In some contexts, the extent of wealth homophily we observe may be readily understood, such as the high wealth homophily in caste-segmented villages in South Asia (UP, AZ, TE, NP) or the low wealth homophily in places where state policies (e.g., in China: YI, YN, LB), or local norms of charity (e.g., PE) or egalitarianism (e.g., BY) encourage cross-wealth connection.
Communities where sharing unit wealth and average alter wealth are not positively associated raise new questions about the likely multiple processes that may impede wealth homophily \citep{stibbard-hawkes_egalitarianism_2025}, and more generally about whether social capital facilitates or impedes the movement of poor households out of poverty \cite{chantarat_social_2012}.
Homophily (or heterophily) may derive from a wide range of underlying mechanisms operating in tandem \cite{mcpherson_birds_2001, shalizi_homophily_2011, angrist_perils_2014,chetty_social_2022-1}. %We note that many of the communities that show a negative association between wealth and AAW_i are also those with low wealth inequality...
Although the general consistency of the within-site associations across such a wide array of cultures and social structures is revealing,
new theoretical frameworks are needed to account for the range and patterning of wealth homophily and its association with inequality.
%The wealth homophily found within most sites could be a key engine in maintaining and amplifying inequality, allowing wealthy households to engage in higher return uses of their wealth consistent with models in \cite{bardhan_chapter_2000}.

Beyond economic connectedness, we considered other community attributes that might relate to wealth inequality.
The extent to which members of a community have well-structured and protected rights to the ownership of private property is positively associated with inequality. It could be that private property rights allow for people to collect (and monopolize) wealth in ways that lead to highly unequal distributions.  Indeed, in a few of the communities (e.g., AR and NI), there are small numbers of sharing units that own extremely high proportions of the assets of the community.
In reverse, it could be that higher inequality in the distribution of assets leads a society to adopt and protect ownership rights, especially if the leadership is governed by the relatively wealthier, consistent with recent interpretations of prehistoric wealth inequality based on archaeological evidence \citep{bowles_origins_2024}.

We also find that the amount of time community members spend in wage or salaried labor (employed by someone else in return for a wage/salary) is positively associated with wealth inequality.  This provides new data for the large literature on labor market access and its effects on inequality \citep[e.g.,][]{shenk_intergenerational_2010,lakner_global_2016}. %Notably, we find no equivalent positive association between wealth inequality and the extent to which a community produces goods for the market.
%The association of wage labor and wealth inequality could arise because in more structured economies tasks become more specialized and people are formally paid for some tasks, enabling a greater variety of compensations. Further, wealthier sharing units may have superior access to these opportunities (e.g., because of investment or caste/cultural barriers) or be more likely to take advantage of that access (due to limited information or emulation) as discussed above.
This could reflect elements of a dual economy, as originally modeled by Arthur Lewis \citep{lewis_economic_1954}.\footnote{  Such an economy consists of a modern capital-intensive sector in which wage labor is the primary form of work, as well as an informal sector that remains labor intensive with lower labor productivity and average income. In a dual economy model, the wages paid by employers exceed the average incomes in the informal sector which is the pool from which wage labor is recruited.  This mechanism is a basis for the inverted U-shaped Kuznets curve whereby economic inequality first increases and then decreases (as the informal sector shrinks) with  economic growth \citep{kuznets_economic_1955}.}  Here, this could arise from the costs of foregoing informal production and finding paid employment. If opportunities for wage/salaried labor are limited (as shown in Figure \ref{fig:site_characteristics}, the mean across the communities is 18\%), and wages exceed the value of production in the informal sector, that can increase income inequality and hence wealth inequality.

Our analyses therefore suggest associations that warrant further investigation of the causal pathways that underlie them \citep{an_causal_2022}. With a second wave of data collection, we are collecting the crucial longitudinal data to help facilitate this task, and we join others in calling for such efforts at both data collection and analysis \citep{yu_local_2024}. Beyond the question of causality, there are a number of other important open questions raised by these analyses.

First, we have focused on an aggregate network that combines a range of types of support; further investigation of the distinct types of support is an important area for future study.
If connections across wealth disparities are of consequence, then it is important to consider which sorts of connections entail such bridging relationships.
Certain types of connections may more readily lead to relationships of dependency, exploitation, or mutuality.
Further, we may reasonably expect variation in which types of support are consequential in different ecological or political contexts.
Our multi-layer network data is a unique affordance to explore this and so is an ongoing area of our research.

Second, our attention has been on connections \emph{within} the community, but connections that extend \emph{beyond} the local community are another important resource \citep{yu_local_2024}. Such connections can provide access to resources and novel information not available locally \citep{granovetter_strength_1973, garip_social_2008}.
They can facilitate access to new markets or opportunities or give access to state and non-governmental organizations \citep{woolcock_social_1998}.
Further, when some shocks (e.g., drought, disease) are spatially distributed, connections to people based elsewhere, rather than neighbors, may help recovery \citep{bollig_risk_2006, blumenstock_airtime_2016}. Connections beyond the community may also be less burdened by normative expectations and obligations that could foreclose some opportunities as readily as they facilitate them.
Future work should investigate the role that connections beyond the immediate community play in wealth accumulation and its distribution within and across communities.

Finally, there may be important intra-household dynamics that are obscured by a focus on the household as the unit of analysis. There is ample work on the household and the (un)reasonableness of seeing it as a singular decision-maker \citep{alderman_unitary_1995, behrman_intrahousehold_1997, haddad_intrahousehold_1997}.
Our concern for how household composition may shape our results \citep{bover_wealth_2010, cowell_accounting_2018} led us to include different weightings, but this should be seen as only a first step. It is not obvious that the benefits of social capital or material wealth will be equally shared among all household members.
Particularly as household composition changes over time, future work should explore the economic mobility not only of households, but also of their individual members.

In summary, we explore the relationships between material wealth and social capital using household-level data on social connections collected by researchers working across a highly diverse sample of communities worldwide. The consistencies between some of our findings and those found in studies of residents of large global economies as well as from analyses of large-scale digitally-mediated data sources are striking and help to substantiate the validity of inferences drawn from earlier work regarding the importance of social structure for inequalities in material wealth. The variation across these diverse communities will nevertheless open new lines of inquiry into how inequities arise from, and can be mitigated by, other features of the community, including transitioning into the global market economy.

\clearpage

\bibliography{ENDOW.bib}

\clearpage

\section{Methods}\label{methods}

\subsection{Sample construction}

The aim of this project has been to compile demographic, economic, and social network data from the widest possible range of human communities. Our research team, therefore, primarily comprises anthropologists who have extensive prior experience in their respective study communities, building the necessary relationships and local knowledge to facilitate this comparative work. The demands of data collection for this project---including the need to have as-close-to-complete coverage of all sharing units---means that the included communities ultimately comprise a convenience sample. Still, the sampled communities encompass much of the diversity of human social and economic arrangements (see Figures~\ref{fig:site_characteristics} and \ref{fig:subsistence}).

Researchers working with each community conducted surveys with representatives of effectively all sharing units in each community. The minimal expectation was one adult representative from each sharing unit, but, where possible, researchers were encouraged to solicit responses from more. Table~\ref{tab:su_summary} shows details of each community and the sampling within it. See Supplementary Section~\ref{supp:databackground} for further details on data collection.

\subsection{Variable definitions}

\subsubsection{Material Wealth}

Our measure of material wealth is an aggregation of the property and assets held by each sharing unit (see Supplementary Section \ref{supp:materialwealthbackground} for details), obtained from conducting extensive inventories of each sharing unit---including land, livestock, household goods, tools, etc.
Using estimates of asset values, we then construct estimates of sharing unit wealth converted into US Dollars for comparison within and across communities.
As noted above, what defines a sharing unit varies across communities. A rich literature has debated how best to account for household composition when considering measures of wealth or consumption \citep[e.g.,][]{deaton_household_2003}, including as constructing per capita or per adult measures \citep[e.g.,][]{piketty_capital_2013}.
Each measure has its own advantages and disadvantages.
As we show in Supplementary Sections~\ref{supp:robustness_within} and \ref{supp:robustness_between}, our results are robust to this choice, as well as to other ways of assessing material wealth.
In the main text, per capita measures are employed because otherwise there is a consistently strong relationship that sharing units with more people have both more network connections and more material wealth. We also use percentile ranks of wealth (however measured) to facilitate cross-site comparisons and mitigate the influence of outlier wealth values.
%\footnote{A per capita normalization could be wrong since a sharing unit's production function can be nonlinear. To investigate this, we also perform an analysis where we estimate this relationship and examine residual wealth and social ties (see Supplementary Sections~\ref{supp:alternative_SUweightings_within} and \ref{supp:alternative_SUweightings_between}), which results in comparable inferences.}
%While we do calculate and present such measures, we give primacy to the overall measure of sharing unit wealth. This is because we see the members of a sharing unit as being inextricably part of its collective ``endowment'' \citep{baulch_economic_2000}, contributing to its ability to undertake productive labor and echoing the idea of ``wealth in people'' \citep{guyer_wealth_1993}. Further, each particular attempt to account for sharing unit composition will, we contend, be more or less appropriate in the various sites, which differ not only in sharing unit composition but also the extent to which sharing unit assets are rivalrous or inclusive. We discuss these points further in Supplementary Section \ref{supp:alternative_SUweightings_within}.

% rationale for construction of material wealth measure
% difficulty of measuring income in these contexts: \citep{deaton_analysis_2018}
%assets-based measurement of poverty: \citep{sahn_exploring_2003, mckenzie_measuring_2005, carter_economics_2006}
%endowments \citep{baulch_economic_2000} -- READ!

\subsubsection{Network Measures of Social Capital}

Our network measures are derived from the social connections reported by members of each sharing unit (see Supplementary Section~\ref{supp:socialnetsbackground} for details). Respondents were asked a series of ``name generators,'' some with standardized phrasing, as well as some tailored by the contributing researchers to fit the context of the respective communities \citep{borgatti2024analyzing}.
These questions cover a range of types of support, from tangible aid to behavioral assistance to affective support:
borrowing a week's wages, %(1, \emph{``monetary loan''}),
borrowing food or other basic household items, % (2, \emph{``household items''}),
receiving behavioral assistance (for both women and men), % (3, \emph{``female behavioral''}) and men (4, \emph{``male behavioral''}) in the sharing unit,
and socializing and information sharing (for both women and men). % (5, \emph{``female information''}) and men (6, \emph{``male information''}) in the sharing unit.
Notably, these prompts capture not just relationships of ``friendship'' (relatively balanced, mutual relationships between nominal peers) \cite{hruschka_friendship_2010}, but also ones that may be forged out of economic necessity, with decidedly different tenor and substance.
Full genealogical records additionally provide measures of kinship relationships between residents.
These collectively give us a representation of the multilayer network of each community \citep{boccaletti_structure_2014, atkisson_why_2020}.

%social network data and measurement, esp. in developing countries: \citep{perkins_social_2015, chuang_social_2015}

%Acknowledge that supportive relationships and kin can also be a burden/obligation \citep{krackhardt_ties_1999,di_falco_dark_2011, jakiela_does_2016, villalonga-olives_dark_2017}

%Measurement and comparison -- \citep{ravallion_poverty_1994}

Because our fundamental unit of measurement is the sharing unit \citep[cf.][]{aguiar_illusion_2025}, we aggregate individual responses to construct networks reflecting connections among the sharing units within a community.
We primarily consider a weighted, directed network based on the survey questions outlined above (the \emph{``composite''} network), constructed as follows. We go through each question in turn, assigning a weight for that question equal to the number of times any individual from sharing unit $i$ nominates any individual from sharing unit $j$, divided by the number of individuals from sharing unit $i$ who were surveyed. %\footnote{So for example, if two separate individuals in sharing unit $i$ are interviewed, and one names 3 people in sharing unit $j$ for the borrowing money question, and another names 2 people, then that question gets a weight of (3+2)/2=2.5.}
We do this for each question and sum across all the relevant questions (see Supplementary Section \ref{supp:socialnetsbackground} for further details). The normalization accounts for that fact that if more people were surveyed in one sharing unit, then that mechanically increases the number of relationships that are included in the weights.
The relevant survey questions are each given equal weight in this composite network, so greater edge weights reflect sharing unit $i$ turning to more members of sharing unit $j$ for more types of support.
%Other aggregations and weightings are plausible, and in Supplementary Section \ref{supp:alternative_networks} we check that our findings are robust to several different representations of the network.

With these data, we can then construct our network measures of sharing units' social capital. ``Support access'' and ``support provisioning'' both draw solely on the weighted, directed network. Note that a connection from $i$ to $j$ reflects $i$'s requests for support, so resources flow in the opposite direction (i.e., from $j$ to $i$). This means that ``support access'' is the out-degree (capturing the sharing units that are supporters) and ``support provisioning'' is the in-degree (capturing the sharing units that are supportees) in the composite network, where we use the edge weights as described above.
``Average Alter Wealth'' ($\text{AAW}_i$) our main measure of economic connectedness, draws not only on the network but also on the wealth data. Briefly, it is the average wealth of a sharing unit's connections. Just as with our first two measures, these connections can be defined through a consideration either of supporters (those provisioning support) and supportees (those receiving support). Consequently, we have two parallel measures: $\text{AAW}_i$ of supporters and $\text{AAW}_i$ of supportees. Here again, we use the network edge weights when constructing these measures, such that alters that are more heavily relied on are given more weight. As with our construction of material wealth measures, we again consider different accountings of wealth (e.g., per capita or absolute) and use the percentile rank of wealth. See Supplementary Section~\ref{supp:definitions-measures} for formal definitions of these measures.

\subsubsection{Community-Level Measures}

From these sharing unit-level measures, we then derive a parallel set of community-level measures.
For material wealth, we simply calculate the Gini, using whatever relevant measure of material wealth (e.g., per capita or absolute).
As an alternative, in Section~\ref{supp:ninety-ten} we construct measures of material inequality for each community based on the ratio of wealth per capita for sharing units at the $90^{th}$ vs $10^{th}$ percentiles of the distribution, as well as the ratio of the $80^{th}$ to $20^{th}$ percentiles of the distribution, which are less susceptible to sharing units with outlying wealth levels \cite{piketty2003income, faggio2010evolution}, finding largely comparable results.
For ``support access'' and ``support provisioning,'' we similarly calculate the Gini of each, as a simple measure of the inequality in these measures.

For ``Average Alter Wealth,'' we follow earlier definitions of community-level economic connectedness \cite{chetty_social_2022} and consider below-median sharing units' access to support. In brief, we take the average of the below-median sharing units' $\text{AAW}_i$ to construct a general measure of the Average Alter Wealth of the Poor. We then normalize this by the Average Alter Wealth of the above-median-wealth sharing units and term this new measure ``Relative Average Alter Wealth'' (RAAW). This normalization means that if the connections of the poor were just as wealthy as those of the rich, this measure would be 1. As with our other measures including wealth, we consider different weightings that variably account for sharing unit composition. Supplementary Section~\ref{supp:definitions-measures} provides formal definitions of these measures and their variants. Our main RAAW measure is based on alters who provide support (i.e., supporters), as that is hypothetically important in generating wealth. In Supplementary Section~\ref{supp:ec_incoming} we show that similar results hold if we consider instead alters to whom support is provided (i.e., supportees).

A potential concern is that the relationship between economic connectedness (RAAW) and wealth inequality could be mechanical if sharing units form relationships within a set band around their own \emph{absolute} material wealth. In that scenario, as the wealth distribution takes on a wider absolute spread (e.g., as the distribution takes on a higher standard deviation and larger support), the difference between the AAW of the poor and that of the wealthy would widen. In Supplementary Section~\ref{supp:mechanical} we show that this is not the case, both in terms of correlation between standard deviation of alters' wealth with respect to overall wealth, and also by contrasting the data with a simulated process in which sharing units form relationships in an absolute band.
%that the average standard deviation in people's friends' wealth shifts with a correlation of .85 (.9 when omitting the top 5 percent of wealthiest sharing units) with respect to the overall standard deviation of wealth (noting that a slope of 1 would be fully relativistic and a slope of 0 would be fully absolute).  We also show via simulations that a network formation process in which sharing units form relationships in an absolute band around their wealth yields correlations with support far from our empirically observed correlation.

\subsubsection{Wealth Modularity}

A possible critique of these economic connectedness measures is that they rely on dividing the community into poor and rich on the basis of a median split. We therefore construct a more flexible measure---(normed) wealth modularity---based on locally-salient wealth classes defined on the basis of the network data and wealth data jointly.
To do this, we iteratively compute network modularity \citep{newman_finding_2004} for different possible thresholds of sharing unit wealth that might differentiate coherent classes.
Modularity is larger when the relative density of connections among some set of nodes (e.g., poor sharing units) is higher than the density of the connections to the other nodes in the network (e.g., rich sharing units). We rank order sharing units by their material wealth and examine the modularity score for each potential split of the network along this sequence.

Comparing this modularity measure at each possible wealth class division allows us to determine the wealth split that leads to the most coherent divisions, without restricting the analysis to a prespecified median split.
To determine whether the identified split is meaningful, we construct a null distribution for the maximum of the wealth modularity that we would observe under uniform randomness. To assess where we find significant splits, we subtract the 95\% quantile of the null distribution from the maximum wealth modularity measured in each community. This is our resulting normed wealth modularity measure. In Supplementary Section~\ref{supp:wealth_modularity} we formally define this measure and show the ``wealth modularity curves" constructed via this procedure.

\subsubsection{Site-Level Variables}

To characterize each community, we compile a wide range of variables drawn from a variety of sources.
First, there are variables that are aggregates of observable characteristics from our own data, such as the fraction of sharing unit's connections that are with close kin.
Second, there are variables that the contributing researchers reported for each community. These include characteristics such as how exposed a community is to various shocks and whether there are norms that preclude certain subsets of people from doing certain tasks.
Third, there are geospatial variables derived from various open datasets that capture characteristics about the immediate area in which a community is situated, such as a measure of the biological productivity of the local area.
Finally, there are variables that represent the national context in which the communities are situated, such as the national inclusiveness of property rights.
Supplementary Section~\ref{supp:sitevars} provides further details on these measures and their sources.
%The list of these variables is in Table \ref{tab:variables}.

\subsection{Correlations}

For our within-site analyses, we use simple correlations between sharing units' attributes.
Throughout, we show the correlations between percentile ranks, to facilitate comparability.
We provide two primary ways to assess the overall trend of the within-site correlations.
First, we calculate the two-sided $p$-value of observing so many positive correlations under the sharp null that the true correlation in each community is zero.
Second, we conduct a random-effects meta-analysis to derive a pooled coefficient and its associated $p$-value.
There is no obvious ``proper'' way to cluster standard errors across communities, as each is more similar to some communities than others across some dimensions, and less on others, and there is no single dimension that is clearly the one that correlates networks and/or wealth across communities. Thus, significance levels should be interpreted cautiously.
See Supplementary Section~\ref{supp:pooled_coefs} for further description of the pooled coefficients and an alternative calculation of them derived from a multilevel model.

For our cross-site analyses, we primarily present simple bivariate correlations. As complements to these, in Supplementary Section~\ref{supp:multilevel_model} we present a multilevel model of network formation based on wealth classes, and in Supplementary Section~\ref{supp:othervariables} we present Lasso and multivariate regressions predicting the Gini of wealth per capita and Relative Average Alter Wealth.

\subsection{Ethics}

Ethical approval for each community was secured by the relevant contributing researcher following local, national, and institutional policies. An overarching ethics review was also conducted by the University of Cincinnati's Institutional Review Board (Study ID 2016-4691). Survey respondents gave informed consent before participating.

\subsection{Data availability}

Given the many varied informed consent processes and associated assurances of privacy and access made by team researchers across the various communities, we cannot currently share the data used herein. Site-level aggregate measures are included in tables in the Supplementary Information. [Note to reader: we are currently drafting a paper presenting the multilayer network data in more detail, which presents many aggregate summary statistics.]

\subsection{Code availability}

The code used to generate the findings shown in this paper are available at \url{https://github.com/endowproject/endow-wealth-corr-rep-code}, with simulated data to facilitate its use.

\clearpage

% -------------------------------------------------------
% INSTITUTIONS (alphabetical by main institution name)
% -------------------------------------------------------

\section*{Author Affiliations}
\footnotesize$^{1}$Australian National University, Canberra, Australia.\and
\footnotesize$^{2}$Boise State University, Boise, ID, USA.\and
\footnotesize$^{3}$Boston University, Boston, MA, USA.\and
\footnotesize$^{4}$Brunel University of London, London, UK.\and
\footnotesize$^{5}$Centre National de la Recherche Scientifique (CNRS), France.\and
\footnotesize$^{6}$Consejo Nacional de Investigaciones Cient\'{i}ficas y T\'{e}cnicas (CONICET), Argentina.\and
\footnotesize$^{7}$Durham University, Durham, UK.\and
\linebreak[0]
\footnotesize$^{8}$German Institute of Development and Sustainability (IDOS), Bonn, Germany.\and
\footnotesize$^{9}$Harvard T.H.\ Chan School of Public Health, Boston, MA, USA.\and
\footnotesize$^{10}$Institut de Recherche pour le D\'{e}veloppement (IRD), France.\and
\footnotesize$^{11}$Institute for Globally Distributed Open Research and Education (IGDORE).\and
\footnotesize$^{12}$International Centre for Diarrhoeal Disease Research, Bangladesh (icddr,b), Dhaka, Bangladesh.\and
\footnotesize$^{13}$iReview Limited, Tanzania.\and
\footnotesize$^{14}$Kiel Institute for the World Economy, Kiel, Germany.\and
\footnotesize$^{15}$London School of Economics and Political Science, London, UK.\and
\footnotesize$^{16}$Masaryk University, Brno, Czech Republic.\and
\footnotesize$^{17}$Max Planck Institute for Evolutionary Anthropology, Leipzig, Germany.\and
\footnotesize$^{18}$North-Eastern Hill University, Shillong, Meghalaya, India.\and
\footnotesize$^{19}$Oregon State University, Corvallis, OR, USA.\and
\footnotesize$^{20}$Passau University, Passau, Germany.\and
\footnotesize$^{21}$The Pennsylvania State University, University Park, PA, USA.\and
\footnotesize$^{22}$Pontificia Universidad Cat\'{o}lica del Per\'{u}, Lima, Peru.\and
\footnotesize$^{23}$Rutgers University, New Brunswick, NJ, USA.\and
\footnotesize$^{24}$Santa Fe Institute, Santa Fe, NM, USA.\and
\footnotesize$^{25}$Stanford University, Stanford, CA, USA.\and
\footnotesize$^{26}$Universidad Nacional de Jujuy, San Salvador de Jujuy, Argentina.\and
\footnotesize$^{27}$Universit\"{a}t Hamburg, Hamburg, Germany.\and
\footnotesize$^{28}$Universit\'{e} de Montpellier, Montpellier, France.\and
\footnotesize$^{29}$Universit\'{e} Mohammed VI Polytechnique, Ben Guerir, Morocco.\and
\footnotesize$^{30}$University College London, London, UK.\and
\footnotesize$^{31}$University of Alabama, Tuscaloosa, AL, USA.\and
\footnotesize$^{32}$University of Bristol, Bristol, UK.\and
\footnotesize$^{33}$University of Calgary, Calgary, Canada.\and
\footnotesize$^{34}$University of California Davis, Davis, CA, USA.\and
\footnotesize$^{35}$University of California Los Angeles, Los Angeles, CA, USA.\and
\footnotesize$^{36}$University of California Merced, Merced, CA, USA.\and
\footnotesize$^{37}$University of California Santa Barbara, Santa Barbara, CA, USA.\and
\footnotesize$^{38}$University of Cincinnati, Cincinnati, OH, USA.\and
\footnotesize$^{39}$University of Georgia, Athens, GA, USA.\and
\footnotesize$^{40}$University of Groningen, Groningen, Netherlands.\and
\footnotesize$^{41}$University of Maine, Orono, ME, USA.\and
\footnotesize$^{42}$University of Maryland, College Park, MD, USA.\and
\footnotesize$^{43}$University of Massachusetts Amherst, Amherst, MA, USA.\and
%\footnotesize$^{44}$University of Missouri, Columbia, MO, USA.\and
\footnotesize$^{45}$University of New Mexico, Albuquerque, NM, USA.\and
\footnotesize$^{46}$University of Sydney, Sydney, Australia.\and
\linebreak[0]
\footnotesize$^{47}$University of Utah, Salt Lake City, UT, USA.\and
\footnotesize$^{48}$University of Washington, Seattle, WA, USA.\and
\footnotesize$^{49}$University of Zurich, Zurich, Switzerland.\and
\footnotesize$^{50}$Washington State University, Pullman, WA, USA.\and
% -------------------------------------------------------
% CORRESPONDING AUTHOR
% -------------------------------------------------------
\linebreak[0]
\footnotesize$^{\dagger}$Corresponding author. Email: \href{mailto:e.a.power@lse.ac.uk}{e.a.power@lse.ac.uk}

\section*{Author Contributions}
E.A.P., M.B.M., S.B., M.O.J., J.K., D.R., T.R., S.S., and J.W. contributed substantially to the conceptualization, organization, analysis, and writing of this paper.
E.A.P., M.B.M., S.B., M.O.J., J.K., and P.L.H. conceptualized the initial study.
E.A.P., M.B.M., J.K., N.A., S.A., A.A., C.A., M.Ba., B.B., C.M.B., M.Br., M.C., W.C., K.C., J.C., S.C., A.D.R., A.L.D., I.D., F.F., J.P.F., D.G., M.G., C.G., G.G., W.B.H., K.K., G.K., B.L., R.L., S.L., S.M., E.M., K.J.M., M.M., R.M.C., K.O., E.P., S.P., A.P.V., C.R., B.S., M.S., E.S., M.K.S., K.E.S., C.S., B.T., B.V., V.V., and J.P.Z. gathered and contributed data for this project and so were involved in investigation, providing resources, and project administration, and all were responsible for data curation for the particular fieldsite(s) where they work. D.M. and M.O.J. also undertook project administration.
E.A.P. and D.R. curated the data.
Methodology was developed by E.A.P., M.O.J., J.K., D.R., T.R., S.S., J.W., P.L.H., and C.R.
Formal analysis was conducted by E.A.P., M.O.J., D.R., T.R., S.S., J.W., and A.S.
Software was developed by E.A.P., D.R., T.R., S.S., J.W., and A.S.
Validation and visualization were carried out by E.A.P., M.O.J., D.R., T.R., and S.S.
Supervision was provided by E.A.P., D.R., S.C., M.M., R.M.C., C.R., and M.K.S.
E.A.P., M.B.M., S.B., M.O.J., and T.R. wrote the original draft.
E.A.P., M.B.M., S.B., M.O.J., J.K., D.R., T.R., S.S., J.W., M.Ba., J.C., S.C., C.G., G.K., M.M., C.R., and J.P.Z. reviewed and edited the manuscript.
Funding was acquired by E.A.P., M.B.M., S.B., M.O.J., J.K., D.R., M.Ba., M.C., S.C., K.K., G.K., D.M., M.S., and P.L.H.

\section*{Competing Interests}
The authors declare no competing interests.

\section*{Acknowledgments}
\footnotesize{
% FUNDING
This paper is based on fieldwork supported by the National Science Foundation under Award Nos. 1743019, 2218860, and 2218861.
We are also grateful for financial support from the Santa Fe Institute and the Omidyar Network through its Emergent Political Economies project.

Michele Barnes' fieldwork was co-funded by the Australian Research Council through a Discovery Early Career Fellowship Grant (DE190101583).
Christine Beitl's fieldwork was also supported by the University of Maine Faculty Summer Research Award.
Siobh\'{a}n Cully's fieldwork was supported by National Science Foundation grant Nos. 1461514, 1920812, and 2222262.
Karen Kramer's fieldwork was partially supported by National Science Foundation grant Nos. BCS-2051264, BCS-1632338.
%Gianluca Grimalda's fieldwork was partially supported by the grant "Research on Peripheries to Strengthen the Resilience of Czech Society'', CZ.02.01.01/00/23_025/0008727 % MUST BE FOR WAVE 2, so not relevant
Fieldwork by Ivan Deschenaux and Komal Chauhan was funded by Leverhulme Research Leadership Award RL-2022-039.
Matthew Jackson was partially supported by NSF grant SES-1629446.
Joshua Cinner was partially supported by Australian Research Council grants FL230100201, FT160100047.
Geoff Kushnick was partially supported by a grant from the ANU College of Arts and Social Science.

% THANKS
We thank all community residents and participants for their support and engagement with our fieldwork.
We further thank the research assistants who helped with data collection, entry, and processing.
Thanks to colleagues in the Department of Human Behavior, Ecology and Culture at the Max Planck Institute for Evolutionary Anthropology, including the student assistants who helped to process the data: Lucy Betke, Anastassiya Bublikova, Nikkin Dev, Gargi Kodgirwar, Enzo Lima, John Newton, Anne Buechner, Leonie Ette, Maria Froehlich, Vincent Kiepsch, Kristina Kunze, Carlotta Rueck, and Constanze Wiedemann.
Thanks to Evgeny Noi and Ella Vacic for compiling various geospatial datasets.

Sheina Lew Levy thanks Prof. Clobite Bouka-Biona from IRSEN, who facilitated the acquisition of research permits and infrastructure, Dzabatou Moise, who served as a community liaison, Mekouno Pau for translation, and Sarah Pope-Caldwell and Adam Boyette for field support.
Geoff Kushnick thanks his Indonesian counterpart, Dr Fikarwin Zuska (Universitas Sumatera Utara), who facilitated research permits through RISTEK (2017) and BRIN (2023-24).
Michele Barnes thanks Jacqueline Lau for her assistance in the field and Sarah Sutcliffe for her assistance with data management.
Karen Kramer thanks Joe Hackman for his concerted help in managing the databases.
Rafael Morais Chiaravalloti thanks Mira\'{i}ra Noal Manfroi.
Eleanor Power thanks the Chella Meenakshi Centre for their support in facilitating fieldwork.
Mary Shenk thanks the International Centre for Diarrhoeal Disease Research, Bangladesh (icddr,b) and Tanima Rashid.
}

%\section*{Author contributions}

%\section*{Competing interests}

\begin{comment}
\clearpage
\section*{Extended Data}

\renewcommand{\figurename}{Extended Data Figure}
\renewcommand{\tablename}{Extended Data Table}
\renewcommand{\thefigure}{\arabic{figure}}
\renewcommand{\thetable}{\arabic{table}}
\setcounter{figure}{0}
\setcounter{table}{0}

%%% EXTENDED DATA TABLES/FIGURES
%%% Subsistence
%%% Site summary table with coverage etc
%%% Summary violins of community size, mean SU size, SU mean support partners, and SU wealth
%%% Aggregate Lorenz Curves
%%% Example Composite Networks
%%% 90/10 and 80/20 instead of Gini
%%% Quartile Splits
%%% Stepwise model results
\end{comment}

\appendix

\clearpage

\renewcommand{\figurename}{Figure}
\renewcommand{\tablename}{Table}
\renewcommand{\thefigure}{A\arabic{figure}}
\renewcommand{\thetable}{A\arabic{table}}

\setcounter{figure}{0}
\setcounter{table}{0}

\section{Supplementary Information}

\subsection{Data Collection} \label{supp:databackground}

This project, called by the acronym ``ENDOW'' (for Economic Networks and the Dynamics Of Wealth inequality), began in 2017, with the aim of better understanding inequality by assembling data to facilitate cross-cultural comparisons \citep{borgerhoff_mulder_intergenerational_2009, bowles_emergence_2010} by involving anthropologists (and other social scientists) who could contribute data from the communities where they work.
Due to the opportunistic nature of this sample (willing researchers with active fieldsites) our inferences are primarily of a wide range of communities transitioning into market-based economies.
An overview of the communities, the lead researcher(s) working in each, and the number of sharing units for each is presented in Table~\ref{tab:su_summary}.

\begin{figure}
    \centering
    \includegraphics[width=0.9\linewidth]{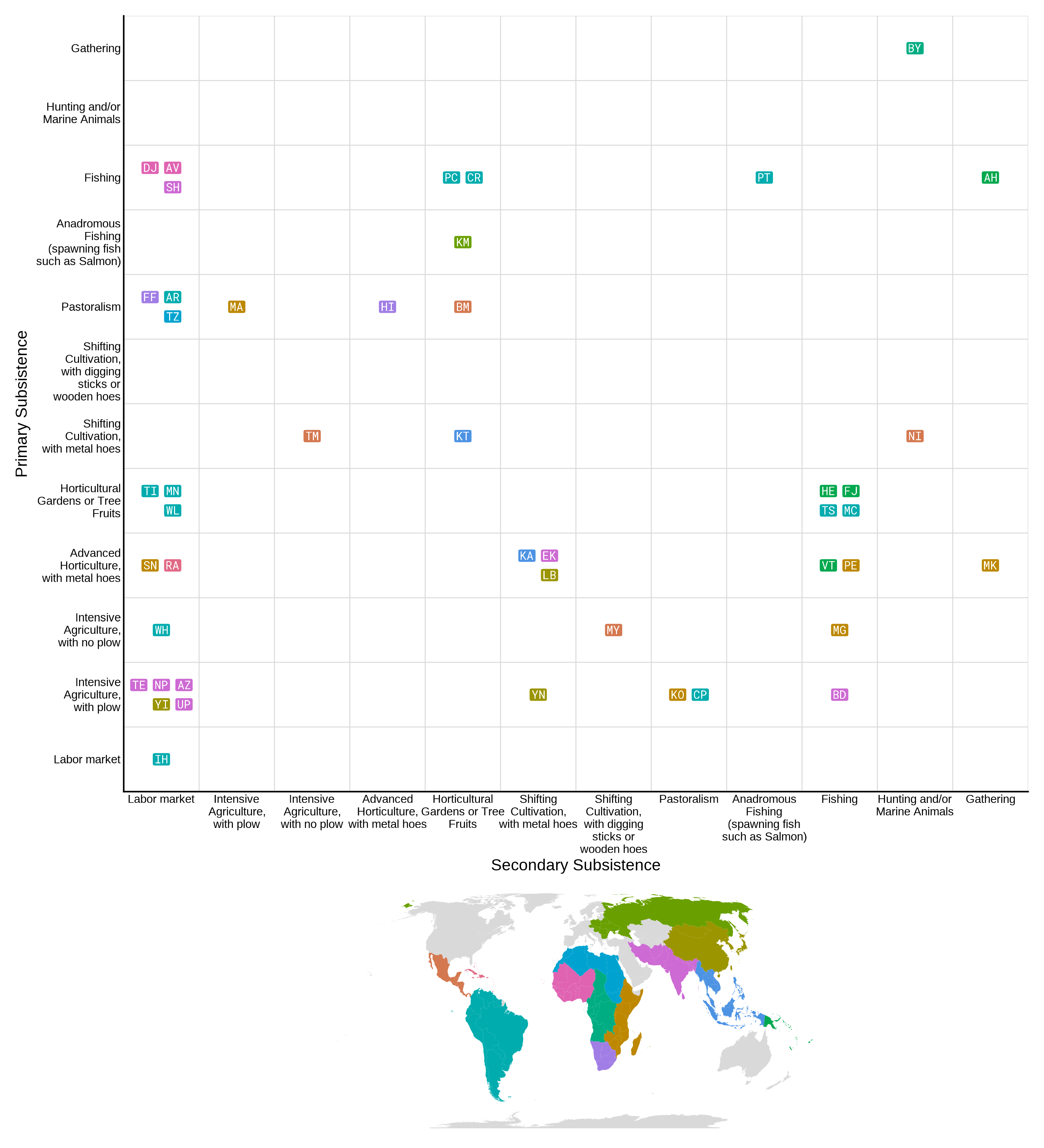}
    \caption{The primary and secondary mode of subsistence practiced in each ENDOW community. Site codes are colored by United Nations geoscheme regions, shown in the world map (with only represented regions colored). Consult Figure~\ref{fig:sites} for more exact locations.}
    \label{fig:subsistence}
\end{figure}

Figure~\ref{fig:subsistence} shows the primary and secondary mode of subsistence practiced in each community, as reported by the contributing ENDOW researcher.

\begin{figure}
    \centering
    \includegraphics[width=\linewidth]{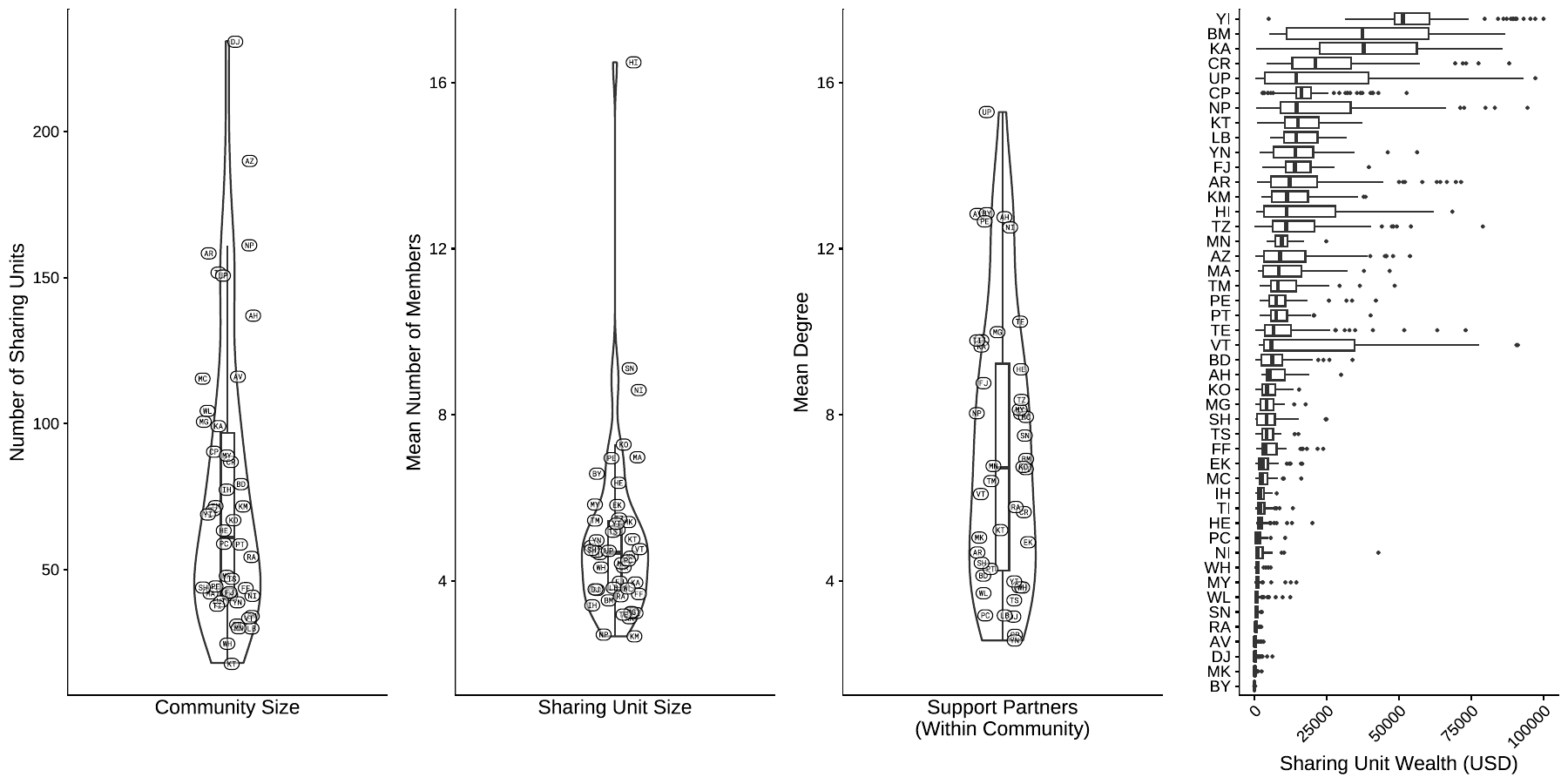}
    \caption{\textbf{Descriptive plots of sharing units, across ENDOW communities.} (Note that the wealth plot has a truncated axis; see the full distributions in Figure~\ref{fig:boxplot-wealth-su}). }
    \label{fig:global-su-summary-stats}
\end{figure}

The ``sharing unit'' is our fundamental unit of analysis, the unit to which resources are (principally) pooled and shared. This may often, but not always, align with the idea of a ``household.''
As shown in Table~\ref{tab:su_summary} and Figure~\ref{fig:global-su-summary-stats}, across all ENDOW communities, each sharing unit has, on average, four or five members.

\begin{landscape}
  \include{tables/su_summary}
\end{landscape}

ENDOW researchers gathered comprehensive information on virtually all sharing units in their communities (coverage is, on average, 96\%).
As summarized in Table \ref{tab:wealthmeasures}, the aim was to get information not just on their material wealth (i.e., land, livestock, household goods, tools, etc.), but also their ``embodied wealth'' (i.e., bodily strength, dexterity, skill, knowledge, etc.) and ``relational wealth'' (i.e., access to and position within a social support network) \citep[cf.][]{borgerhoff_mulder_intergenerational_2009}.
Prompts were devised collaboratively by the full team of researchers, with the aim of being as culturally- and site-specific as appropriate, while also facilitating cross-cultural comparison (see Table \ref{tab:wealthmeasures} for a summary).
For material wealth, this entailed detailed inventories of sharing unit assets and estimates of their values, used to create singular monetary estimates that are comparable within and across sites \citep[cf.][]{hruschka_estimating_2015}.
For embodied wealth, this entailed standardized questions on food security and individuals' ability to work, alongside questions on educational attainment and knowledge in other locally valuable domains, such as fluency in national languages and other beneficial or marketable skills \citep{godoy2007language, hadley2012coping, mattison2023market, piperata2023measuring}.

\begin{table}[ht]
\centering
\scalebox{0.75}{
\begin{tabular}{llcc}
  \hline
Type & Domain & Level & Tailored? \\
  \hline
\multirow{9}{4em}{Relational} & Money borrowing (double sampled) & SU &  N \\
& Borrowing/sharing goods (double sampled) & SU & Y \\
& Female-oriented labor/behavioral support & SU & Y \\
& Male-oriented labor/behavioral support & SU & Y \\
& Female-oriented informative/affective support & SU &  Y \\
& Male-oriented informative/affective support & SU & Y \\
& Ties to government and NGO employees & SU & N \\
& Support outside the community & SU & Y \\
& Kinship and marital ties & I & N \\
   \hline
Material & Inventory of sharing unit property and assets & SU & Y \\
  \hline
\multirow{4}{4em}{Embodied} & Food security & SU & N \\
 & Ability to work & I & N \\
 & Years of formal education & I & N \\
 & Locally valuable knowledge & I & Y \\
   \hline
\end{tabular}
}
\caption{Measures of relational, material, and embodied wealth gathered at all ENDOW communities. Some were queried at the sharing unit level (SU), while others were documented for each individual (I). ``Tailored'' questions have a set domain, but the exact prompt and phrasing is ethnographically tailored to each community.}
\label{tab:wealthmeasures}
\end{table}

Data collection was timed and conducted at each community based on the availability of community members (e.g., seasonal constraints), in alignment with the scheduling constraints of the contributing ENDOW researcher(s). Data began to be collected at some sites in 2017, but others joined the project at a later date, so our resulting data set was collected between 2017 and 2025. Contributing researchers were given a description of the data requirements and a template that they could adapt to their particular setting. Many decisions were left to the researchers(s), shaped by their knowledge of and discussions with the community, as well as various logistical constraints. Accordingly, while many researchers gathered network data with pen and paper, others used tablets. Some had existing demographic and kinship data whereas others needed to newly ask about these domains. This leads to heterogeneity in the resulting data.

Ethical approval for each community was secured by the relevant contributing ENDOW researcher following local, national, and institutional policies. An overarching ethics review was also conducted by the University of Cincinnati's Institutional Review Board (Study ID 2016-4691).

Following data collection, contributing researchers followed guidelines on the structure of the data files for submission. These files were then subjected to a series of checks for completeness and possible data errors (e.g., implausible age differences between mothers and their children). In response to the questions that emerged when collating these reports, the collaborating researchers provided additional details to clarify ambiguities or missingness in the contributed data.

\subsection{Material Wealth Data} \label{supp:materialwealthbackground}

%Importance of considering household composition / structure / life cycle for its impact on both production and consumption \citep{cain_household_1978}
% for consumption side, there's work on ``equivalence scales'' as potential refinement of per capita measures (e.g., Deaton and Muellbauer 1986, but then also pushback see e.g. in Lahoti et al GCIP 2016)
% Chayanov consumer/producer ratio -- extension by Hammel PNAS and his earlier work; e.g. application in quanjer_homemakers_2019
% Hammel and Laslett typology; see use in e.g. bengtsson_life_2004
% Netting "Some Home Truths" 1982
% Household production of health berman_household_1994

As our measure of material wealth, we draw from detailed enumerations of each sharing unit's property and assets, and their associated valuations. ENDOW team researchers (primarily anthropologists) working in each community determined the set of relevant items through conversations with residents and visits to a range of homes; the aim was to be exhaustive of property and material assets, with a focus on those that would be economically productive (e.g., clothing was not enumerated in detail). The number and granularity of the items vary from community to community, reflecting the local context.

In order to reduce the burden on respondents, ENDOW researchers were directed to identify any items that virtually all sharing units would possess, and so did not need individual enumeration (this typically included a small set of items, for example, perhaps every sharing unit might possess a machete, three cooking pots, and a bowl and spoon for each member). These comprised a `baseline' set of items for each sharing unit. All other items were enumerated. ENDOW researchers and/or research assistants visited each sharing unit and determined the count of each such item through direct observation and reports by sharing unit members. %The draft survey shown in Section~\ref{draft_survey} gives a sense of the detail of the enumerated items.

Item values were determined by the ENDOW researchers, through conversations with residents and by visiting local markets. The aim was to determine current market values of items. For items that are not regularly traded or brought to market, values were typically determined by what it would cost to replace the item anew (e.g., this was often used to determine house valuation: what would it cost to rebuild a house of equivalent size and material). In some cases, ENDOW researchers determined that some assets could not be understood as a wealth holding (e.g., in contexts where land was fully usufruct, it was not given a monetary value). In Section~\ref{sec:wealth-sensitivity-noise} we evaluate the robustness of our results to these valuations.

A sharing unit's total wealth is then determined through a simple summation of the multiplication of their enumerated items (plus the `baseline' items) and their associated values.

So, the primary wealth measure for each sharing unit is the total value of enumerated assets, computed as
$w_i = \sum_k q_{ik} \cdot p_k$,
where $q_{ik}$ is the observed quantity of item $k$ owned by unit $i$ and
$p_k$ denotes the value of a unit of item $k$.
Per-capita wealth is $w_i^{\text{pc}} = w_i / n_i$, where $n_i$ is
sharing unit size.

In the small number of cases where items were not fully enumerated (i.e., where we have some missing data) we conducted multivariate imputation via chained equations using the \texttt{mice} package in \textsf{R} \citep{mice}. We generate 50 imputed datasets and took the median value from this set as the baseline value used in the paper. In Section~\ref{sec:wealth-sensitivity-imputation} we evaluate the robustness of our results to this approach.

\begin{figure}
    \centering
    \includegraphics[width=0.9\linewidth]{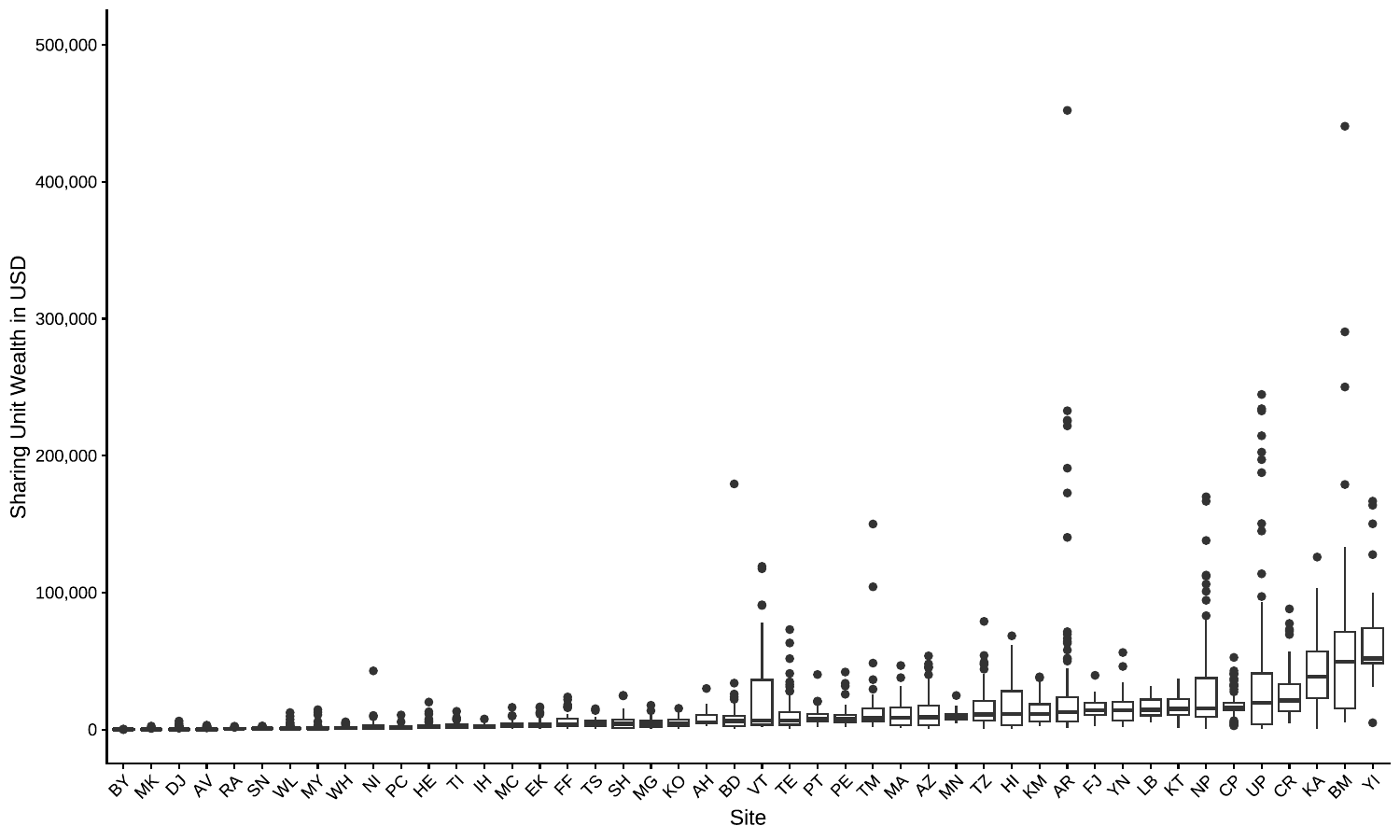}
    \includegraphics[width=0.9\linewidth]{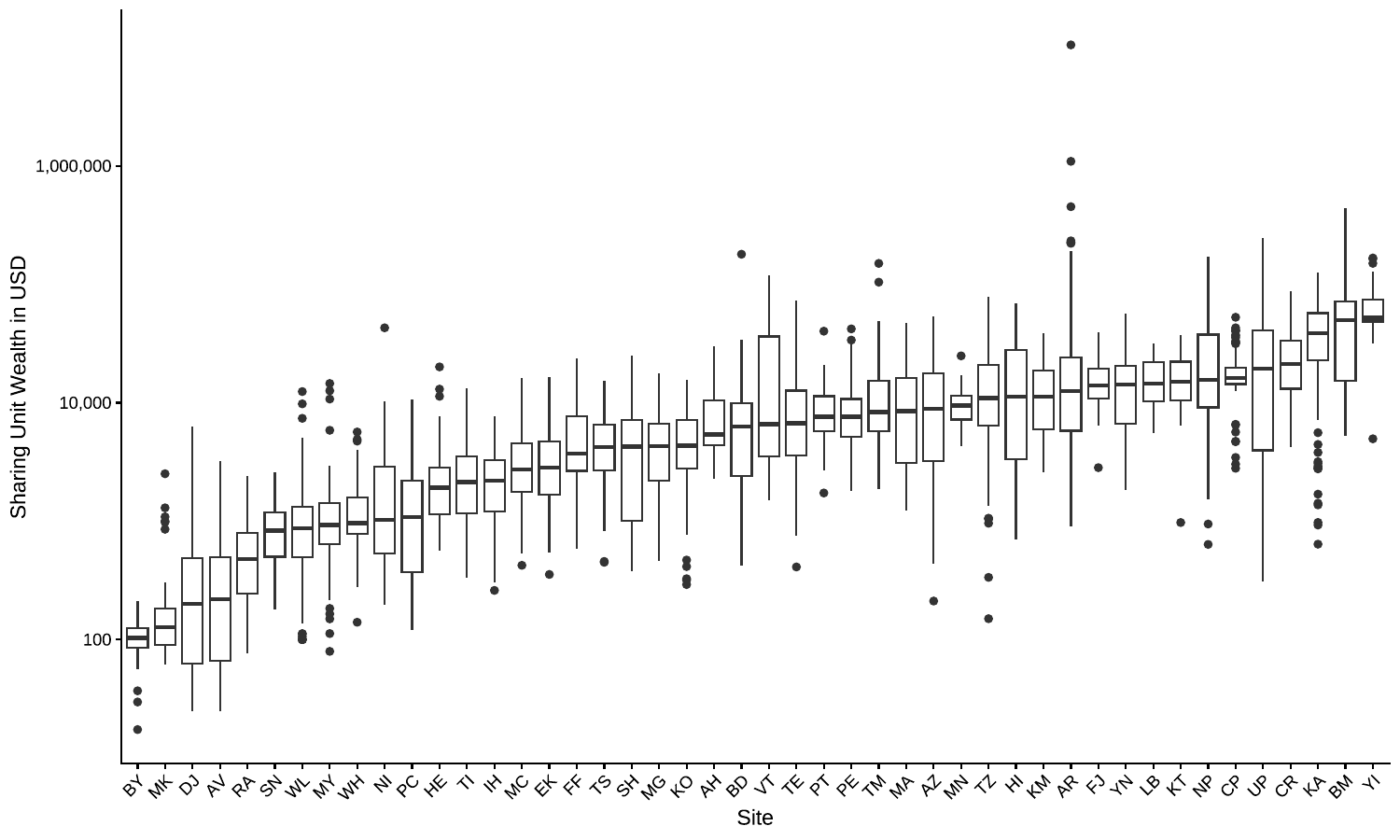}
    \caption{\textbf{Boxplots by community for sharing unit wealth.} The top plot shows the raw distributions but has a truncated axis, removing a handful of outliers. The bottom plot shows the same distribution with a log transformed axis, so that all outliers can be included.}
    \label{fig:boxplot-wealth-su}
\end{figure}

\begin{figure}
    \centering
    \includegraphics[width=0.9\linewidth]{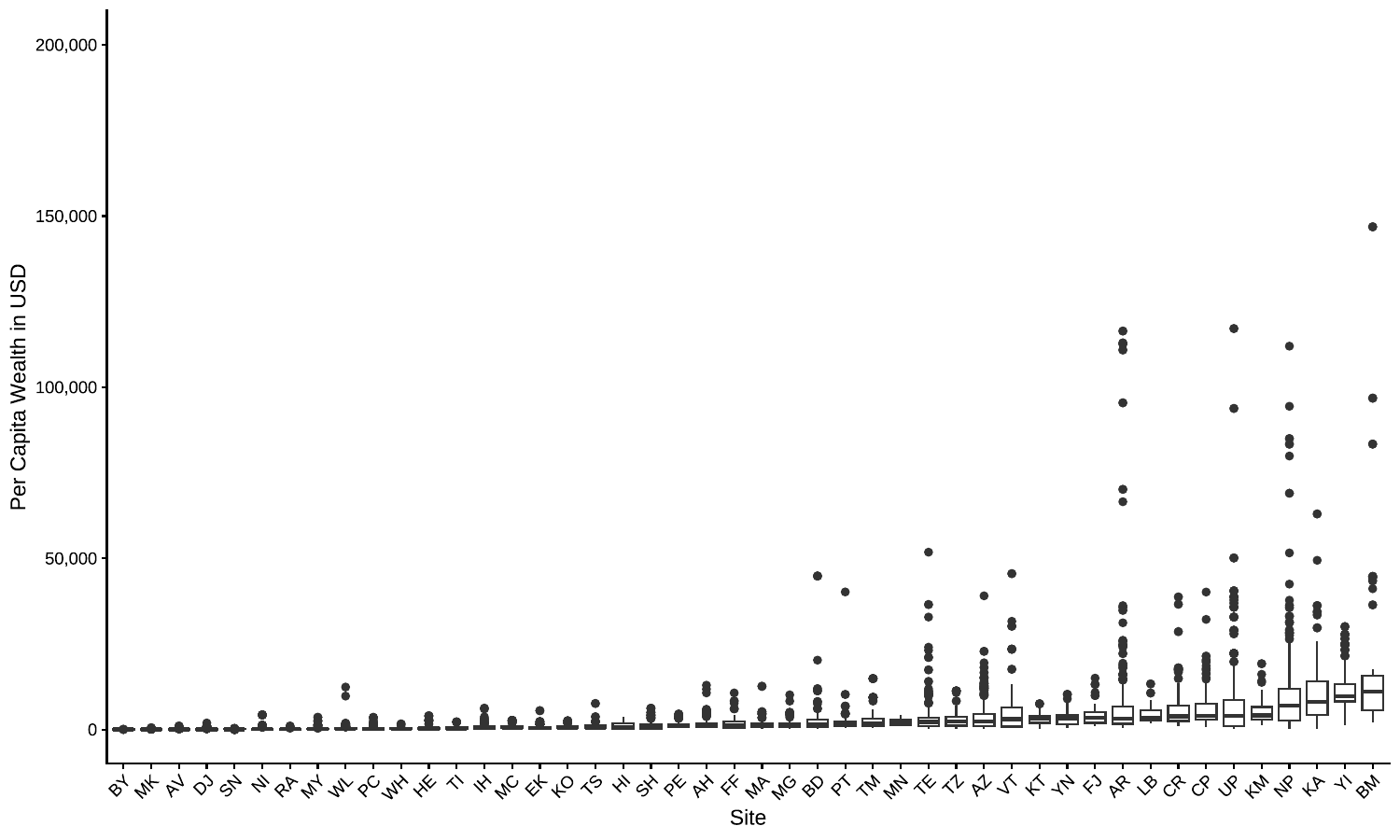}
    \includegraphics[width=0.9\linewidth]{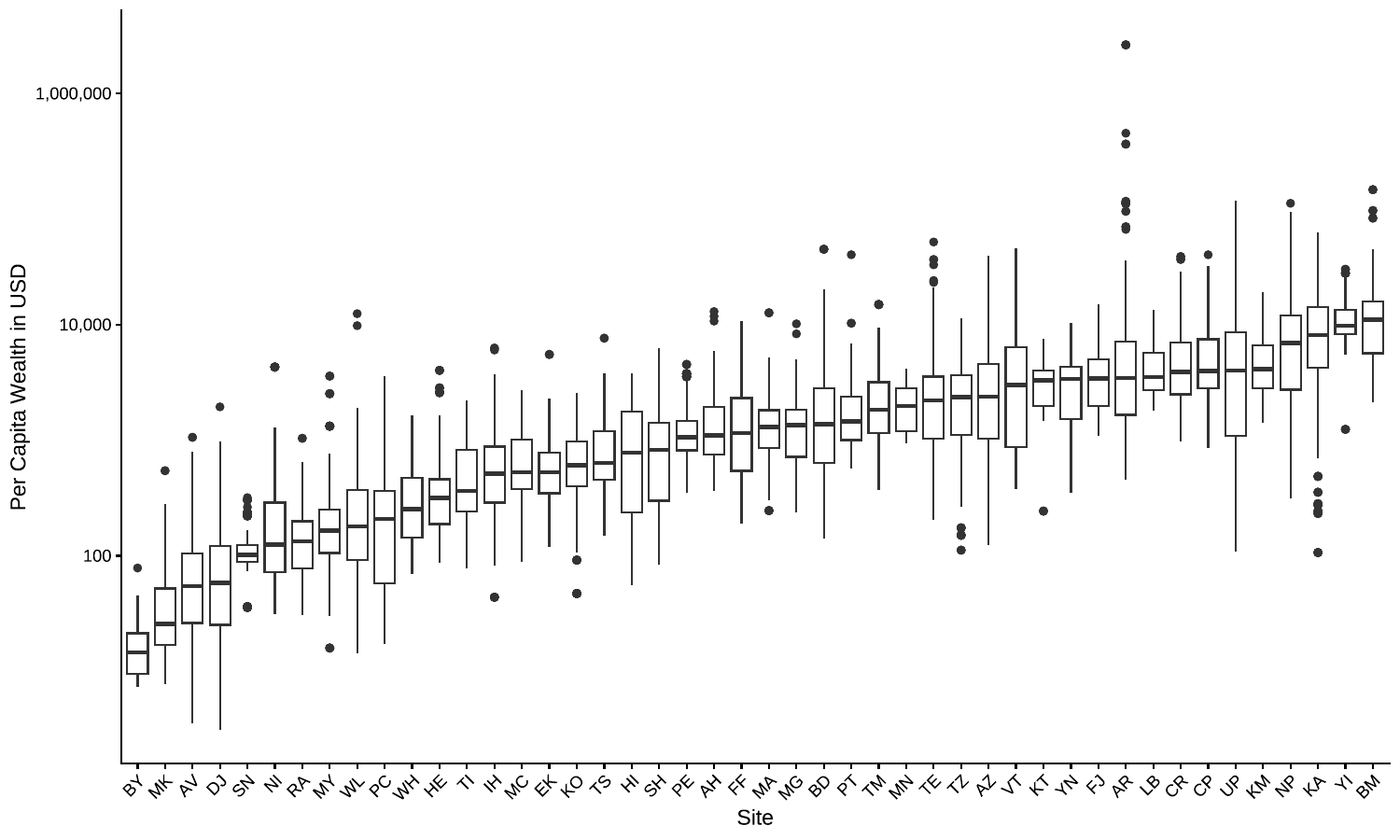}
    \caption{\textbf{Boxplots by community for wealth per capita.} The top plot shows the raw distributions but has a truncated axis, removing a handful of outliers. The bottom plot shows the same distribution with a log transformed axis, so that all outliers can be included.}
    \label{fig:boxplot-wealth-person}
\end{figure}

\begin{figure}
    \centering
    \includegraphics[width=0.9\linewidth]{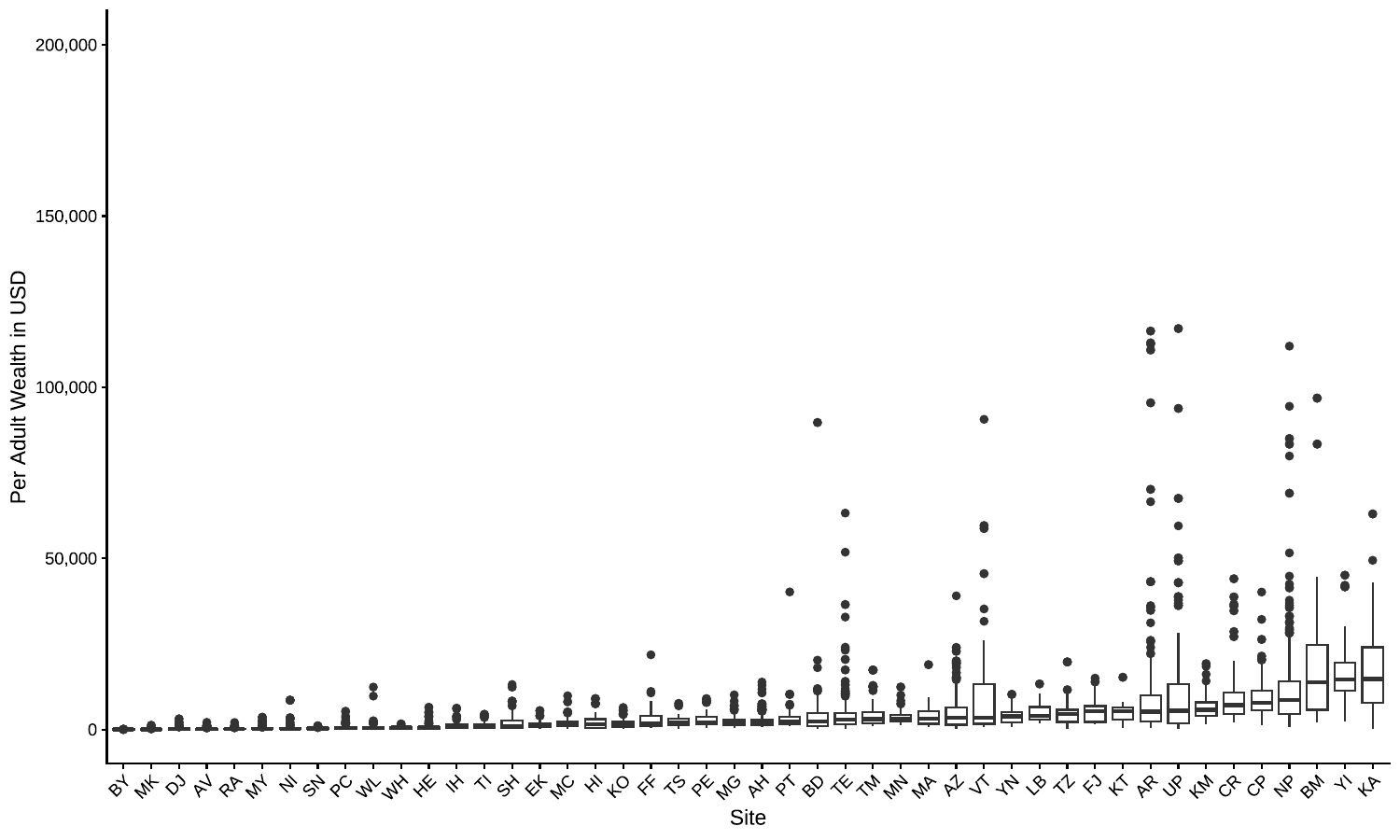}
    \includegraphics[width=0.9\linewidth]{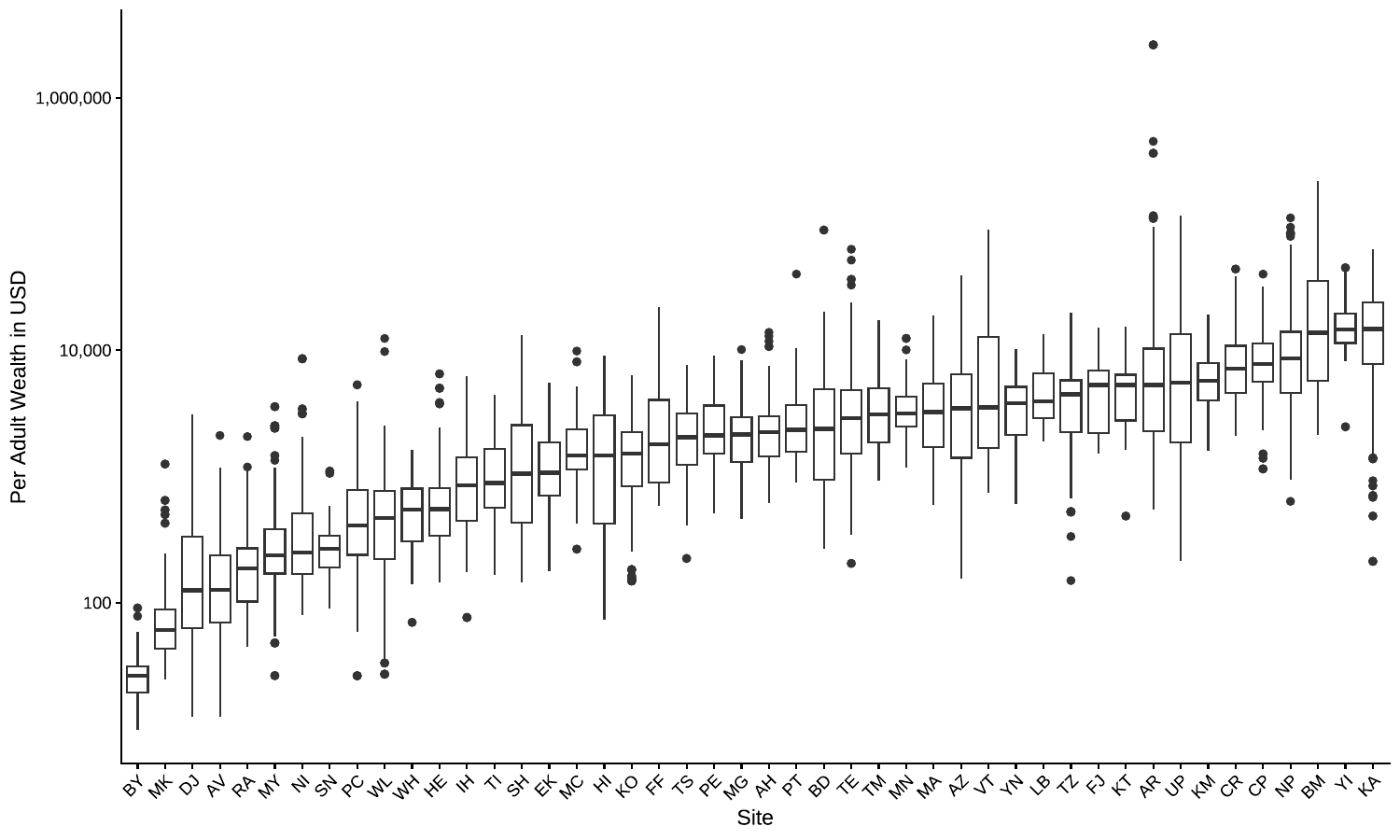}
    \caption{\textbf{Boxplots by community for wealth per adult.} The top plot shows the raw distributions but has a truncated axis, removing a handful of outliers. The bottom plot shows the same distribution with a log transformed axis, so that all outliers can be included.}
    \label{fig:boxplot-wealth-adult}
\end{figure}

It is possible to conceptualize material wealth in different ways---the overall material value of a sharing unit's assets (Figure~\ref{fig:boxplot-wealth-su}), the per capita value (Figure~\ref{fig:boxplot-wealth-person}), the per adult value (Figure~\ref{fig:boxplot-wealth-adult}), among others. We provide more discussion of these distinctions in Section~\ref{supp:alternative_SUweightings_within}.

\begin{figure}
    \centering
    \includegraphics[page=1, width=\linewidth]{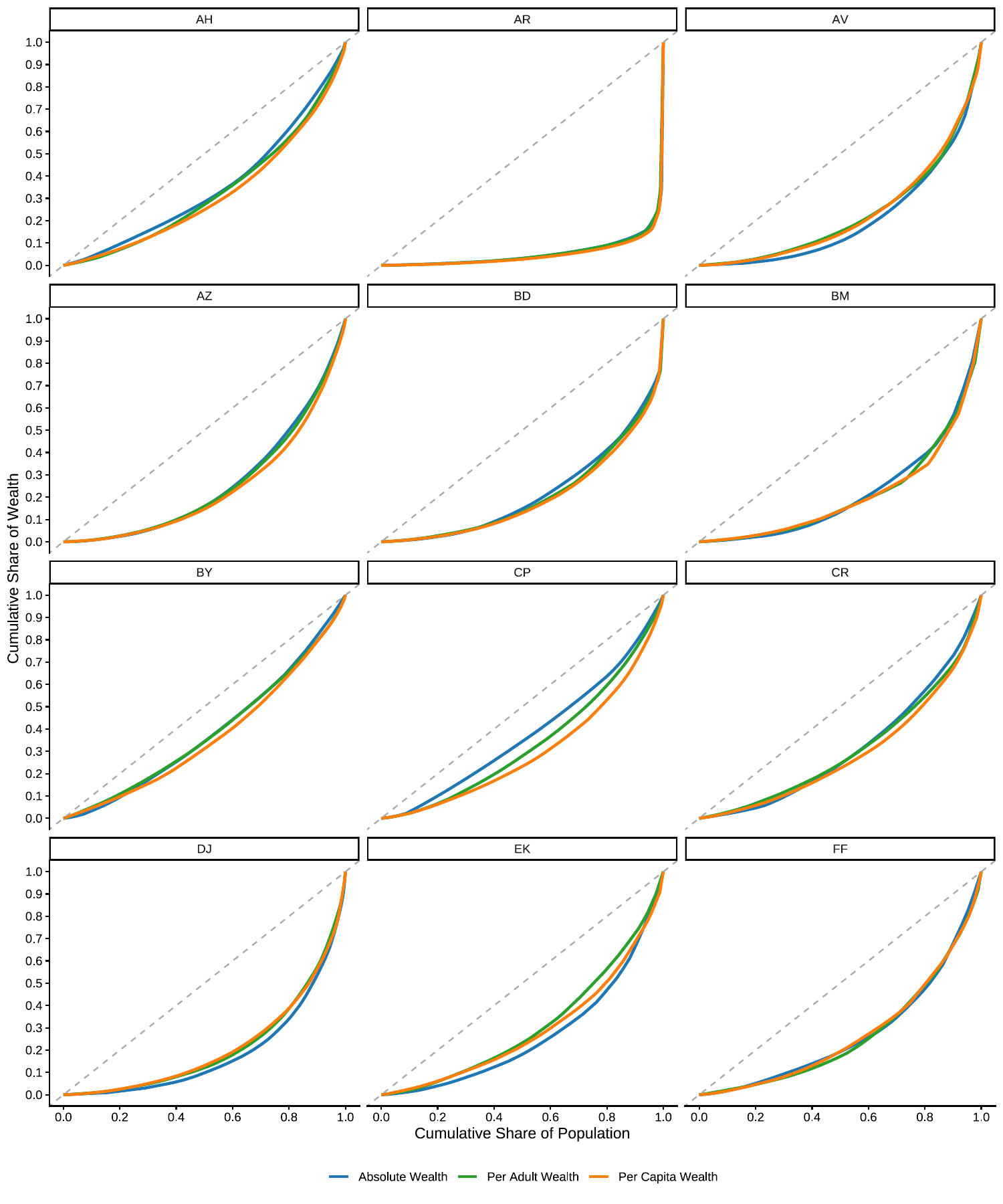}
    \caption{Lorenz Curves}
    \label{fig:lorenz1}
\end{figure}

\begin{figure}
    \centering
    \includegraphics[page=2, width=\linewidth]{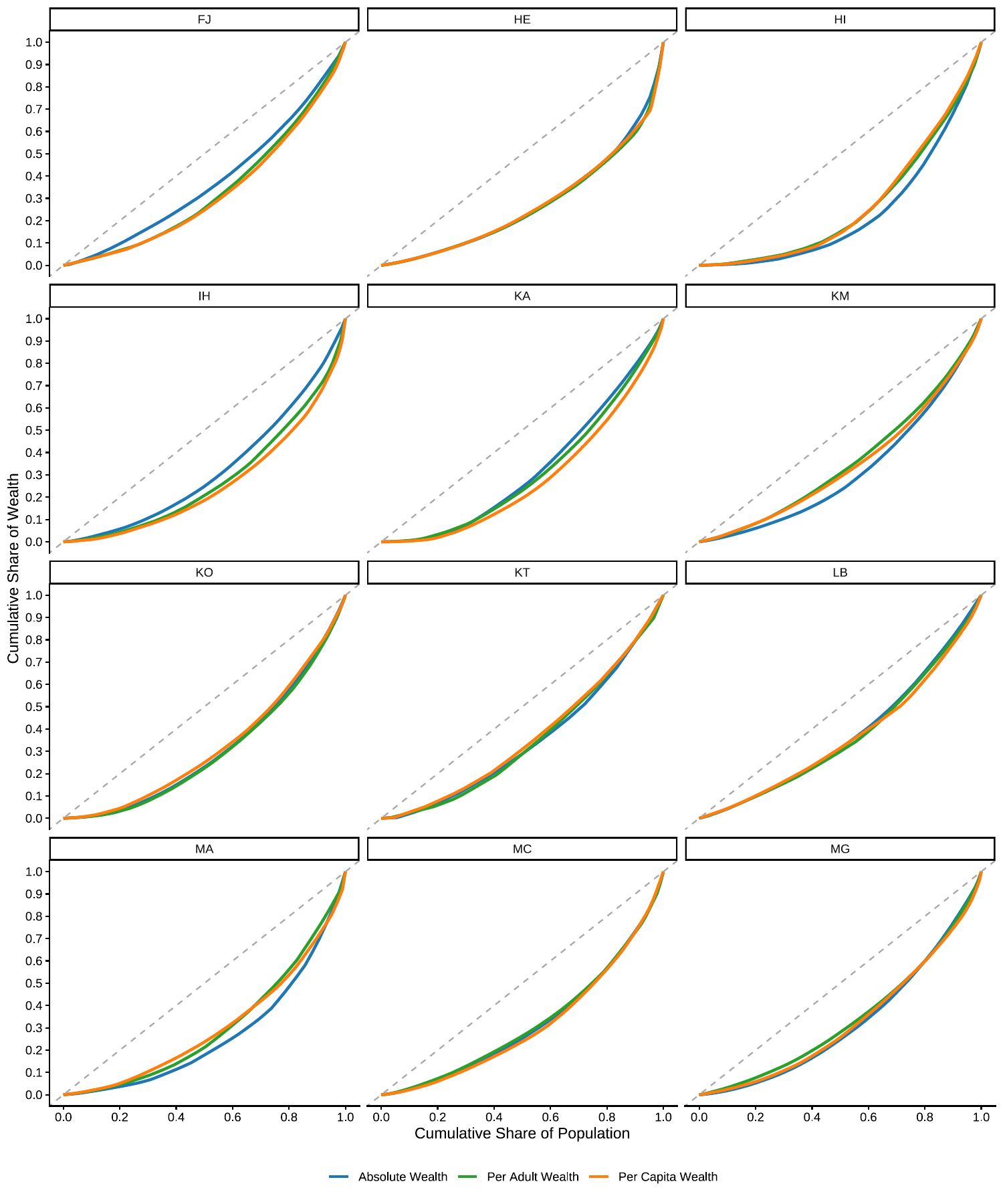}
    \caption{Lorenz Curves}
    \label{fig:lorenz2}
\end{figure}

\begin{figure}
    \centering
    \includegraphics[page=3, width=\linewidth]{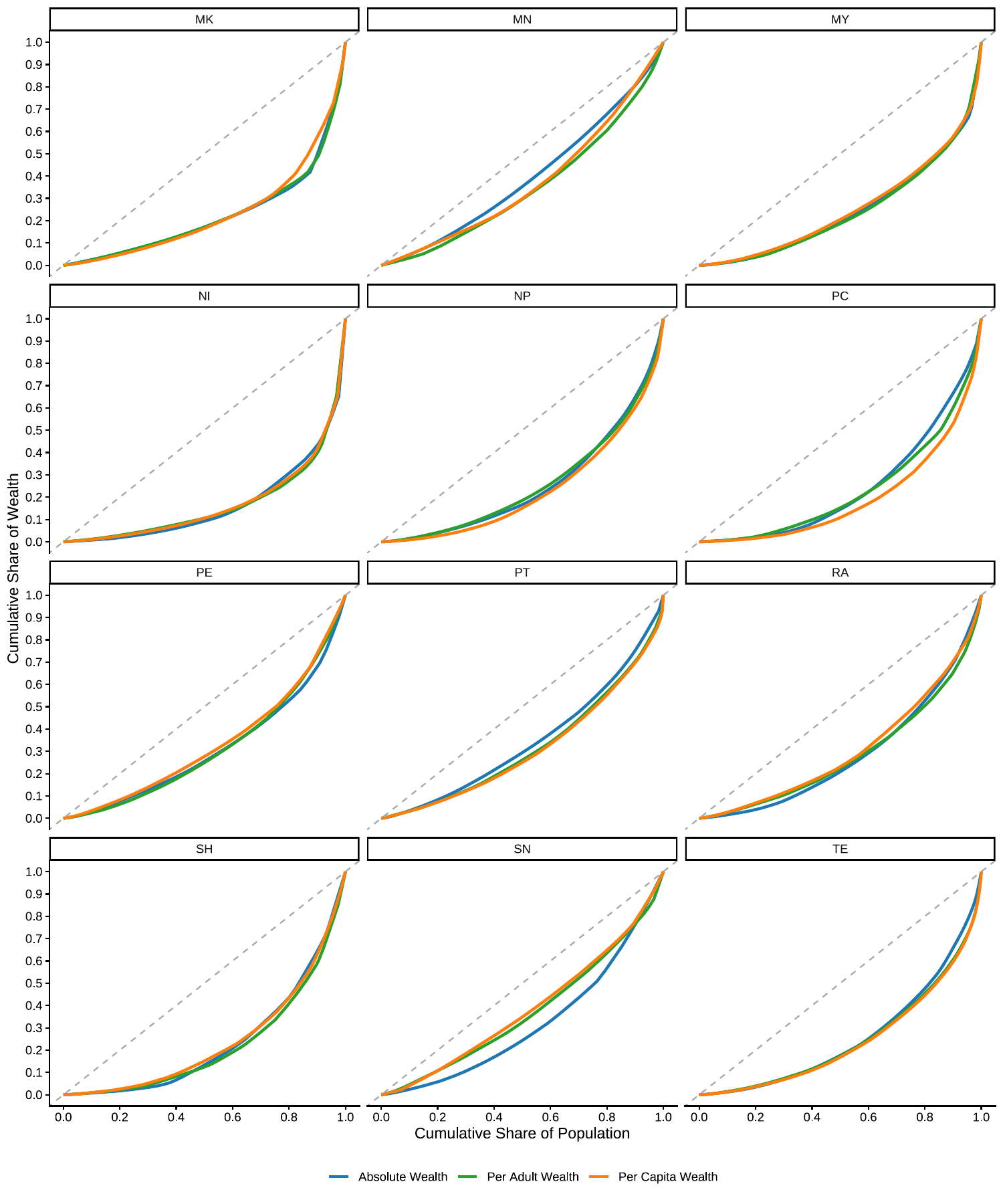}
    \caption{Lorenz Curves}
    \label{fig:lorenz3}
\end{figure}

\begin{figure}
    \centering
    \includegraphics[page=4, width=\linewidth]{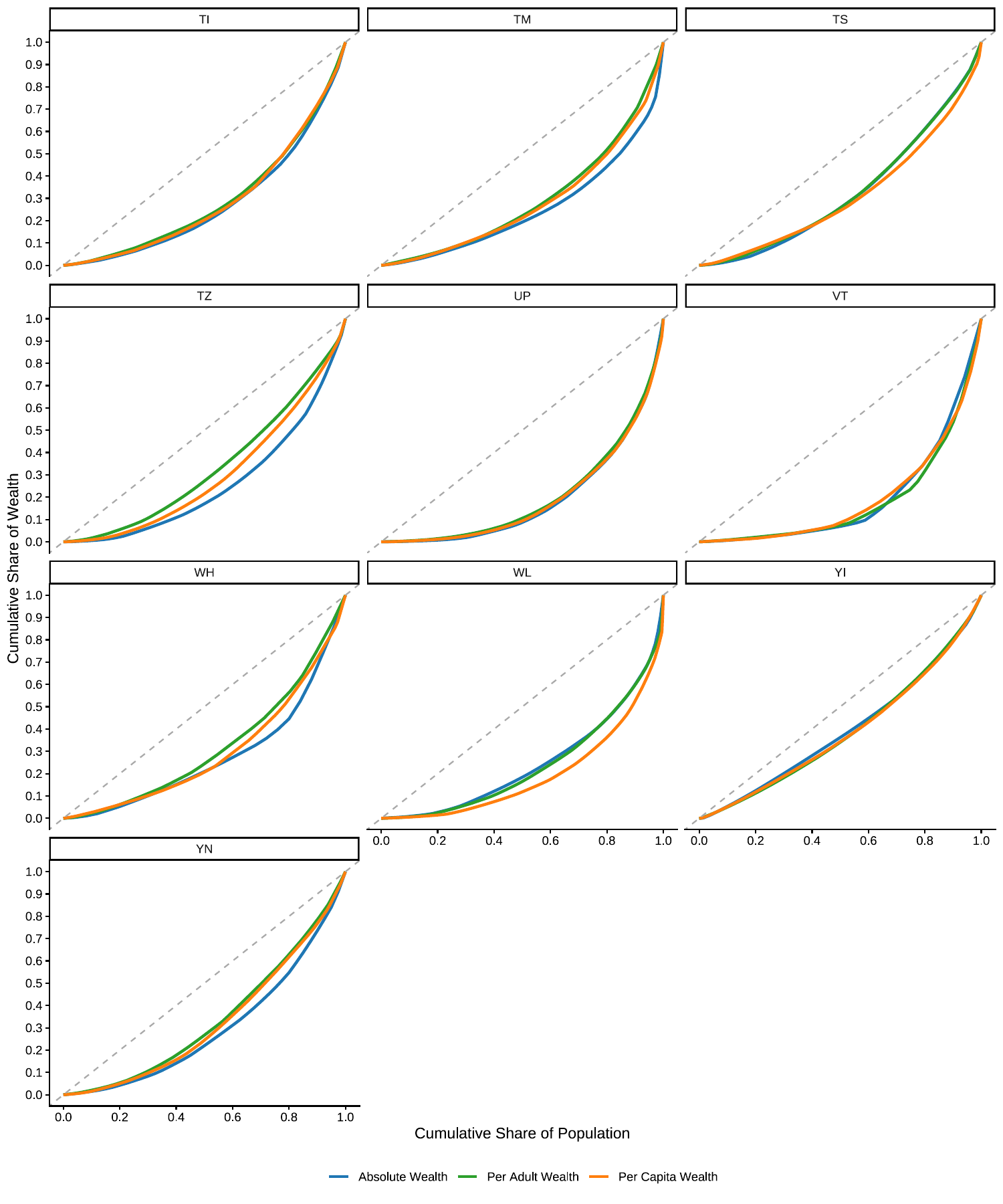}
    \caption{Lorenz Curves}
    \label{fig:lorenz4}
\end{figure}

\begin{figure}
    \centering
    \includegraphics[width=0.8\linewidth]{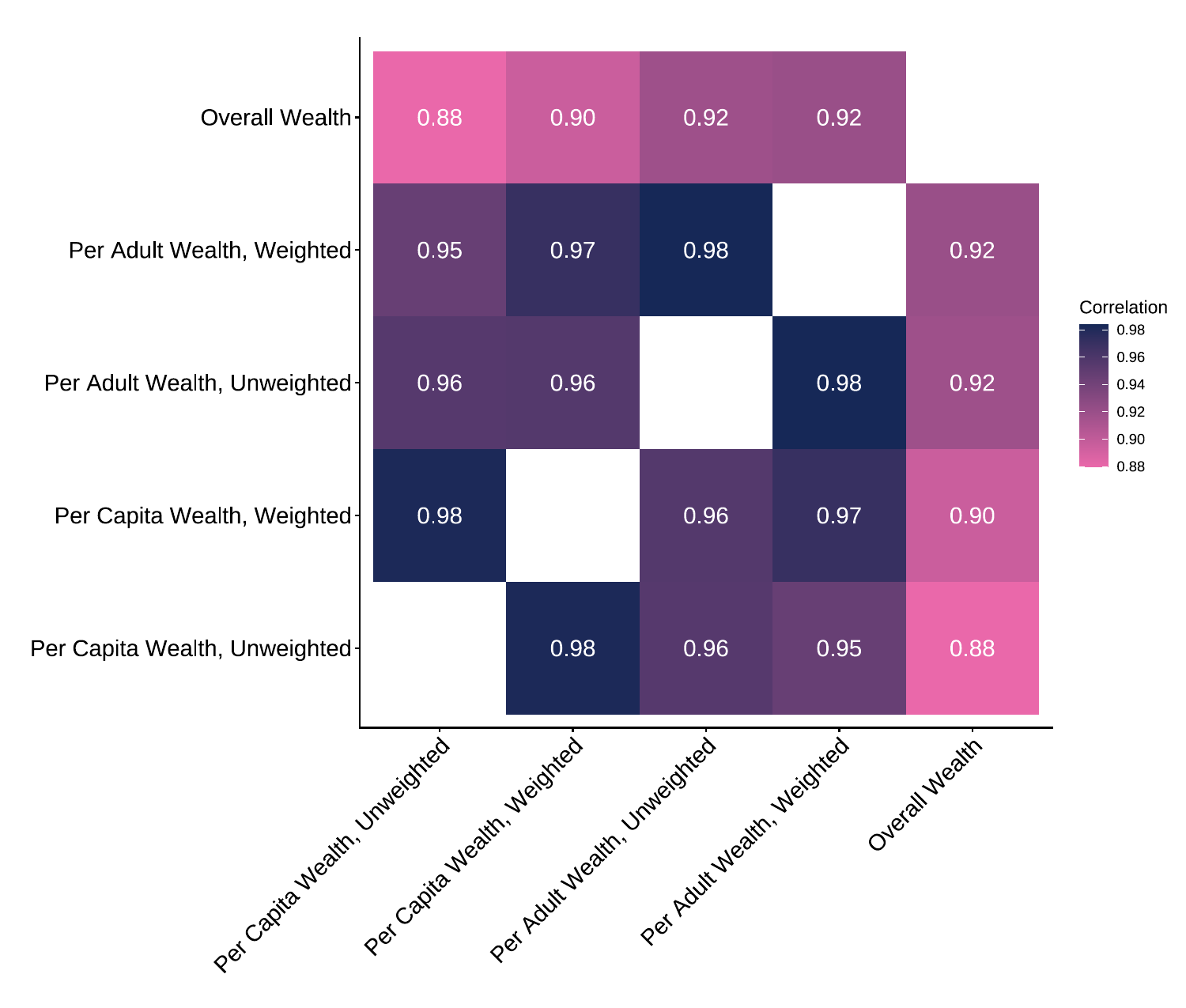}
    \caption{\textbf{Cross-site correlations between Ginis using different measures of sharing unit wealth} Each entry presents the cross-site correlation between a pair of Gini measures. The axis labels indicate the form of wealth used when computing the Gini and whether sharing-unit weights are used when calculating the Gini coefficient. For example, when calculating the weighted Gini using per adult wealth, we weight each sharing unit by the number of adults in the unit, so that the Gini can be thought of as being calculated at the level of adults, as opposed to the level of sharing units.}
    \label{fig:corrs-various-ginis}
\end{figure}

The cross-site patterns of inequality are present regardless of whether wealth is conceptualized as total sharing unit-wealth, per-capita wealth, or per-adult wealth. Those patterns are presented in Figures~\ref{fig:lorenz1}--\ref{fig:lorenz4}. Each figure shows a set of three Lorenz curves for each site, one for the sharing-unit-level wealth distribution, one for the adult-level wealth distribution (where we generally consider people aged 18 or older as ``adults''), and one for the person-level wealth distribution.
Figure~\ref{fig:corrs-various-ginis} shows that, at the site level, the Gini coefficients stemming from these various wealth measures are well-correlated with each other.

\begin{comment}
\begin{figure}
    \centering
    \includegraphics[width=\linewidth]{figures/lorenz/Lorenz_Curves_AllSites_Panel.png}
    \caption{Lorenz curves of sharing unit wealth. Showing per capita and absolute wealth, respectively. Site-wise curves are shown in Figures~\ref{fig:lorenz1}--\ref{fig:lorenz4}. }
    \label{fig:lorenz_agg}
\end{figure}
\end{comment}

\clearpage
\subsection{Social Network Data}\label{supp:socialnetsbackground}

As measures of relational wealth, we use a series of ten name generator questions, some with standardized phrasing and some tailored to fit the context of the respective communities (see Table \ref{tab:wealthmeasures}). We provide further detail in a forthcoming companion paper. % Further details on the exact prompts, etc. can be found in [[LINK/REFERENCE SOMEHOW TO DATA PAPER]]. xx
Questions were asked in a set order.

First were two questions about money lending, first about \emph{borrowing} a week's wages (the specific amount in the local currency determined by the contributing ENDOW researchers) and then about \emph{lending} a week's wages.
Next were two questions about sharing unit-level exchanges of common goods (the particular phrasing was site-specific) again ``double sampled'' (i.e., asking both about asking/borrowing and giving/lending).
These were followed by two questions on behavioral assistance and labor support, with one question focused on women's assistance and one on men's. This bifurcation by gender was done in recognition of the gendered division of labor in many sites, meaning that the types of behavioral assistance could plausibly differ for men and women sharing unit members. This was further done to help respondents think holistically of the relationships of their entire sharing unit, not just their own personal connections. The particular prompts were specific to each site (as the relevant types of behavioral assistance vary based on the set of livelihoods in the respective communities).
The same approach was taken for the two subsequent questions on communication and affective support (i.e., gender-specific questions, specific to each site).
The two final questions were aimed at capturing support from beyond the community, particularly to organizations and people who can facilitate the navigation of bureaucracy and economic opportunities. First was a question on government and NGO-employees who were well known and so could render assistance, which was standard across sites. Second was a site-specific question on other important sources of support \emph{outside} the community, with a particular focus on economic opportunities (e.g., accessing markets).
Finally, some ENDOW researchers included additional name generators at the end, as they desired (not studied here).
Alongside these prompts, ENDOW researchers gathered information on kinship, allowing us to identify consanguineal and affinal relatedness between all community residents, including both adults and children. ENDOW researchers were tasked with gathering information on existing partnerships and on descent relationships that would suffice to identify relatives to the level of first cousins (this could be done, for example, by gathering information on all community residents' parents and grandparents), though for our purposes here we consider only close kin relationships (i.e., parent-child and full siblings) as we can be more confident in our coverage.
These collectively give us a comprehensive representation of the multilayer network of the communities \citep{boccaletti_structure_2014, atkisson_why_2020}.

%Tables \ref{tab:prompts-standardized-money-lending-question}--\ref{tab:prompts-male-oriented-K-network} contain details of the prompts used in each site for each layer.

%\include{tables/standardized money-lending question}
%\include{tables/household-level exchange question}
%\include{tables/female-oriented Q network}
%\include{tables/male-oriented Q network}
%\include{tables/female-oriented K network}
%\include{tables/male-oriented K network}

\subsubsection{Network Construction}\label{supp:net_construct}

A key point of variation across communities is the sampling strategy of respondents for the social support network name generator prompts. The decisions on sampling strategies were made by the contributing ENDOW researchers in light of practical considerations, primarily which residents would most consistently be available to provide answers and logistical constraints on data collection (time available for fieldwork, etc). For example, the majority of men in some sites are away from the community for extended periods of time, making them impractical to survey. Consequently, in some sites, it was the female household head who was the intended respondent, in others the male household head, in others both male and female household heads, and in others all adult residents.

Depending on the sampling strategy and the size of the sharing unit, sharing units consequently had variable opportunities to name support partners.
There are several different ways that the person-level responses can be aggregated into sharing-unit-level networks, in light of this. The two key features that need to be accounted for when moving from person-level to sharing-unit-level networks are (1) the fact that since the questions were usually asked to specific people, some sharing units received more opportunities to nominate connections than others, and (2) in some sharing units, multiple people from the same sharing unit nominated a particular person in another sharing unit. There is no obvious best way to handle these features of our data, so we check that our findings are robust to several different ways of aggregating the network to the sharing-unit level.

\begin{figure}
    \centering
    \includegraphics[width=\linewidth]{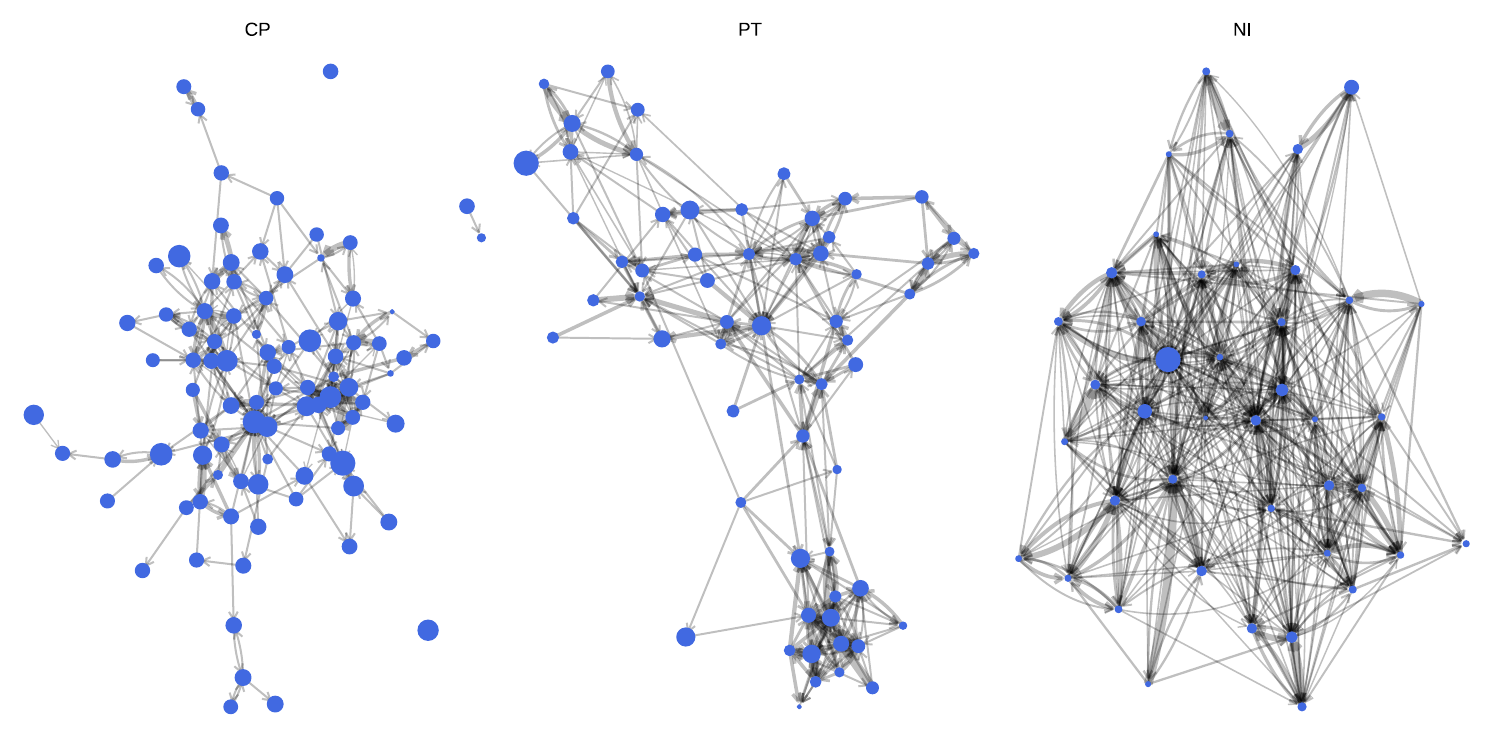}
    \caption{Examples of the composite network for three communities. Nodes (sharing units) are sized by their absolute wealth (rescaled in each plot to cover the same size range). Edges are directed and follow requests for support (so, go from supportee to supporter). Thicker edges reflect higher edge weight, meaning that the requesting sharing unit more strongly relies on the supporting sharing unit for support (i.e., requests support across a greater number of domains). Curved edges reflect a reciprocal relationship; straight edges reflect unreciprocated relationships. }
    \label{fig:eg_nets}
\end{figure}

In our main analysis, we use the composite network, which is weighted and directed with sharing units being the nodes, defined as follows. We go through each question in turn.  We assign a weight for that layer equal to the number of times any individual from sharing unit $i$ nominates any individual from sharing unit $j$, divided by the number of individuals from sharing unit $i$ who were surveyed. So for example, if two separate individuals in sharing unit $i$ are interviewed, and one names $3$ people in sharing unit $j$ for the borrowing money question, and another names $2$ people, then the loan question layer gets a weight of $(3+2)/2=2.5$. Then we do this for each layer and sum across all the layers. Figure~\ref{fig:eg_nets} shows the composite network for three example communities.

Below, we show that our results are similar if we use two other networks---the ``proportional composite'' network and the ``raw composite'' network.

The ``proportional composite'' network is similar to the composite network except the edge weights for each question are equal to the proportion of nominating individuals in sharing unit $i$ who nominate any member of sharing unit $j$.  For each respondent in sharing unit \textit{i}, that is, there is a binary value of $0$ or $1$ depending on the presence of \textit{any} nominations of individuals who reside in sharing unit \textit{j}. For example, if two separate individuals in sharing unit $i$ are interviewed, and one names 3 people in sharing unit $j$ for the borrowing money question, and another names 2 people, then the loan question layer gets a weight of $(1+1)/2=1$. When extended to multiple network layers, for all of the respective respondents in sharing unit $i$, these binary indicators are summed across layers and divided by the potential number of nominations of individuals in sharing unit $j$ that collectively could have been reported by those respondents.

The  ``raw composite'' network, which is weighted and directed with sharing units being the nodes, is defined as follows. We go through each question in turn.  We assign a weight for that layer equal to the number of times any individual from sharing unit $i$ nominates any individual from sharing unit $j$. So for example, if two separate individuals in sharing unit $i$ are interviewed, and one names $3$ people in sharing unit $j$ for the borrowing money question, and another names $2$ people, then the loan question layer gets a weight of $3+2=5$.   Then we do this for each layer and sum across all the layers. The key difference from the ``composite network'' is that we do not divide by the number of individuals from sharing unit $i$ who were surveyed when determining an edge weight.  % These are the collapse_sum and main_supp networks.

Note that due to the complexities already introduced here with the weightings, we do not additionally consider the ``double sampled'' name generators (i.e., those where people were asked to report whom they would \emph{lend} a week's wages or a range of household items to), but just the prompts that are consistent with all other name generators (i.e., where people were asked to report who they would make requests of). In future work, we plan to incorporate these responses.

\clearpage
\subsection{Measures Constructed from Network and Wealth Data} \label{supp:definitions-measures}

Among the network measures of social capital that we construct, the most predictive of wealth inequality are features that combine data on edges in the network with the attributes of sharing units (nodes). Specifically, we consider the wealth attribute of a node $i$, $w_i$. We note that in the paper we consider various metrics of wealth: absolute wealth, wealth per adult, wealth per capita, and size-adjusted wealth. Here, we will just consider a wealth value $w_i$ that could be a stand-in for any of the above measures. Due to the skewness of wealth within sites, we transform our wealth values to percentile ranks, such that $w_i \in [0, 1]$ for all $i$. We sometimes use weighted percentile ranks on the basis of sharing unit weights such as the number of individuals living in the sharing unit or the number of adults living in the sharing unit.

Let $\mathcal{V}$ denote the set of nodes. In our applications, this set will always lie \textit{within} a given study community. Given a vector of node characteristics $\bm{w} = (w_i)_{i \in \mathcal{V}}$ and directed and weighted edges $\{e_{ij}\}_{i, j \in \mathcal{V}}$ between all pairs of nodes $i$ and $j$,
we define the \textit{average alter wealth} ($\text{AAW}_i$) of a sharing unit $i$ as the edge-weighted average of the wealth percentile ranks of its connections:
\begin{align*}
\text{AAW}_i = \frac{\sum_{j \in \mathcal{V}} e_{ij} w_j}{\sum_{j \in \mathcal{V}} e_{ij}}
\end{align*}
where $e_{ij}$ represents the weight of the directed edge from $i$ to $j$, and $w_j \in [0, 1]$ is the wealth rank of alter $j$.

We then define the \textit{average alter wealth of the poor} (AAWP) in a community as: \begin{align*}
\text{AAWP} = \frac{\sum_{i: w_i \le 0.5} a_i \cdot \text{AAW}_i}{\sum_{i: w_i \le 0.5} a_i}
\end{align*}
where $a_i$ is a weight variable. For example, when constructing measures on the basis of wealth per capita, we weight sharing units by the number of individuals living in the sharing unit.\footnote{
More specifically, our weighting proceeds like this:
When constructing measures on the basis of wealth per capita, for example, denote the number of individuals in sharing unit $i$ as $n_i$, and total wealth of sharing unit $i$ as $w_i$:
\begin{enumerate}
\item Calculate $w_i$/$n_i$
\item Rank sharing units in the within-site distribution after weighting each unit by $n_i$. Equivalently, a unit with five residents is represented five times in the percentile distribution.
\item Classify the entire sharing unit as below-median if its weighted rank is at most $0.5$, and above-median otherwise.
\item Calculate its $\text{AAW}_i$ using only the network edge weights \(e_{ij}\).
\item Average $\text{AAW}_i$ within each half using $n_i$ as a weight again. So a five-person sharing unit receives five times the weight of a one-person sharing unit.
\end{enumerate}
}
Similarly, the \textit{average alter of the rich} (AAWR) in a community is:
\begin{align*}
\text{AAWR} = \frac{\sum_{i: w_i > 0.5} a_i \cdot \text{AAW}_i}{\sum_{i: w_i > 0.5} a_i}
\end{align*}
The measure of \textit{relative average alter wealth} (RAAW) in a community that we make use of throughout the paper is AAWP normalized by AAWR:
\begin{align*}
\text{RAAW} = \frac{\text{AAWP}}{\text{AAWR}}
\end{align*}

Similarly, we can construct a measure of \textit{unit economic connectedness} (UEC) for each sharing unit, similar to \textit{individual economic connectedness} defined by \cite{chetty_social_2022}:
    \begin{align*}
        \text{Unit Economic Connectedness}_i = 2 \frac{\sum_{j \in \mathcal{V}} e_{ij} \mathbb{I}(w_j > 0.5)}{\sum_{j \in \mathcal{V}} e_{ij}}
    \end{align*}
    The factor of 2 is simply a normalization so that a $\text{UEC}_i$ value of 1 indicates that a given node is connected to an equal number of above- and below median-wealth nodes.

We then construct \textit{economic connectedness} (EC) for a community as:
\begin{align*}
\text{Economic Connectedness} = \frac{\sum_{i: w_i \le 0.5} a_i \cdot \text{UEC}_i}{\sum_{i: w_i \le 0.5} a_i}
\end{align*}
where $a_i$ is again a weight variable.

Similarly, the \textit{economic connectedness for high-wealth (i.e., ``rich'') sharing units} (ECR) in a community is constructed as:
\begin{align*}
    \text{ECR} = \frac{\sum_{i: w_i > 0.5} a_i \cdot \text{UEC}_i}{\sum_{i: w_i > 0.5} a_i}
\end{align*}
From these, we can form a measure of \textit{relative connectedness} (RC) for a community:
\begin{align*}
    \text{RC} = \frac{\text{EC}}{\text{ECR}}
\end{align*}

When we consider quartiles of the wealth distribution, as in Figure~\ref{fig:cross-site-two-panel}, we normalize by $\text{AAW}_i$ or $\text{UEC}_i$ of \emph{all} sharing units, as opposed to normalizing by the $\text{AAW}_i$ or $\text{UEC}_i$ of above-median-wealth sharing units as in the binary split case.

Note that we can construct distinct measures based on each tie direction, one considering $i \rightarrow j$ and one considering $i \leftarrow j$. For example, given the meaning of tie directionality in our networks---where ties go from requester to requested---$\text{AAW}_i$ constructed using $i \rightarrow j$ represents the Average Alter Wealth for alters who provide support (i.e., supporters), while $\text{AAW}_i$ constructed using $i \leftarrow j$ represents the Average Alter Wealth for alters to whom support is provided (i.e., supportees). All that is done to construct the latter is to reverse the direction of the $ij$ edge. Unless otherwise stated, the measures we use are constructed considering alters who \emph{provide} support (i.e., supporters).

\clearpage
\subsection{Community-Level Information}\label{supp:sitevars}

Production processes, environmental factors, and work arrangements differ across communities. These differences may be of interest on their own, but for the purposes of our exploration of the relationship between material inequality and network structure across communities, we mainly view these variables as confounders.

\subsubsection{National Attributes}

We compiled a range of national-level variables, drawing from existing datasets.

Our national attributes include:
\begin{itemize}
    \item GDP per capita, the national Gini coefficient, and national adult literacy rate from the World Bank World Development Indicators \citep{worldbank2025wdi} gathered with \citep{scheuch2026wbwdi}.
    \item national indices of corruption, socioeconomic exclusion, private property rights, and freedom of domestic movement from the Varieties of Democracy (V-Dem) project \citep{vdem2025} gathered with \citep{vdem_package}. Specifically, the corruption index we use is ``$v2x\_corr$'', the private property rights index we use is ``$v2xcl\_prpty$'', the exclusion index is ``$v2xpe\_exlecon$'', and the freedom of domestic movement index is ``$v2xcl\_dmove$''.
    \item state authority characteristics from Polity5 \citep{marshall_polityv_2020}.
\end{itemize}

\subsubsection{Local Geospatial Attributes}

We construct local geospatial attributes as raster-based measures using coordinates of the center of each community and (where the underlying dataset has temporal resolution) the year of data collection, using a mix of R and Python. Based on the centroid for each site, we then construct a radius around that to construct a buffer from which we sample the underlying data and aggregate. Data are gathered using the Google Earth Engine API or separate existing open datasets. These measures include:

\begin{itemize}
    \item local population density from the Gridded Population of the World \citep{ciesin_gridded_2017}
    \item nightlight radiance from VIIRS \citep{elvidge_annual_2021}
    \item the Human Influence Index \citep{sanderson_march_2022} (for which we use a buffer of 250m)
    \item land-based travel time to the nearest densely populated area \citep{weiss_global_2018} (for which we use a buffer of 1000m)
    \item biologically-effective degree days, mean diurnal temperature range, and precipitation \citep{nobakht_agroclimatic_2019}
\end{itemize}

(This work was done with the help of research assistants Evgeny Noi and Ella Vacic).

\subsubsection{Researcher-Provided Attributes}

Drawing on their qualitative ethnographic experience, ENDOW researchers were asked to provide some variables to characterize the community where they work, largely derived from the Standard Cross-Cultural Sample \cite{murdock_standard_1969}. This included basic information on subsistence strategies, residence patterns, inheritance practices, local political organization, etc. These followed variables asked in an earlier cross-cultural study led by some of the same team \cite{borgerhoff_mulder_intergenerational_2009}.

In addition to this, we designed a brief questionnaire with a series of questions tailored to the ENDOW project and its aims:

\begin{enumerate}
    \item What proportion of all produced output goes toward:
    \begin{enumerate}
        \item subsistence consumption
        \item non-monetary exchange within the site
        \item non-monetary exchange outside the site
        \item cash market exchange within the site
        \item cash market exchange outside the site
        \item communal/collective allocation
    \end{enumerate}
    \item What proportion of all non-leisure time goes toward:
    \begin{enumerate}
        \item household subsistence (for example, cooking meals)
        \item non-monetary cooperative or reciprocated labor within the site (for example, providing childcare for another family)
        \item non-monetary cooperative or reciprocated labor outside the site
        \item non-monetary communal labor (for example, working on community projects relating to irrigation)
        \item wage or salaried labor within the community (for example farming or herding)
        \item wage or salaried labor outside the community
        \item producing or acquiring goods/services for intended sale (within the community)
        \item producing or acquiring goods/services for intended sale (outside the community)
    \end{enumerate}
    \item What proportion of sharing units are involved in the most common livelihood?
    \item How often would the following shocks radically affect the day-to-day livelihood of a typical sharing unit in the absence of risk-sharing/insurance. [with options for response of ``weekly'', ``monthly'', ``more than yearly'', ``less than yearly'', and ``never'']:
    \begin{enumerate}
        \item climate-related problems
        \item human diseases or accident-related problems
        \item violence from state or non-state actors
        \item market-related events
    \end{enumerate}
    \item Are there strong norms (such as those based on caste, but excluding those based on gender, age, or wealth) that put substantial limits on the type of work people can perform and who can access or control property or other means of production?
\end{enumerate}

\clearpage

\section{Robustness of Within-Community Results}\label{supp:robustness_within}

\subsection{Different Network Constructions}\label{supp:alternative_networks}

Our main results focus on the ``composite network'', whereby each edge is associated with a weight reflecting the number of alters named in each sharing unit and the number of layers in which the two sharing units are related. In this section, we show that our results are similar when using the ``proportional composite'' and ``raw composite'' networks (described above in Supplementary Section~\ref{supp:socialnetsbackground}).

\begin{figure}[h]
\centering
\begin{subfigure}{0.49\textwidth}
    \centering
    \includegraphics[width=\textwidth]{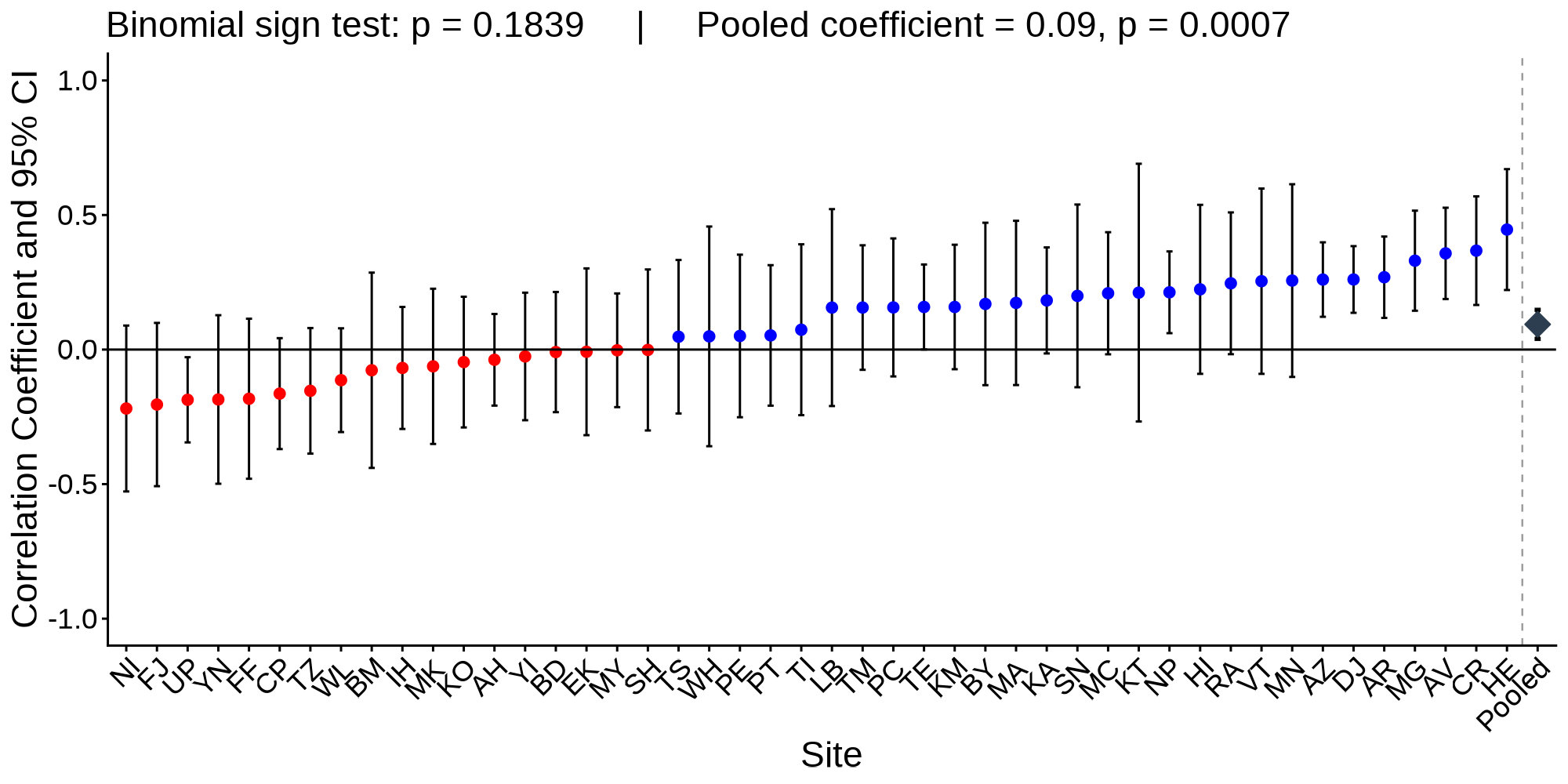}
    \caption{Support access vs wealth.}
\end{subfigure}
%\hspace{2em}
\begin{subfigure}{0.49\textwidth}
    \centering
    \includegraphics[width=\textwidth]{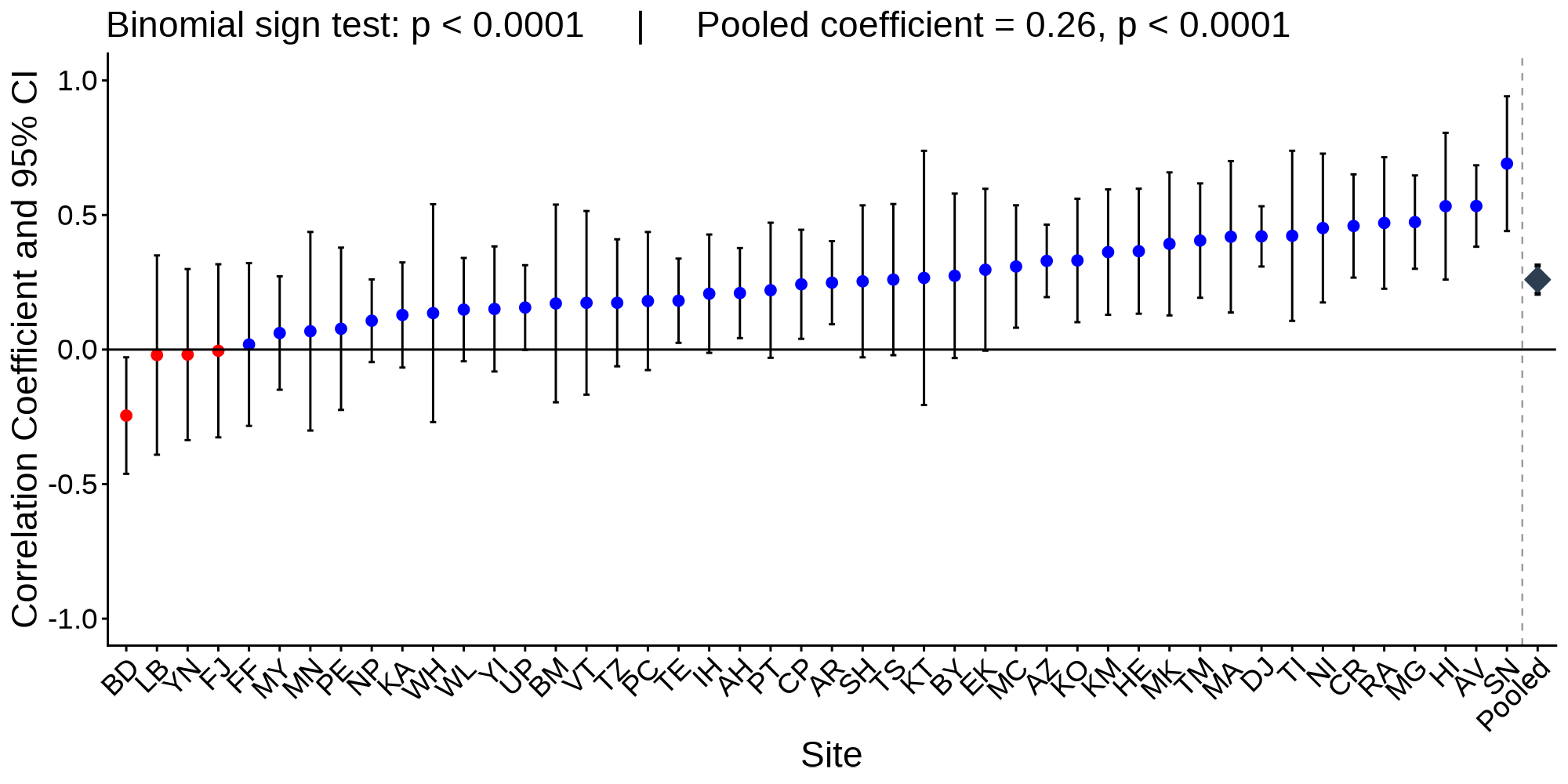}
    \caption{Support provisioning vs wealth.}
\end{subfigure}
\vspace{1em} % Space between the top and bottom rows
\begin{subfigure}{0.49\textwidth}
    \centering
    \includegraphics[width=\textwidth]{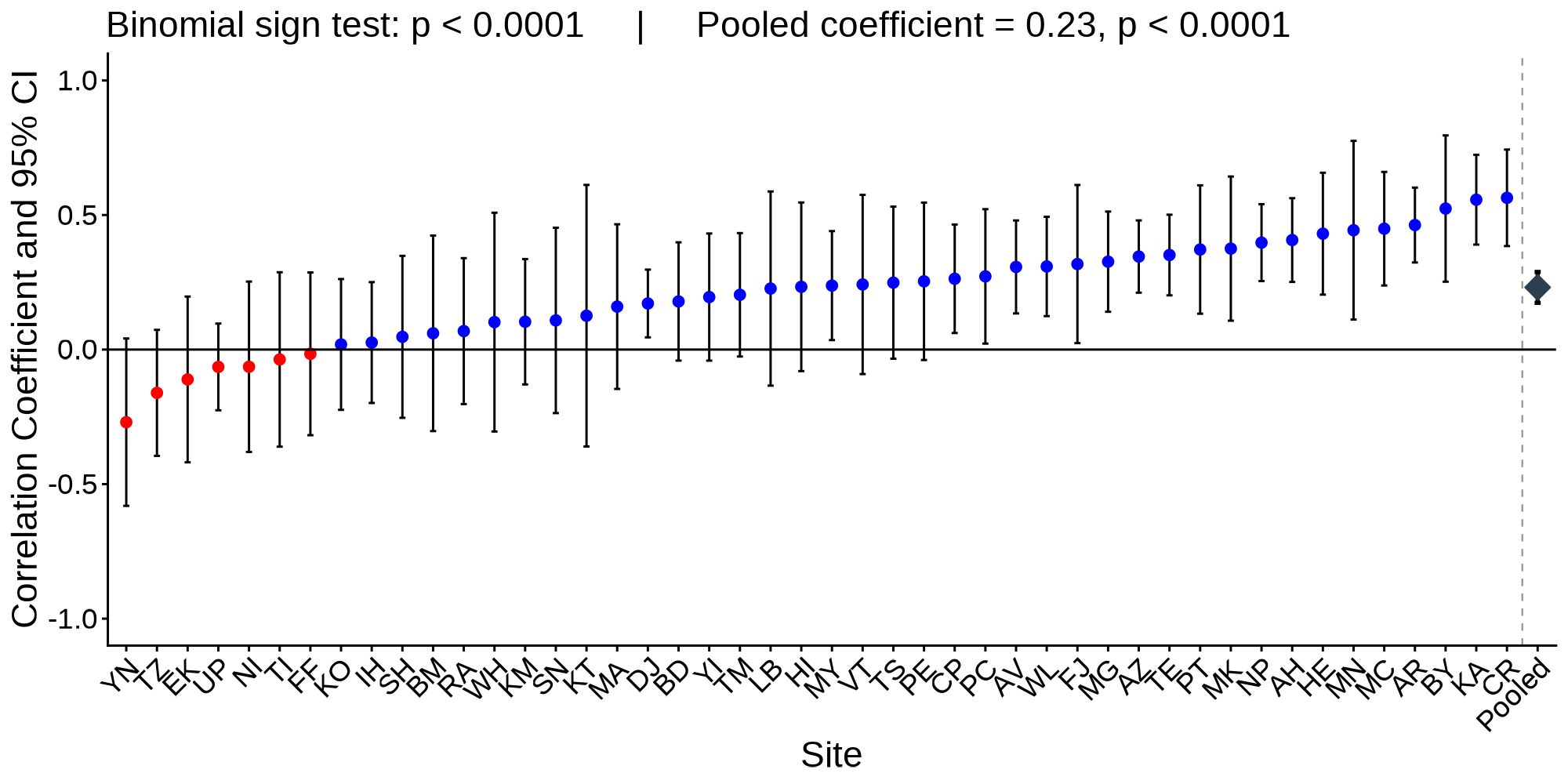}
    \caption{Support access p.c. vs wealth p.c.}
\end{subfigure}%
%\hspace{2em}
\begin{subfigure}{0.49\textwidth}
    \centering
    \includegraphics[width=\textwidth]{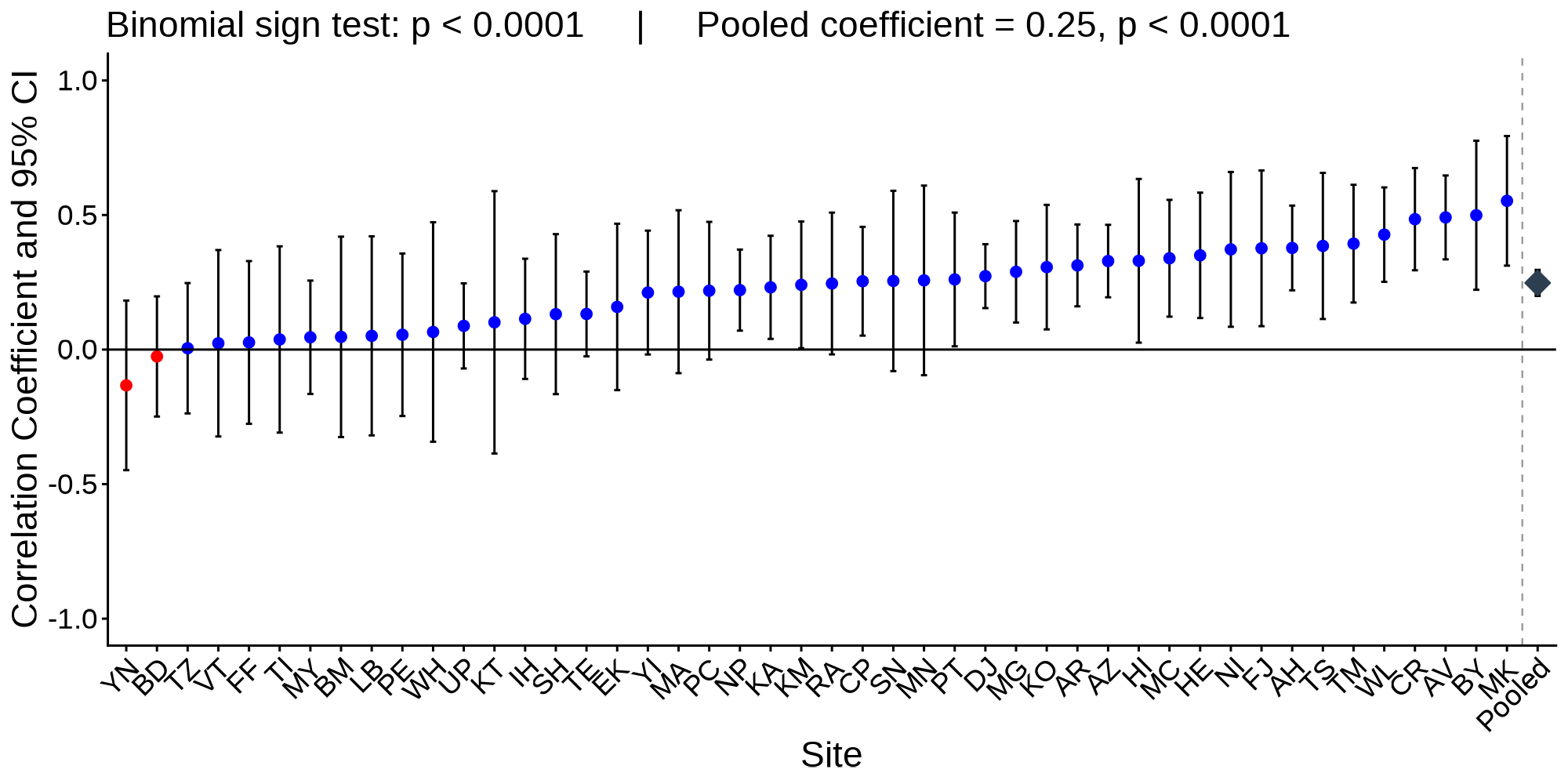}
    \caption{Support provisioning p.c. vs wealth p.c.}
\end{subfigure}
%\vspace{1em} % Increased space before the main figure caption
\caption{\textbf{Correlations between access and provisioning of support and material wealth metrics}. Access to support (i.e., out-degree) and provisioning of support (i.e., in-degree) are calculated using the ``proportional composite'' network. }
\label{fig:within-site-multipanel-collapsed-sum}
\end{figure}

\begin{figure}[t]
\centering
\begin{subfigure}{0.49\textwidth}
    \centering
    \includegraphics[width=\textwidth]{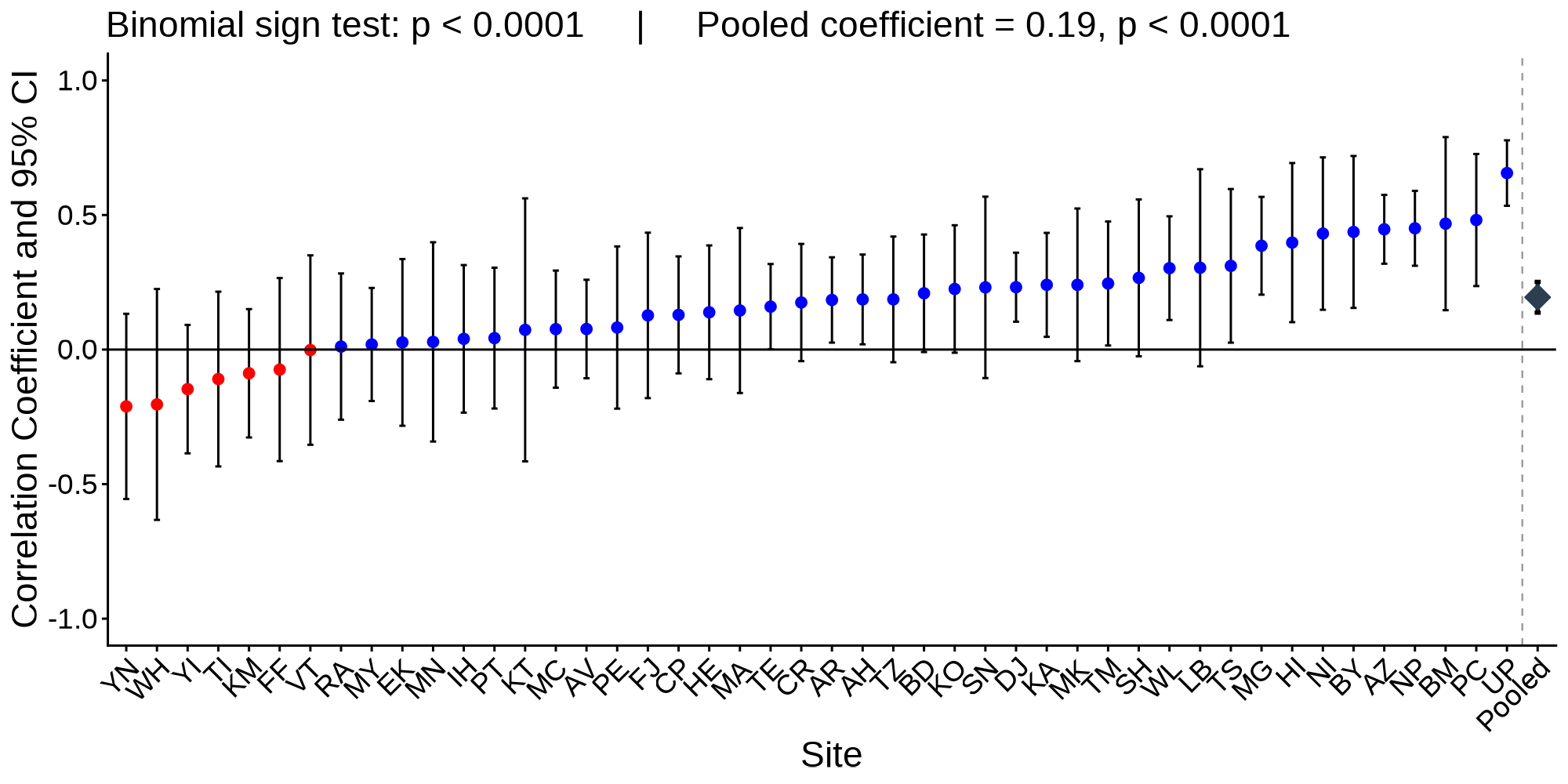}
    \caption{Average Alter Wealth vs wealth.}
\end{subfigure}
\begin{subfigure}{0.49\textwidth}
    \centering
    \includegraphics[width=\textwidth]{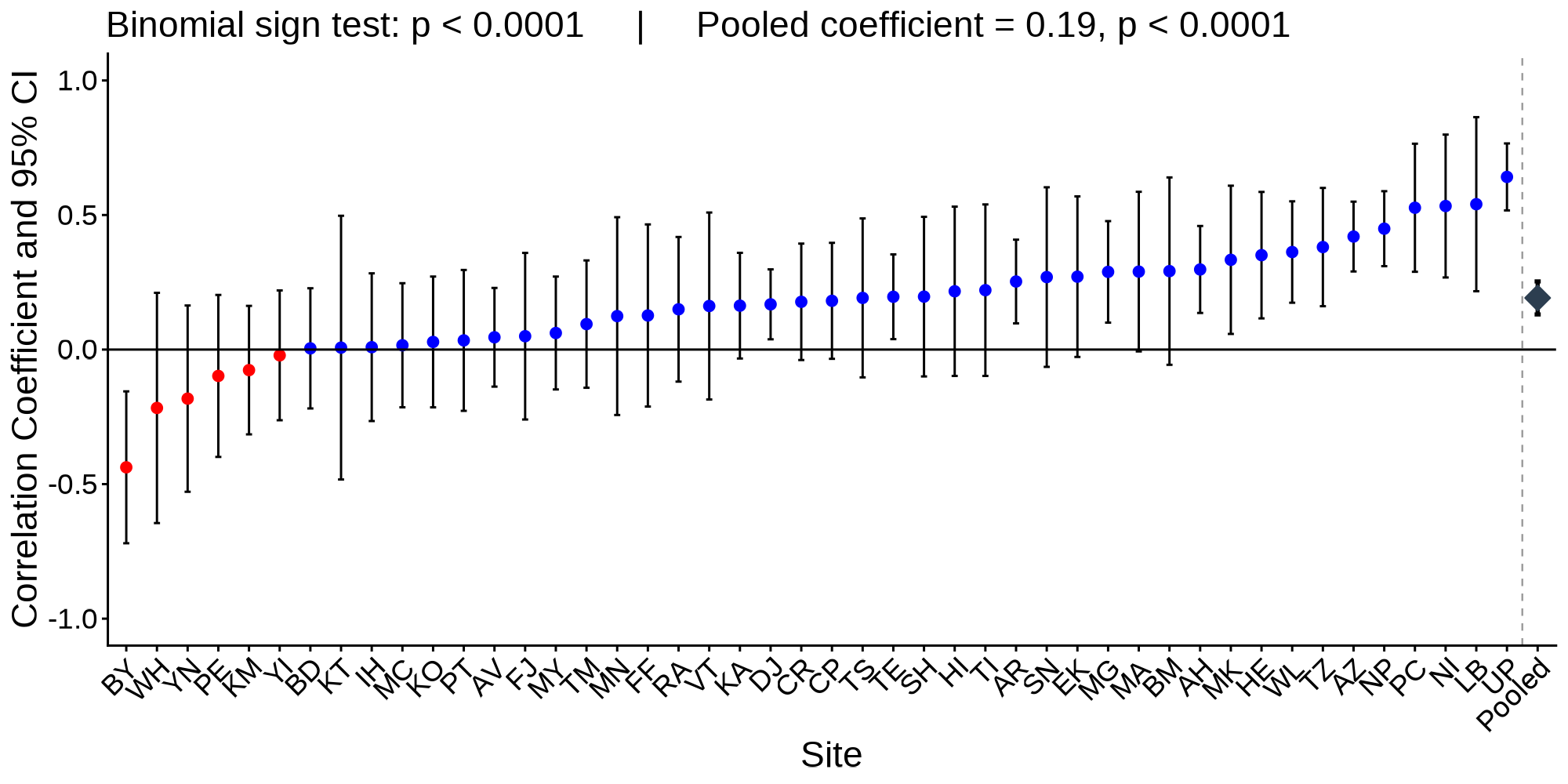}
    \caption{Average Alter Wealth vs wealth per capita.}
\end{subfigure}
\caption{\textbf{Correlations between Average Alter Wealth and material wealth metrics}. Average Alter Wealth is calculated using the ``proportional composite'' network (considering alters who provide support, i.e., supporters) and a wealth distribution using wealth (a) or wealth per capita (b). }
\label{fig:iec-wealth-corr-collapse-sum}
\end{figure}

\begin{figure}[t]
\centering
\begin{subfigure}{0.49\textwidth}
    \centering
    \includegraphics[width=\textwidth]{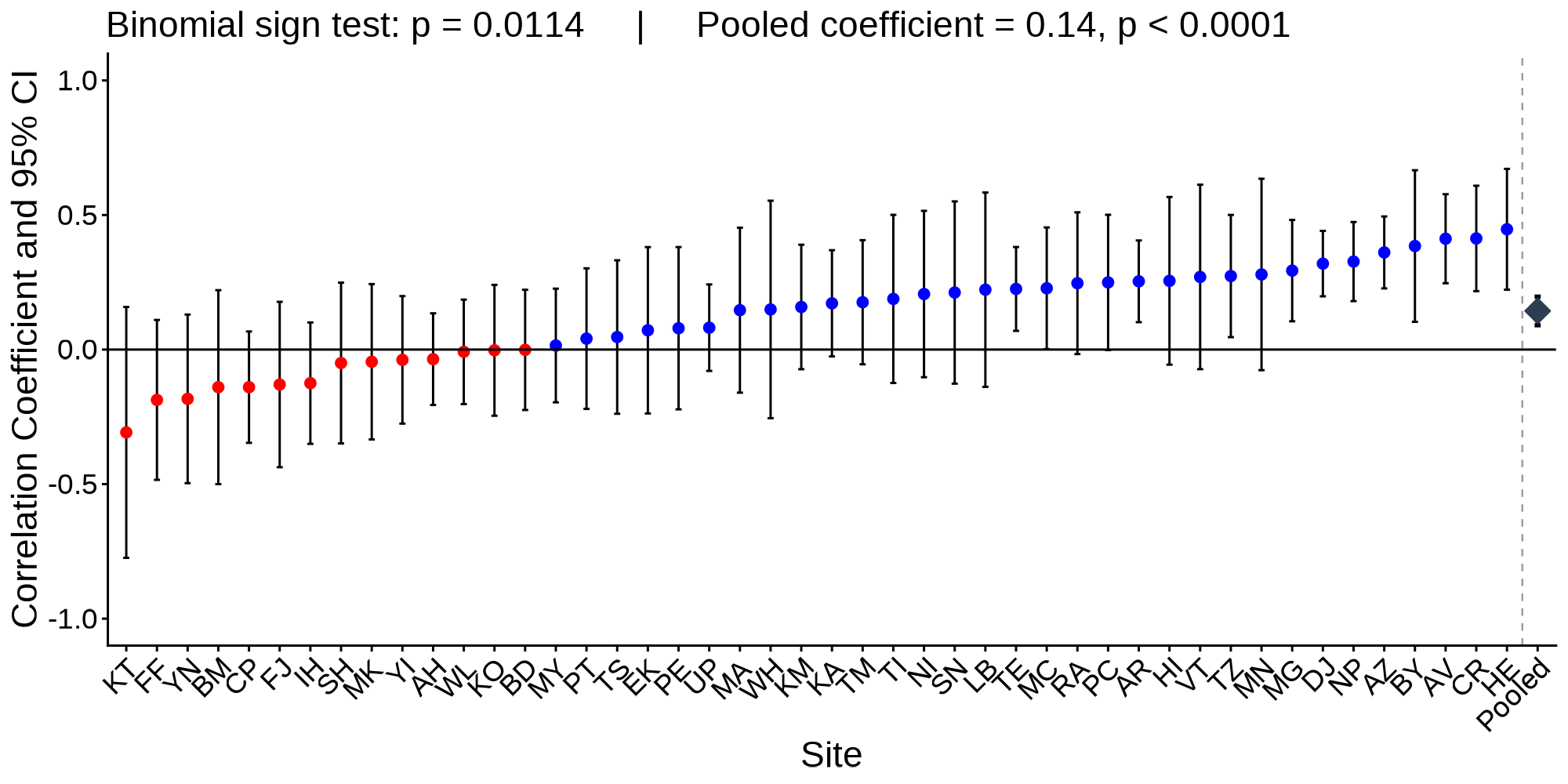}
    \caption{Support access vs wealth.}
\end{subfigure}
\vspace{1em} % Space between the top and bottom rows
\begin{subfigure}{0.49\textwidth}
    \centering
    \includegraphics[width=\textwidth]{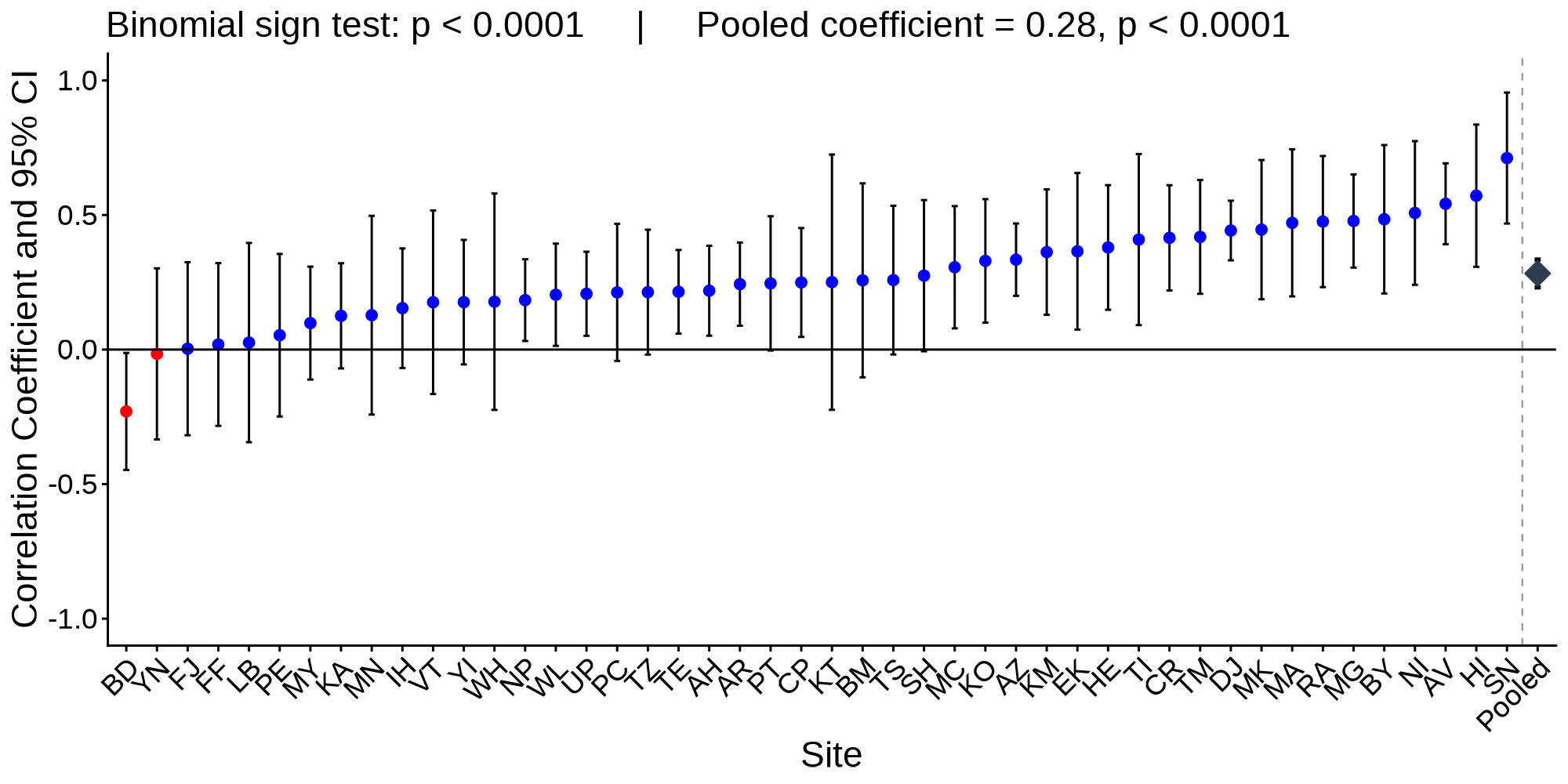}
    \caption{Support provisioning vs wealth.}
\end{subfigure}
\begin{subfigure}{0.49\textwidth}
    \centering
    \includegraphics[width=\textwidth]{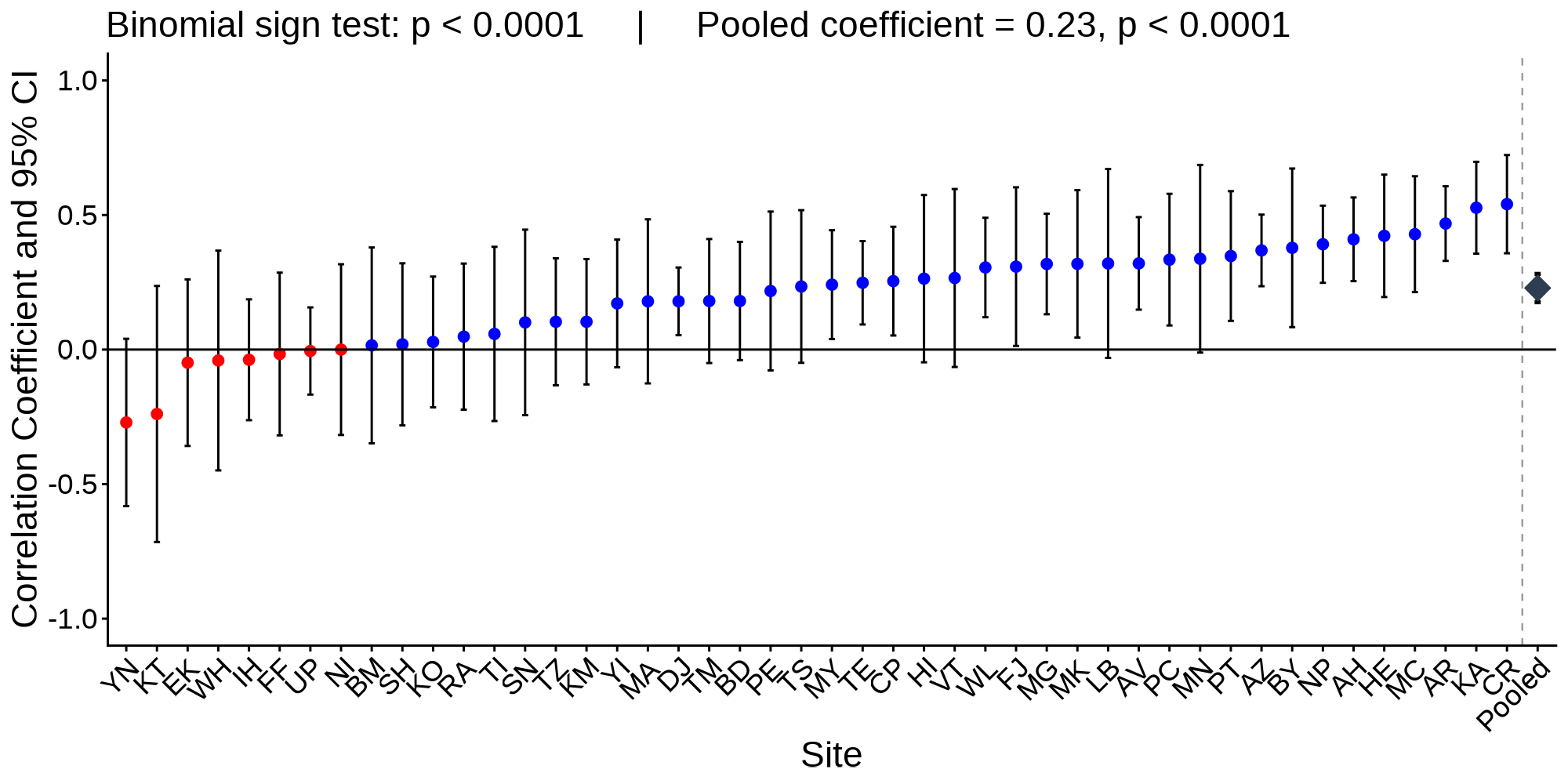}
    \caption{Support access p.c. vs wealth p.c.}
\end{subfigure}%
%\hspace{2em}
\begin{subfigure}{0.49\textwidth}
    \centering
    \includegraphics[width=\textwidth]{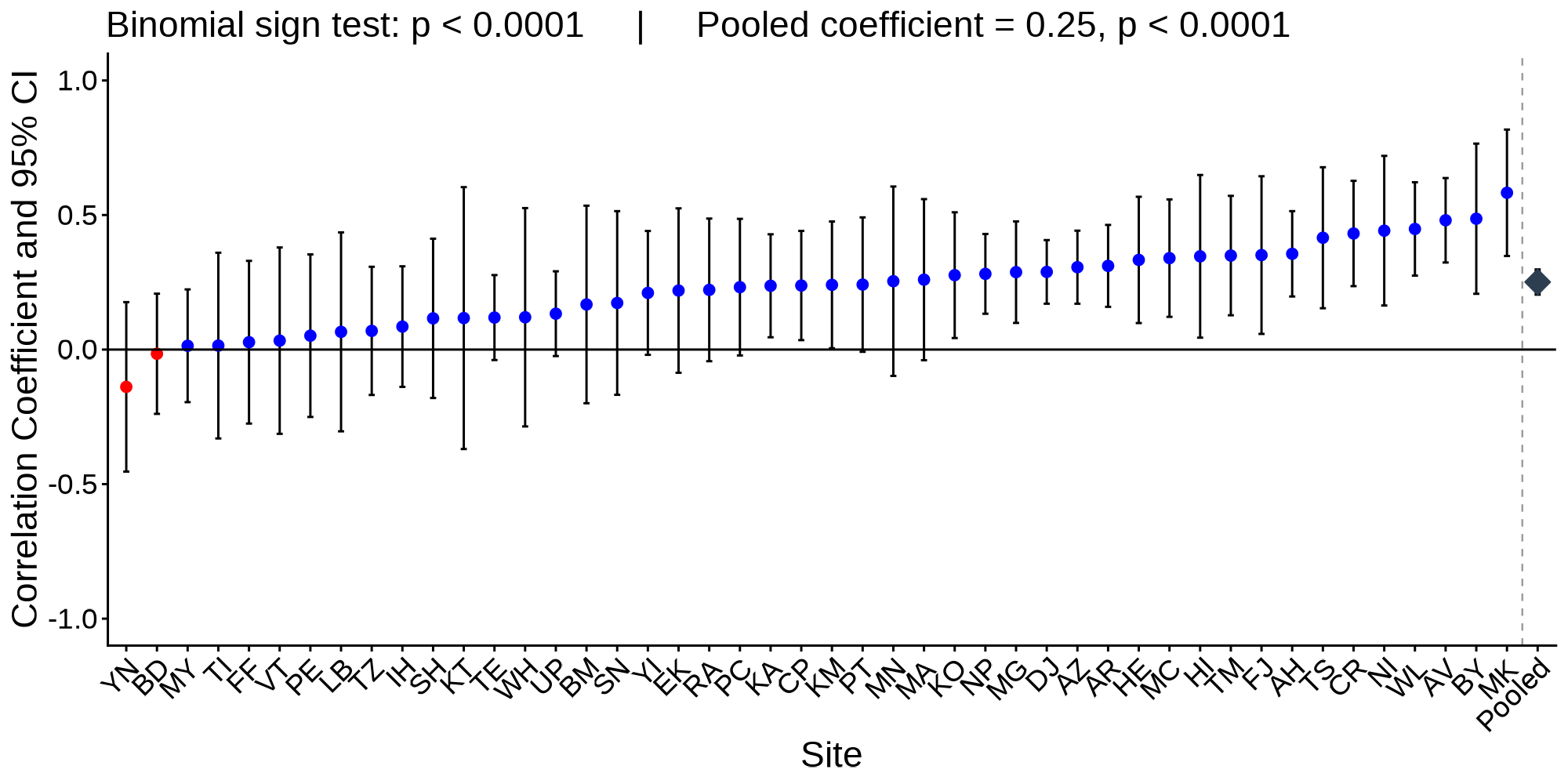}
    \caption{Support provisioning p.c. vs wealth p.c.}
\end{subfigure}
\caption{\textbf{Correlations between access and provisioning of support and material wealth metrics}. Access to support (i.e., out-degree) and provisioning of support (i.e., in-degree) are calculated using the ``raw composite'' network. }
\label{fig:within-site-multipanel-union}
\end{figure}

\begin{figure}[t]
\centering
\begin{subfigure}{0.49\textwidth}
    \centering
    \includegraphics[width=\textwidth]{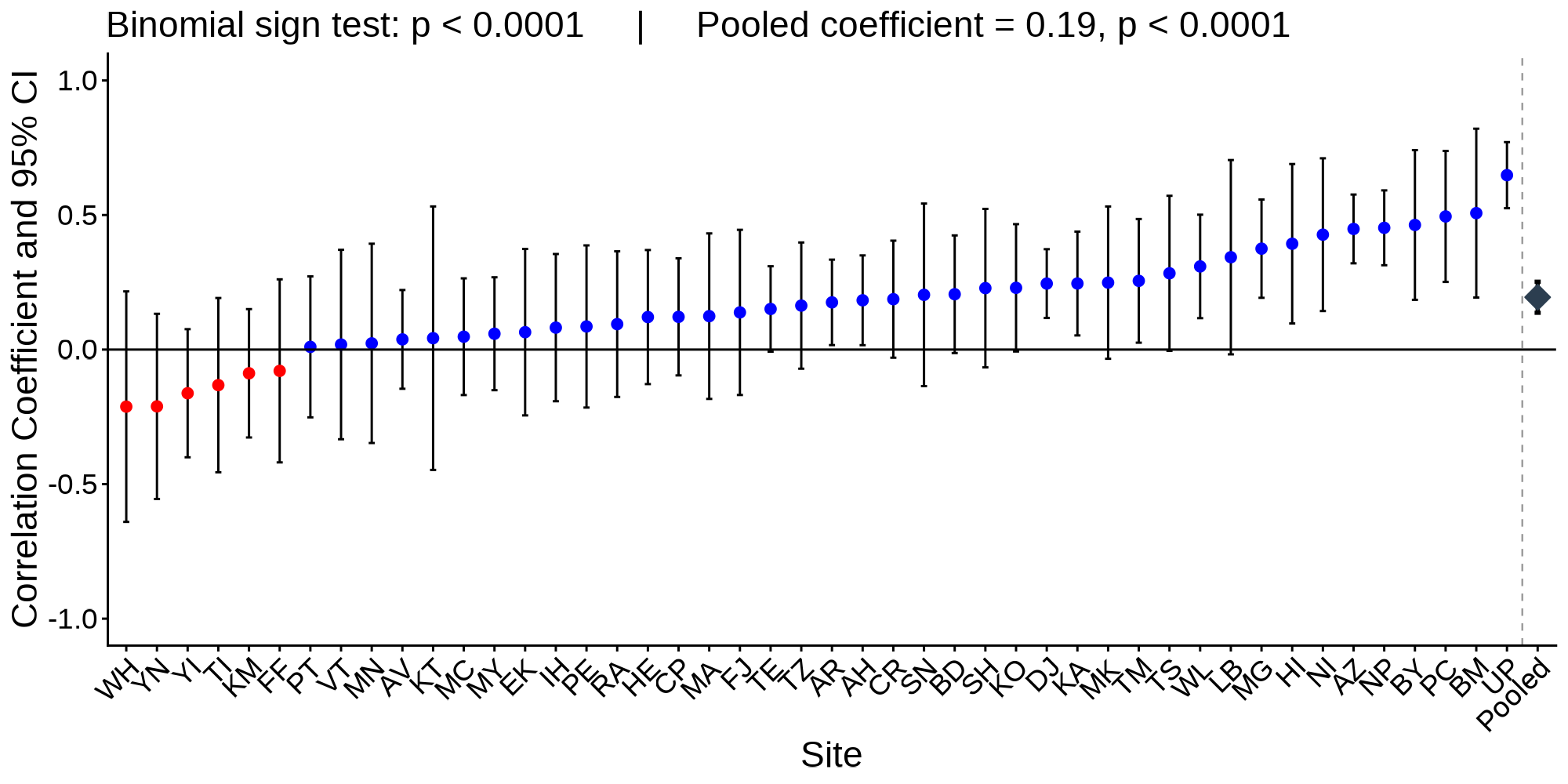}
    \caption{Average Alter Wealth vs wealth.}
\end{subfigure}
\begin{subfigure}{0.49\textwidth}
    \centering
    \includegraphics[width=\textwidth]{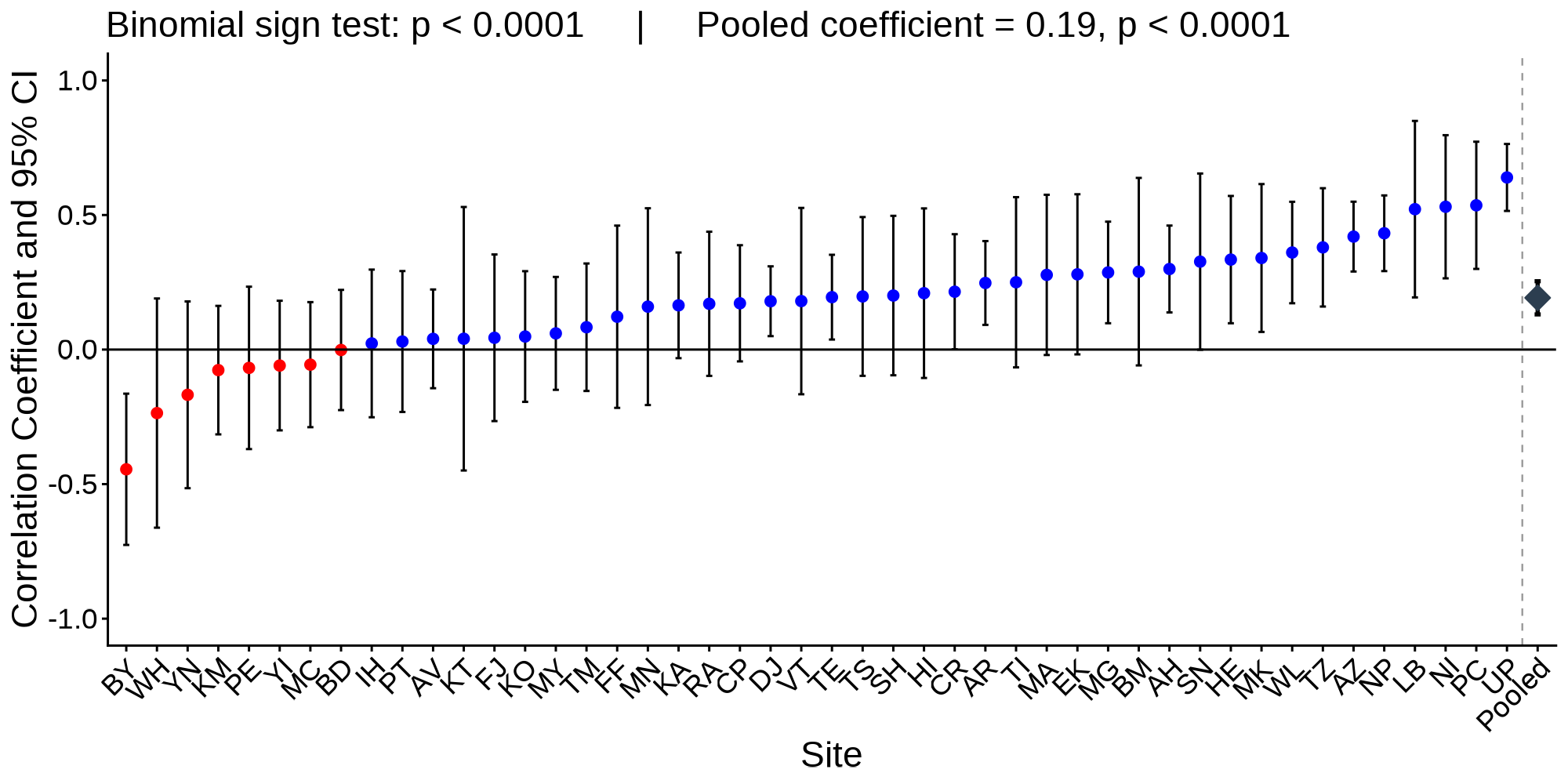}
    \caption{Average Alter Wealth vs wealth per capita.}
\end{subfigure}
\caption{\textbf{Correlations between Average Alter Wealth and material wealth metrics}. Average Alter Wealth is calculated using the ``raw composite'' network (considering alters who provide support, i.e., supporters) and a wealth distribution using wealth (a) or wealth per capita (b). }
\label{fig:iec-wealth-corr-union}
\end{figure}

Figures~\ref{fig:within-site-multipanel-collapsed-sum} and~\ref{fig:iec-wealth-corr-collapse-sum} show our results for the proportional composite network. Figures~\ref{fig:within-site-multipanel-union} and~\ref{fig:iec-wealth-corr-union} show our results for the raw composite network. We see in both cases that they are similar to our main results in Figure~\ref{fig:within-site-multipanel}.

Note that, unless stated otherwise in the figure caption, in the plots showing within-site correlations each (sub-)figure presents a set of correlation coefficients and 95\% confidence intervals, one for each site, for a pair of variables. Correlation coefficients are calculated as the slope coefficient (and the corresponding standard error) of a regression of the $y$-variable, ranked across sharing units in a site, on the $x$-variable ranked across sharing units in a site. Bars denote 95\% confidence intervals. We display the two-sided $p$-value in each (sub-)figure of observing so many positive correlations under the sharp null that the true correlation in each site is zero, as well as a pooled coefficient (and associated p-value) resulting from a random-effects meta-analysis. (The pooled coefficient is described further in Section~\ref{supp:pooled_coefs}).

\clearpage
\subsection{Different Accountings for Sharing Unit Composition}\label{supp:alternative_SUweightings_within}

In our main analyses we give primacy to per capita adjustments to the total material wealth of sharing units, in order to correct for sharing unit size. We do this to be consistent with most comparable studies conducted by economists. We do, however, note that there are contexts, particularly those studied by anthropologists, where this correction might be inappropriate, which is why we also include results here using total sharing unit wealth.

First, in many societies where economic and social security are tied to kinship and labor rather than money, material wealth is tightly linked to household size. The members of a sharing unit can then be seen as being inextricably part of its collective ``endowment'' \citep{baulch_economic_2000}. The concept of ``wealth in people'' was initially developed for the West African context, where the expansion of households through marriage and fostering provides clear economic benefits \citep{goody2005contexts, guyer_wealth_1993}; the concept gained much traction throughout subsequent anthropological, historical, and sociological literature, not only in Africa but other parts of the world where economic well-being and security are strongly linked to labor. For example, in southeast Asian peasant societies large family units ensure resilience against material losses caused by economic shocks \citep{scott1977moral}, among urban poor in Mexico large households provide mutual support and resource-sharing across different domains of the informal economy \citep{lomnitz2014networks}, and in many contemporary contexts where there is little state welfare provision large households enable migration strategies where some members work abroad and send remittances home.

Second, sharing unit assets will consist of items from which all sharing unit members benefit (the house structure, a plow, a grinding stone), items that entail some degree of rivality (livestock, acres under cultivation), and items that accumulate with the number of sharing unit members (potentially clothing, jewelry, mobile phones, vehicles). On the assumption that items with some sort of rivality predominate, analysts typically correct measures of household wealth for household size, calculating various indices of household wealth per individual. Recognizing the heterogeneity of our sample, and the fact that our populations will differ significantly in the proportion of non-rival goods, rival goods, and private goods held within sharing units, we opted to make the most minimalist assumptions by simply using the summed value.

However, we recognize that sharing unit size and composition is an important point of variation both within and across communities, and therefore consider different ways of accounting for sharing unit composition, to see how they consequently shape the conclusions. The concern is that the relationship between sharing unit wealth and sharing unit connections could plausibly be driven by a mechanical relationship, wherein sharing units with more (productive) members are both wealthier \citep{netting1982some} and also have more supportive relationships. This would be a particular concern for our provisioning and accessing of support (in- and out-degree) measures, as they are focused on the number of supportive relationships, moreso than for Average Alter Wealth.

\begin{figure}
    \centering
    \begin{subfigure}{0.65\textwidth}
        \centering
        \includegraphics[width=\textwidth]{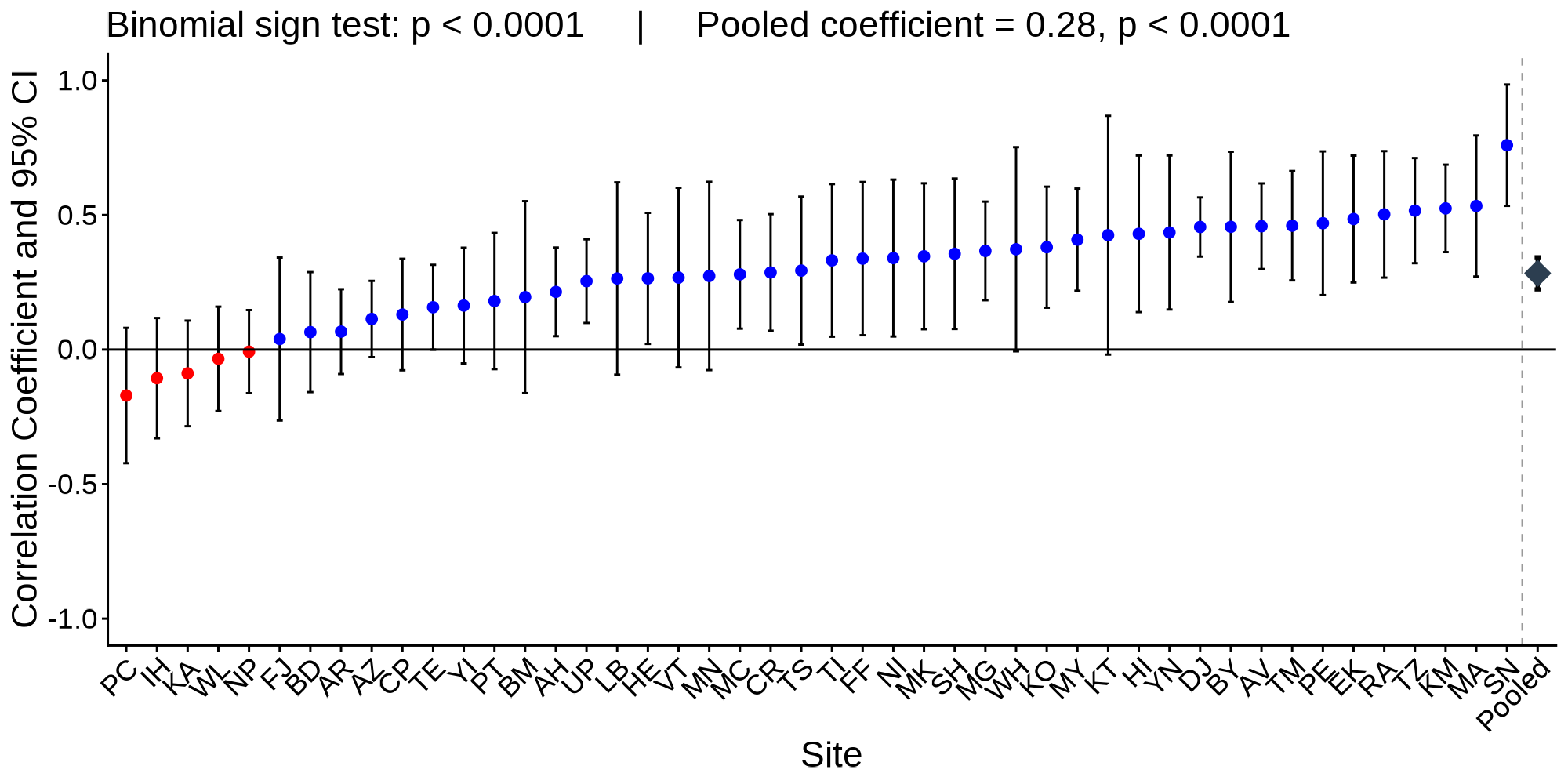}
        \caption{Unit wealth vs size.}
    \end{subfigure}
        \begin{subfigure}{0.65\textwidth}
        \centering
        \includegraphics[width=\textwidth]{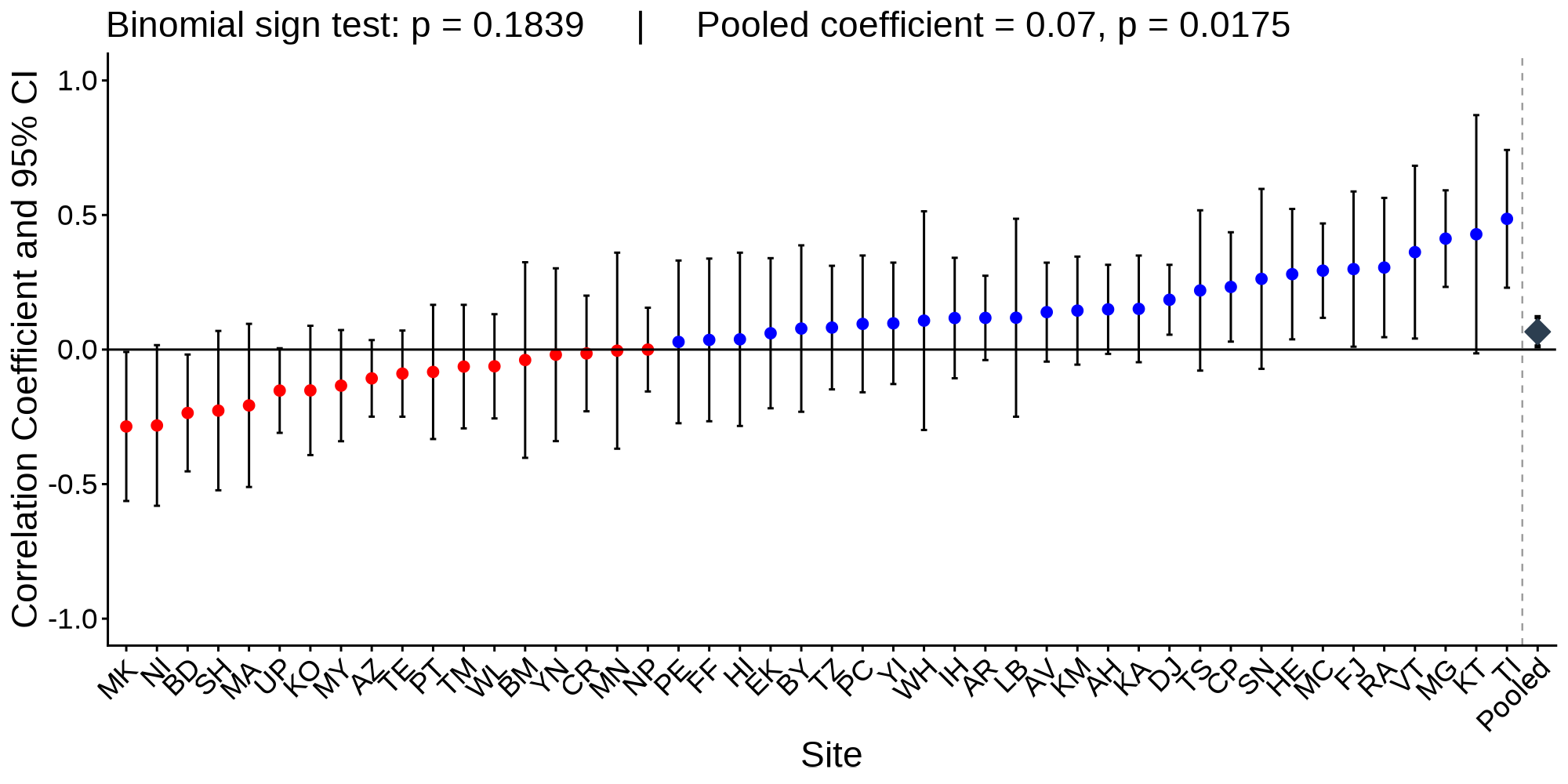}
        \caption{Unit support access (out-degree) vs size.}
    \end{subfigure}
    \begin{subfigure}{0.65\textwidth}
        \centering
        \includegraphics[width=\textwidth]{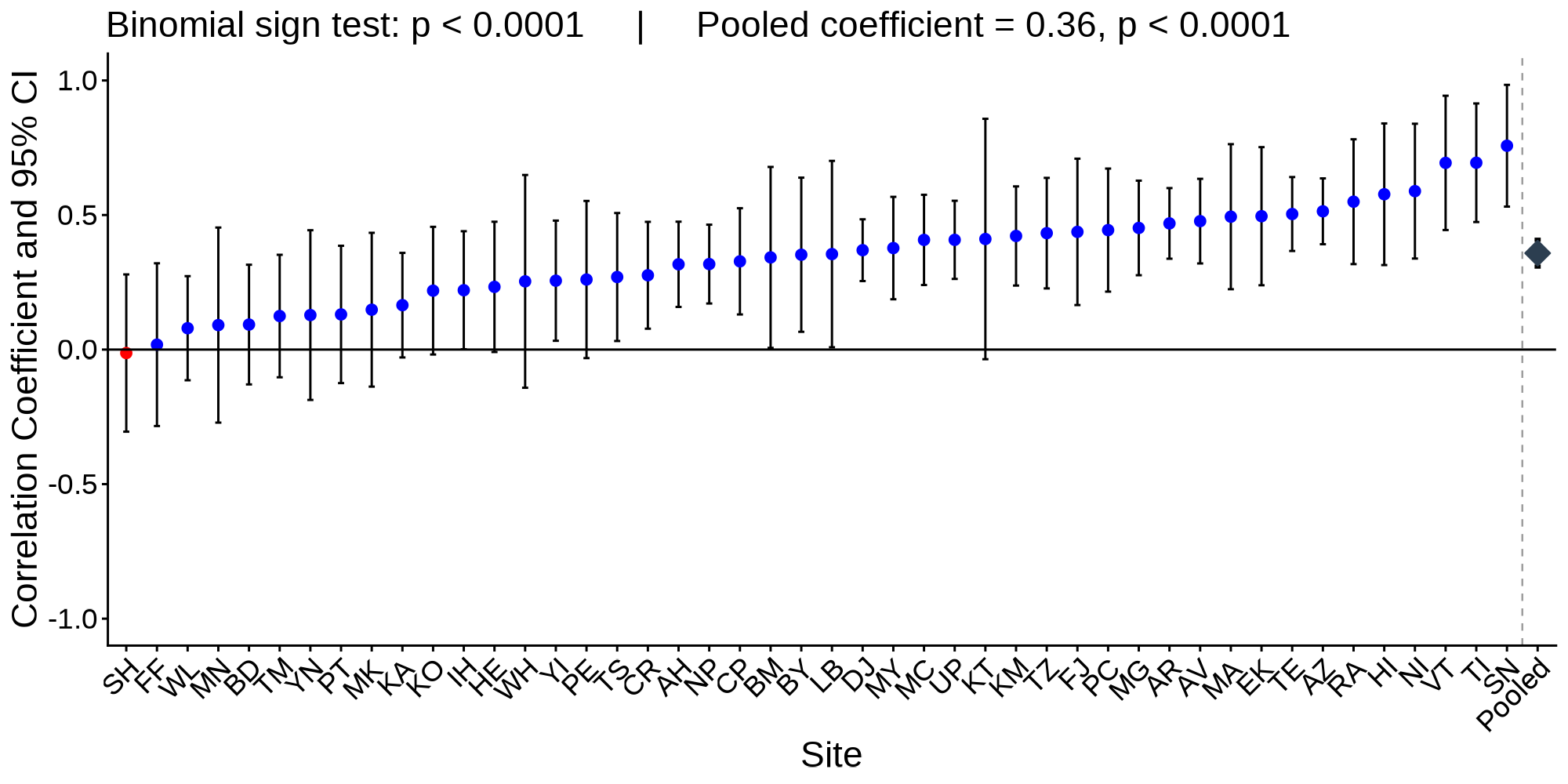}
        \caption{Unit support provisioning (in-degree) vs size.}
    \end{subfigure}
    \vspace{1em}
    \caption{\textbf{Correlations between sharing unit size and wealth, support access, and support provisioning.} Access to support (i.e., out-degree) and provisioning of support (i.e., in-degree) are calculated using the ``composite'' network.}
    \label{fig:sitewise-size}
\end{figure}

In Figure~\ref{fig:sitewise-size} we show the correlations between sharing unit size, wealth, provisioning of support (i.e., in-degree) and access to support (i.e., out-degree).

\begin{figure}[t]
\centering
\begin{subfigure}{0.49\textwidth}
    \centering
    \includegraphics[width=\textwidth]{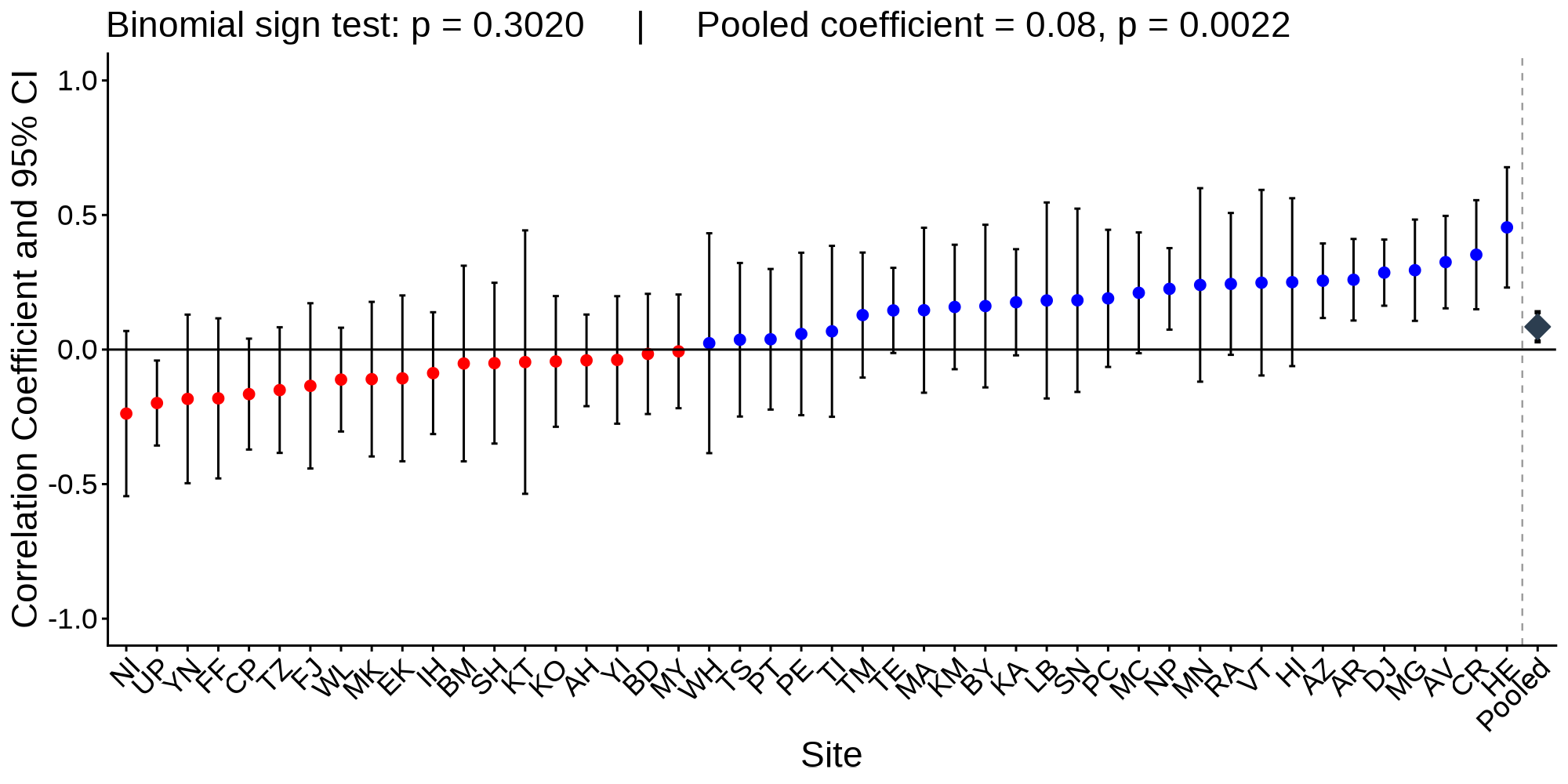}
    \caption{Support access vs wealth}
\end{subfigure}
\vspace{1em}
\begin{subfigure}{0.49\textwidth}
    \centering
    \includegraphics[width=\textwidth]{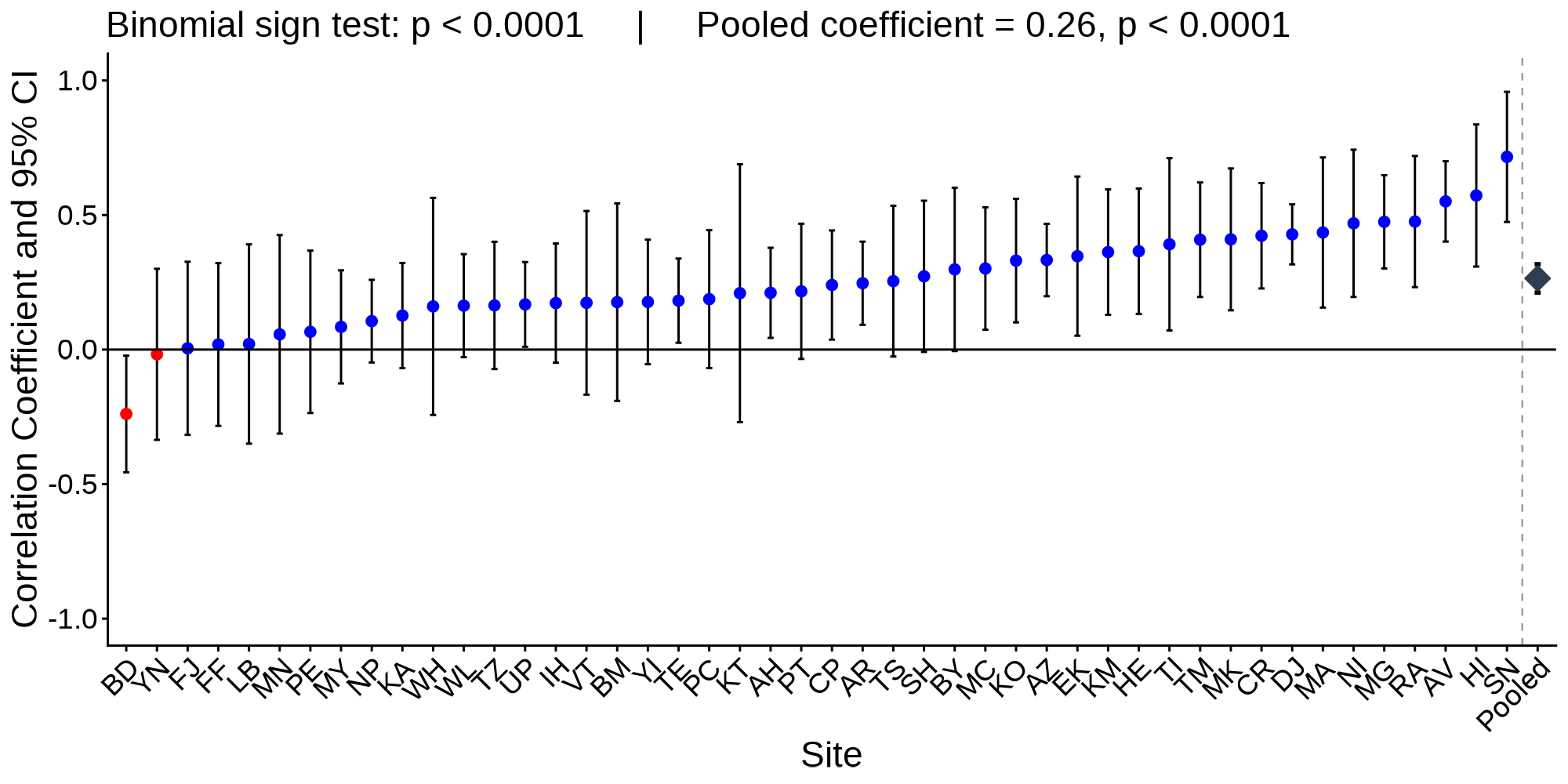}
    \caption{Support provisioning vs wealth}
\end{subfigure}
\vspace{2em}
\begin{subfigure}{0.49\textwidth}
        \centering
        \includegraphics[width=\textwidth]{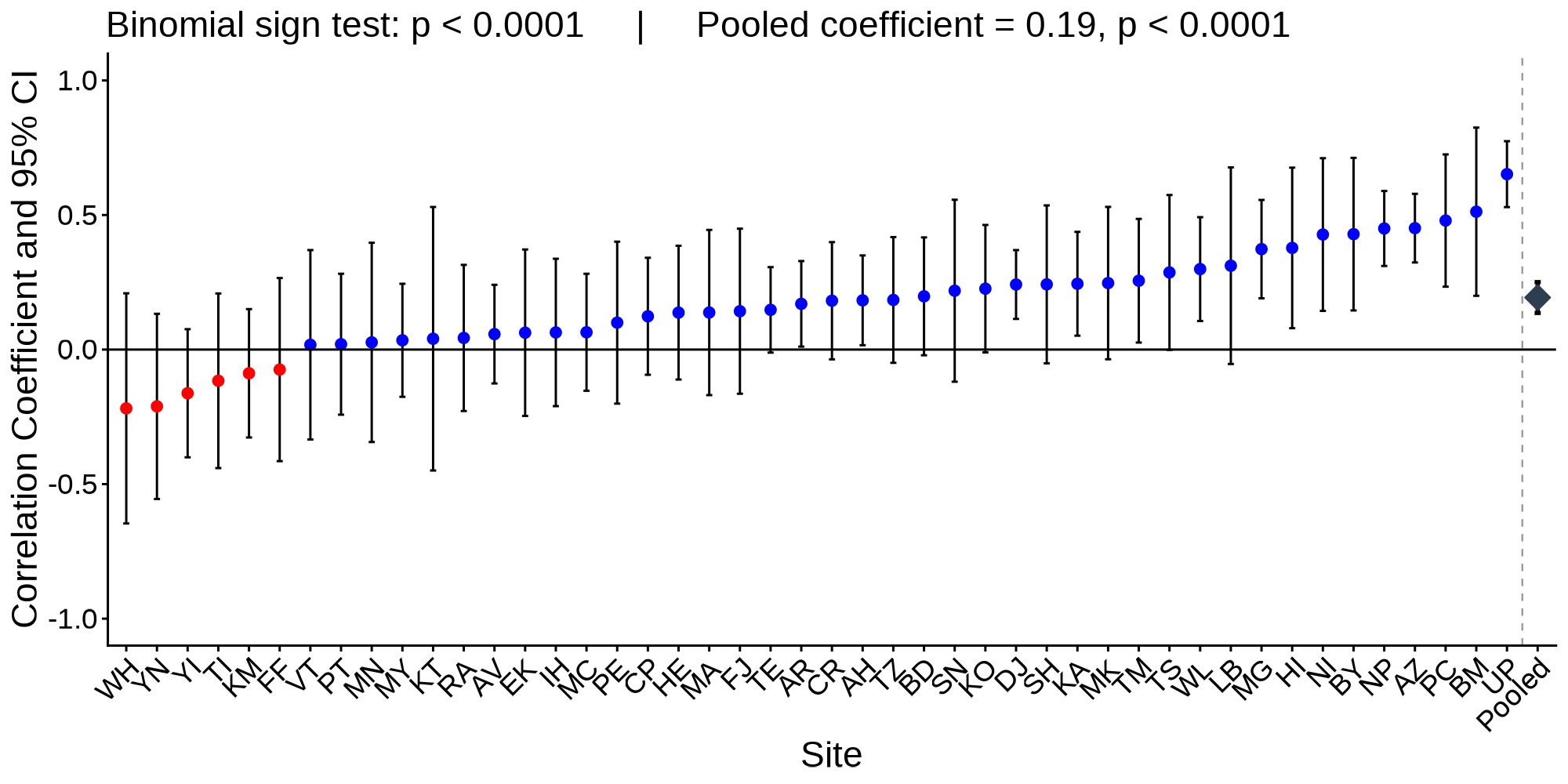}
        \caption{$\text{AAW}_i$ (of supporters) vs wealth}
    \end{subfigure}
    %\hspace{2em}
    \begin{subfigure}{0.49\textwidth}
        \centering
        \includegraphics[width=\textwidth]{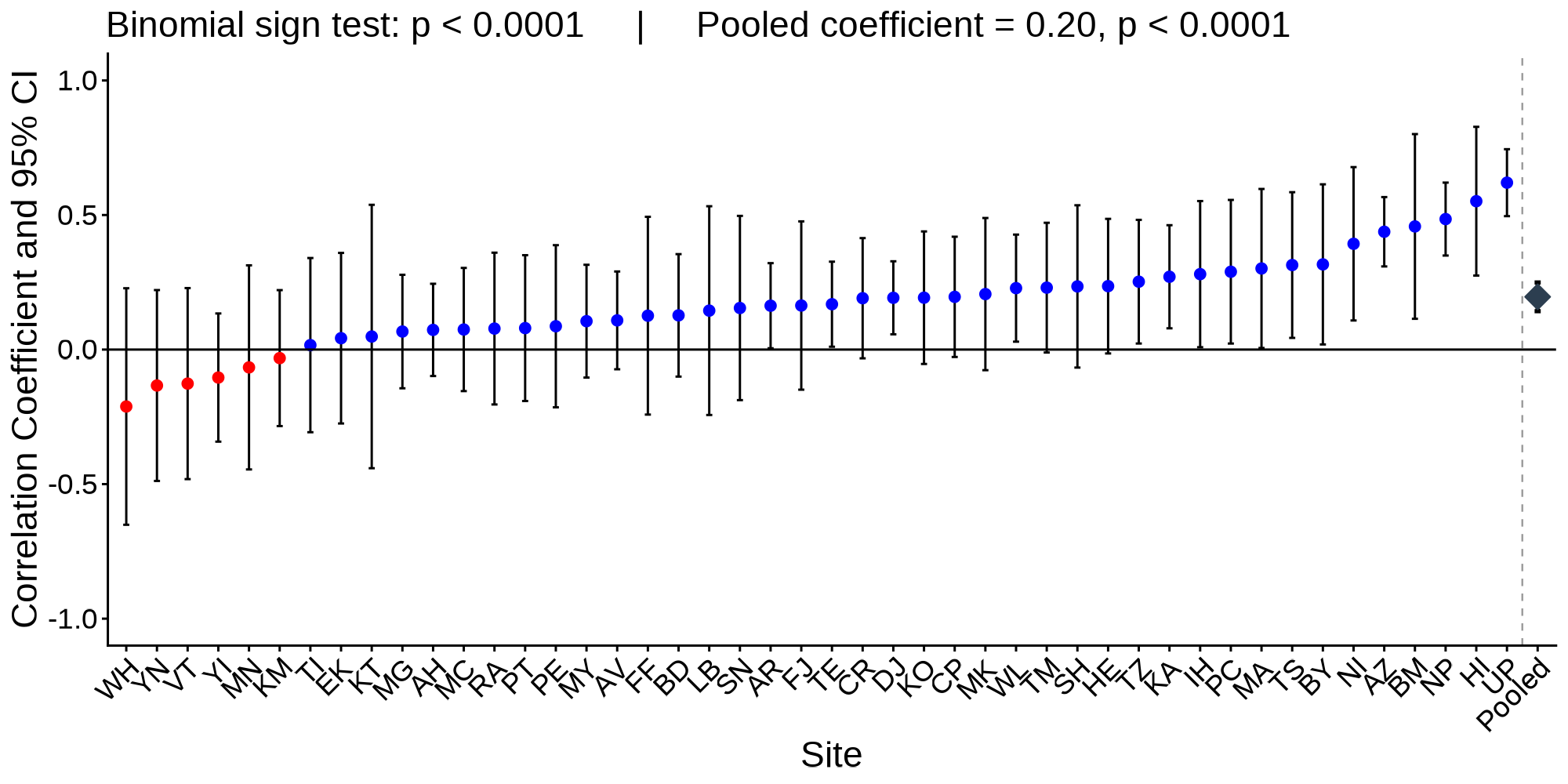}
        \caption{$\text{AAW}_i$ (of supportees) vs wealth}
    \end{subfigure}
\caption{\textbf{Correlations between network connections and material wealth, using absolute measures}. Measures are defined and calculated as described in Figure~\ref{fig:within-site-multipanel}, but using absolute instead of per capita measures. } %Supplementary Figure~\ref{fig:within-site-multipanel-adults} shows that our findings in panels (c) and (d) are similar if we use per adult as opposed to per capita measures.}
\label{fig:within-site-multipanel-absolute}
\end{figure}

Figure~\ref{fig:within-site-multipanel-absolute} presents equivalent results to what is shown in Figure~\ref{fig:within-site-multipanel} in the main text, without any weighting by sharing unit size (i.e., no per capita weighting on our measures of material wealth, or our measures of support access or provisioning (i.e., in-/out-degree) or Average Alter Wealth). The results are broadly comparable, though somewhat weaker for the relationship between out-degree and wealth.

\begin{figure}[p]
\centering
\begin{subfigure}{0.49\textwidth}
    \centering
    \includegraphics[width=\textwidth]{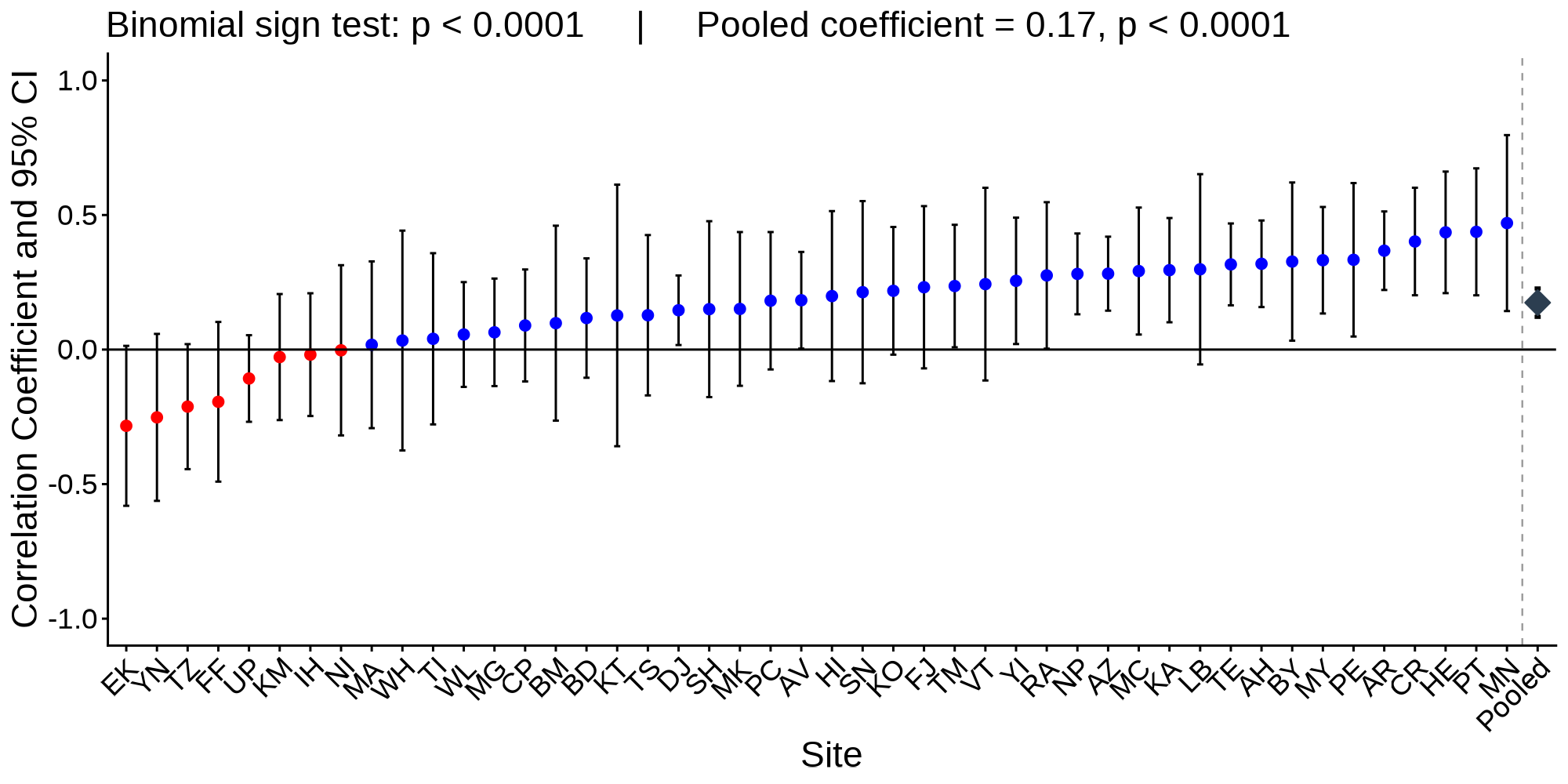}
    \caption{Support access per adult vs wealth per adult.}
\end{subfigure}
\vspace{1em}
\begin{subfigure}{0.49\textwidth}
    \centering
    \includegraphics[width=\textwidth]{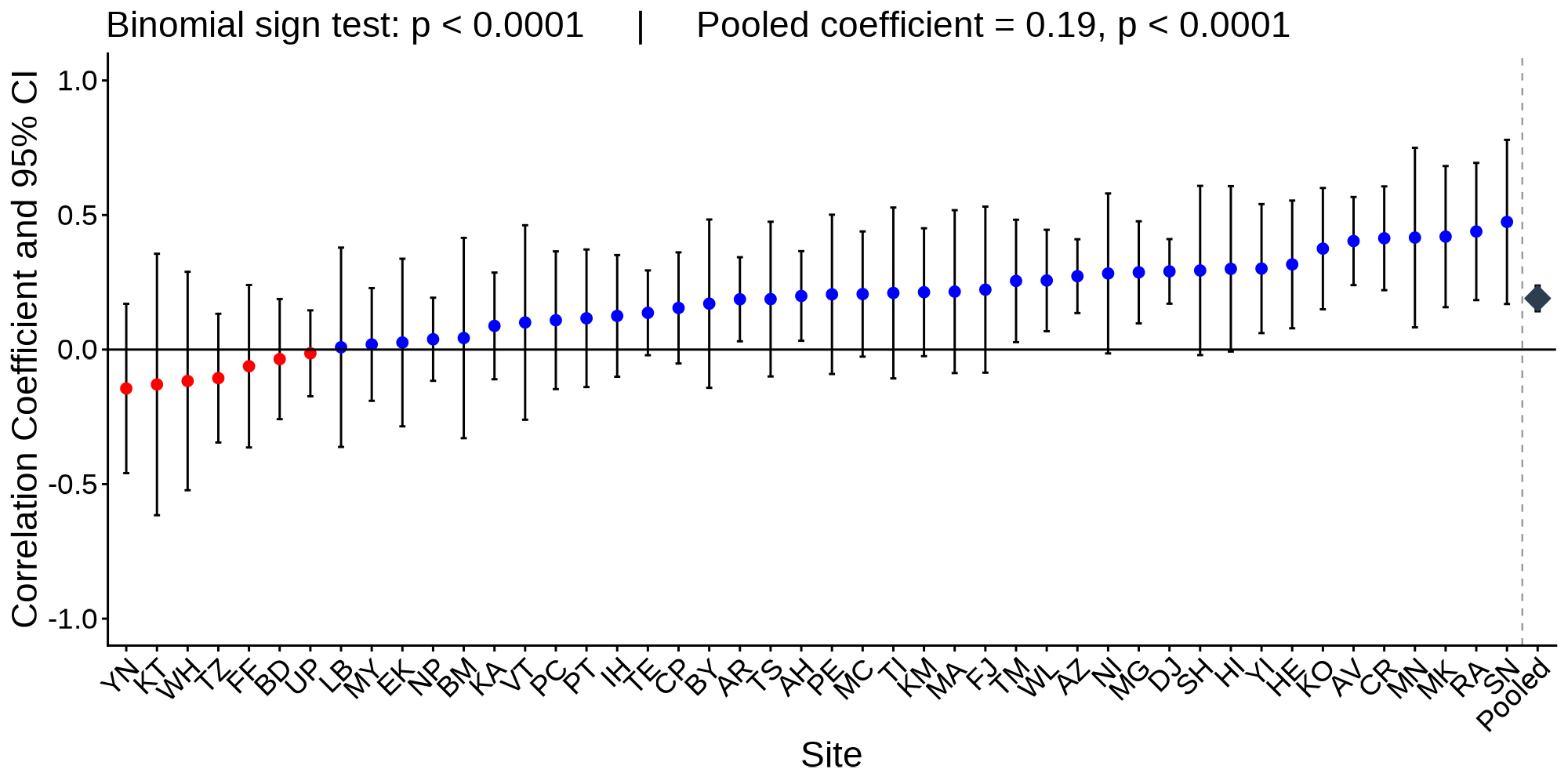}
    \caption{Support provisioning per adult vs wealth per adult.}
\end{subfigure}
\begin{subfigure}{0.49\textwidth}
    \centering
    \includegraphics[width=\textwidth]{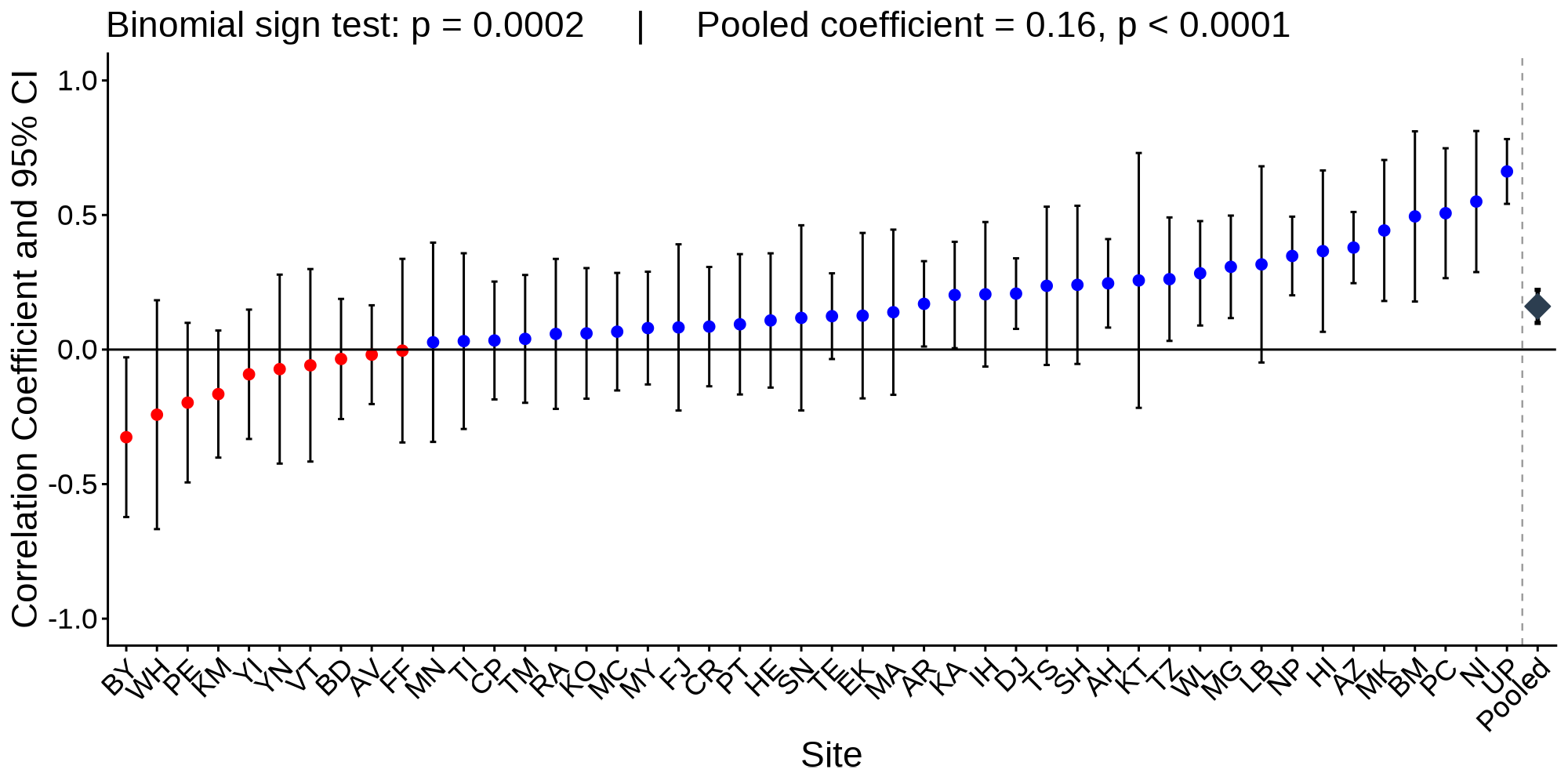}
    \caption{$\text{AAW}_i$ per adult (of supporters) vs wealth per adult}
\end{subfigure}
\begin{subfigure}{0.49\textwidth}
    \centering
    \includegraphics[width=\textwidth]{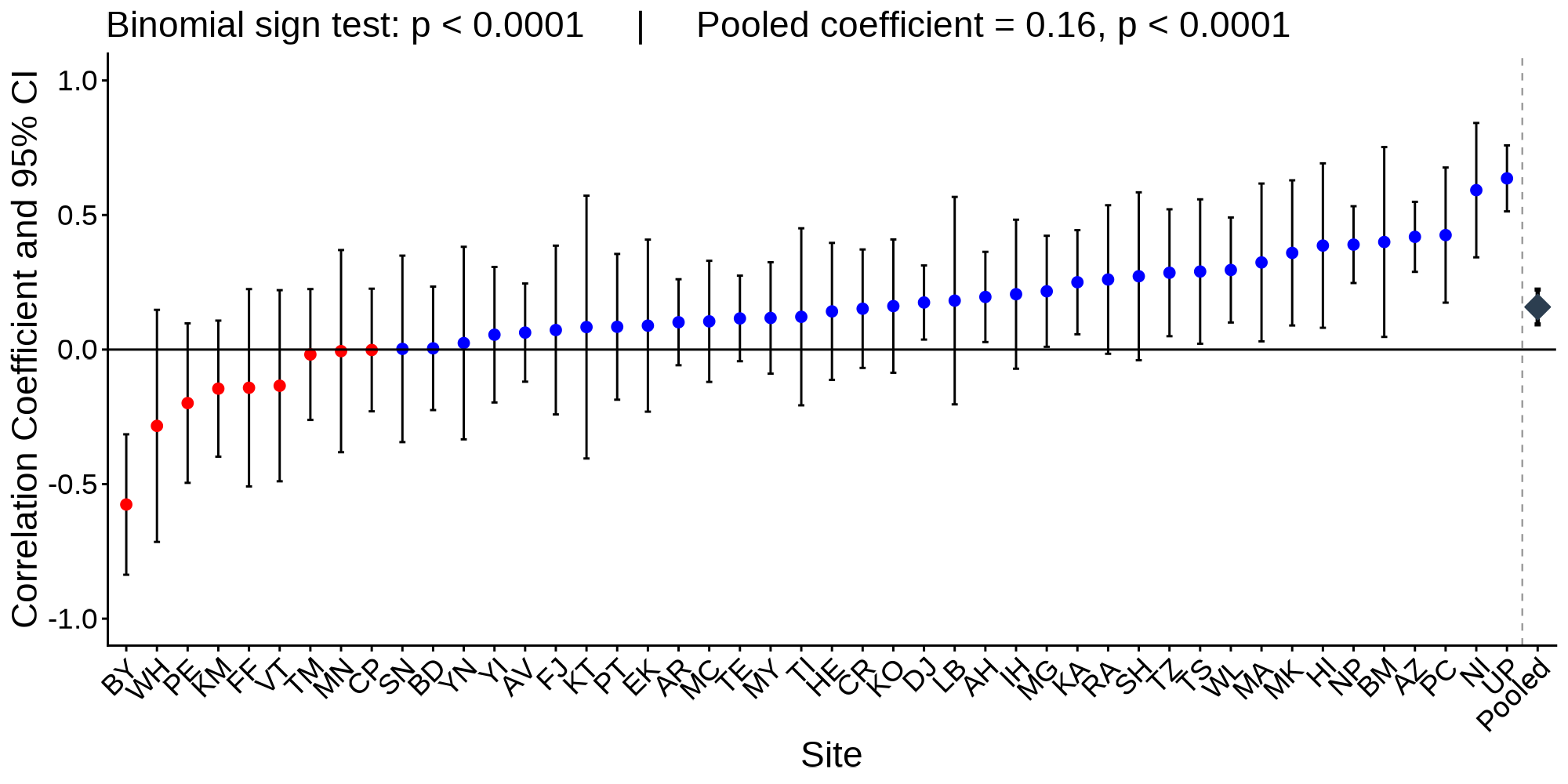}
    \caption{$\text{AAW}_i$ per adult (of supportees) vs wealth per adult}
\end{subfigure}
\caption{\textbf{Correlations between network connections and material wealth, using per adult weights}. Measures are defined and calculated as described in Figure~\ref{fig:within-site-multipanel}, but using per adult instead of per capita measures. }
\label{fig:within-site-multipanel-adults}
\end{figure}

Figure~\ref{fig:within-site-multipanel-adults} again presents the equivalent results to what is shown in Figure~\ref{fig:within-site-multipanel} in the main text, now with measures being weighted per adult rather than per capita (where we generally define ``adults'' as those age 18 or older). The broad pattern remains similar to the main text.

\begin{figure}[p]
\centering
\begin{subfigure}{0.49\textwidth}
    \centering
    \includegraphics[width=\textwidth]{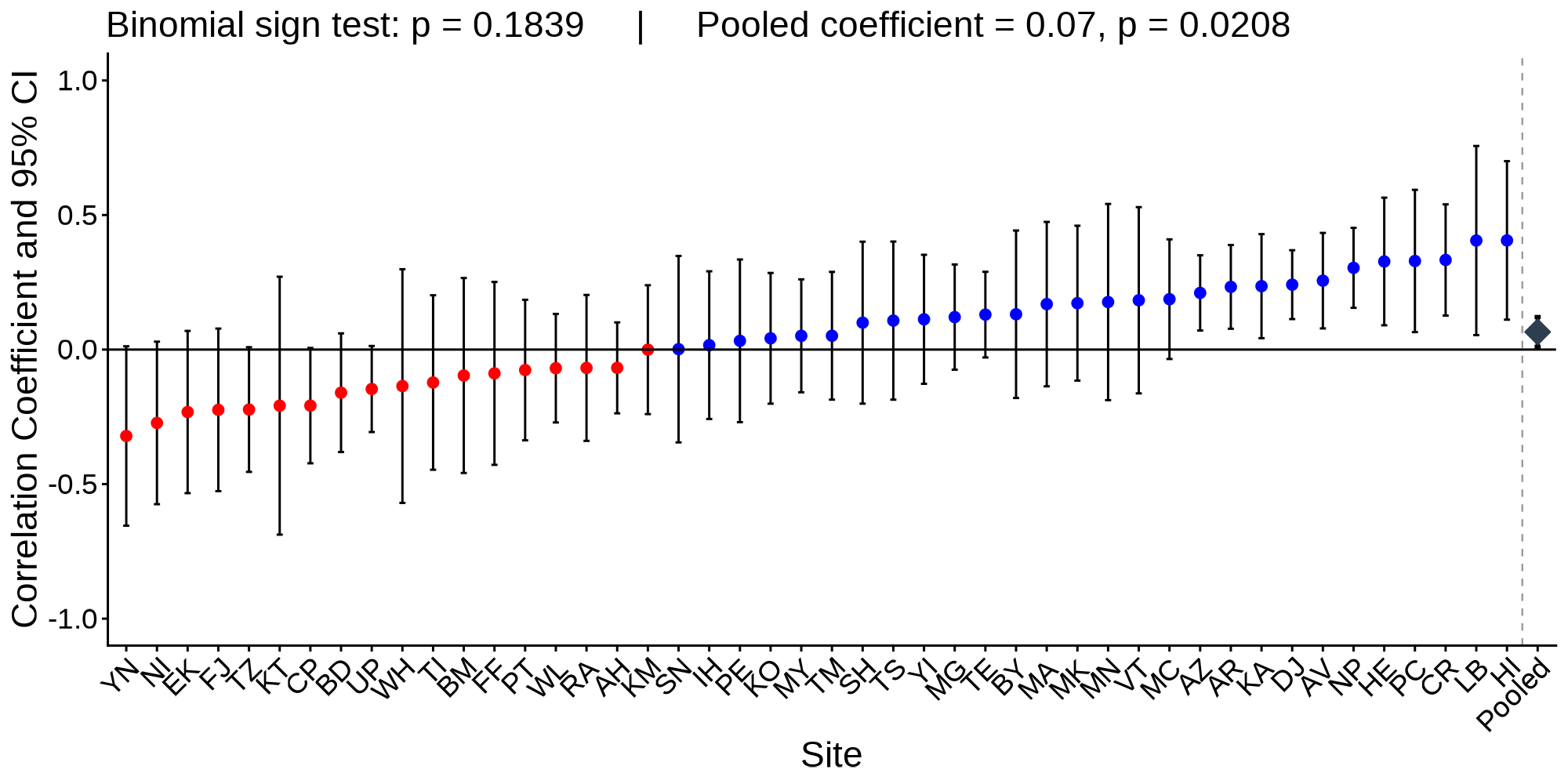}
    \caption{Support access vs wealth.}
\end{subfigure}%
\begin{subfigure}{0.49\textwidth}
    \centering
    \includegraphics[width=\textwidth]{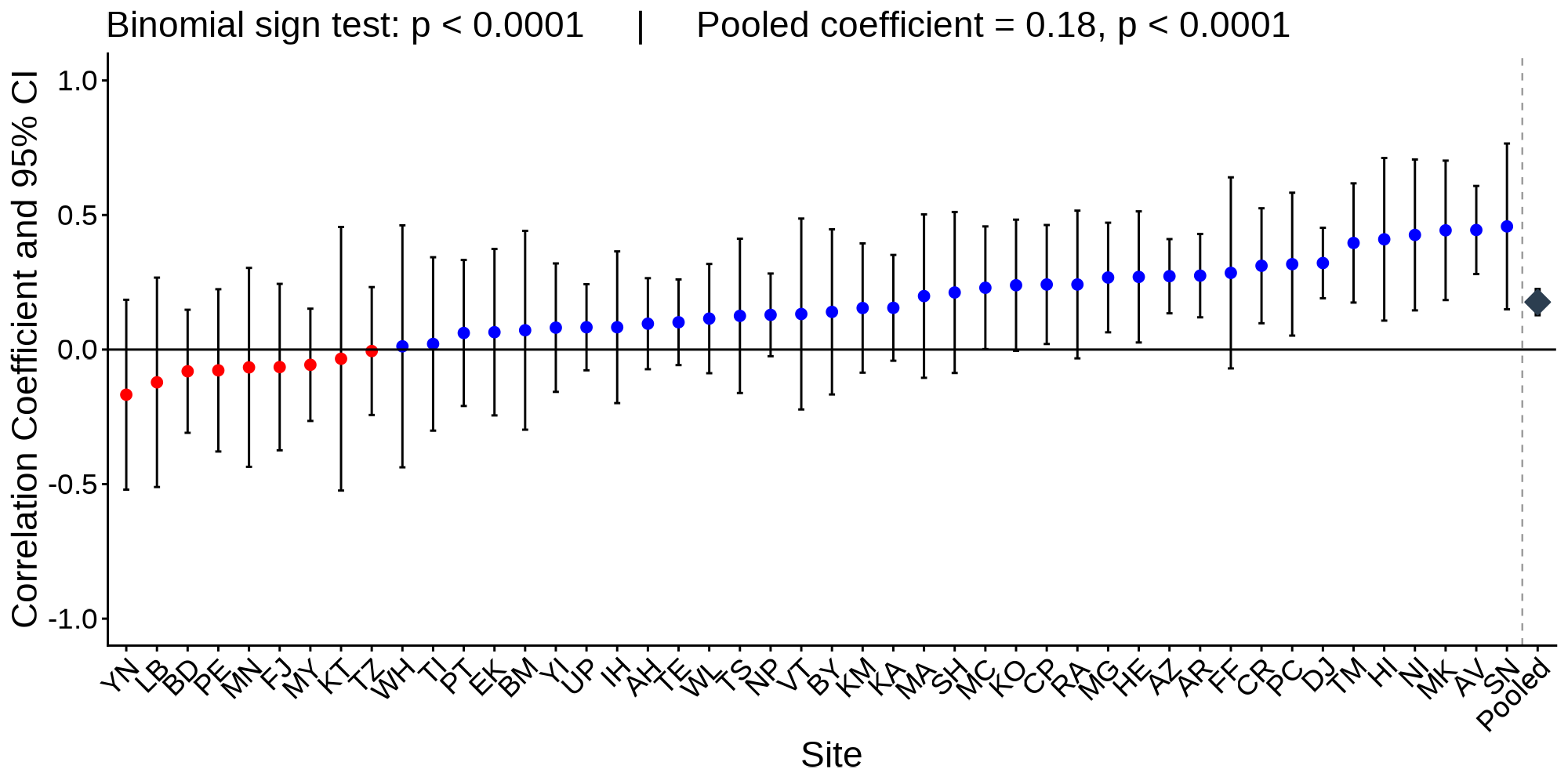}
    \caption{Support provisioning vs wealth.}
\end{subfigure}
\caption{\textbf{Correlations between access and provisioning of support and material wealth metrics, controlling for sharing unit size}. Support access (i.e., out-degree) and support provisioning (i.e., in-degree) are calculated using the ``composite'' network. Correlation coefficients, and their standard errors, are calculated as the slope coefficient (and the corresponding standard error) of a regression of the residualized $y$-variable, ranked across sharing units in a community, on the residualized $x$-variable ranked across sharing units in a community. Variables are residualized on log(sharing unit size). }
\label{fig:within-site-size-control}
\end{figure}

\begin{figure}[p]
\centering
\begin{subfigure}{0.49\textwidth}
    \centering
    \includegraphics[width=\textwidth]{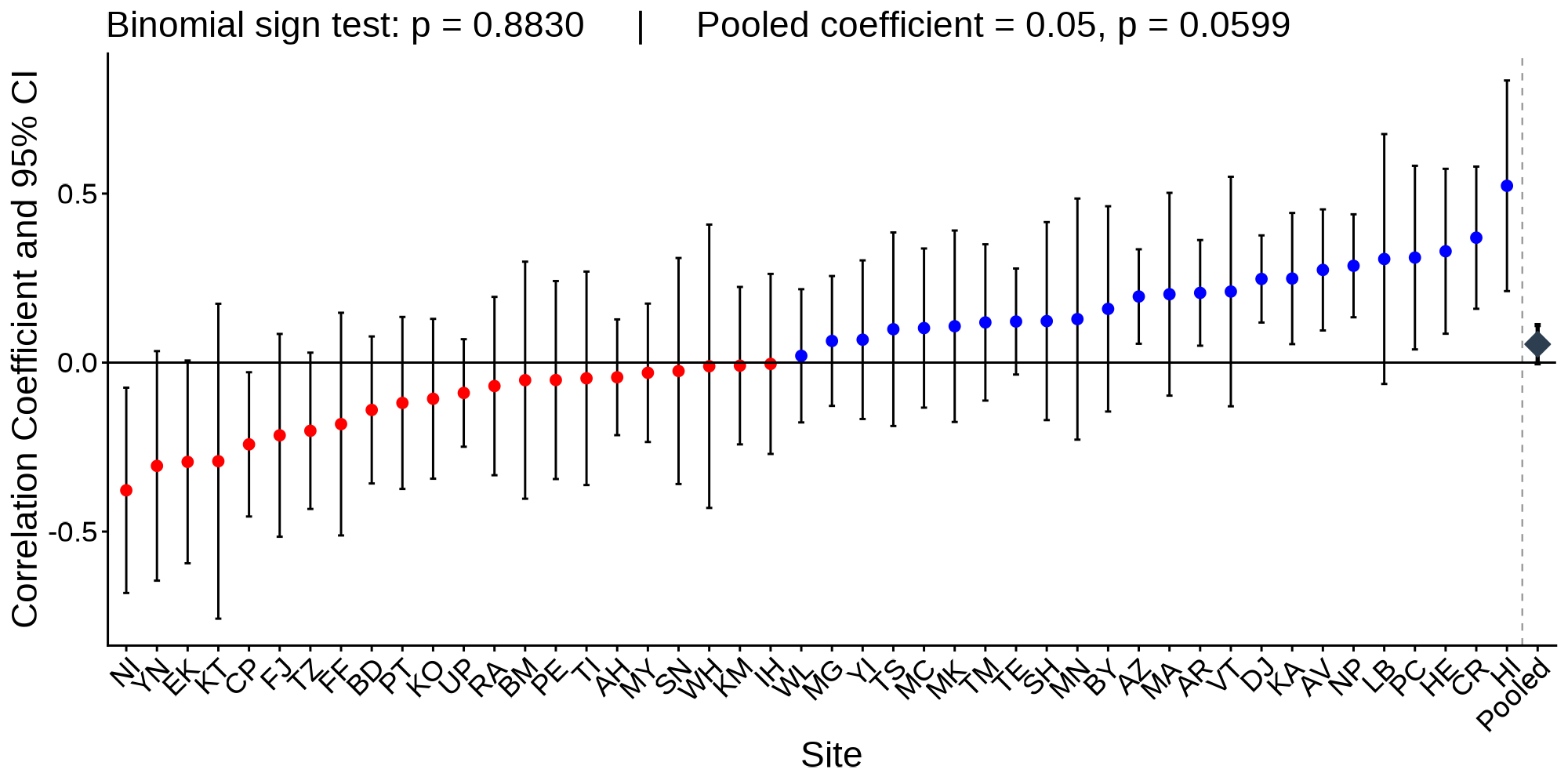}
    \caption{Support access vs wealth.}
\end{subfigure}%
\begin{subfigure}{0.49\textwidth}
    \centering
    \includegraphics[width=\textwidth]{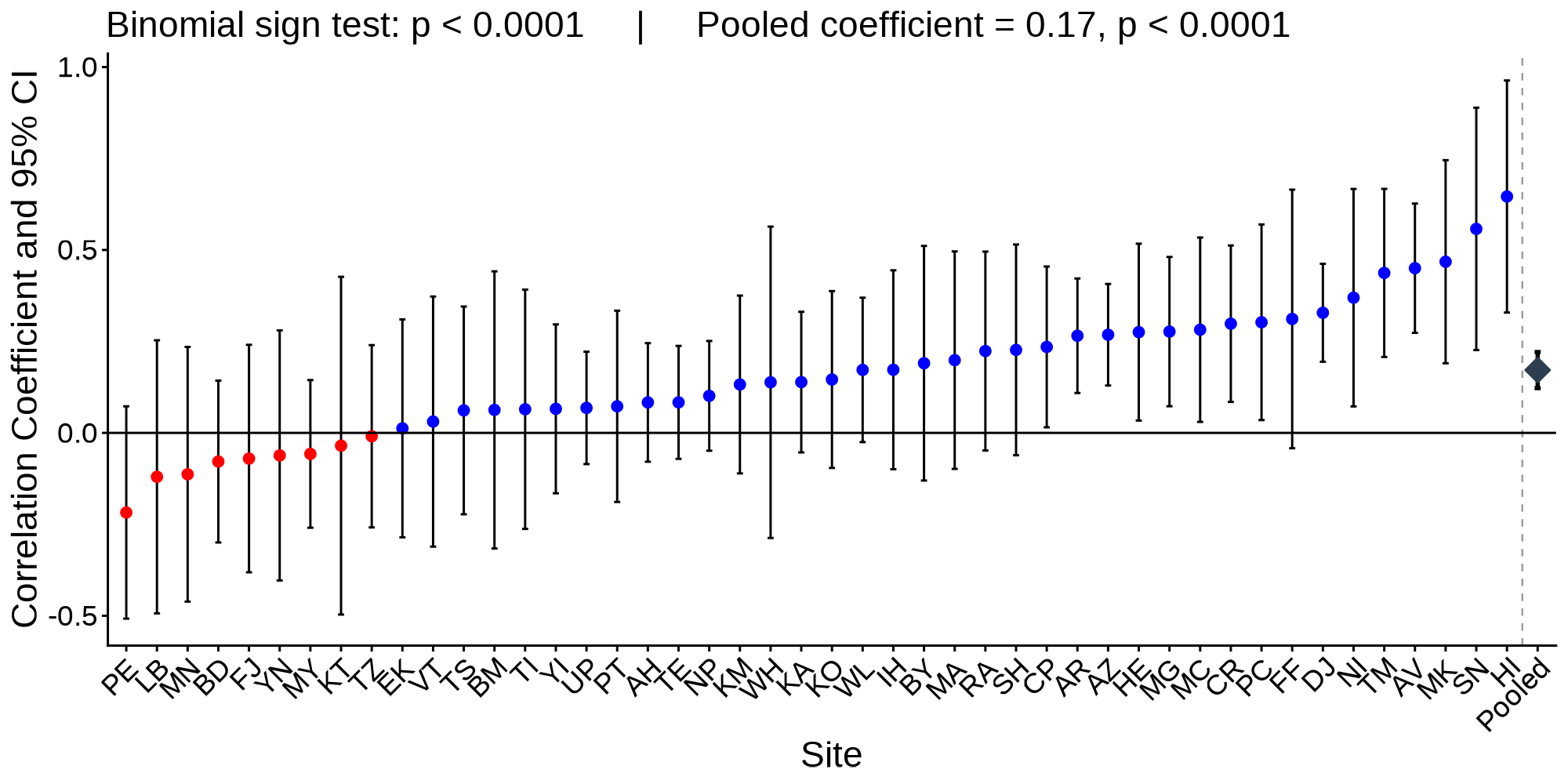}
    \caption{Support provisioning vs wealth.}
\end{subfigure}
\caption{\textbf{Correlations between access and provisioning of support and material wealth metrics, with sharing unit size fixed effects}. Support access (i.e., out-degree) and support provisioning (i.e., in-degree) are calculated using the ``composite'' network. Correlation coefficients, and their standard errors, are calculated as the slope coefficient (and the corresponding standard error) of a regression of the residualized $y$-variable, ranked across sharing units in a community, on the residualized $x$-variable ranked across sharing units in a community. Variables are residualized on a full set of dummies for every sharing unit size.}
\label{fig:within-site-size-fixed-effects}
\end{figure}

Figures~\ref{fig:within-site-size-control} and \ref{fig:within-site-size-fixed-effects} take a different approach.
In Figure~\ref{fig:within-site-size-control}, we
calculate correlations on variables that are residualized on log(sharing unit size), while in Figure~\ref{fig:within-site-size-fixed-effects} this is done on variables residualized on a full set of dummies for each observed sharing unit size. In both, we see that while the results for provisioning support (i.e., in-degree) continue to hold, they do not for accessing support (i.e., out-degree).

\clearpage
\subsection{Different Measures of Economic Connectedness}\label{supp:sum_wealth_within}

In our analyses, we use the average percentile rank of alters' wealth (however weighted) when constructing our measures of economic connectedness. First, we consider as an alternative to Average Alter Wealth the Unit Economic Connectedness measure (following \cite{chetty_social_2022}).

\begin{figure}[h]
\centering
\begin{subfigure}{0.49\textwidth}
    \centering
    \includegraphics[width=\textwidth]{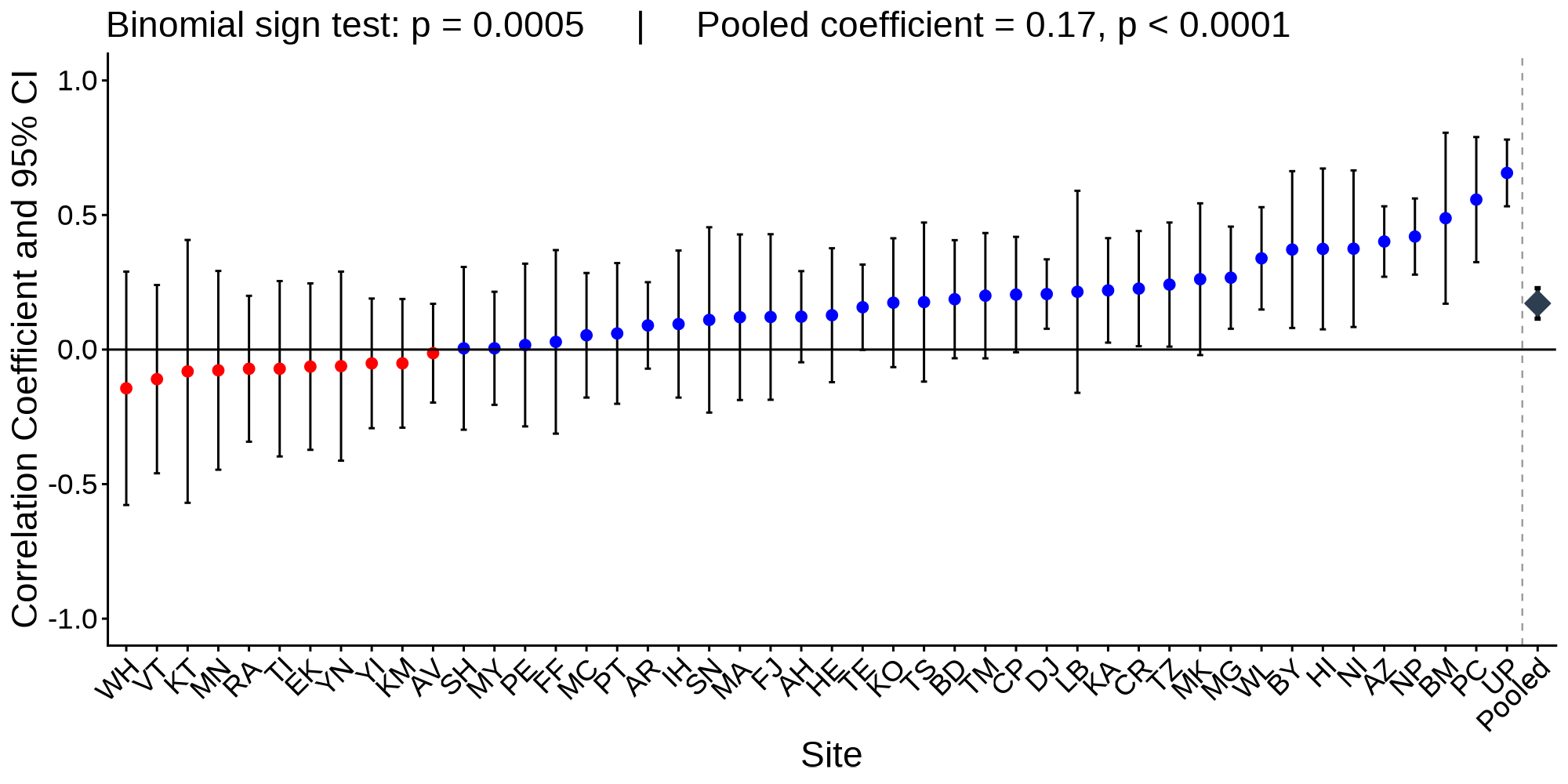}
    \caption{Unit economic connectedness (to supporters) vs wealth.}
\end{subfigure}
\vspace{1em}
\begin{subfigure}{0.49\textwidth}
    \centering
    \includegraphics[width=\textwidth]{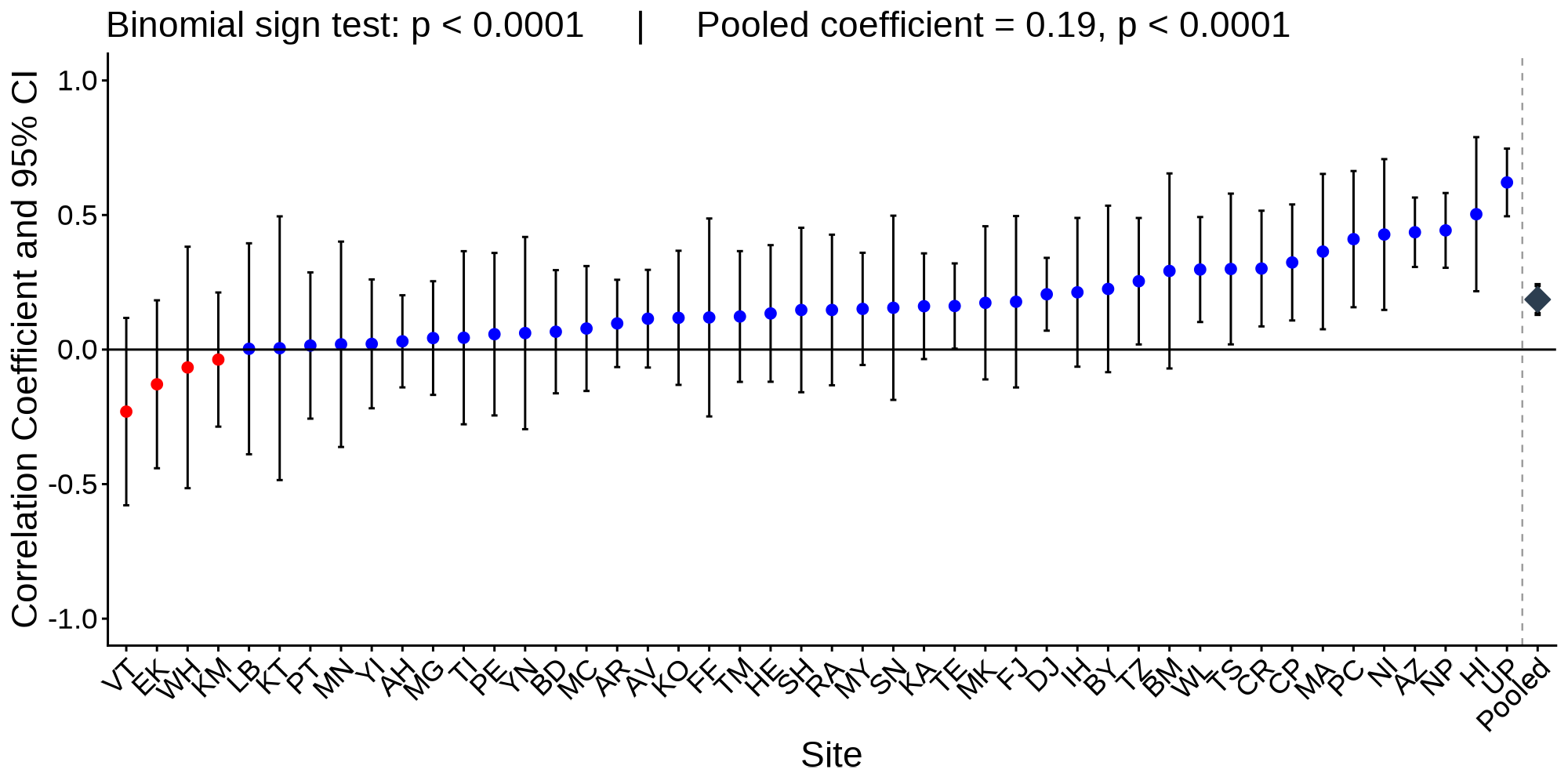}
    \caption{Unit economic connectedness (to supportees) vs wealth.}
\end{subfigure}
\begin{subfigure}{0.49\textwidth}
    \centering
    \includegraphics[width=\textwidth]{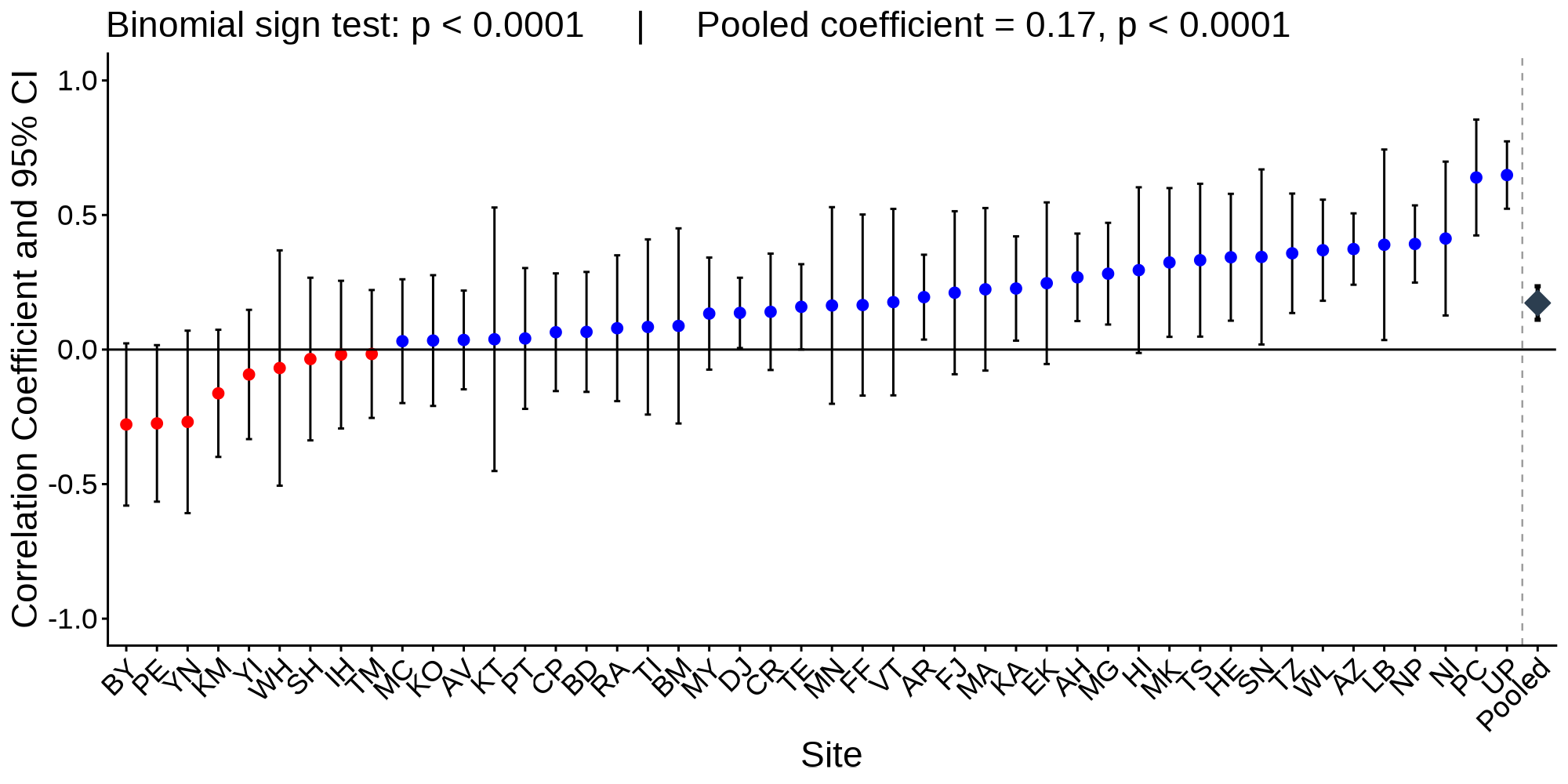}
    \caption{Unit economic connectedness (to supporters) vs wealth per capita.}
\end{subfigure}
\begin{subfigure}{0.49\textwidth}
    \centering
    \includegraphics[width=\textwidth]{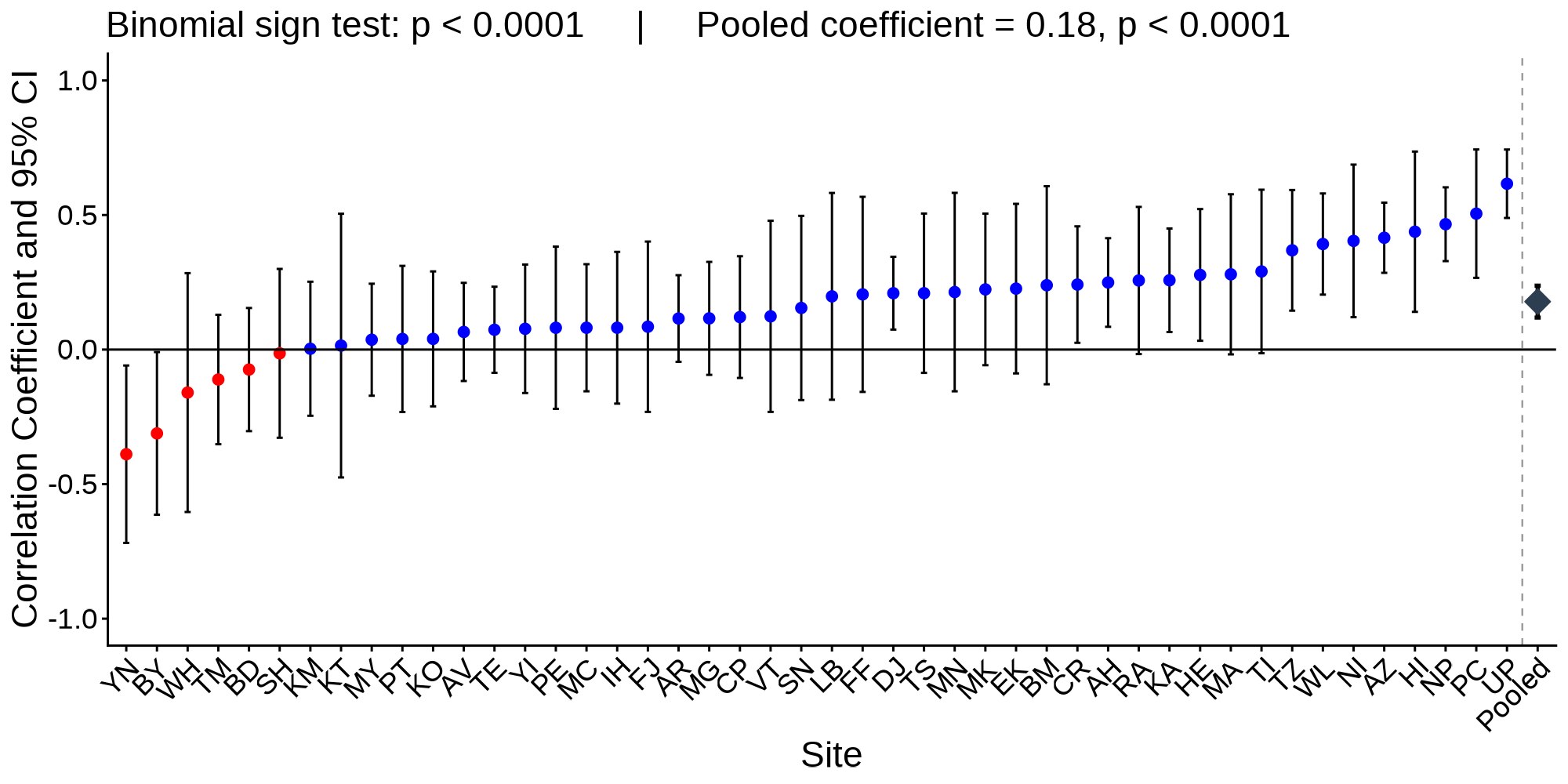}
    \caption{Unit economic connectedness (to supportees) vs wealth per capita.}
\end{subfigure}
\caption{\textbf{Correlations between Unit Economic Connectedness and material wealth}. Unit economic connectedness is calculated using the ``composite'' network and considering alters who provide support (a, c) or alters to whom support is provided (b, d) and a wealth distribution using absolute (a, b) or per capita (c, d) measures.}
\label{fig:iec-wealth-corr}
\end{figure}

Figure~\ref{fig:iec-wealth-corr} shows that Unit Economic Connectedness, constructed considering either alters who provide support (i.e., supporters) or alters to whom support is provided (i.e., supportees), is positively associated with both absolute wealth and wealth per capita across the majority of communities.

\begin{figure}[h]
\centering
\begin{subfigure}{0.49\textwidth}
    \centering
    \includegraphics[width=\textwidth]{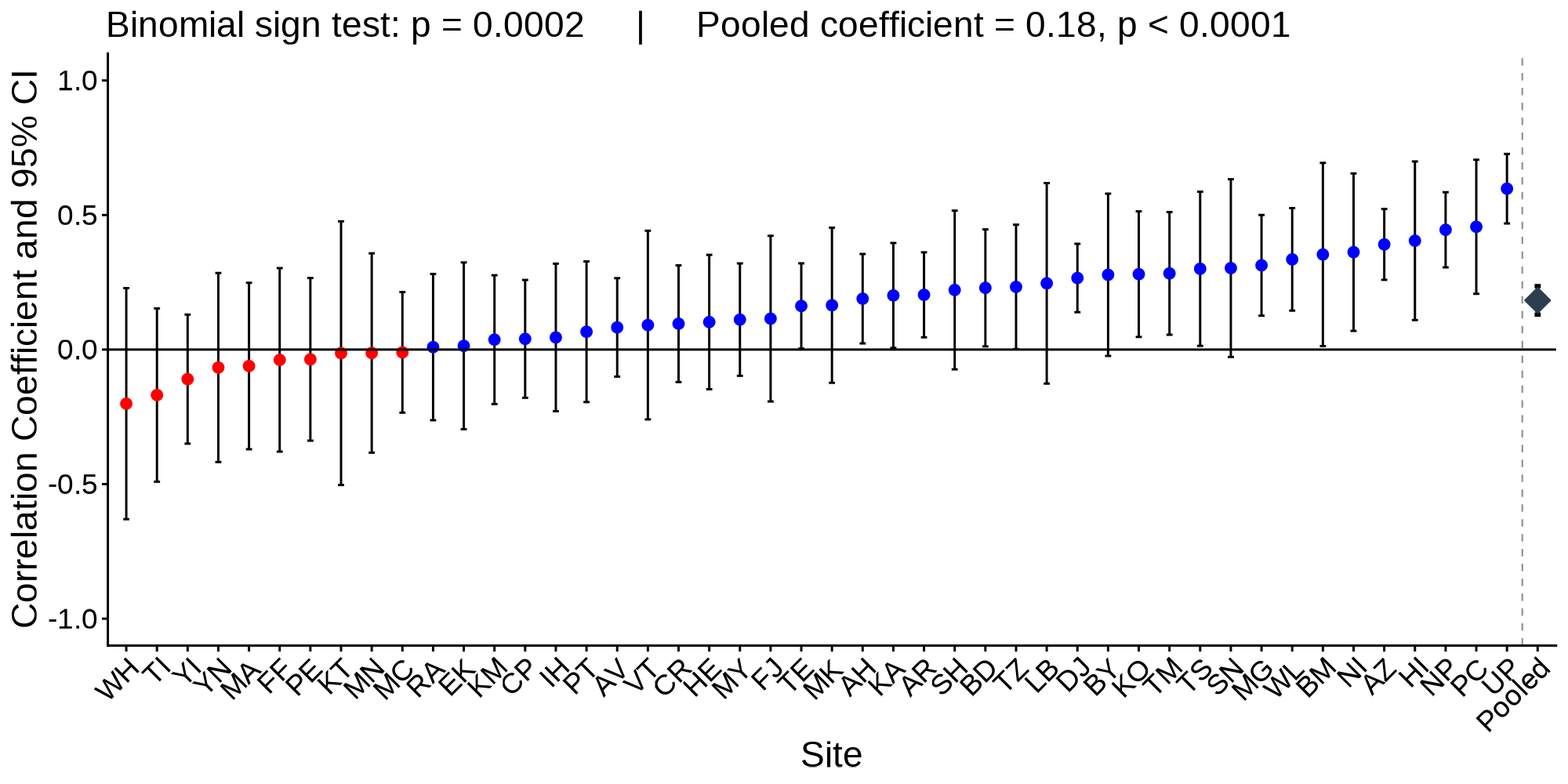}
    \caption{Median alter wealth (of supporters) vs wealth.}
\end{subfigure}
\begin{subfigure}{0.49\textwidth}
    \centering
    \includegraphics[width=\textwidth]{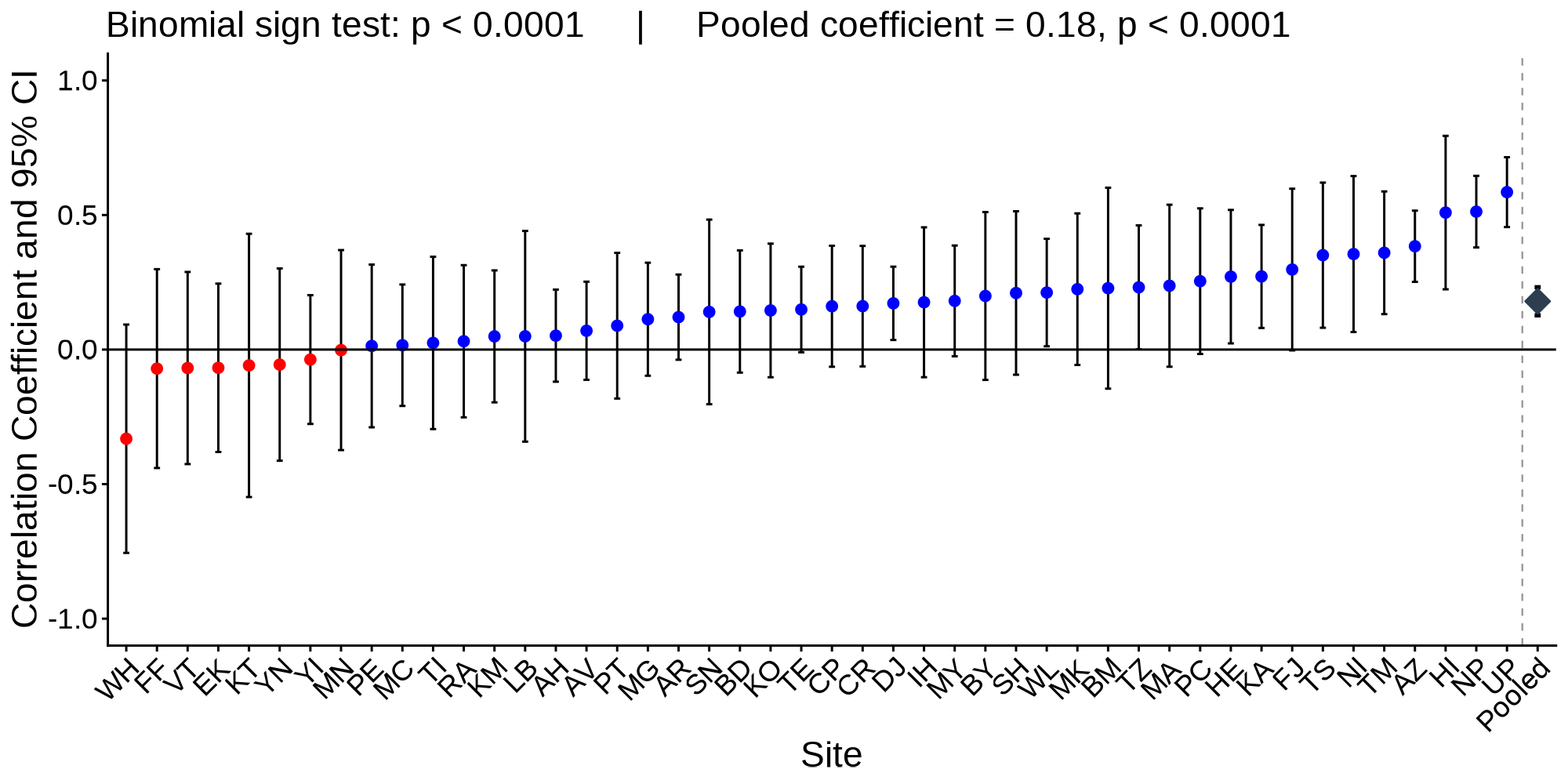}
    \caption{Median alter wealth (of supportees) vs wealth.}
\end{subfigure}
\caption{\textbf{Correlations between the percentile rank of median alter wealth and percentile rank of sharing unit wealth}. The median of alter wealth is calculated using the ``composite'' network and based on either (a) alters who provide support (i.e., supporters) or (b) alters to whom support is provided (i.e., supportees). }
\label{fig:iec-median_wealth}
\end{figure}

\begin{figure}[h]
\centering
\begin{subfigure}{0.49\textwidth}
    \centering
    \includegraphics[width=\textwidth]{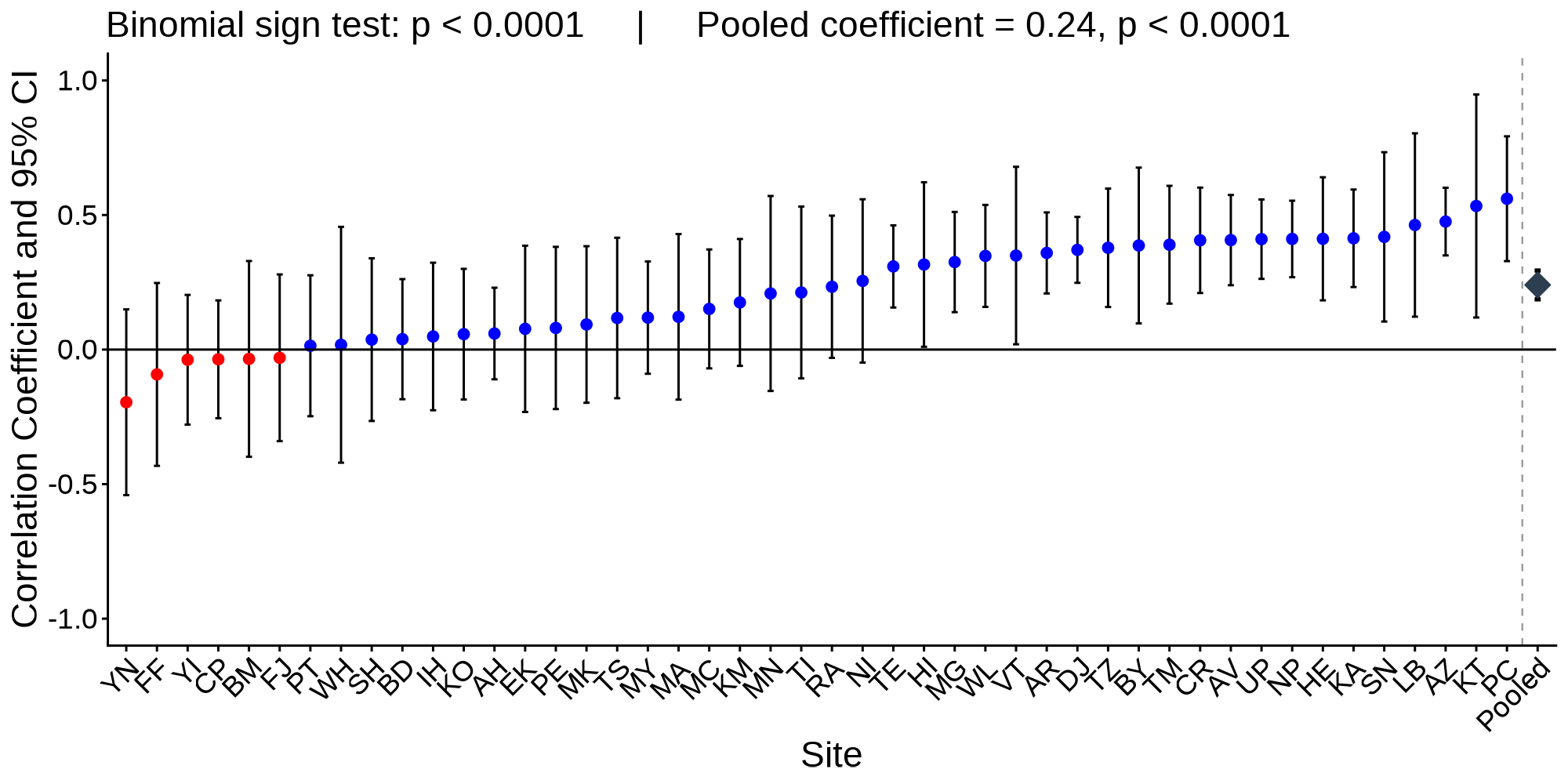}
    \caption{Sum of alter wealth (of supporters) vs wealth.}
\end{subfigure}
\begin{subfigure}{0.49\textwidth}
    \centering
    \includegraphics[width=\textwidth]{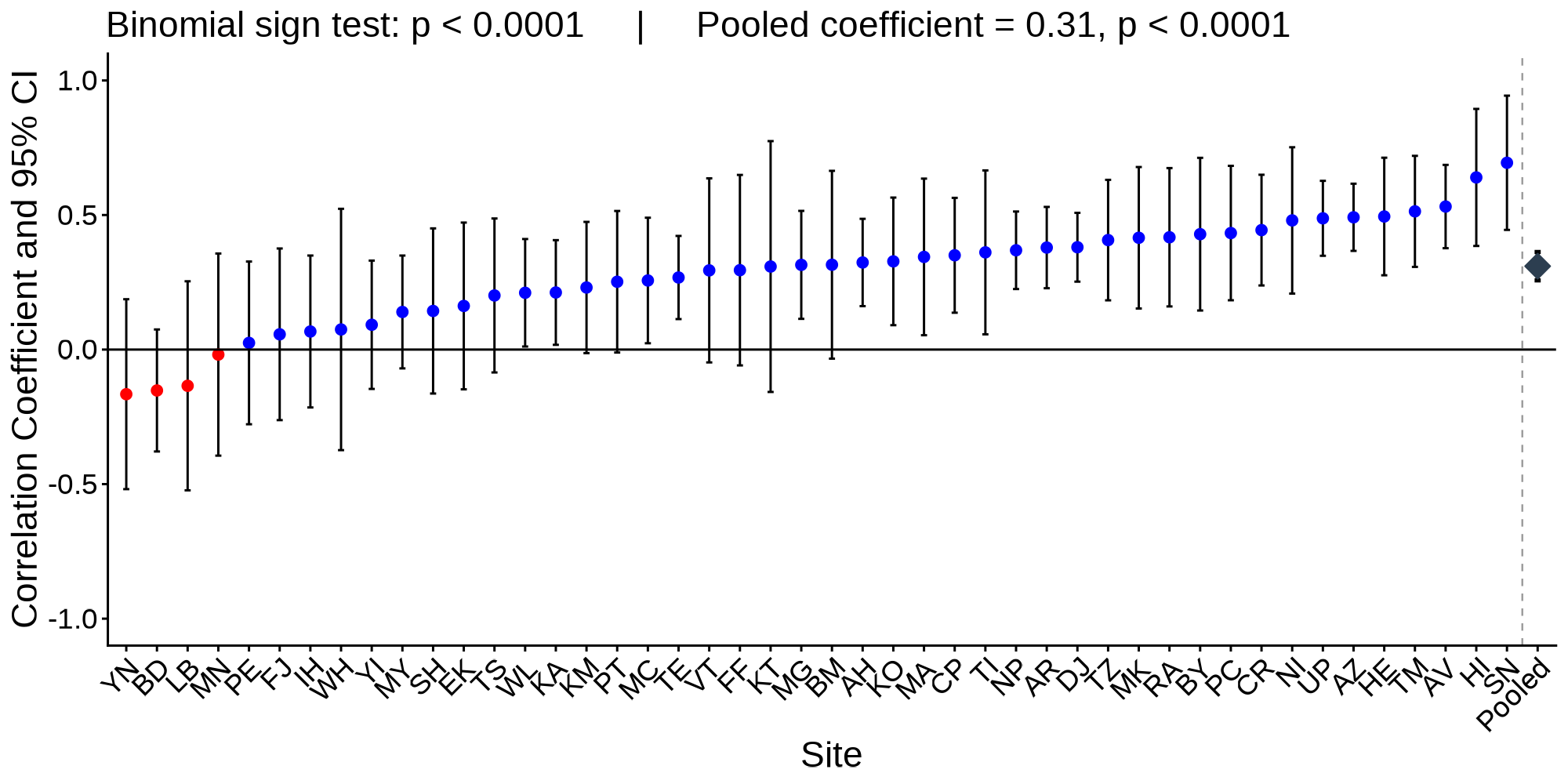}
    \caption{Sum of alter wealth (of supportees) vs wealth.}
\end{subfigure}
\caption{\textbf{Correlations between the percentile rank of the sum of alter wealth and percentile rank of sharing unit wealth}. The sum of alter wealth is calculated using the ``composite'' network and based on either (a) alters who provide support (i.e., supporters) or (b) alters to whom support is provided (i.e., supportees). }
\label{fig:iec-sum_wealth}
\end{figure}

We may also be interested in access to the absolute wealth implied by network connections. We therefore consider the median (Figure~\ref{fig:iec-median_wealth}) and the sum (Figure~\ref{fig:iec-sum_wealth}) of the wealth of a sharing unit's alters.

Similar to what we see for Average Alter Wealth constructed using percentile ranks (as shown in Figure~\ref{fig:within-site-multipanel}), we see consistent positive associations across communities between a sharing unit's wealth and the median wealth or the sum of the wealth of their alters.

\clearpage
\subsection{Multilevel Models of Degree, AAW, and Wealth}\label{supp:pooled_coefs}

In Figure~\ref{fig:within-site-multipanel}, and related figures, we provide a site-by-site visualization of the within-community relationship between network connections and wealth. To summarize the average relationship across communities, we estimate a random-effects meta-analytic coefficient using the site-specific regression slopes and their standard errors. This is the ``pooled'' coefficient displayed on the figure. Formally, we estimate:
\[
\hat{\beta}_s = \beta + u_s + \epsilon_s,
\]
where $\epsilon_s$ captures sampling uncertainty using the estimated standard error of $\hat{\beta}_s$, and $u_s \sim N(0,\tau^2)$ captures between-site heterogeneity. The pooled coefficient reported in the figure is the estimated average effect $\hat{\beta}$ from this random-effects model, with $\tau^2$ estimated by restricted maximum likelihood.

We see that our pooled coefficient is positive and significant across communities, showing that on average support access and support provisioning are positively correlated with wealth across communities, and that similarly AAW (measured using either alters who provide support or alters to whom support is provided) and wealth are positively correlated. To further support this interpretation, in this section we estimate a multilevel model demonstrating these relationships.

Let $i$ index sharing units and $s$ index sites. For each panel, we construct the same within-site percentile-rank variables used in Figure~\ref{fig:within-site-multipanel}, and then \textit{z}-score standardize both variables within site. We then estimate a varying-slope multilevel model of the form
\[
\widetilde{Y}_{is} = \left(\beta + b_s\right)\widetilde{X}_{is} + \varepsilon_{is},
\]
where $\widetilde{Y}_{is}$ and $\widetilde{X}_{is}$ are within-site standardized ranks, $b_s \sim N(0,\sigma^2_b)$ is a site-specific deviation from the average slope, and $\varepsilon_{is} \sim N(0,\sigma^2)$ is the sharing-unit-level residual. We omit random intercepts because both variables are standardized within community, so each community's mean is zero by construction.

We estimate this model separately for the four panels of Figure~\ref{fig:within-site-multipanel}. Throughout, the dependent variable is wealth per capita rank. In Panels A and B and the predictors are support access (i.e., out-degree) per capita rank and support provisioning (i.e., in-degree) per capita rank, respectively. For panels C and D, the predictor variable is Average Alter Wealth per capita rank; Panel C considers alters who provide support (i.e., supporters), while Panel D considers alters to whom support is provided (i.e., supportees).

The coefficient $\beta$ is the multilevel analogue of the pooled relationship across communities. It estimates the average within-site association while allowing each site to have its own slope. The estimated standard deviation $\sigma_b$ summarizes how much the site-specific slopes vary around this average. This model therefore uses the underlying sharing-unit-level data directly, rather than first estimating site-level slopes and then combining them.

Tables~\ref{tab:within-site-multilevel-degree} and~\ref{tab:within-site-multilevel-aaw} report the resulting estimates. In all four specifications, the average slope is positive, with coefficients that are almost identical to the pooled coefficients shown in Figure~\ref{fig:within-site-multipanel}. Alongside this overall effect, there is variability across study communities. With the exception of Panel B, negative slopes are within two standard deviations ({$2\tau$}) of the pooled slope ($\beta$).

\include{tables/within_site_multilevel_tables}

\clearpage
\section{Robustness of Across-Community Results}\label{supp:robustness_between}

% \subsection{Economic Connectedness versus Site Size}

% \begin{figure}
%     \centering
%     \includegraphics[width=\linewidth]{figures/EC_vs_Size.pdf}
%     \caption{\textbf{Economic Connectedness vs Number of Sharing Units.} Economic connectedness is constructed on the basis of wealth per capita reweighted by sharing unit size, and using the  composite network.}
%     \label{fig:ec-vs-size}
% \end{figure}

%\begin{figure}
%    \centering
%    \includegraphics[width=\linewidth]{figures/Relative_Connectedness_vs_Gini.pdf}
%    \caption{\textbf{Relative connectedness vs Gini of wealth per capita.} Economic connectedness is constructed on the basis of wealth per capita reweighted by sharing unit size, and using the  composite network for both above-median wealth per capita and below-median wealth per capita sharing units. Relative connectedness is the ratio of the economic connectedness of below-median wealth per capita sharing units to above-median wealth per capita sharing units. The Gini of wealth per capita is constructed from the distribution of wealth per capita reweighted by the number of individuals in each sharing unit.}
%    \label{fig:connectedness-vs-gini}
%\end{figure}

\subsection{Different Accountings for Sharing Unit Composition}\label{supp:alternative_SUweightings_between}

\begin{figure}
    \centering
    \includegraphics[width=\linewidth]{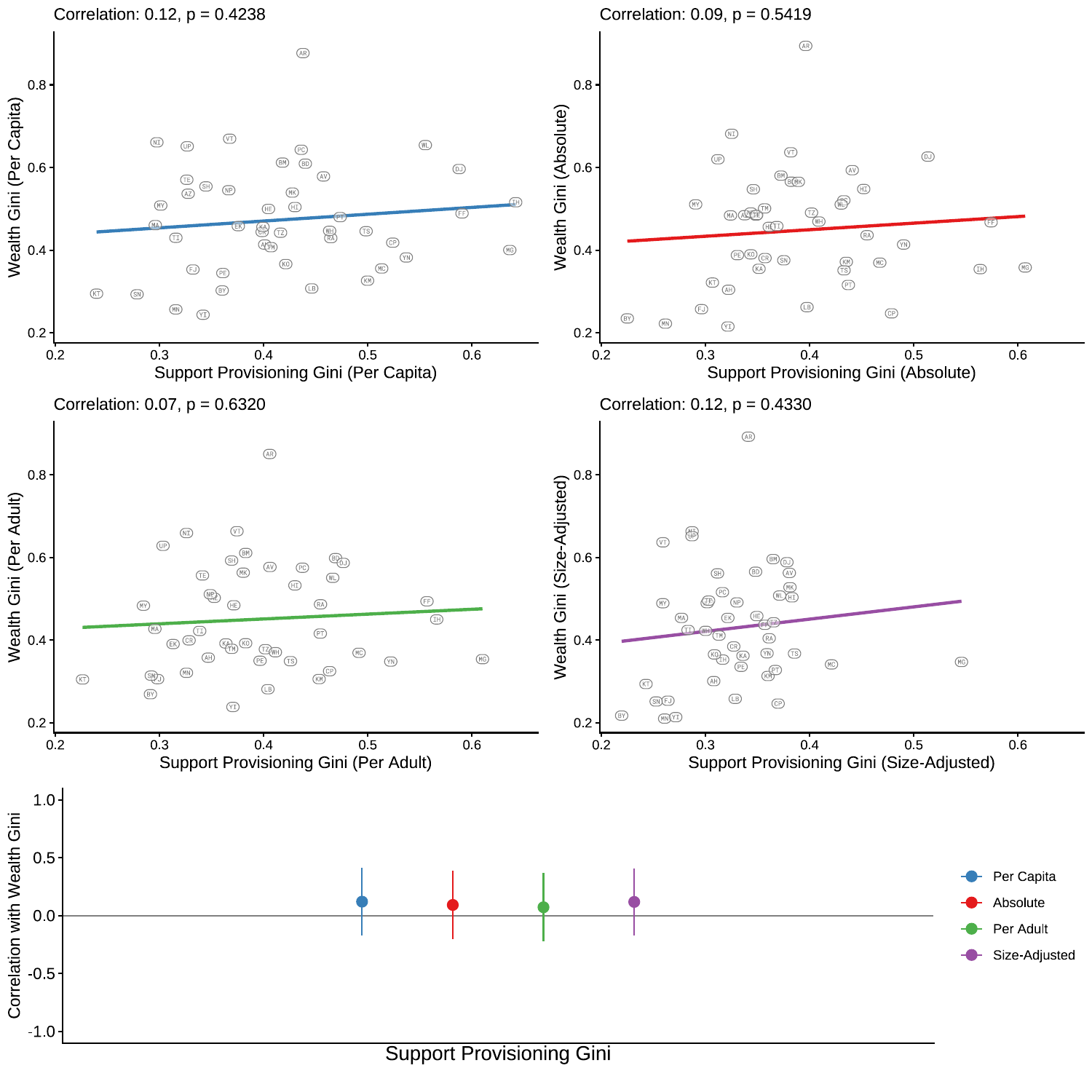}
    \caption{\textbf{Wealth Ginis vs Support provisioning (i.e., in-degree) Ginis.} Top left, we use per capita measures. Top right, we use absolute measures. Middle left, we use the per adult measures. Middle right, we use size-adjusted measures (i.e., residualized on log(sharing unit size)). The bottom figure shows the correlation coefficient (i.e., the slopes shown in the panels above) and its 95\% confidence interval for each accounting of sharing unit wealth.}
    \label{fig:wealth-ginis-ideg}
\end{figure}

\begin{figure}
    \centering
    \includegraphics[width=\linewidth]{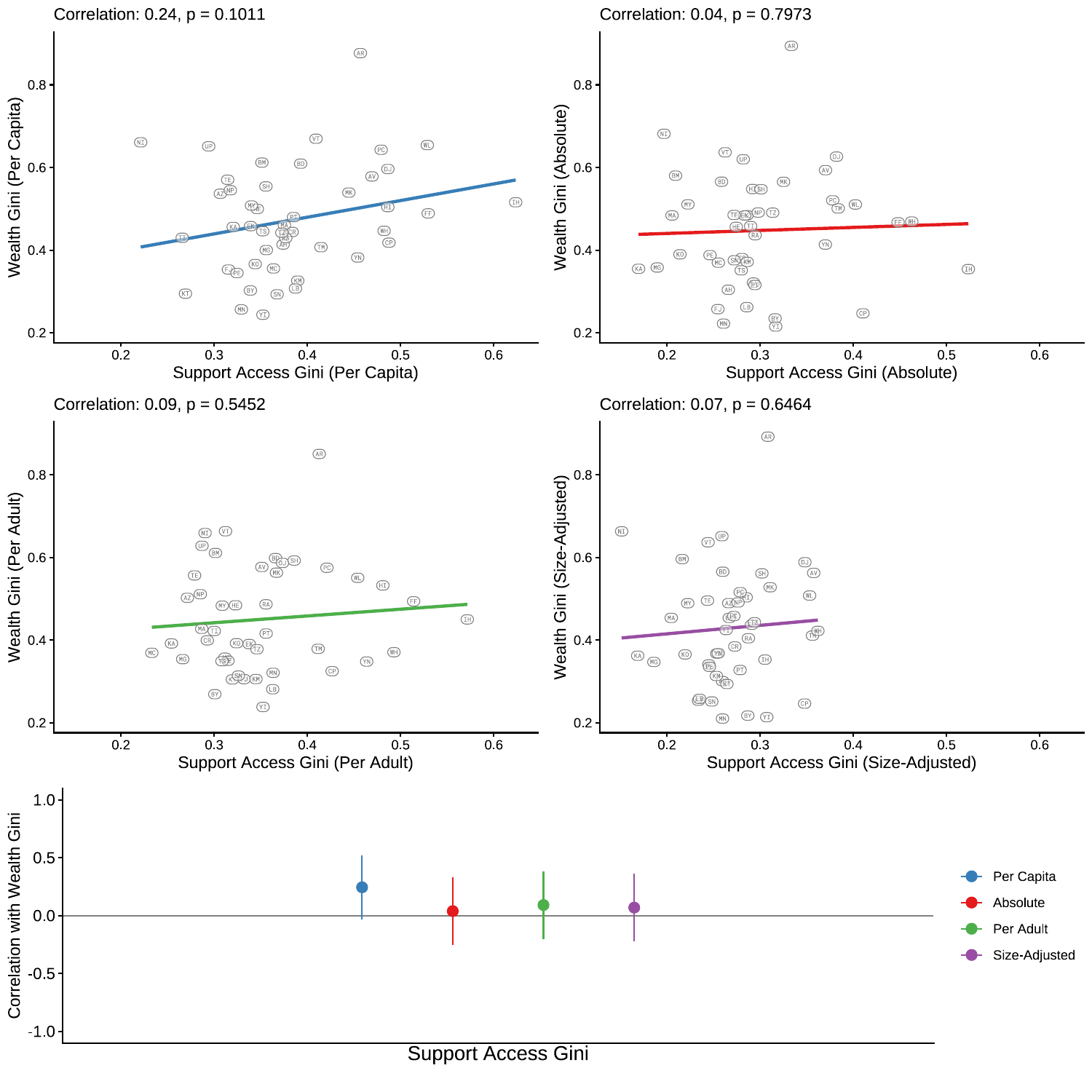}
    \caption{\textbf{Wealth Ginis vs Support access (i.e., out-degree) Ginis.} Top left, we use per capita measures. Top right, we use absolute measures. Middle left, we use the per adult measures. Middle right, we use size-adjusted measures (i.e., residualized on log(sharing unit size)). The bottom figure shows the correlation coefficient (i.e., the slopes shown in the panels above) and its 95\% confidence interval for each accounting of sharing unit wealth.}
    \label{fig:wealth-ginis-odeg}
\end{figure}

\begin{figure}
    \centering
    \includegraphics[width=\linewidth]{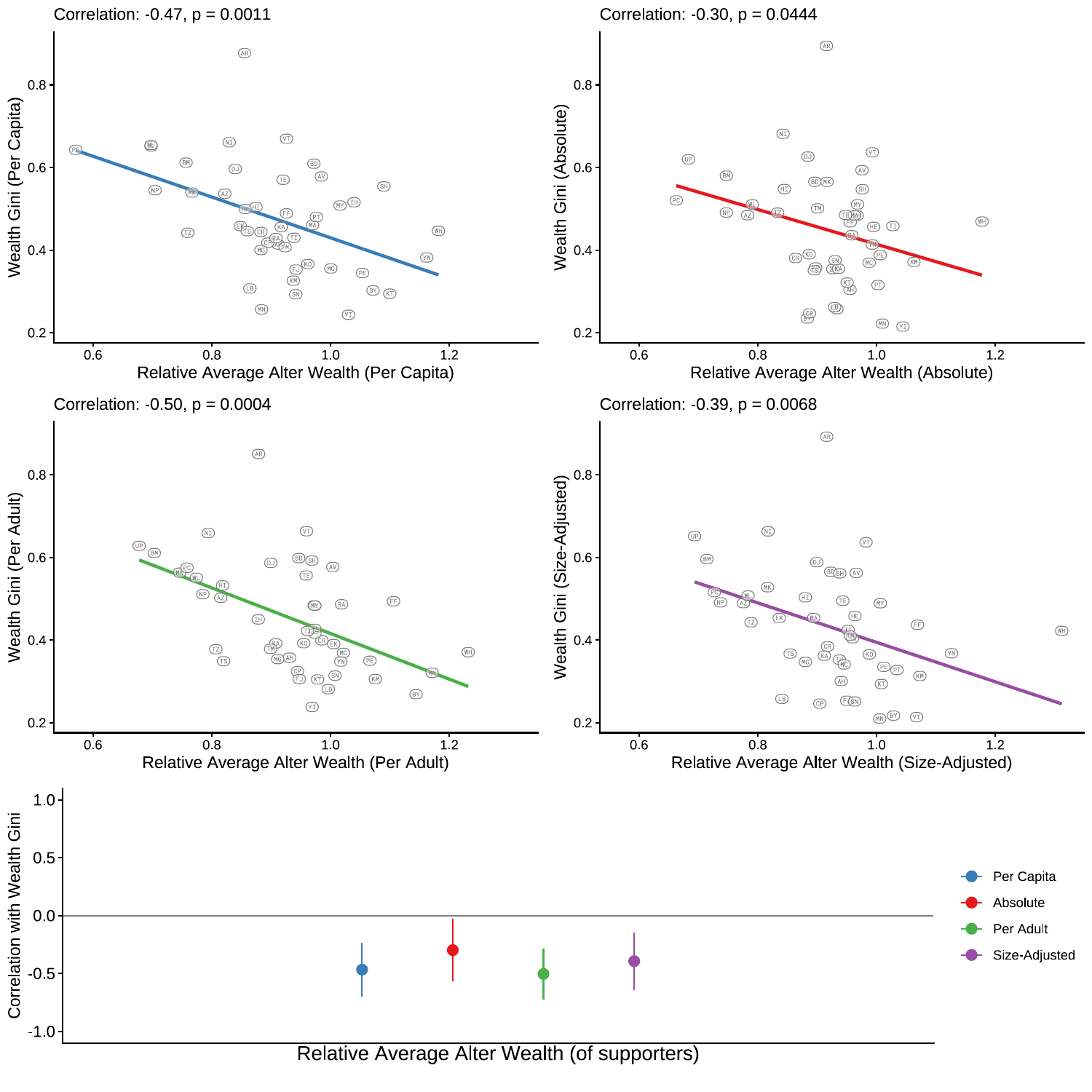}
    \caption{\textbf{Wealth Ginis vs Relative Average Alter Wealth (of supporters).} Top left, we use per capita measures. Top right, we use absolute measures. Middle left, we use the per adult measures. Middle right, we use size-adjusted measures (i.e., residualized on log(sharing unit size)). The bottom figure shows the correlation coefficient (i.e., the slopes shown in the panels above) and its 95\% confidence interval for each accounting of sharing unit wealth.}
    \label{fig:wealth-ginis-friend-rank-ratios}
\end{figure}

\begin{figure}
    \centering
    \includegraphics[width=\linewidth]{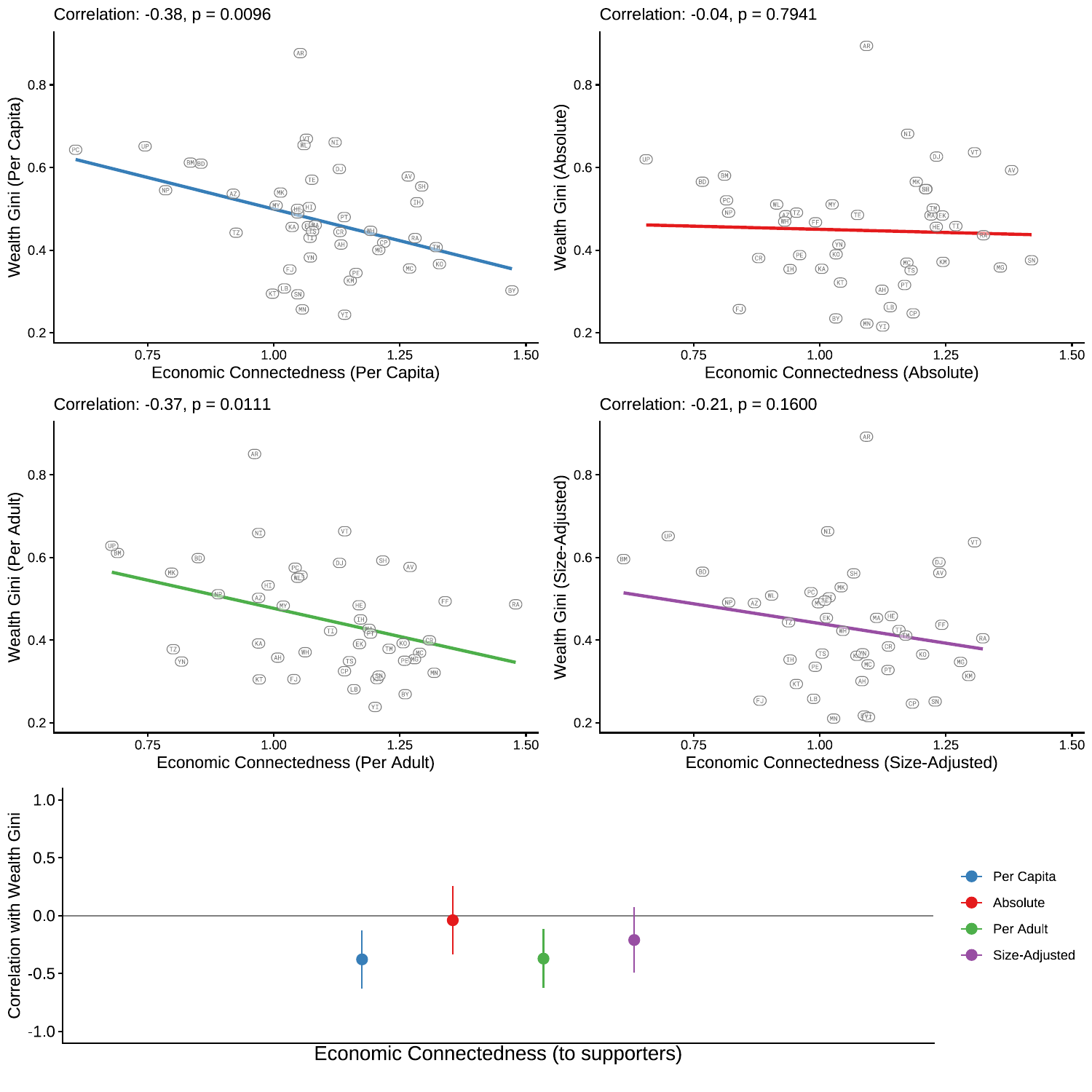}
    \caption{\textbf{Wealth Ginis vs Economic Connectedness (to supporters).} See Section~\ref{supp:definitions-measures} for definition of Economic Connectedness. Top left, we use per capita measures. Top right, we use absolute measures. Middle left, we use the per adult measures. Middle right, we use size-adjusted measures (i.e., residualized on log(sharing unit size)). The bottom figure shows the correlation coefficient (i.e., the slopes shown in the panels above) and its 95\% confidence interval for each accounting of sharing unit wealth.}
    \label{fig:wealth-ginis-ec}
\end{figure}

\begin{figure}
    \centering
    \includegraphics[width=\linewidth]{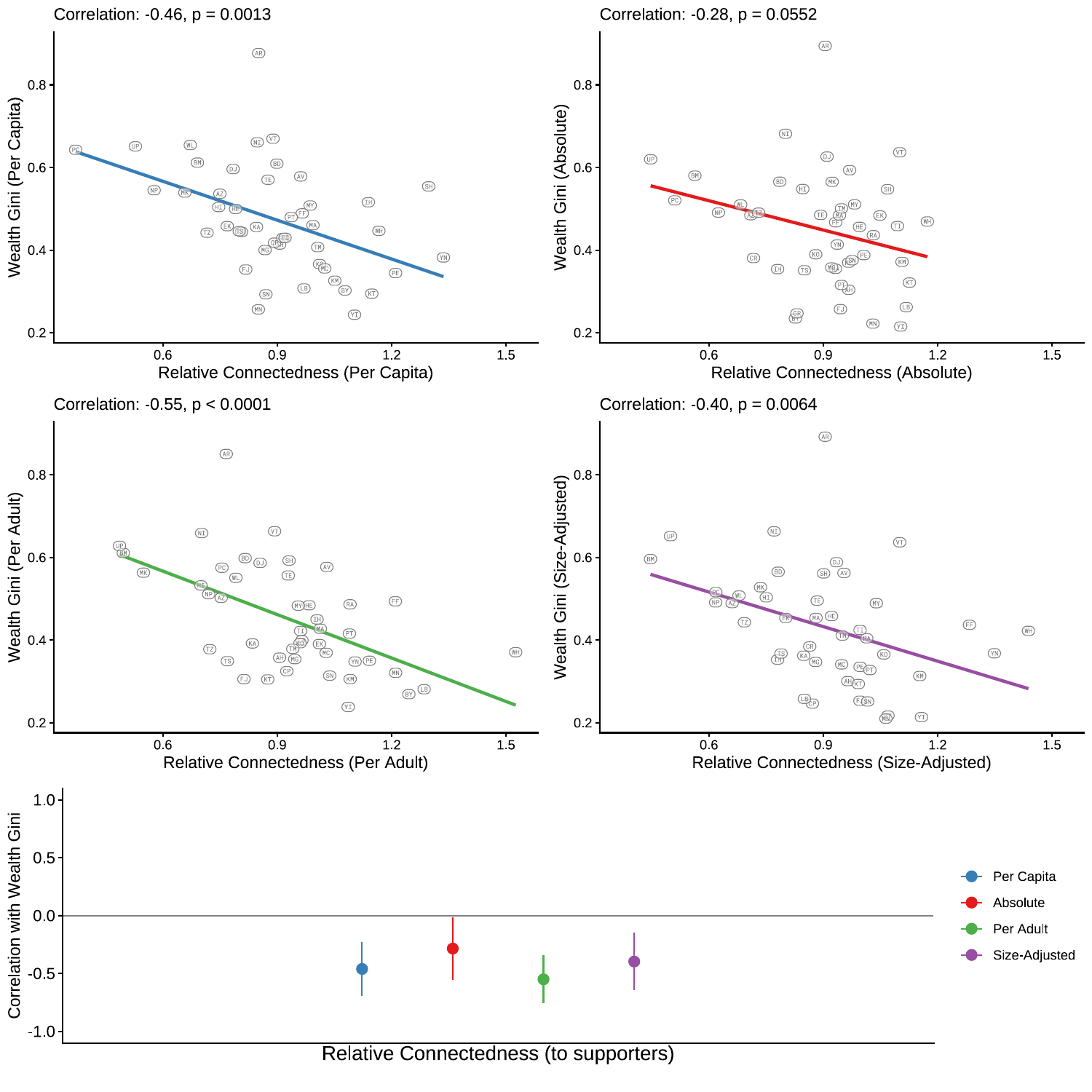}
    \caption{\textbf{Wealth Ginis vs Relative Connectedness (to supporters).} See Section~\ref{supp:definitions-measures} for definition of Relative Connectedness. Top left, we use per capita measures. Top right, we use absolute measures. Middle left, we use the per adult measures. Middle right, we use size-adjusted measures (i.e., residualized on log(sharing unit size)). The bottom figure shows the correlation coefficient (i.e., the slopes shown in the panels above) and its 95\% confidence interval for each accounting of sharing unit wealth.}
    \label{fig:wealth-ginis-rc}
\end{figure}

As was done above when looking within communities in Section \ref{supp:alternative_SUweightings_within}, here again for our cross-site analyses we consider different accountings for sharing unit composition. Figures~\ref{fig:wealth-ginis-ideg}-\ref{fig:wealth-ginis-rc} show bivariate associations between wealth Ginis and each of our measures of social capital (support provisioning in Figure~\ref{fig:wealth-ginis-ideg}, support access in Figure~\ref{fig:wealth-ginis-odeg}, Relative Average Alter Wealth in Figure~\ref{fig:wealth-ginis-friend-rank-ratios}, Economic Connectedness in Figure~\ref{fig:wealth-ginis-ec}, and Relative Connectedness in Figure~\ref{fig:wealth-ginis-rc}) using four different ways of accounting for sharing unit composition.

\clearpage
\subsection{Using Incoming Nominations}\label{supp:ec_incoming}

\begin{figure}[h]
    \centering
    \includegraphics[width=\linewidth]{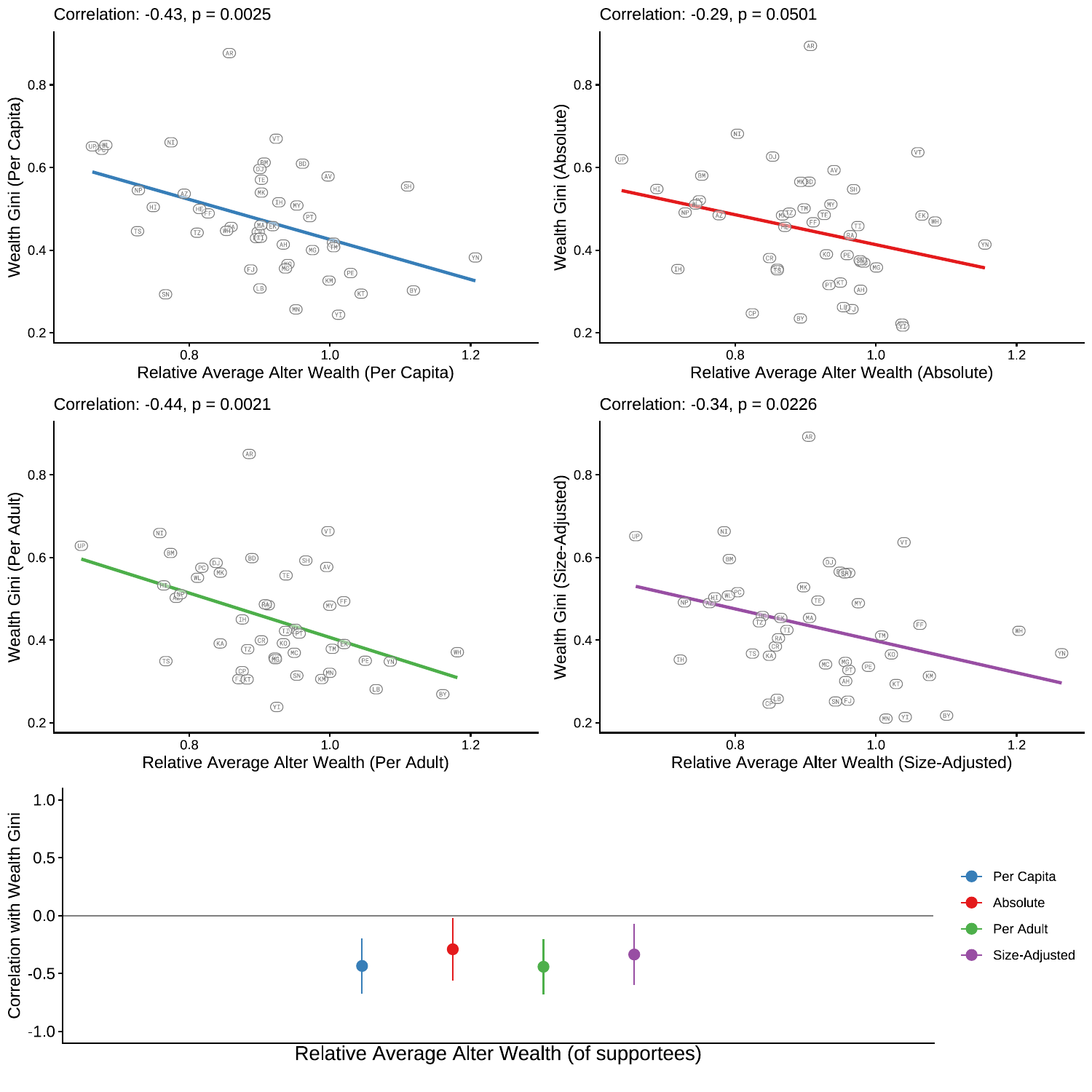}
    \caption{\textbf{Wealth Ginis vs Relative Average Alter Wealth of supportees (contra Figure~\ref{fig:wealth-ginis-friend-rank-ratios})}. Top right, we use absolute measures. Middle left, we use the per adult measures. Middle right, we use size-adjusted measures (i.e., residualized on log(sharing unit size)). The bottom figure shows the correlation coefficient (i.e., the slopes shown in the panels above) and its 95\% confidence interval for each accounting of sharing unit wealth.}
    \label{fig:wealth-ginis-friend-rank-ratios-rev}
\end{figure}

\begin{figure}[h]
    \centering
    \includegraphics[width=\linewidth]{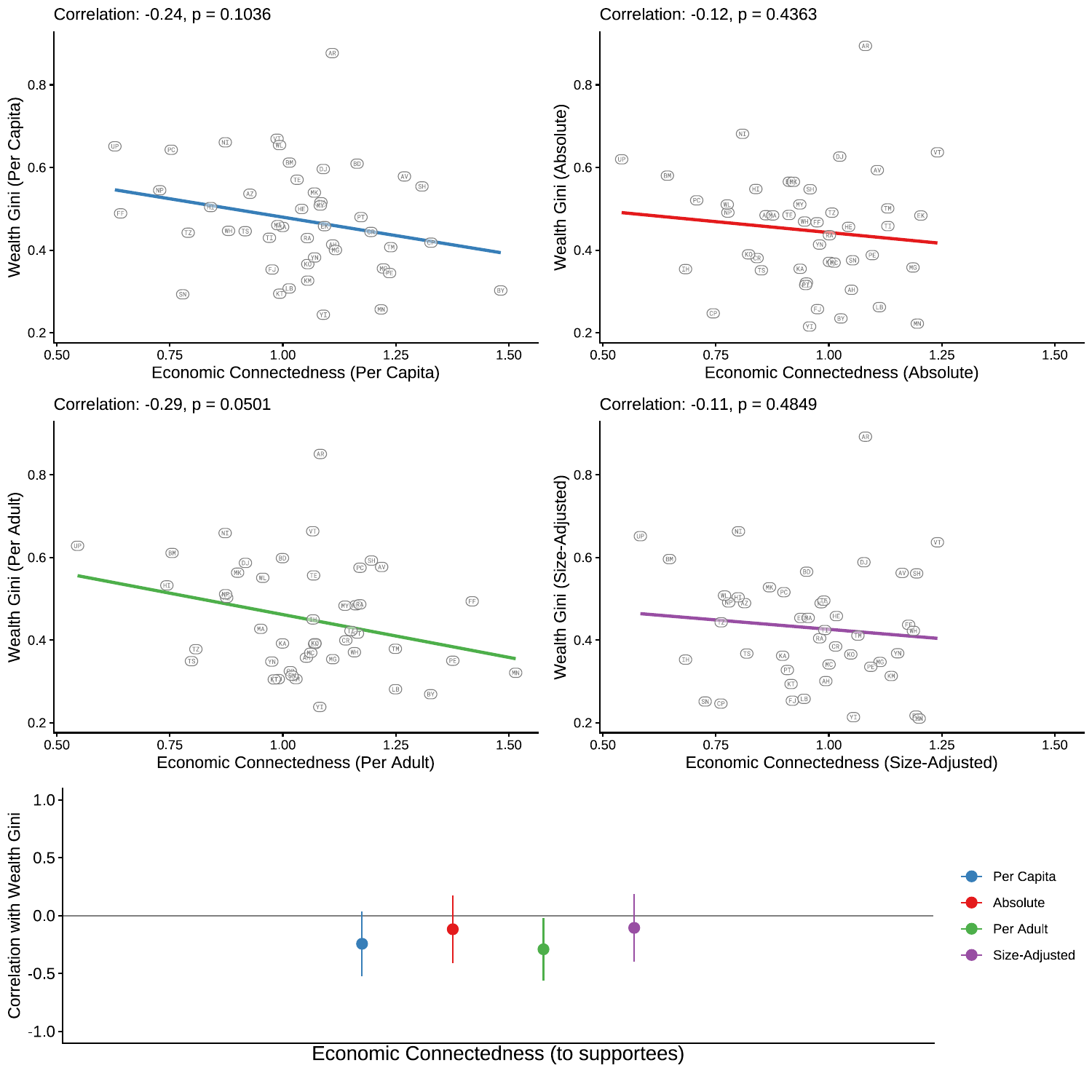}
    \caption{\textbf{Wealth Ginis vs Economic Connectedness to supportees (contra Figure~\ref{fig:wealth-ginis-ec}).} See Section~\ref{supp:definitions-measures} for definition of Economic Connectedness. Top right, we use absolute measures. Middle left, we use the per adult measures. Middle right, we use size-adjusted measures (i.e., residualized on log(sharing unit size)). The bottom figure shows the correlation coefficient (i.e., the slopes shown in the panels above) and its 95\% confidence interval for each accounting of sharing unit wealth.}
    \label{fig:wealth-ginis-ec-rev}
\end{figure}

\begin{figure}[h]
    \centering
    \includegraphics[width=\linewidth]{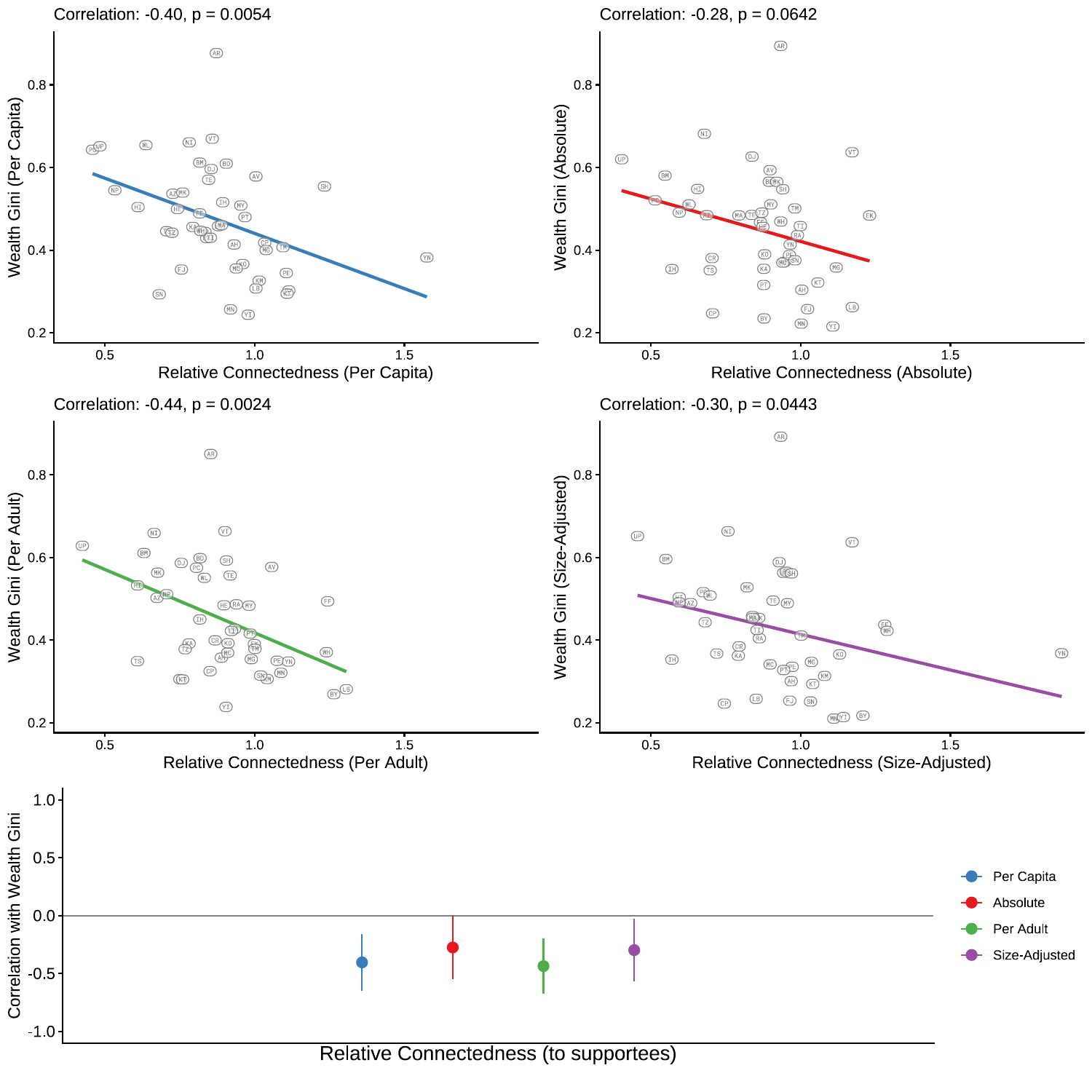}
    \caption{\textbf{Wealth Ginis vs Relative Connectedness to supportees (contra Figure~\ref{fig:wealth-ginis-rc}).} See Section~\ref{supp:definitions-measures} for definition of Relative Connectedness. Top right, we use absolute measures. Middle left, we use the per adult measures. Middle right, we use size-adjusted measures (i.e., residualized on log(sharing unit size)). The bottom figure shows the correlation coefficient (i.e., the slopes shown in the panels above) and its 95\% confidence interval for each accounting of sharing unit wealth.}
    \label{fig:wealth-ginis-rc-rev}
\end{figure}

For measures of economic connectedness, we see outgoing nominations (i.e., sharing unit's access to support) as being most pertinent for generating wealth, as it captures the alters who provide support. But, throughout we have also noted that provisioning support can also plausibly be productive. In Figures~\ref{fig:wealth-ginis-friend-rank-ratios-rev}, \ref{fig:wealth-ginis-ec-rev}, and \ref{fig:wealth-ginis-rc-rev} we therefore show equivalent cross-site associations between wealth inequality and measures of economic connectedness based on incoming nominations (so, considering alters to whom support is provided, i.e., supportees), and using different accountings for sharing unit wealth. The association is quite similar to what is found using outgoing nominations (shown in Figures~\ref{fig:wealth-ginis-friend-rank-ratios}, \ref{fig:wealth-ginis-ec}, and \ref{fig:wealth-ginis-rc}).

\clearpage
\subsection{Using Absolute Wealth}\label{supp:sum_wealth_between}

\begin{figure}[h]
\centering
\begin{subfigure}{0.48\textwidth}
    \centering
    \includegraphics[width=\textwidth]{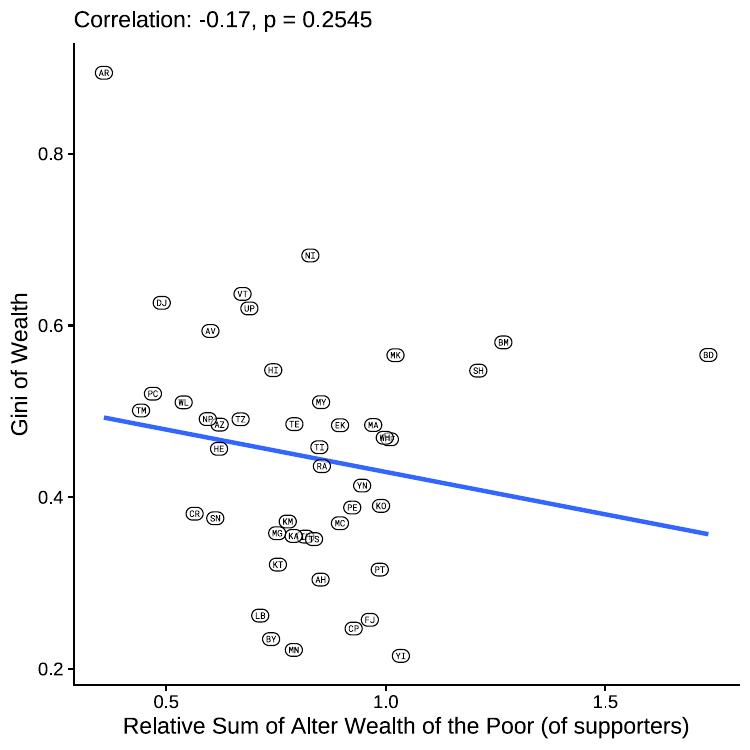}
\end{subfigure}
\hfill
\begin{subfigure}{0.48\textwidth}
    \centering
    \includegraphics[width=\textwidth]{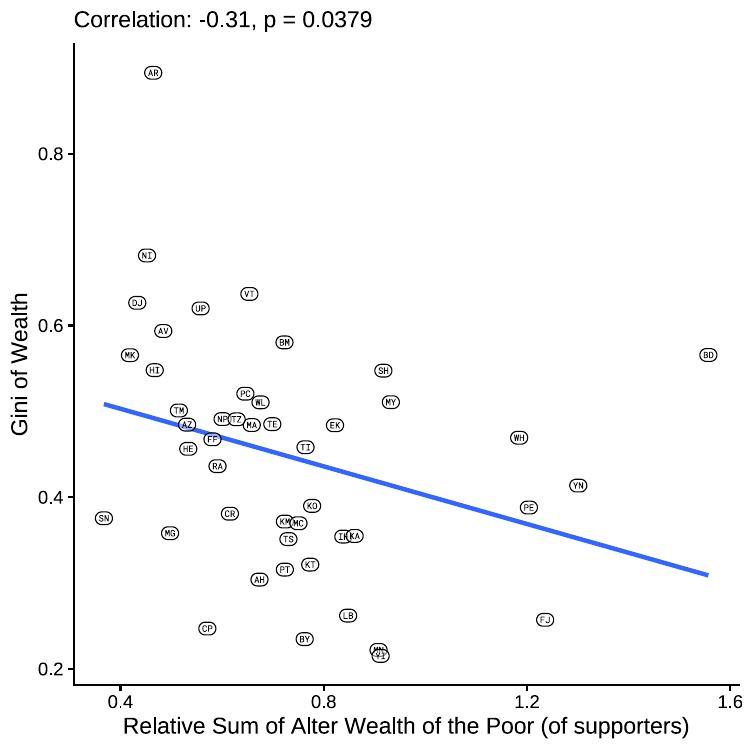}
\end{subfigure}
\caption{\textbf{Wealth Gini vs Relative Sum of Alter Wealth.} Cross-site correlations between the Gini of sharing unit wealth and the relative sum of alter wealth, based on (left) alters who provide support and (right) alters to whom support is provided.}
\label{fig:sum-wealth-gini}
\end{figure}

\begin{figure}[h]
\centering
\begin{subfigure}{0.48\textwidth}
    \centering
    \includegraphics[width=\textwidth]{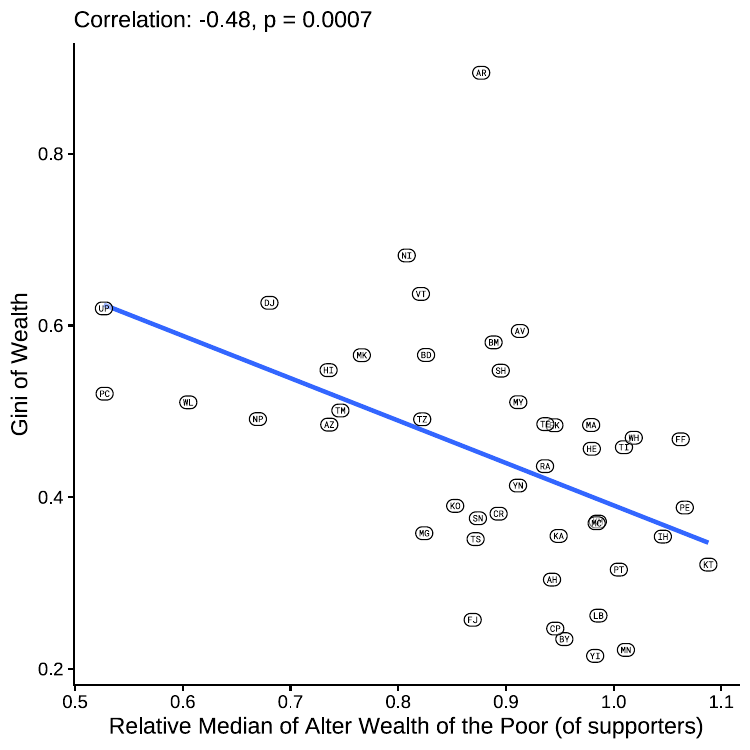}
\end{subfigure}
\hfill
\begin{subfigure}{0.48\textwidth}
    \centering
    \includegraphics[width=\textwidth]{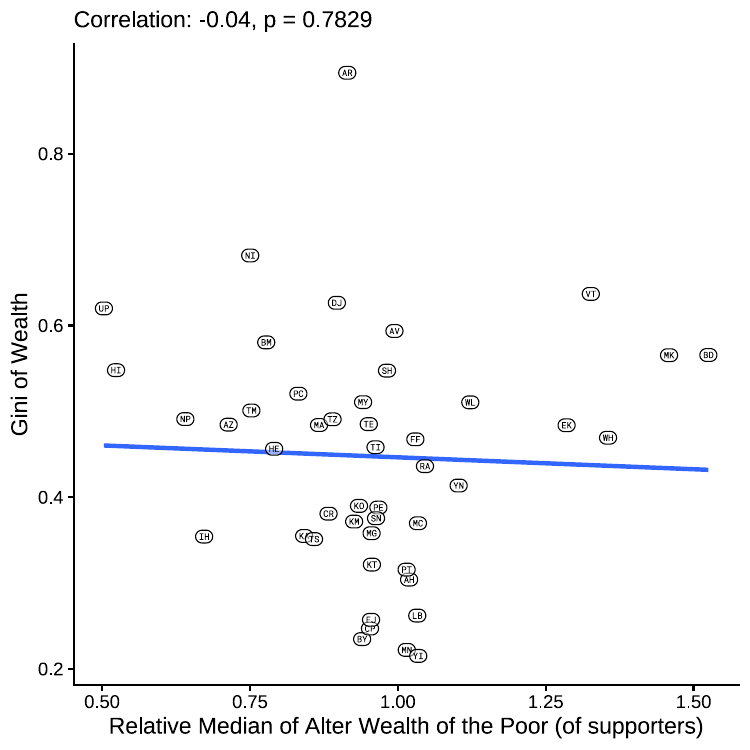}
\end{subfigure}
\caption{\textbf{Wealth Gini vs Relative Median of Alter Wealth.} Cross-site correlations between the Gini of sharing unit wealth and the relative median of alter wealth, based on (left) alters who provide support and (right) alters to whom support is provided.}
\label{fig:median-wealth-gini}
\end{figure}

In Section~\ref{supp:sum_wealth_within} we considered the absolute access to wealth implied by network connections and its association with sharing unit wealth within communities. Here now we construct community-level measures akin to measures of Relative Average Alter Wealth, by constructing the ratio of the average summed alter wealth for below-median wealth sharing units to above-median wealth sharing units (Figure~\ref{fig:sum-wealth-gini}). Similarly, we construct the ratio of the average median alter wealth for below-median sharing units to above-median wealth sharing units (Figure~\ref{fig:median-wealth-gini}).

\clearpage
\subsection{Using Different Measures of Inequality}\label{supp:ninety-ten}

\begin{figure}[ht]
\centering
\begin{subfigure}{0.48\textwidth}
    \centering
    \includegraphics[width=\textwidth]{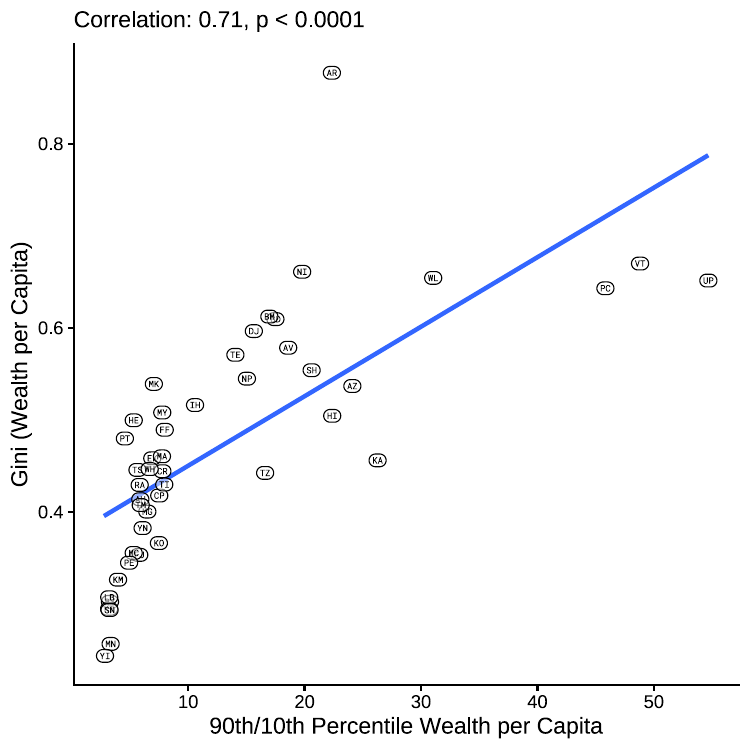}
%    \caption{In-Degree Per Capita Gini vs Wealth per Capita Gini}
\end{subfigure}
\hfill
\begin{subfigure}{0.48\textwidth}
    \centering
    \includegraphics[width=\textwidth]{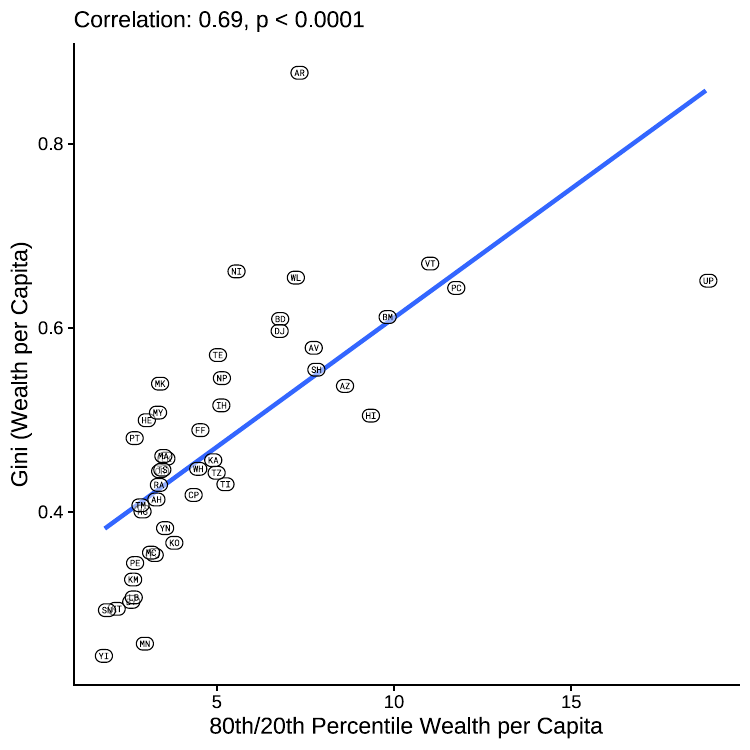}
%    \caption{In-Degree Per Capita Gini vs Wealth per Capita Gini}
\end{subfigure}
\vspace{1em}
\begin{subfigure}{0.48\textwidth}
    \centering
    \includegraphics[width=\textwidth]{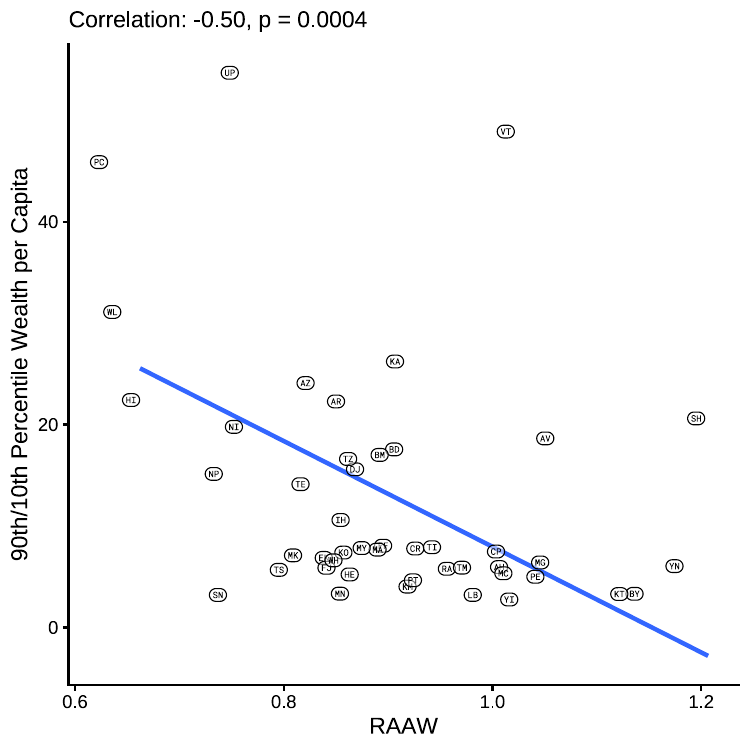}
%    \caption{RAAW vs Wealth per Capita Gini.}
\end{subfigure}
\hfill
\begin{subfigure}{0.48\textwidth}
    \centering
    \includegraphics[width=\textwidth]{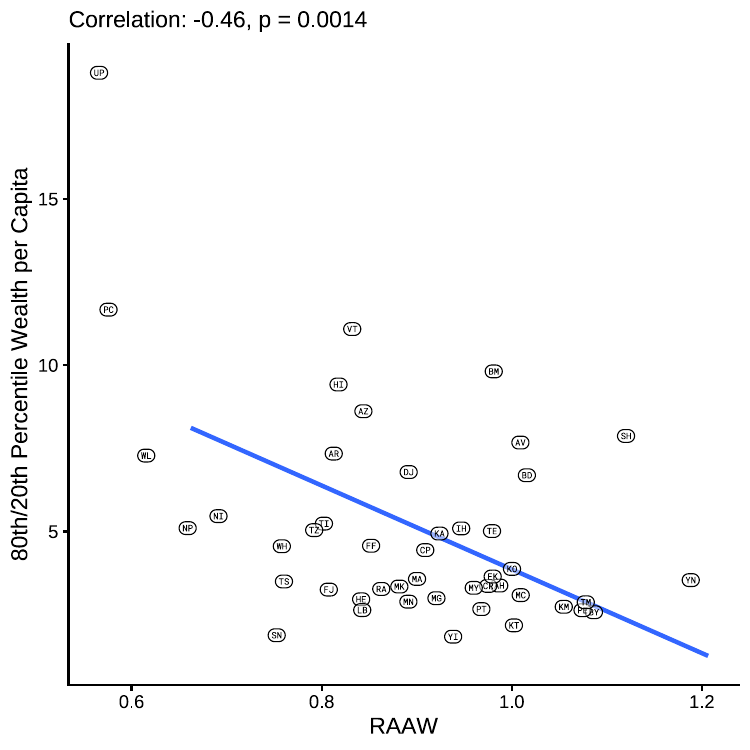}
%    \caption{In-Degree Per Capita Gini vs Wealth per Capita Gini}
\end{subfigure}
\caption{Cross-site correlations between the ratio of wealth per capita for the 90th percentile sharing unit vs the 10th percentile sharing unit (left), the ratio of wealth per capita for the 80th percentile sharing unit vs the 20th percentile sharing unit (right), the Gini of wealth per capita (top), and Relative Average Alter Wealth (bottom). The top plots show the alignment of these alternative measures of inequality to the Gini, while the bottom plots are equivalent to what we show in the bottom left panel of Figure~\ref{fig:wealth-ginis-vs-scatters}. }
\label{fig:ninety-tens}
\end{figure}

We may also be concerned that our results could be driven by outliers with extreme wealth (per capita) which may disproportionately be influencing the Gini coefficients. To check this, we construct as well the ratio of wealth per capita for the 90th percentile versus the 10th percentile, and the ratio of the 80th percentile versus the 20th percentile \cite{piketty2003income, faggio2010evolution}. As shown in Figure~\ref{fig:ninety-tens}, those measures are clearly positively correlated with the Gini coefficient, and both are significantly negatively correlated with Relative Average Alter Wealth.

\clearpage
\subsection{Ruling Out Wealth-Band Tie Formation}\label{supp:mechanical}

As mentioned in the text, a potential concern is that the association between RAAW and wealth inequality could be a driven by sharing units forming ties in an absolute manner around their own wealth. Then as the distribution of wealth overall is widened, inequality increases while RAAW would decrease. For example, if everyone randomly friends people within X wealth units from themselves and the wealth distribution shifts from being uniform on [0,X] to being uniform on [0,10X], then friendships go from being uniform at random to being highly segregated by wealth. In this way RAAW would decrease as the distribution was spread out more broadly, and so that would be an explanation for the relationship between RAAW and inequality.  If instead people's friending tendencies are relativistic (widening with the widening distribution), then this mechanical result would not hold.  We first explore this directly by comparing the standard deviation of alters' wealth with the overall standard deviation of wealth, and then via comparison to a simulated model.

\begin{figure}[h]
\centering
\includegraphics[width=0.49\linewidth]{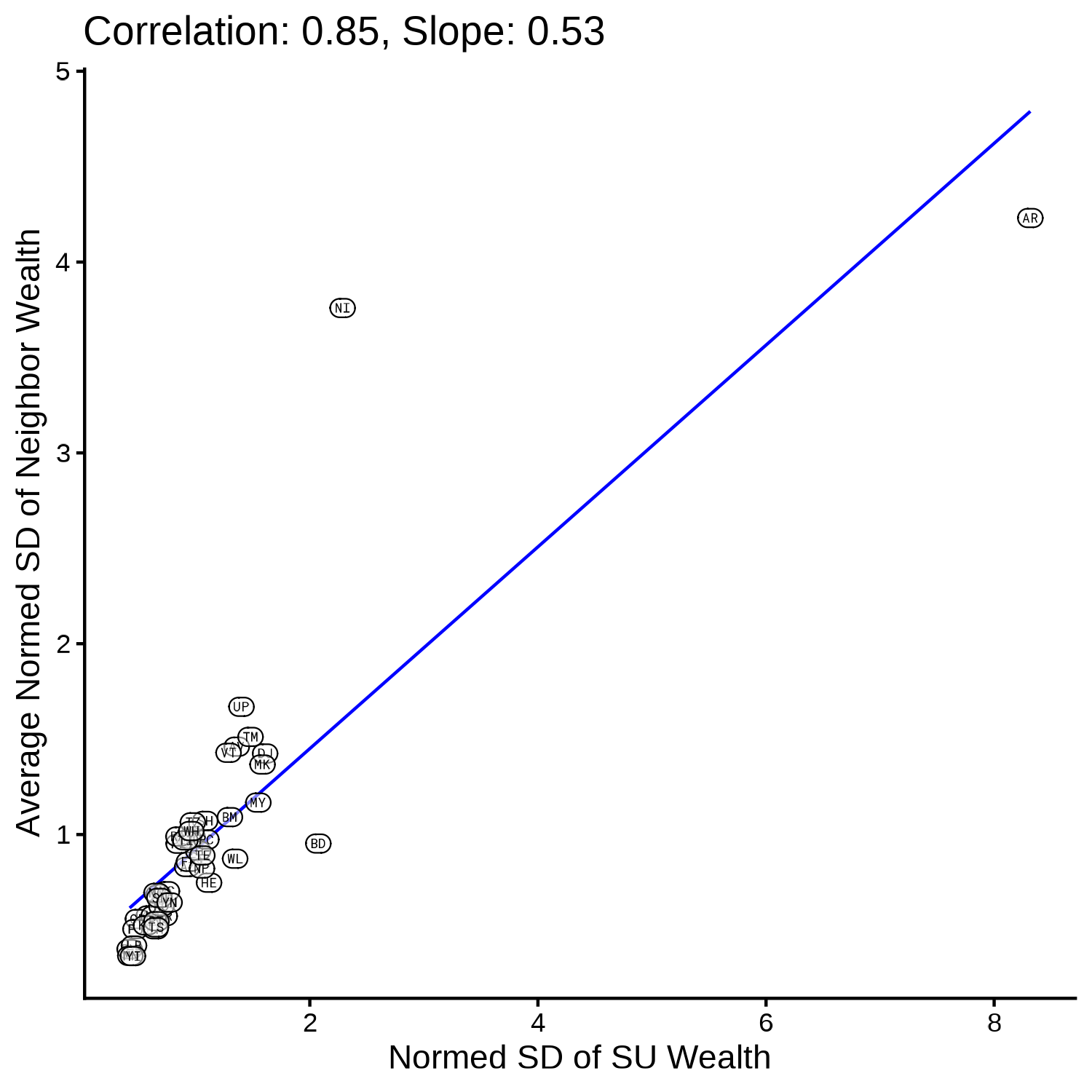}
\includegraphics[width=0.49\linewidth]{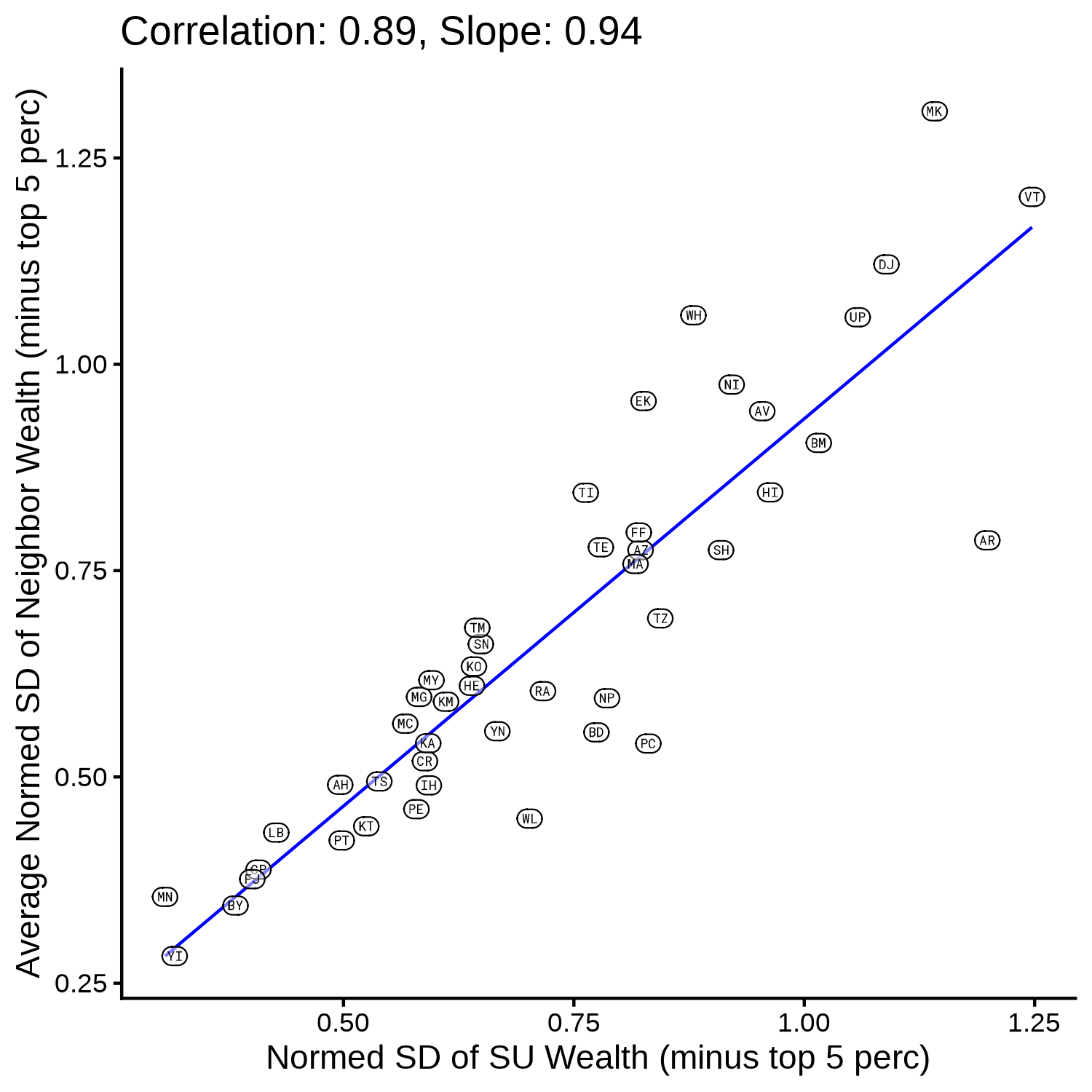}
\caption{Correlation between a community's standard deviation of sharing unit wealth (normalized by mean sharing unit wealth -- that is, a Coefficient of Variation) and the  average across a community's sharing units of the standard deviation of its alters' (i.e., the set of sharing units named as providing support) wealth, again normalized by mean wealth. The right plot is equivalent to the left, but removes sharing units in the top 5\% of the wealth distribution in each community. }
\label{fig:neighborwealth}
\end{figure}

As shown in Figure~\ref{fig:neighborwealth} we find that sharing units' standard deviation of alters' wealth grows with the standard deviation of wealth in the community, with a correlation of .85 ( .89 when omitting the top 5 percent of wealthiest sharing units). Thus, the association between economic connectedness and material wealth inequality is not driven in this mechanical manner.

To further rule out this mechanical channel, we simulate networks where sharing units formed connections in an absolute manner around their own wealth. For these simulations, we generate placebo networks separately within each community, holding fixed the observed set of sharing units, their wealth per capita, and each sharing unit's
observed out-degree (i.e., the number of alters named as providing support) in the ``composite'' network. For each community and each bandwidth rule, we run 500 simulations.

We consider two sets of rules: (i) ``fixed-band" rules and (ii) ``mean-scaled'' rules. Under the fixed-band rules, eligible alters for a sharing unit are all other
sharing units whose wealth per capita is within \$1,000, \$2,000, or \$5,000 of the ego sharing unit's wealth per capita. Under the mean-scaled rules, the same
procedure is used, but the band is set to 20\%, 40\%, or 60\% of the
community's mean wealth per capita. In each simulated network, we preserve each sharing
unit's observed number of outgoing ties, and sample alters
uniformly without replacement from the eligible set. If fewer alters fall within
the band than are needed to match the observed out-degree, all eligible alters
are included and the remaining ties are assigned to the nearest out-of-band
sharing units on the basis of absolute wealth-per-capita differences. We randomly permute observed edge weights for
each ego sharing unit across their simulated alters, preserving weighted nomination volumes.

We then recompute RAAW on each simulated network, using the same procedure as in the observed data.

\begin{figure}
\centering
  \includegraphics[width=0.49\linewidth]{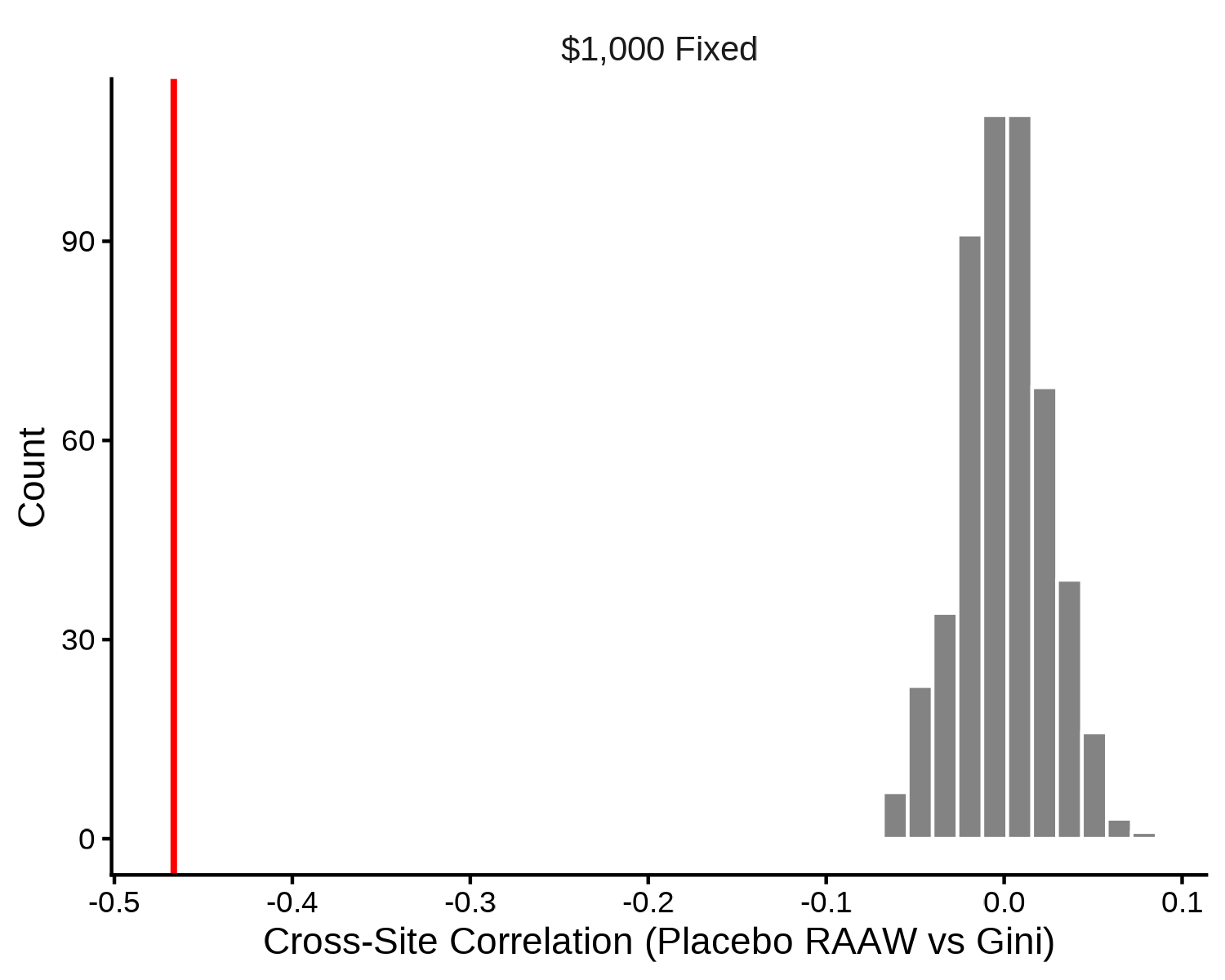} \\
  \includegraphics[width=0.49\linewidth]{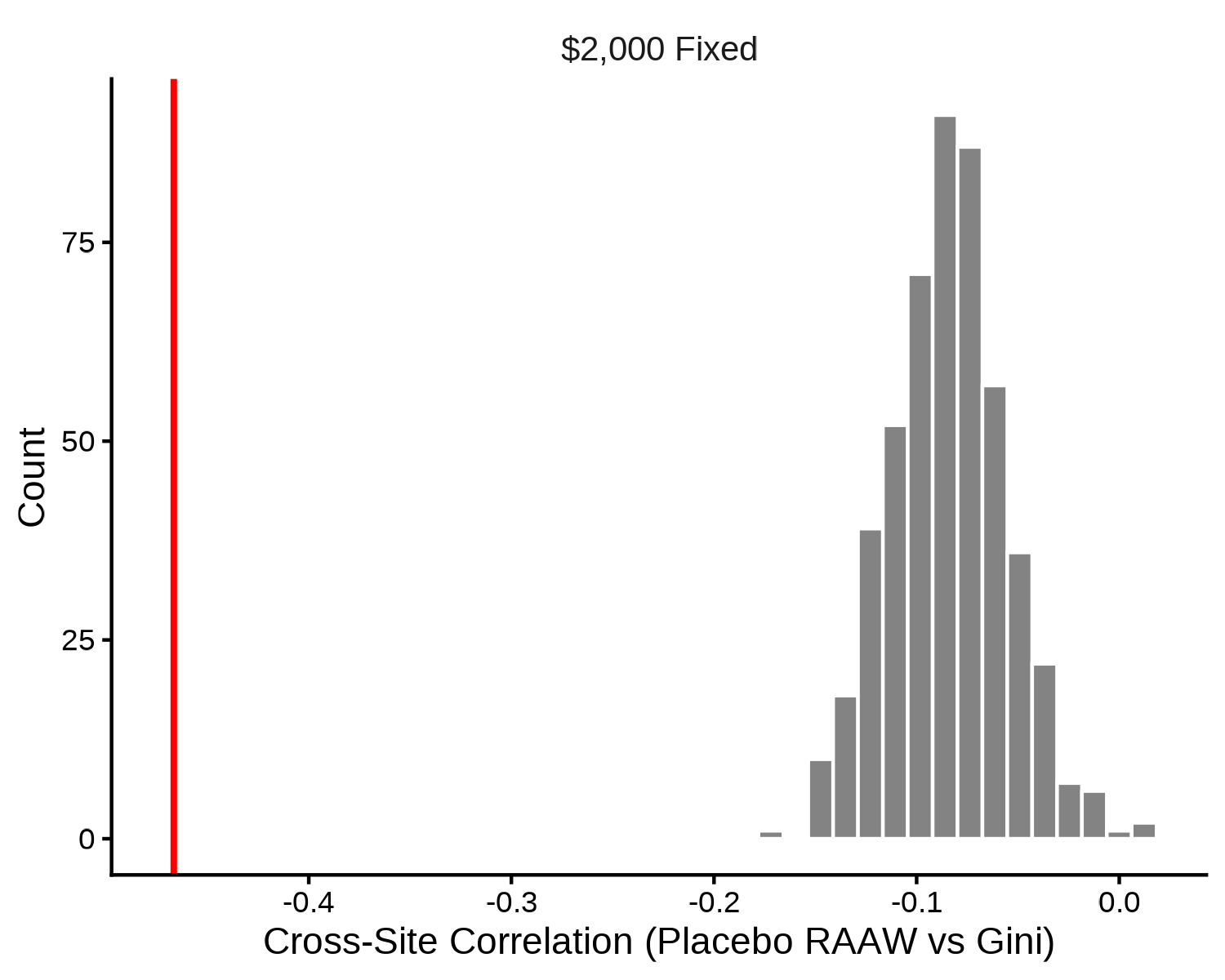} \\
  \includegraphics[width=0.49\linewidth]{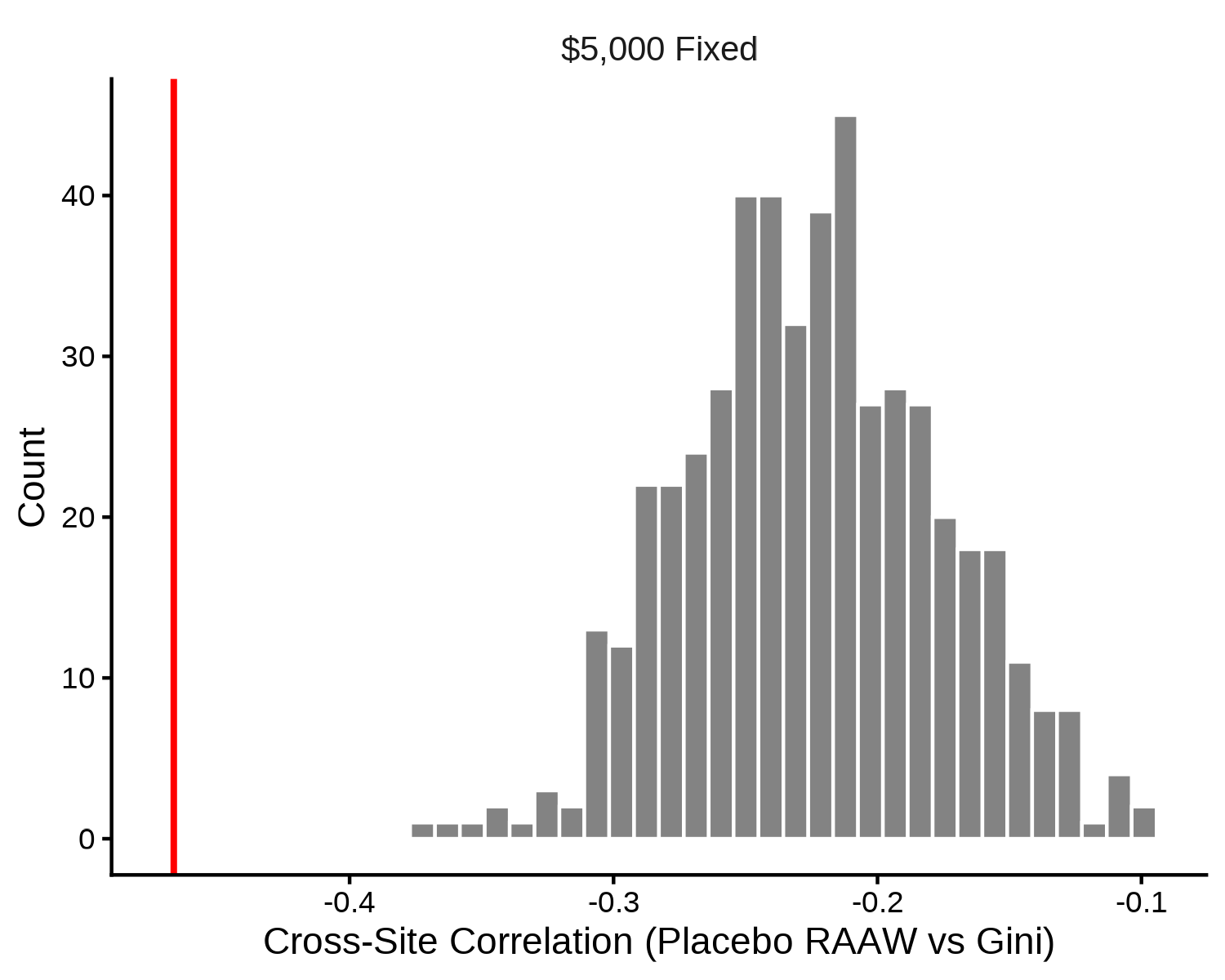}
\caption{\textbf{Distribution of cross-site correlations for simulated networks with absolute-wealth-band connecting rules.} The correlation we observe in the data is marked by the red line.}
\label{fig:friend-band}
\end{figure}

\begin{figure}
\centering
  \includegraphics[width=0.49\linewidth]{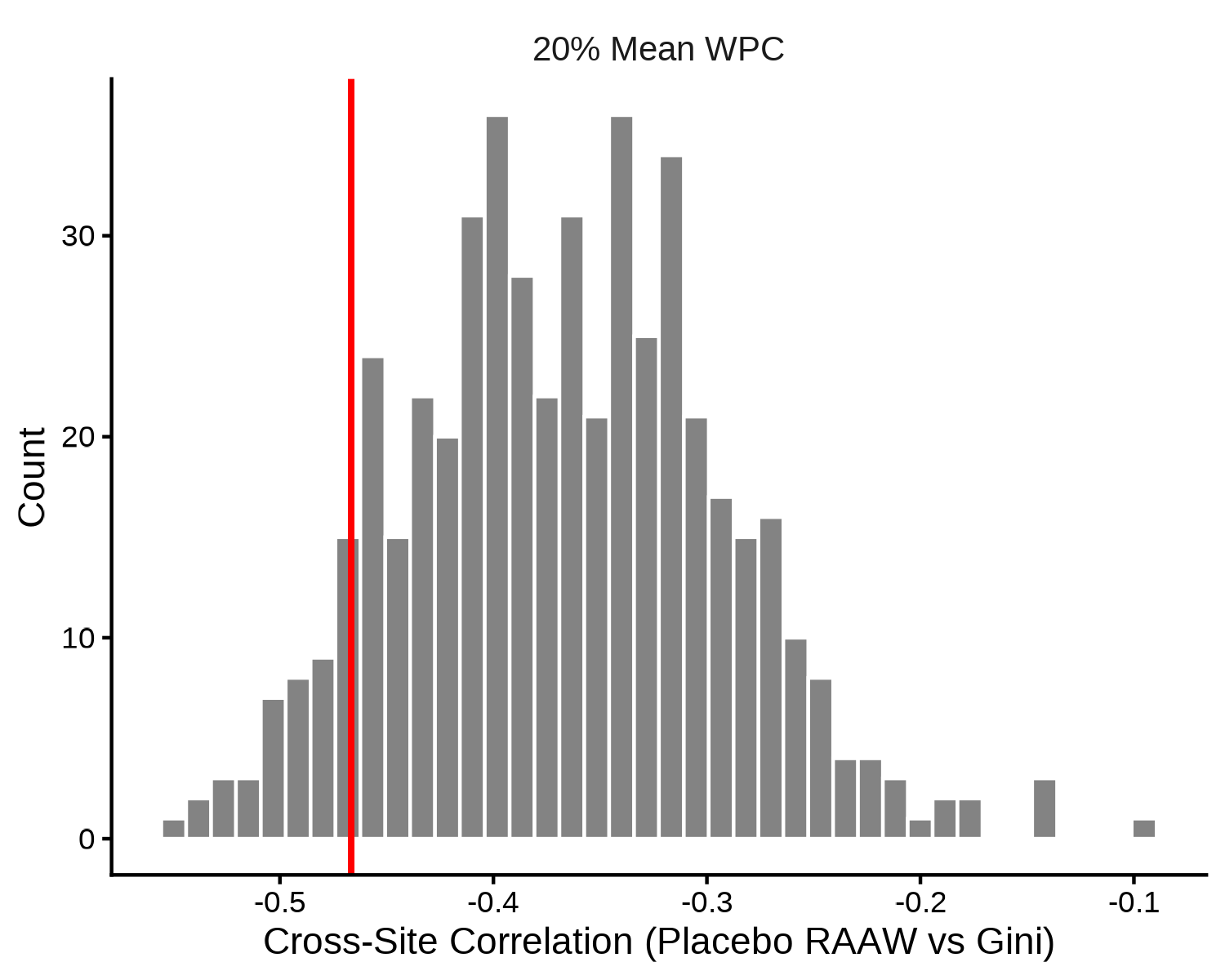} \\
  \includegraphics[width=0.49\linewidth]{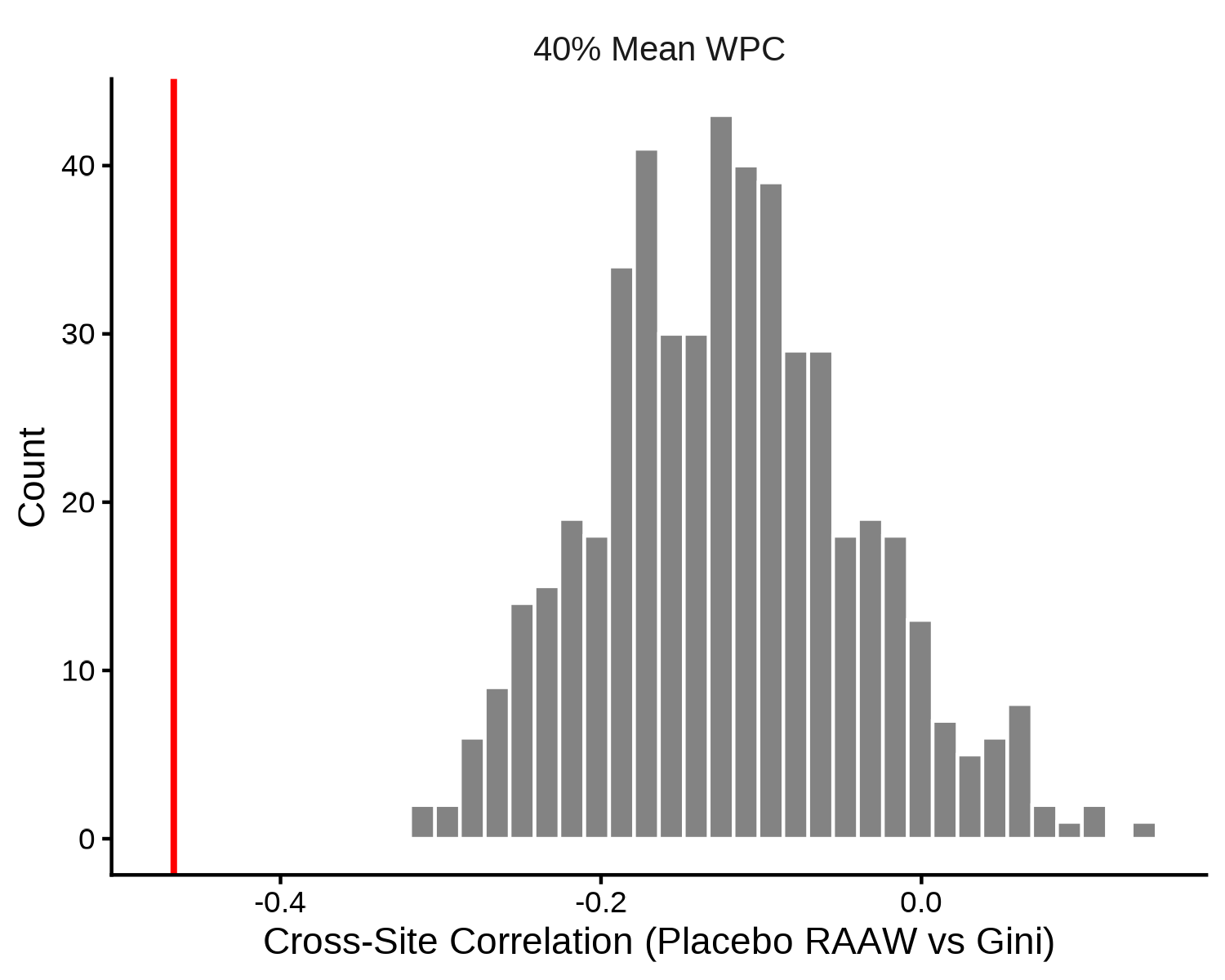} \\
  \includegraphics[width=0.49\linewidth]{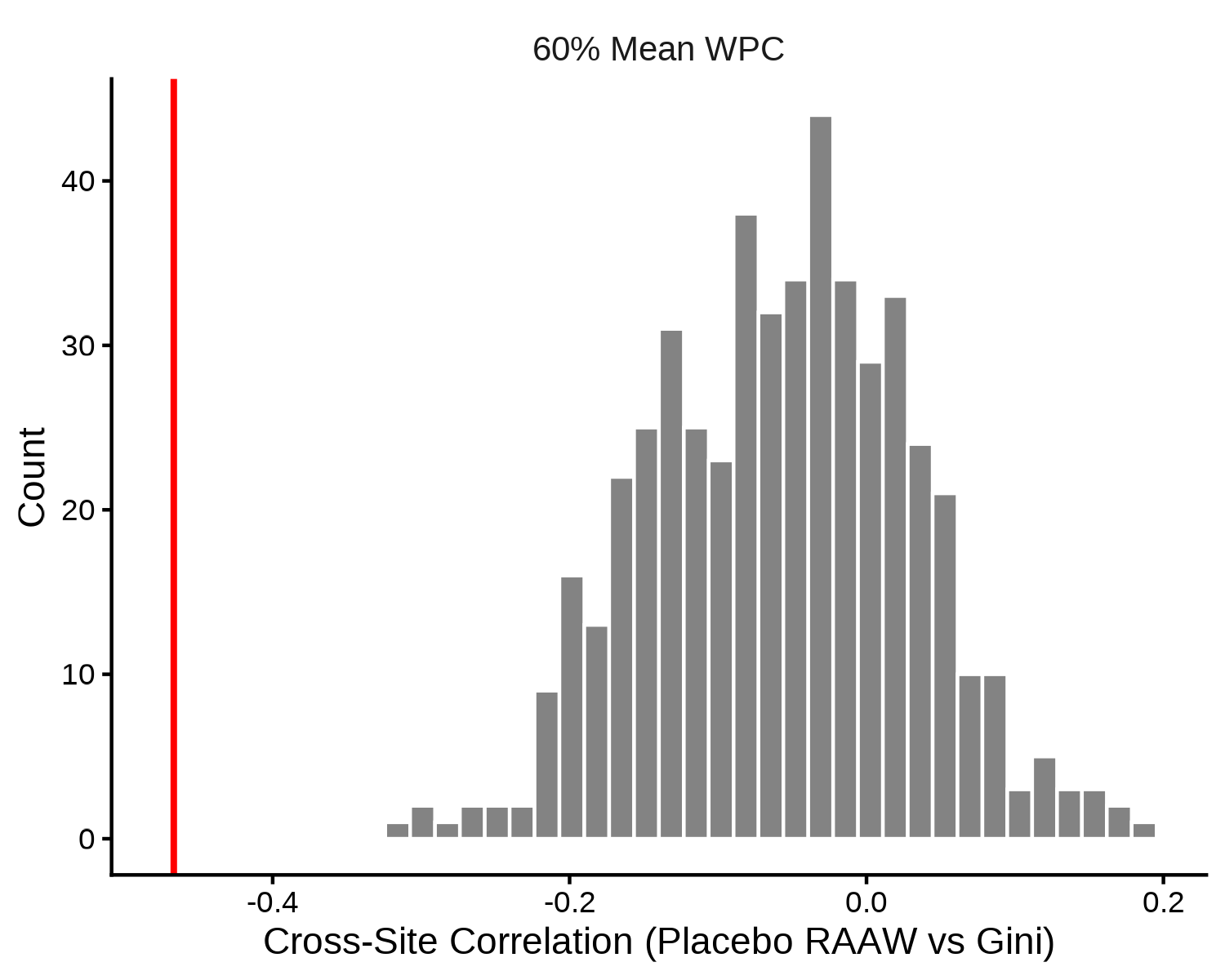}
\caption{\textbf{Distribution of cross-site correlations for simulated networks with within-band connecting rules, where the band varies by linearly by site with mean wealth per capita.} The correlation we observe in the data is marked by the red line.}
\label{fig:friend-band-relative}
\end{figure}

We see in Figure~\ref{fig:friend-band} that the correlations between RAAW and the Gini of wealth per capita on these simulated networks are generally distributed far from our observed correlation. For each bandwidth, the distribution displayed in
Figure~\ref{fig:friend-band} and Figure~\ref{fig:friend-band-relative} is the
distribution of the cross-site Pearson correlations between RAAW on the 500 simulated networks and
the observed wealth-per-capita Gini. The red line denotes the empirically observed correlation, and lies far from the distribution of simulated correlations in each case (with the exception of the 20\% band, though the mass of the distribution is still far off).

\begin{figure}
    \centering
    \includegraphics[width=0.9\linewidth]{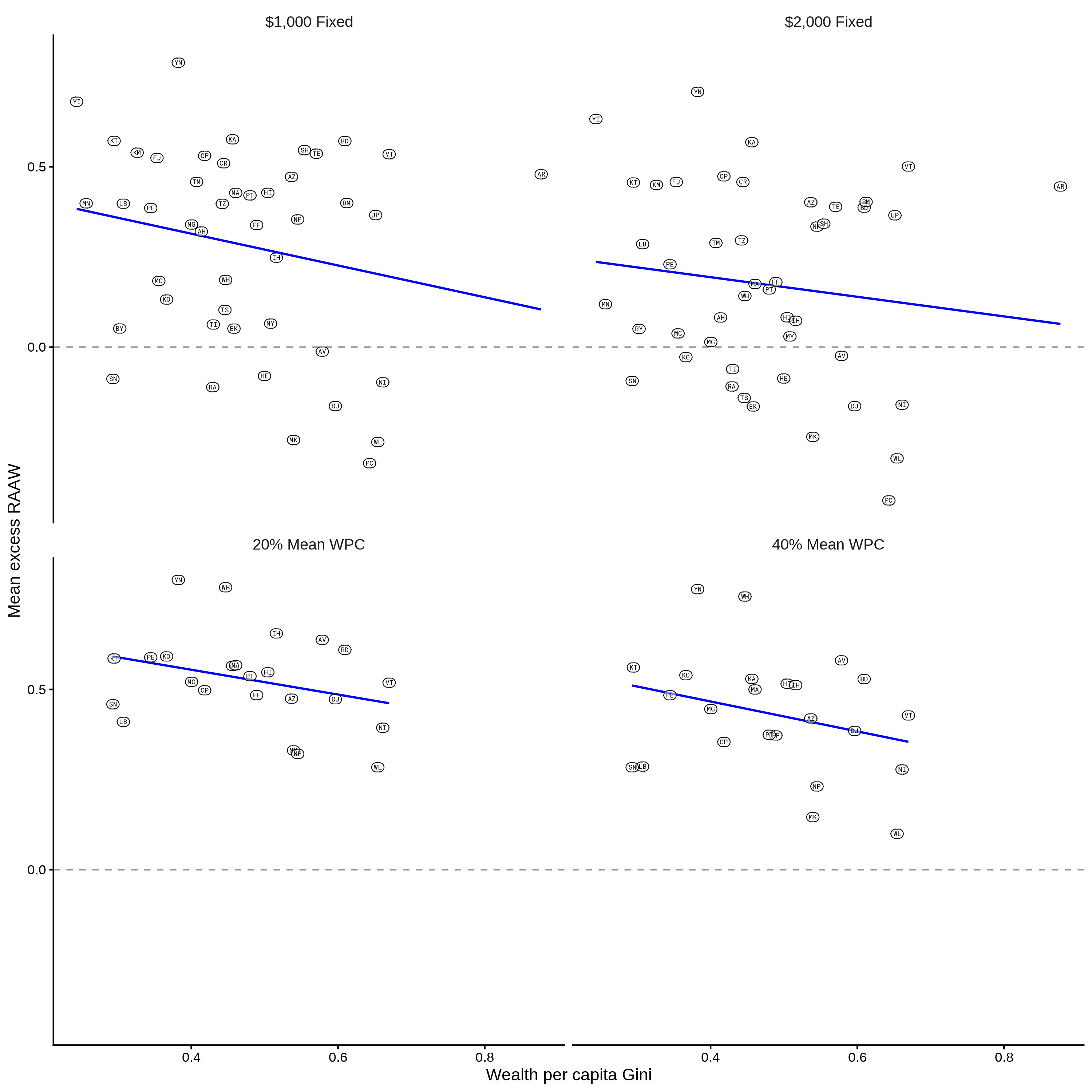}
    \caption{\textbf{Wealth per capita Gini versus mean excess RAAW.} Mean excess RAAW is the empirically observed RAAW minus the average simulated RAAW when sharing units follow a heuristic where they connect to sharing units within a particular wealth band of themselves.}
    \label{fig:excess-raaw}
\end{figure}

As a further check, we also construct a measure of mean excess RAAW beyond what we would expect to see under average-wealth-per-capita-scaled connecting rules. To do this, we take the empirically observed RAAW in each community and subtract the average RAAW under the 500 simulated networks. In Figure~\ref{fig:excess-raaw}, we see that the Gini of wealth per capita continues to be negatively correlated with this measure of excess RAAW.

\begin{comment}
\begin{figure}
\centering
  \includegraphics[width=0.6\linewidth]{figures/Average Normed SD of Neighbor Wealth.png}
\caption{Correlation between communities' material wealth Gini and the average across a community's sharing units of the standard deviation of its neighbors' (i.e., the set of sharing units named as providing support) wealth, normalized by mean wealth. The gray regression line includes all sites; the black regression line excludes NI and AR.}
\label{fig:neighborwealth_old}
\end{figure}
\end{comment}

\clearpage
\subsection{Relative Average Alter Wealth versus Assortativity}\label{supp:ec_vs_assort}

\begin{figure}[h]
\centering
\begin{subfigure}{0.48\textwidth}
    \centering
    \includegraphics[width=\textwidth]{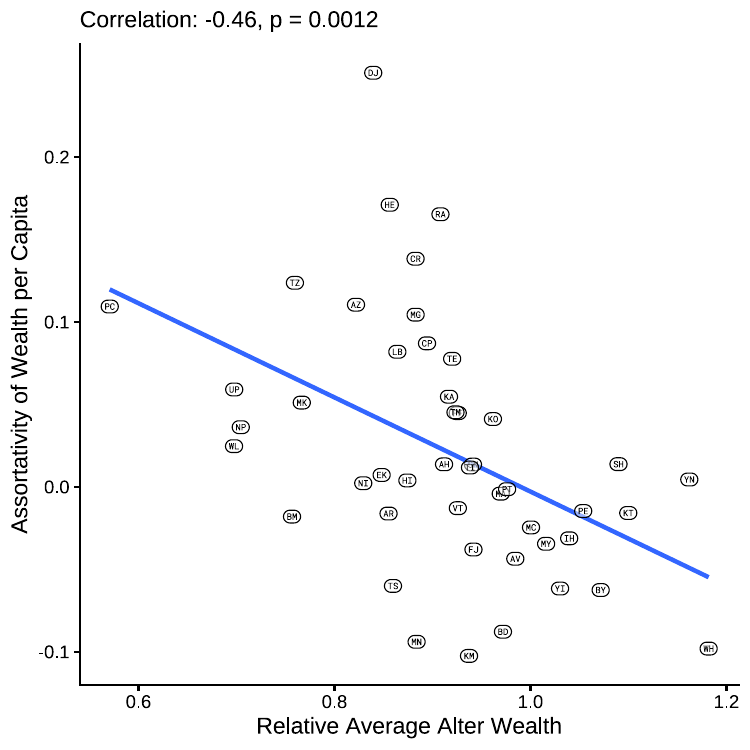}
\end{subfigure}
\hfill
\begin{subfigure}{0.48\textwidth}
    \centering
    \includegraphics[width=\textwidth]{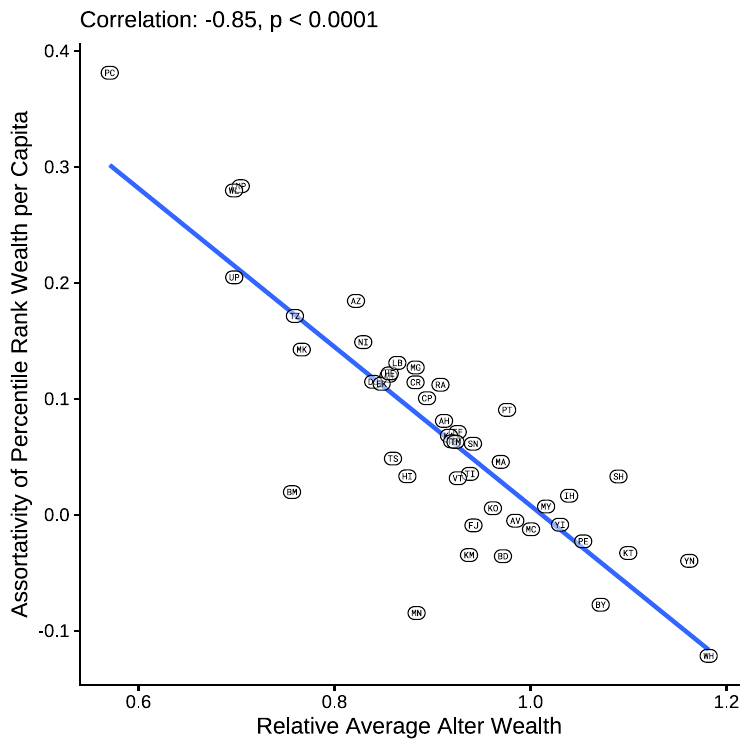}
\end{subfigure}
\caption{\textbf{Relative Average Alter Wealth vs Assortativity by Wealth.} Cross-site correlations between the Relative Average Alter Wealth (using wealth per capita) and assortativity of wealth per capita (left) and assortativity of percentile rank wealth per capita (right).}
\label{fig:raaw-vs-assort}
\end{figure}

We present our various measures of economic connectedness as measures of wealth homophily. Another commonly-used measure of homophily is assortativity \citep{newman_mixing_2003}, which constructs one overall measure for the extent to which vertices are connected with those that are alike (or unalike) in some way. In Figure~\ref{fig:raaw-vs-assort}, we show the bivariate associations between our Relative Average Alter Wealth measure (using wealth per capita) and assortativity on wealth per capita (left) and assortativity on percentile rank of wealth per capita (right). In both cases, we can see a clear, strong negative association between the two.

\clearpage
\subsection{Quartile Splits}\label{supp:quartiles}

\begin{figure}[ht]
    \centering
\begin{subfigure}{\textwidth}
        \centering
        \includegraphics[width=0.49\textwidth]{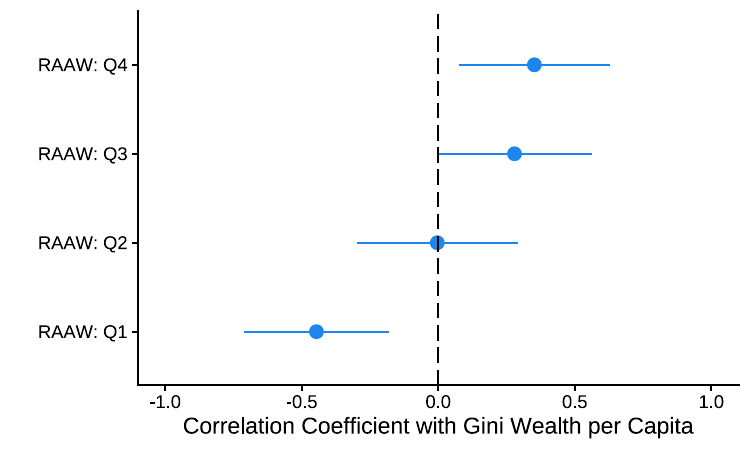}
        \includegraphics[width=0.49\textwidth]{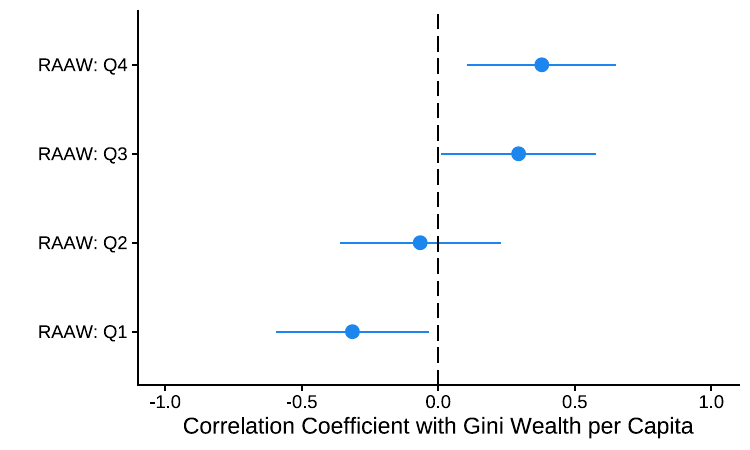}
        \caption{RAAW vs wealth per capita Gini, by ego quartile, based on (left) alters who provide support (i.e., using outgoing nominations that capture accessing support) and (right) alters to whom support is provided (i.e., using incoming nominations that capture provisioning support).}
    \end{subfigure}
    \vspace{1em} % Space between the top and bottom rows
    \begin{subfigure}{\textwidth}
        \centering
        \includegraphics[width=\textwidth]{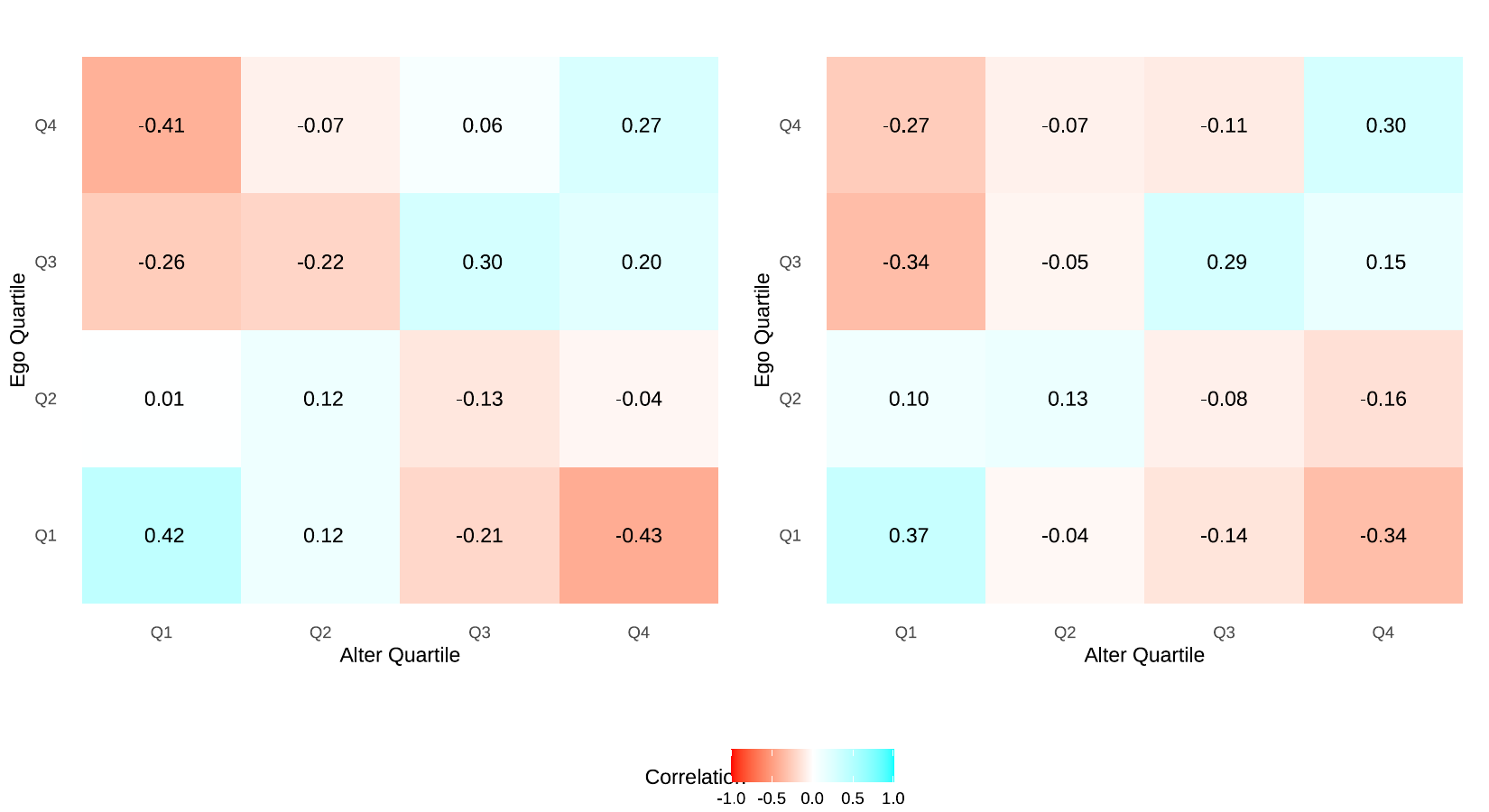}
        \caption{RC vs wealth per capita Gini, based on (left) alters who provide support and (right) alters to whom support is provided.}
    \end{subfigure}
\vspace{1em}
\caption{Correlations of each community's wealth per capita Gini with (a) each wealth quartile's Average Alter Wealth and (b) the relative frequency of quartile to quartile connections (Relative Connectedness). Q1 represents the poorest and Q4 the wealthiest quartile.
The RAAW here is normalized by the $\text{AAW}_i$ of all sharing units in the community. The relative connectedness of a quartile to another is calculated by normalizing the frequency of connections from quartile $x$ to alter quartile $y$ by the frequency of connections from all sharing units to alter quartile $y$. }
\label{fig:cross-site-two-panel}
\end{figure}

In the main text, we use the median split or the split driven by our wealth modularity calculations, effectively always splitting each community into two. But, it may be that the relevant wealth classes are more granular. Here, then, we split each community into wealth quartiles and calculate the Average Alter Wealth of each, as well as the relative frequency of connections across the quartiles.
We find that in communities where sharing units are relatively more connected to those in a similar wealth level (i.e., wealth homophily) there is greater wealth inequality; this is shown in Figure~\ref{fig:cross-site-two-panel} by the positive correlation coefficients for the wealthier quartiles and across the lower diagonal of the heatmaps (from bottom left to top right). Conversely, in communities where poorer sharing units are relatively more connected to wealthier sharing units, there is less material wealth inequality (as is shown in the negative correlation coefficients of the connectedness of the poorer quartiles to the wealthiest quartile). Connections from wealthier to poorer sharing units are similar in effect to connections from poorer to wealthier.

\clearpage
\subsection{Multilevel Model for Economic Connectedness}\label{supp:multilevel_model}

We construct a multilevel generative model of network formation based on wealth classes.
From the parameters of this model, we formulate approximate measures of economic connectedness.
We test the robustness of our findings in Section~\ref{sec:wealth_ineq_ec} to a different quantification of uncertainty by evaluating the posteriors of these measures.

For any $n \in \mathbb{N}$, we define $[n] = \{1,2,\ldots, n \}$. Label the total number of communities $J \in \mathbb{N}$, and the number of sharing units within each community as $n_j \in \mathbb{N}$ for $j \in [J]$. For community $j \in [J]$, we label the wealth per capita of sharing unit $i \in [n_j]$ as $y_{ij} \in \mathbb{R}$. Additionally, we categorize each sharing unit by their position in the wealth distribution of their community. For community $j \in [J]$ and sharing unit $i \in [n_j]$, $z_{ij} \in \{1,2,3,4\}$ is equal to the quartile of wealth that sharing unit $i$ belongs to. Lastly, for community $j \in [J]$, we label
%the size of sharing unit $i \in [n_j]$ as $s_{ij}$ and
the number of people surveyed in sharing unit $i \in [n_j]$ as $w_{ij}$.
%we can divide into more quantiles later
Label the adjacency matrix for community $j \in [J]$ as $\mathcal{A}^j$ such that $\mathcal{A}^j_{pq}$ is equal to the number of directed edges between sharing units $p$ and $q$ in community $j$. We model the probability of an edge between two sharing units as as a function of their positions in the community wealth distribution and the number of people surveyed in the ego sharing unit,
%and the size of the alter sharing unit
%in one of the following ways,
%\begin{align}
%    \mathcal{A}^j_{pq} \sim \mathrm{Poisson} \left [ \exp\left \{ \sum_{k = 1}^4 \sum_{l = 1}^4 \beta_{jkl} \mathbb{I}(z_{pj} = l) \mathbb{I}(z_{qj} = k) + \gamma_j w_{pj}    \right \} \right ], \ \mathrm{for} \ j \in [J], p,q \in  [n_j],
%    \label{eq:ecmodel2}
%\end{align}
%or,
\begin{align}
    \mathcal{A}^j_{pq} \sim \mathrm{Poisson} \left [ \exp\left \{ \sum_{k = 1}^4 \sum_{l = 1}^4 \beta_{jkl} \mathbb{I}(z_{pj} = l) \mathbb{I}(z_{qj} = k)w_{pj}    \right \} \right ], \ \mathrm{for} \ j \in [J], p,q \in  [n_j],&
    \label{eq:ecmodel3}
\end{align}
where $\bbeta_j \triangleq \left [\left \{ \beta_{jkl} \right \}_{k,l = 1}^4 \right ] \in \mathbb{R}^{16}$, $\gamma_j \in \mathbb{R}$,
%$\etab_j = \left (\eta_{1j}, \eta_{2j} \right ) \in \mathbb{R}^2$
and ``$\mathrm{exp}$" is the natural exponential function. We also assume a random effects structure for the regression coefficients,
%This model analyzes how the probability of directed connections between sharing units depends on the wealth and sizes of the two sharing units. %$p,q \in \left \{1,2, \ldots , n_j \right \}$ We model the site level coefficients as random effects,
\begin{align*}
    &\bbeta_j^\top \sim \mathrm{Normal} \left ( \mu_\bbeta, \Sigma_{\bbeta} \right ), \ \mathrm{for} \ j \in  [J], \\
    &\gamma_j \sim \mathrm{Normal} \left ( \mu_{\gamma}, \sigma_{\gamma} \right ), \ \mathrm{for} \ j \in [J], %\\
    %&\etab_j \sim \mathrm{Normal} \left ( \mu_{\etab}, \Sigma_{\etab} \right ), \ \mathrm{for} \ j \in [J],
\end{align*}
where $\mu_{\gamma},\sigma_{\gamma} \in \mathbb{R}$, % $\mu_{\etab} \in \mathbb{R}^2$, $\Sigma_{\etab} \in \mathbb{R}^{2\times 2}$,
$\mu_\bbeta \in \mathbb{R}^{16}$, and $\Sigma_{{\bbeta}} \in \mathbb{R}^{16 \times 16}$.
Suppose we are interested in the relationship between wealth inequality and the connectedness of quartile $k_*$ to $l_*$ across community.
For each community, we assume there are  $C \in \mathbb{N}$ site-level confounders.
%from the survey and other sources
We label the confounders for community $j \in [J]$ as $\bx_{j} \in \mathbb{R}^C$ and the Gini coefficient as $g_j$. We model the Gini coefficient as
%\begin{equation*}
%        \log \left (y_{ij} \right ) \sim \mathrm{Normal} \left (\mu_j, \sigma^2_j \right ), \ \mathrm{for} \ j \in [J], \ \mathrm{for} \ i \in [n_j],
%\end{equation*}
%\vspace{-0.3cm}
\begin{equation}
    \label{equation:eqreg}
    \frac{\mathrm{exp}(g_j)}{1+ \mathrm{exp}(g_j)} \sim \mathrm{Normal} \left ( \alpha_0 + \alpha_1 \frac{\beta_{jk_*l_*}}{\sum_{l=1}^{4} \beta_{jk_*l} }  + {\balpha}_2^\top \bx_j, \sigma_{\balpha}^2 \right ), \ \mathrm{for} \ j \in [J],
\end{equation}
where $\{ \mu_j \}_{j=1}^J, \{ \sigma^2_j \}_{j=1}^J, \alpha_0, \alpha_1, \sigma_{\balpha}^2  \in \mathbb{R}$ and $\balpha_2 \in \mathbb{R}^C$. Note that $\beta_{jk_*l_*}/ \left ( \sum_{l=1}^{4} \beta_{jk_*l} \right )$ is our approximation to the economic connectedness between quartiles $k_*$ and $l_*$.
Equation~\ref{equation:eqreg} relates wealth inequality to economic connectedness (quantified by $\alpha_1$)
while accounting for site-level confounders.

\begin{figure}
    \centering
    \includegraphics[width=\linewidth]{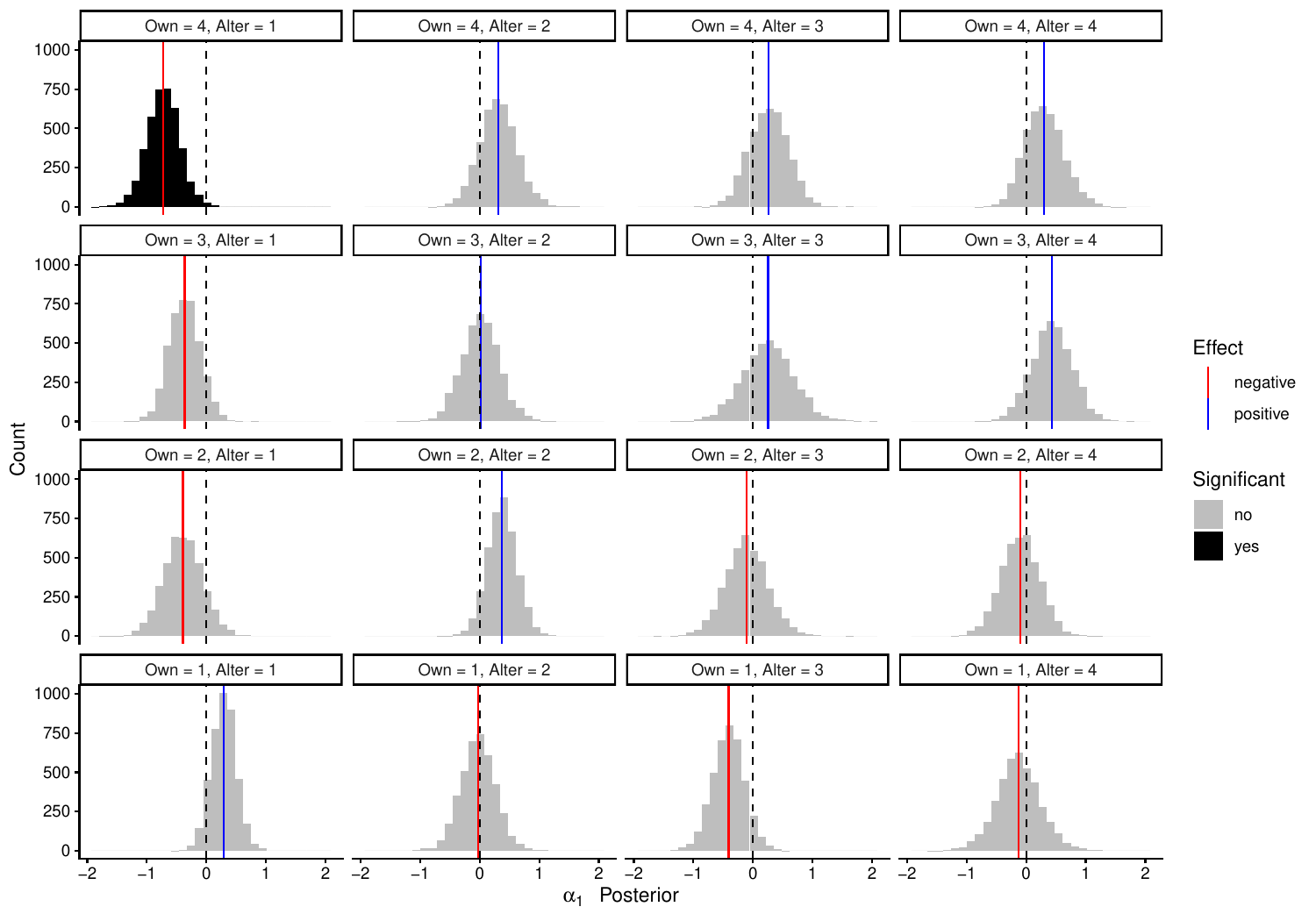}
    \caption{Model-based reproduction of quartile economic connectedness, showing the model posteriors across the wealth quartile combinations, using wealth per capita. 1 represents the poorest and 4 the wealthiest quartile. }
    \label{fig:multilevel_posteriors}
\end{figure}

In Figure~\ref{fig:multilevel_posteriors}, we show the resulting model posteriors, using wealth per capita. In the top left, we can see that communities where the wealthiest quartile is better connected to the poorest quartile have lower wealth inequality. %Looking at the top right, we can see that communities where the wealthiest quartile is better connected to itself generally have greater wealth inequality, though this effect is less clear.

\paragraph{Priors and computation}
We take $\Sigma_\bbeta$ to be diagonal, $\Sigma_\bbeta = \mathrm{diag}\left(\sigma_{\bbeta,1}^2, \ldots, \sigma_{\bbeta,16}^2\right)$; that is, we model the $\beta_{jkl}$ as independent across quartile pairs $(k,l)$ conditional on the hyperparameters. We adopt this simplification because our data include $L \approx 50$ communities, too few to reliably identify the $\binom{16}{2} = 120$ off-diagonal correlations of an unstructured $16 \times 16$ covariance matrix; in preliminary analyses, an unstructured $\Sigma_\bbeta$ produced unstable posterior geometry (frequent divergent transitions) and did not mix reliably even under a shrinkage (LKJ) prior on the correlation matrix. We give each scale $\sigma_{\bbeta,d}$ an independent half-Cauchy prior, $\sigma_{\bbeta,d} \sim \mathrm{Cauchy}^+(0, 2.5)$, for $d = 1, \ldots, 16$. For each $d$, the components of $\mu_\bbeta$ are given independent priors $\mu_{\bbeta,d} \sim \mathrm{Normal}(0, 100^2)$. We further set $\alpha_0 \sim \mathrm{Normal}(0, 100^2)$, $\alpha_1 \sim \mathrm{Normal}(0, 100^2)$, and $\sigma_\balpha \sim \mathrm{Cauchy}^+(0, 2.5)$.

We fit the model separately for each of the $16$ choices of focal quartile pair $(k_*, l_*)$. Posterior inference is carried out via Hamiltonian Monte Carlo (the No-U-Turn sampler) as implemented in Stan (CmdStan 2.39.0, via \texttt{cmdstanr}), running $4$ chains of $1{,}000$ warmup and $1{,}000$ sampling iterations each, yielding $4{,}000$ post-warmup draws per model. We use a target average proposal acceptance probability of $0.95$ and a maximum treedepth of $10$. For computational efficiency given the size of the edge-level dataset ($N \approx 3.5 \times 10^5$ directed pairs per model), $\bbeta_j$ is sampled using a non-centered parameterization, the one-hot quartile-pair design matrix is replaced with an equivalent categorical index to avoid redundant computation, and the edge-level Poisson log-likelihood is evaluated using within-chain multithreading (\texttt{reduce\_sum}, $2$ threads per chain).

Across all $16$ fits, we observed no divergent transitions and $\hat{R} \leq 1.004$ for all reported parameters, with effective sample sizes for $\alpha_1$ exceeding $1{,}100$ (bulk) and $1{,}300$ (tail) in every case out of $4{,}000$ post-warmup draws. A subset of fits exhibited a non-trivial proportion of transitions reaching the maximum treedepth (sometimes around $25\%$), indicating reduced sampling efficiency for some quartile pairs; as this did not co-occur with divergences or with  elevated $\hat{R}$ or reduced effective sample size, we do not consider it to  compromise the reliability of the reported posterior summaries.

%\begin{figure}
%    \centering
%    \includegraphics[width=0.65\linewidth]{figures/Hierarchical Model/Num_Surv_Effect.pdf}
%    \caption{This depicts the variability of the $\eta_{1j}$ in Equation~\ref{eq:ecmodel1}.}
%    \label{fig:heteta1}
%\end{figure}
%\begin{figure}
%    \centering
%    \includegraphics[width=0.65\linewidth]{figures/Hierarchical Model/HH_Size_Effect.pdf}
%    \caption{This depicts the variability of the $\eta_{2j}$ in Equation~\ref{eq:ecmodel1}. The effect of household size is much less variable across sites than the effect of number surveyed.}
%    \label{fig:heteta2}
%\end{figure}
%Lastly, we model the mean parameter of the income distribution for each site as a random effect,
%\begin{equation*}
%    \mu_j \sim \mathrm{Normal} \left ( \mu, \sigma_{\mu}^2 \right ),  \ \mathrm{for} \ j \in [J].
%\end{equation*}

%Figure~\ref{fig:hetgamma} depicts the heterogeneity in the effect of household size on link probabilities, $\gamma_j$ for $j \in [J]$ (coefficients from Equation~\ref{eq:ecmodel1}).

%\begin{figure}
%    \centering
%    \includegraphics[width=0.8\linewidth]{figures/Hierarchical Model/HH_Size_Effect.pdf}
%    \caption{There is a lot of variability in the effect of household size on the probability of connections between sharing units.}
%    \label{fig:hetgamma}
%\end{figure}

\clearpage
\subsection{Wealth Modularity}\label{supp:wealth_modularity}

In this subsection, we construct an additional measure of how segregated the network is on the basis of material wealth, and relate it to measures of material wealth inequality.

Label the total number of sites $J \in \mathbb{N}$, and the number of sharing units within each site as $n_j \in \mathbb{N}$ for $j \in [J]$. Define adjacency matrix $\mathcal{A}^j$ such that $\mathcal{A}^j_{vw}$ is $1$ if there is a directed edge between sharing units $v$ and $w$ in community $j$ and $0$ if not. Additionally, define the degree of sharing unit $v \in [n_j]$ as $k_v \triangleq \sum_{w=1}^{n_j}\mathcal{A}^j_{vw}$ and the total number of edges in $\mathcal{A}^j$ as $m \triangleq \sum_{v =1}^{n_j} \sum_{w=1}^{n_j} \mathcal{A}^j_{vw}$.
If the edges between sharing units were determined at random given their degree, then the probability of an edge between sharing units $v$ and $w$ would be $k_v k_w/2m$.
For classes $\{1,2,\ldots , K\}$, we define $\mathbf{c}$ as a class membership vector such that for $v \in [n_j]$, $c_v = k$ if sharing unit $v$ belongs to class $k$. We evaluate the correspondence between $\mathbf{c}$ and $\mathcal{A}^j$ by comparing the density of edges within to the density of edges between the communities specified by $\mathbf{c}$.
This comparison can be quantified using the modularity metric \citep{newman_network_2018}, which is defined as the number of edges falling within communities minus the expected number of edges if they were placed at random given sharing units' degrees:
\begin{equation*}
    Q_{\mathbf{c}}^j = \frac{1}{2m} \sum_{v=1}^{n_j} \sum_{w =1}^{n_j} \left [A^j_{vw} - \frac{k_v k_w}{2m} \right ] \delta (c_v, c_w),
\end{equation*}
where $\delta(c_v, c_w)$ is $1$ if sharing units $v$ and $w$ are in the same community and $0$ if not.

We are interested in whether and to what extent wealth class network structure exists in sites. For $v \in [n_j]$, we define $y_{vj}$ as the wealth per capita of sharing unit $v$ and $r_{vj} = \sum_{w = 1}^{n_j} \mathbb{I} \left (y_{vj} \geq y_{wj} \right )$ as the rank of sharing unit $v$ in the wealth distribution. To determine the presence of wealth class structure, we focus on $K=2$ and evaluate the set of community structures represented by $\mathcal{C}^j = \left \{ \mathbf{c}^j(v): v \in [n_j] \right \}$ such that $c^j_w(v) = \mathbb{I}\left (r_{wj} \geq r_{vj} \right )$. This allows us to compare modularity at every possible division of the wealth distribution. The maximum wealth modularity and most modular wealth division of site $j$ are therefore $W^j \triangleq \max_{\mathbf{c} \in \mathcal{C}^j} Q_{\mathbf{c}}^j$ and $\mathbf{c}_*^j \triangleq \arg \max_{\mathbf{c} \in \mathcal{C}^j} Q_{\mathbf{c}}^j$ respectively. The magnitude of $W^j$ will allow us to evaluate the significance of the wealth classes specified by $\mathbf{c}_*^j$.
%Defining $\mathbf{s}^j = [s_1, s_2, \ldots, s_{n_j}]$, we make the dependence of $W_j$ on the wealth distribution explicit by denoting $W^j = W^j(\mathbf{s}^j)$.

\begin{figure}
    \centering
    \includegraphics[page=1, width=\linewidth]{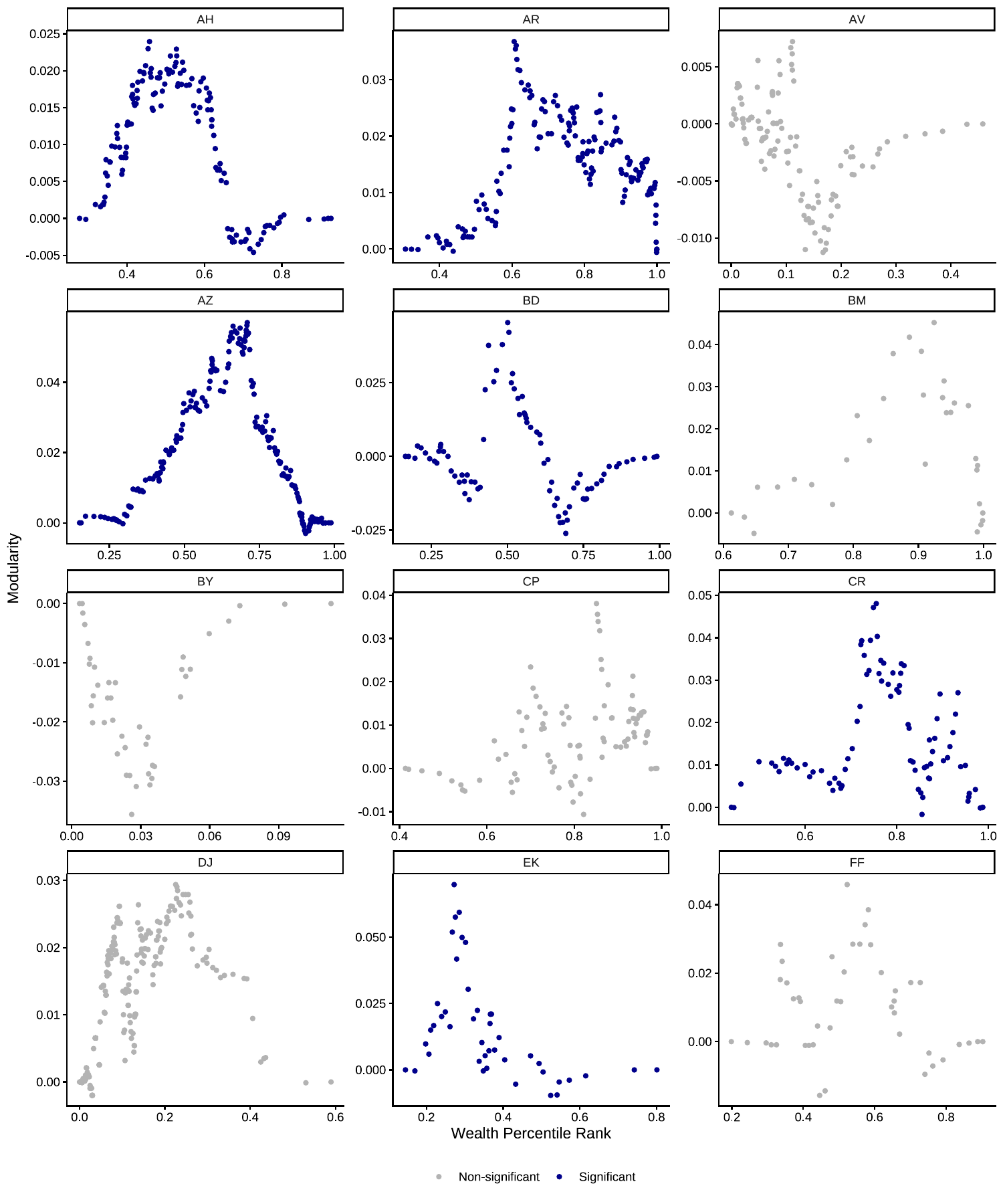}
    \caption{Modularity evaluated at every possible wealth class division for each community, done on the basis of wealth per capita. Dark blue denotes communities with a significant maximum modularity.}
    \label{fig:sig_wealth_mod_1}
\end{figure}

\begin{figure}
    \centering
    \includegraphics[page=2, width=\linewidth]{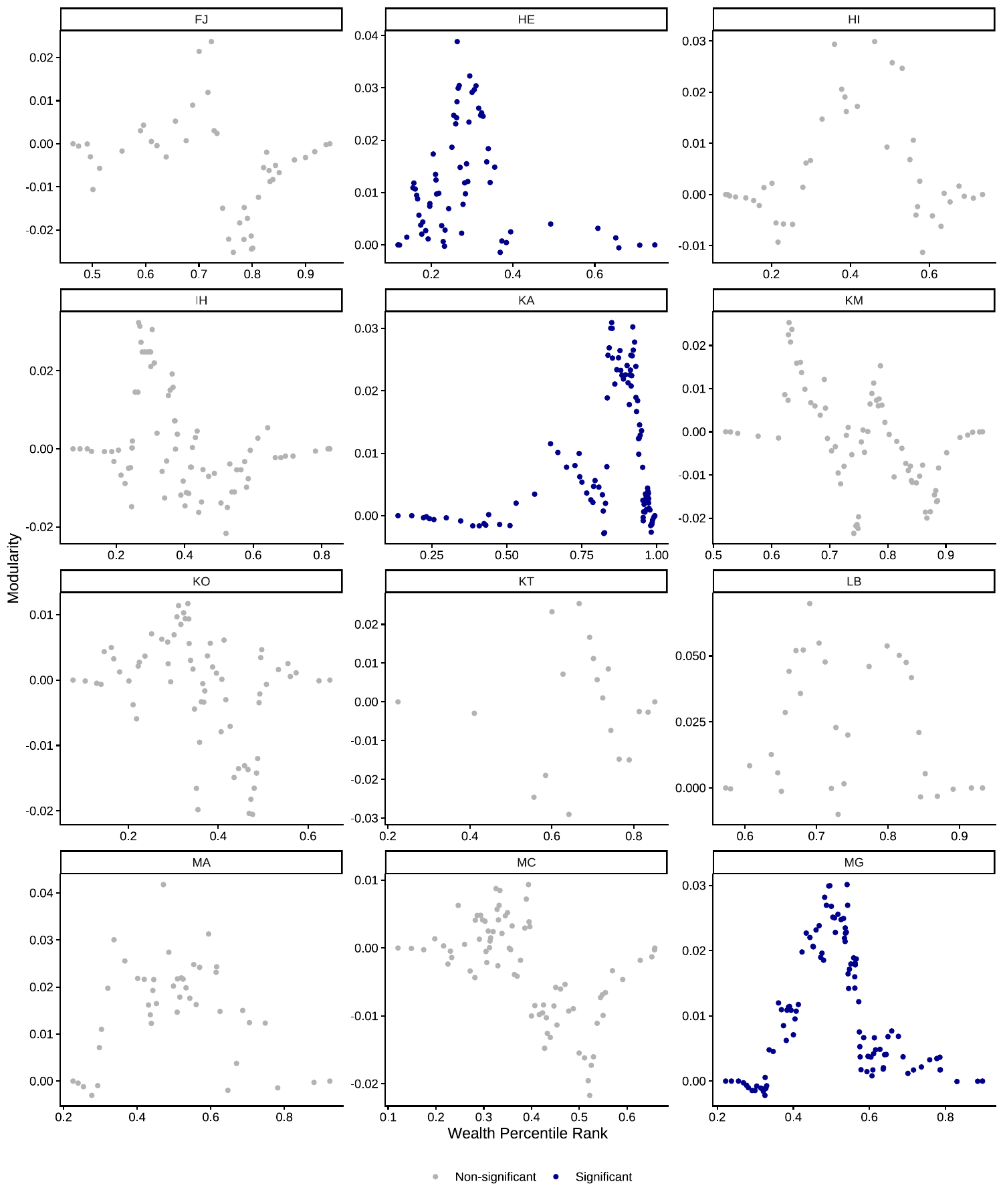}
    \caption{Modularity evaluated at every possible wealth class division for each community, done on the basis of wealth per capita. Dark blue denotes communities with a significant maximum modularity.}
    \label{fig:sig_wealth_mod_2}
\end{figure}

\begin{figure}
    \centering
    \includegraphics[page=3, width=\linewidth]{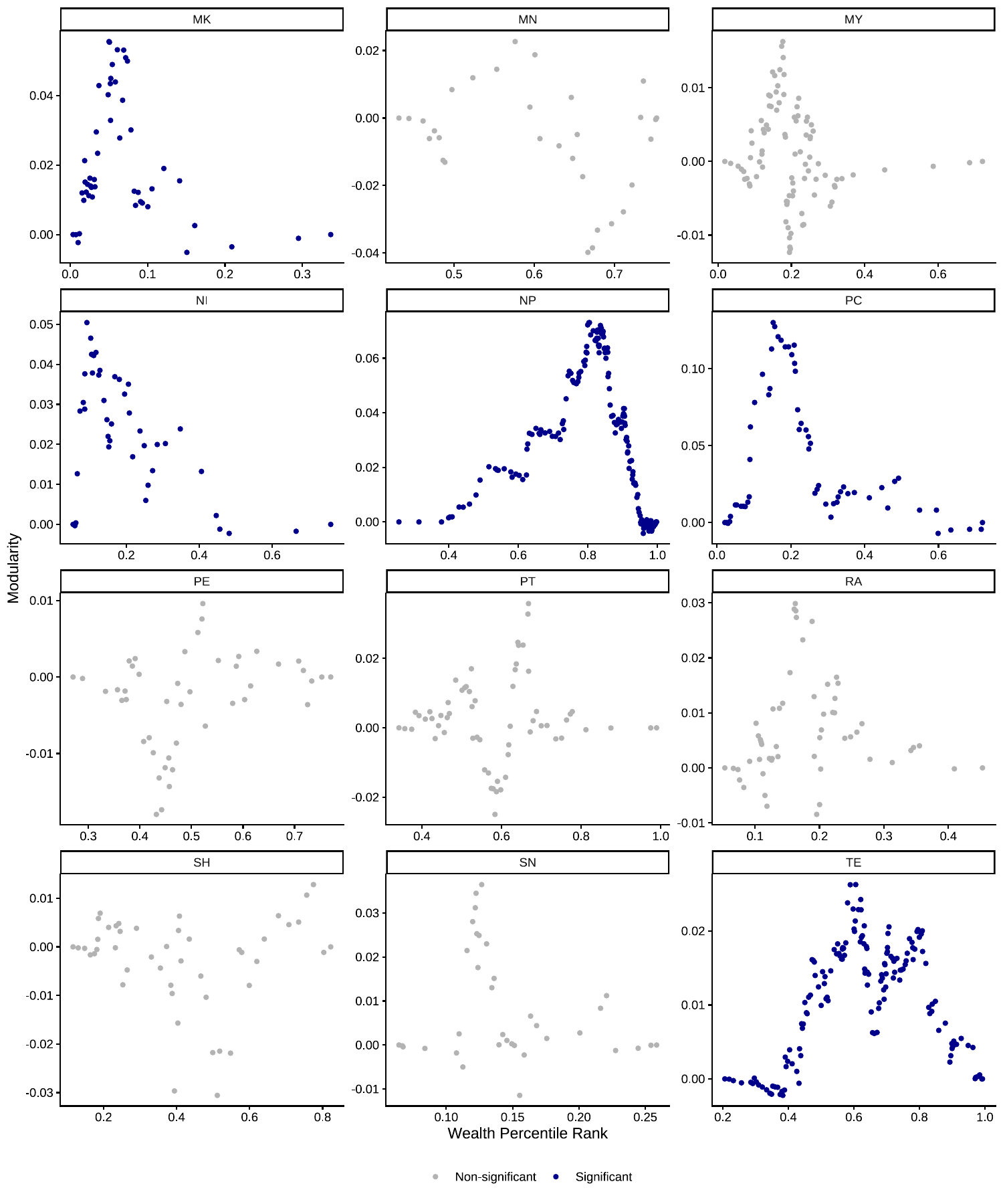}
    \caption{Modularity evaluated at every possible wealth class division for each community, done on the basis of wealth per capita. Dark blue denotes communities with a significant maximum modularity.}
    \label{fig:sig_wealth_mod_3}
\end{figure}

\begin{figure}
    \centering
    \includegraphics[page=4, width=\linewidth]{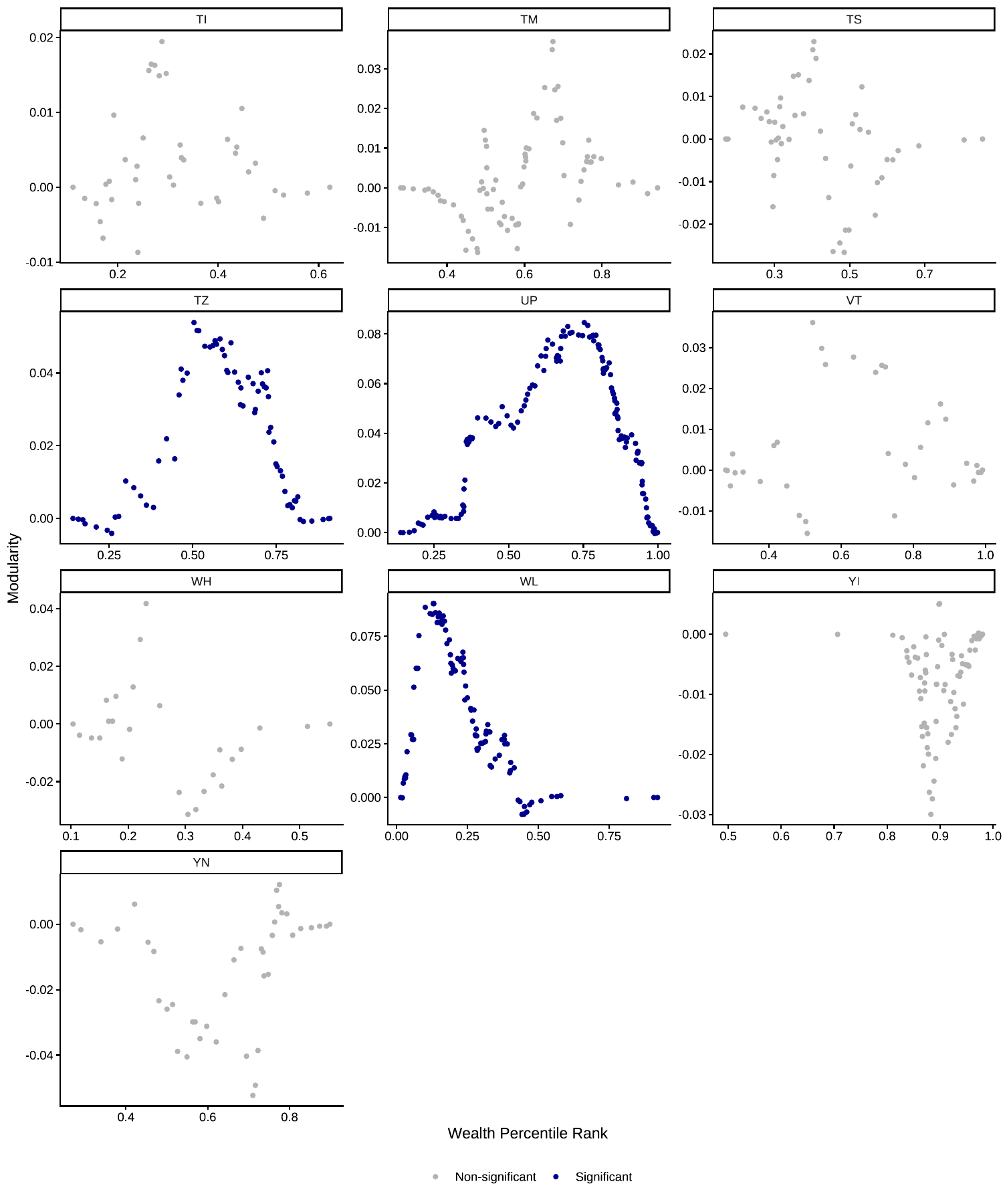}
    \caption{Modularity evaluated at every possible wealth class division for each community, done on the basis of wealth per capita. Dark blue denotes communities with a significant maximum modularity.}
    \label{fig:sig_wealth_mod_4}
\end{figure}

To determine whether the wealth class structure of $\mathbf{c}_*^j$ is significantly different from what we would expect under randomness, we compare $W^j$ to its null distribution under a permutation of sharing unit wealth.
%If we label $\sigma$ as a permutation function, we determine
%\begin{equation*}
%     W^j \left \{ \sigma (\mathbf{s}^j) \right \}.
%\end{equation*}
Figures~\ref{fig:sig_wealth_mod_1}-\ref{fig:sig_wealth_mod_4} depicts the wealth modularity at each possible wealth class division (every member of $\mathcal{C}^j$) for all the communities, denoting those for which the maximum modularity is significant. Lastly, we wish to determine if there is an association between wealth class modularity and wealth inequality. To facilitate a comparison of community structure across sites, we ``normalize" the modularity measure by subtracting the 95\% quantile of the null distribution from $W^j$ in each site. As shown in the bottom right panel of Figure~\ref{fig:wealth-ginis-vs-scatters} in the main text, we see that there is a strong association between this normalized maximum modularity metric and the Gini coefficient of each site.

\clearpage
\section{Consequences of Wealth Mis-measurement}\label{sec:wealth-sensitivity}

Although we took great care to accurately tally up and value the assets of every sharing unit across the communities, our material wealth estimates are not perfect. For example, we often have to make judgments on how many units of one particular asset have equivalent value to a unit of another asset in the absence of market prices (since these assets are not frequently traded or marked to market). Additionally, it is possible that we do not include every single asset a sharing unit owns that constitutes wealth, broadly defined. As a result, our wealth estimates for each sharing unit may be viewed as imperfect proxies for the ``true wealth'' of a sharing unit.

In an ideal world, we would be able to run some kind of cross-validation on our wealth estimates against a sharing unit's true wealth. However, since our wealth estimates are the best estimates of wealth for each sharing unit available, this approach is infeasible.

However, we expect that any measurement issues would be more impactful in lower-wealth communities than in high-wealth communities, since when the absolute level of wealth is lower missing a single asset has more of a relative effect.

\begin{figure}[h]
\centering
\begin{subfigure}{0.49\textwidth}
    \centering
    \includegraphics[width=\linewidth]{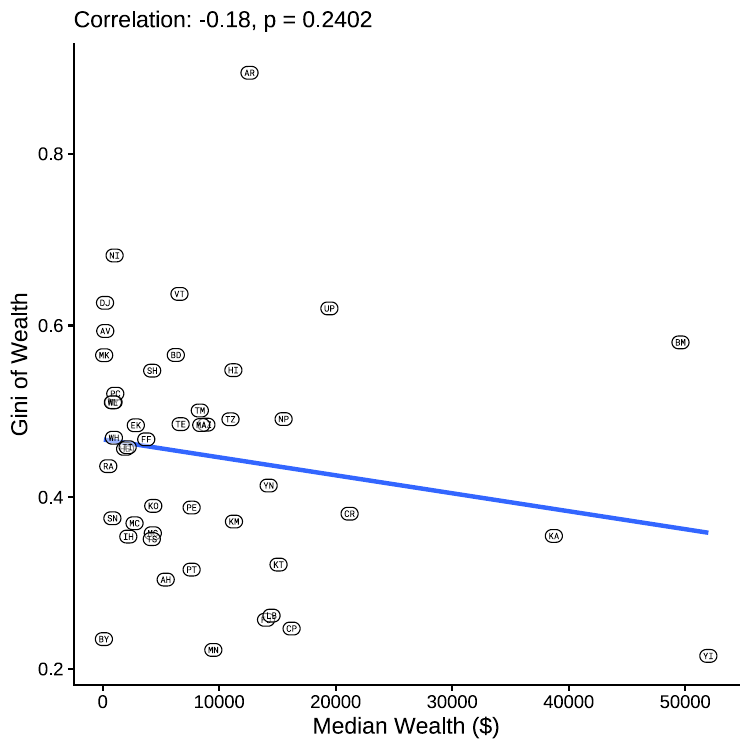}
\end{subfigure}
\vspace{1em}
\begin{subfigure}{0.49\textwidth}
    \centering
    \includegraphics[width=\linewidth]{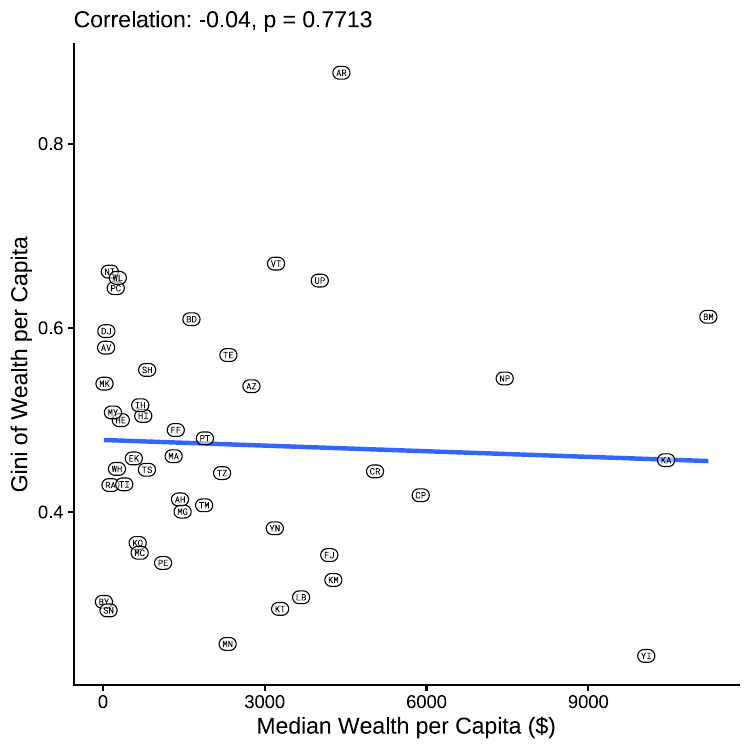}
\end{subfigure}
\vspace{1em}
\caption{The cross-site relationship between (left) median wealth and the wealth Gini and (right) median wealth per capita and the wealth per capita Gini. }
\label{fig:wealth-vs-wealth-gini}
\end{figure}

Figure~\ref{fig:wealth-vs-wealth-gini} shows that the Gini coefficient for wealth is not related across communities to the mean value of wealth at the community, using either absolute or per capita wealth. This abates some of our concerns as to how measurement error in wealth may affect our estimates of wealth inequality.

\begin{figure}
    \centering
    \includegraphics[width=0.5\linewidth]{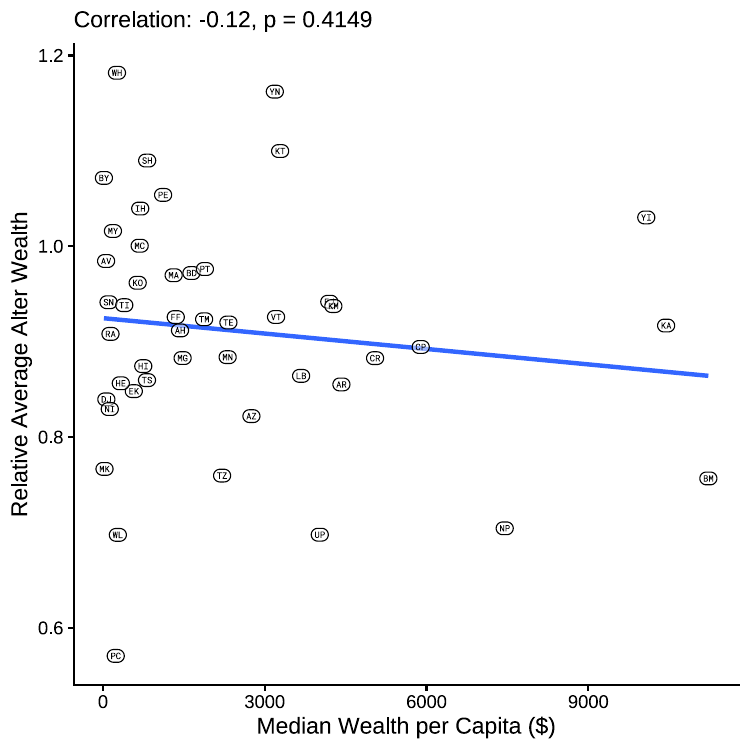}
    \caption{The cross-site relationship between median wealth per capita and Relative Average Alter Wealth.}
    \label{fig:wealth-vs-ec}
\end{figure}

Similarly, Figure \ref{fig:wealth-vs-ec} shows that our estimates of economic connectedness (shown here using Relative Average Alter Wealth, constructed using wealth per capita) are similarly not related to median wealth in the community, further abating the concern.

To further understand the consequences of wealth mis-measurement, we conduct exercises where we deliberately perturb sharing unit wealth and study how it affects our downstream statistics.
Because the accuracy of this measure depends on both the completeness of item inventories and the precision of item-level prices, we conduct two complementary sensitivity analyses. First, we apply random noise perturbations to item values, and second we conduct multiple imputation of missing item-level data.

\clearpage
\subsection{Noise Perturbation of Item Values}
\label{sec:wealth-sensitivity-noise}

To assess robustness to measurement error in item values, we generated 50 perturbed wealth datasets for every community.

\begin{figure}[h]
    \centering
    \includegraphics[width=0.75\linewidth]{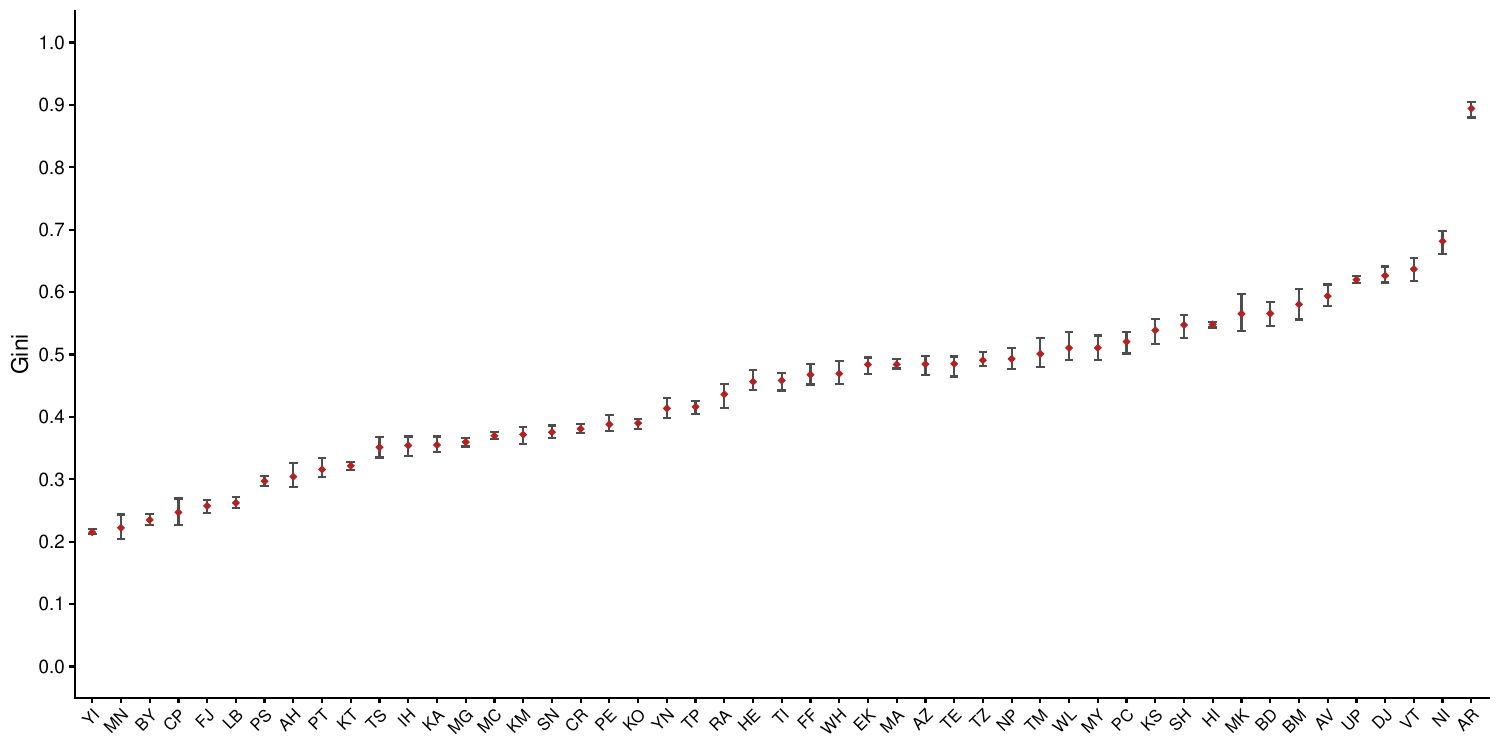}
    \caption{\textbf{Wealth Ginis with Noisy Perturbations.} Points show the median wealth Gini across the 50 trials. Bars show the range from the second-lowest to second-highest wealth Ginis across the trials, effectively representing a 95\% confidence interval.}
    \label{fig:ginis-with-noise}
\end{figure}

\begin{figure}[t]
\centering
\begin{subfigure}{0.49\textwidth}
    \centering
    \includegraphics[width=\textwidth]{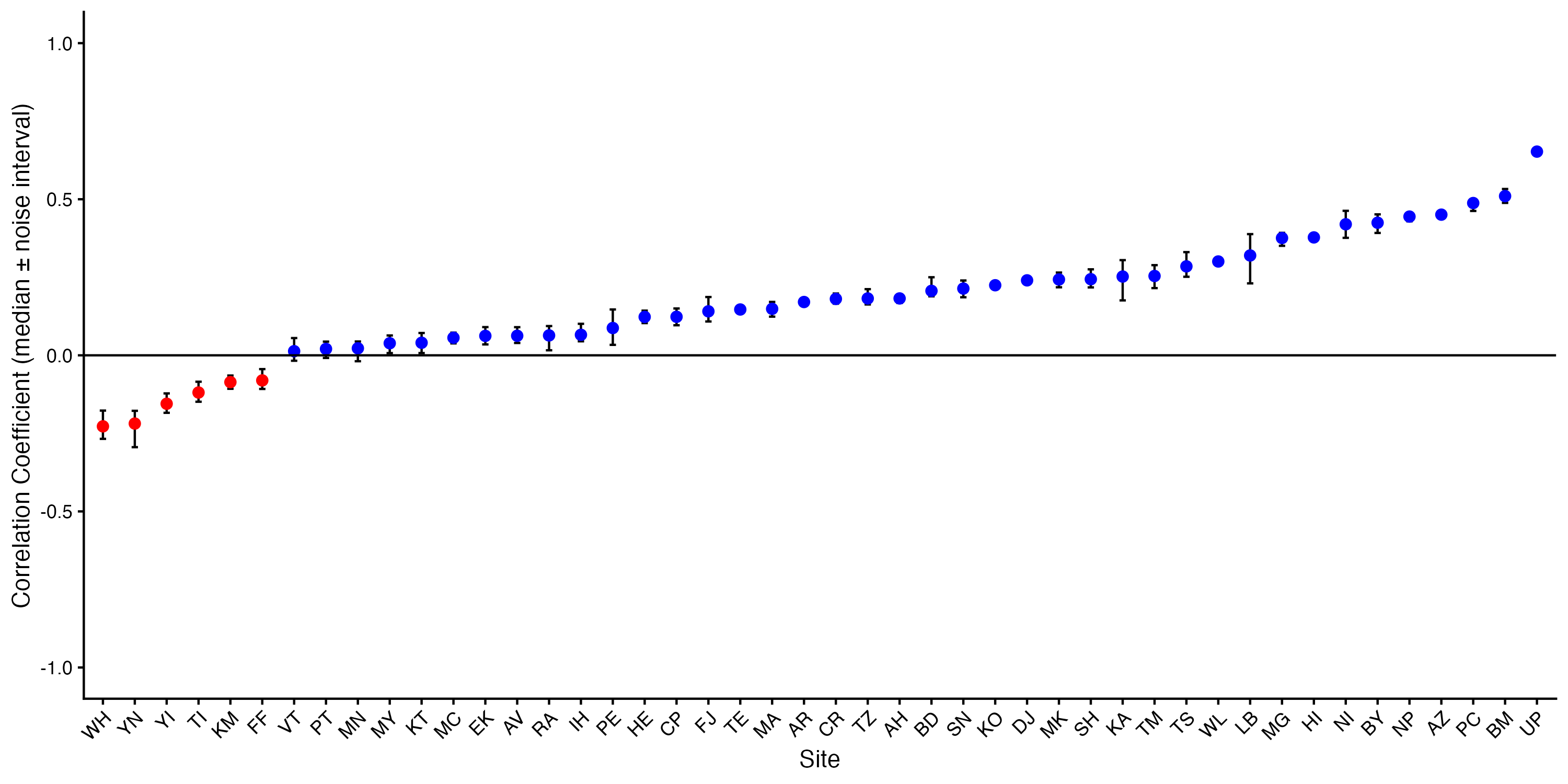}
    \caption{Support access vs wealth}
\end{subfigure}
%\hspace{2em}
\begin{subfigure}{0.49\textwidth}
    \centering
    \includegraphics[width=\textwidth]{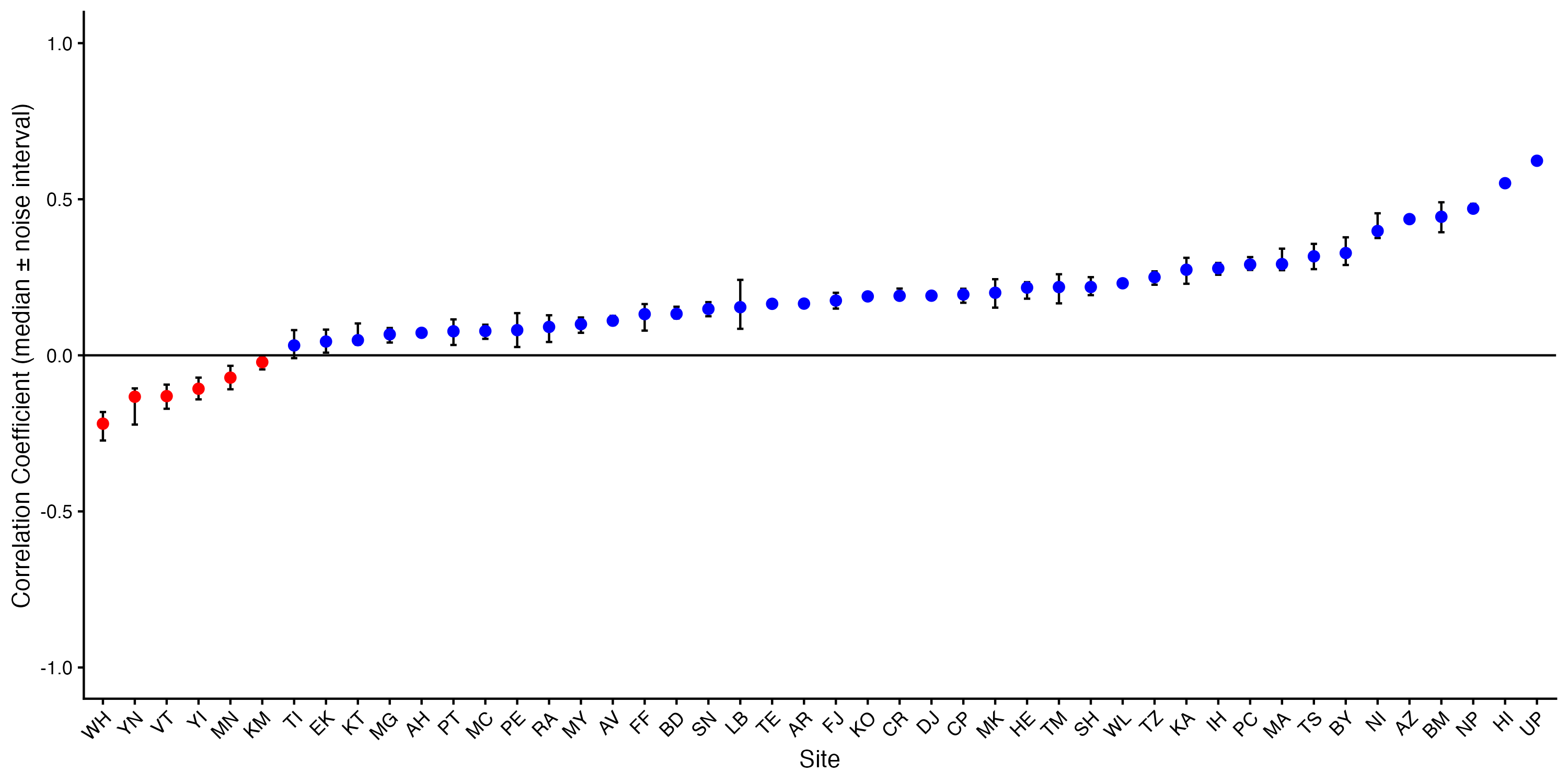}
    \caption{Support provisioning vs wealth}
\end{subfigure}
\vspace{1em} % Space between the top and bottom rows
\begin{subfigure}{0.49\textwidth}
    \centering
    \includegraphics[width=\textwidth]{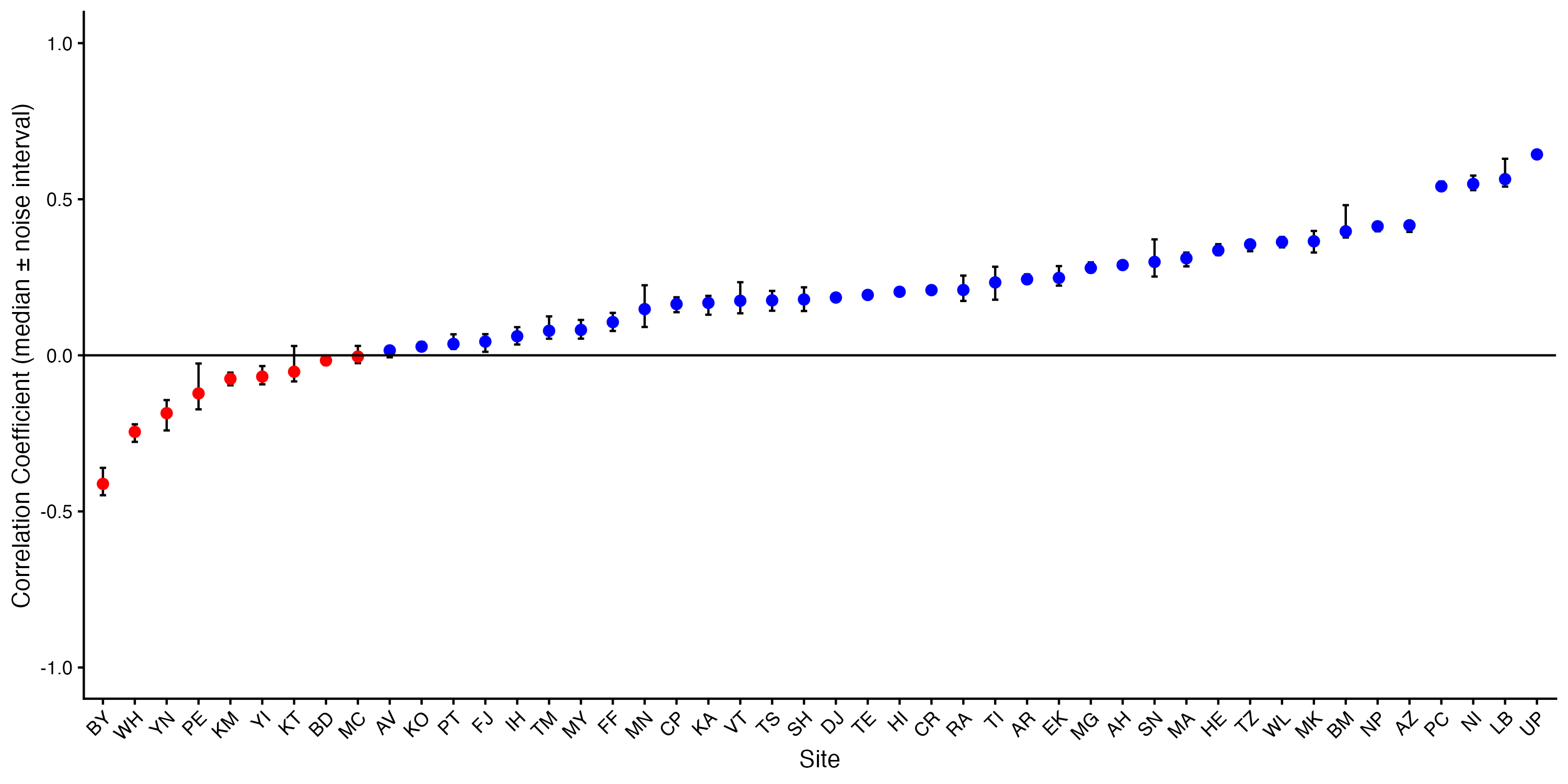}
    \caption{Support access p.c. vs wealth p.c.}
\end{subfigure}%
%\hspace{2em}
\begin{subfigure}{0.49\textwidth}
    \centering
    \includegraphics[width=\textwidth]{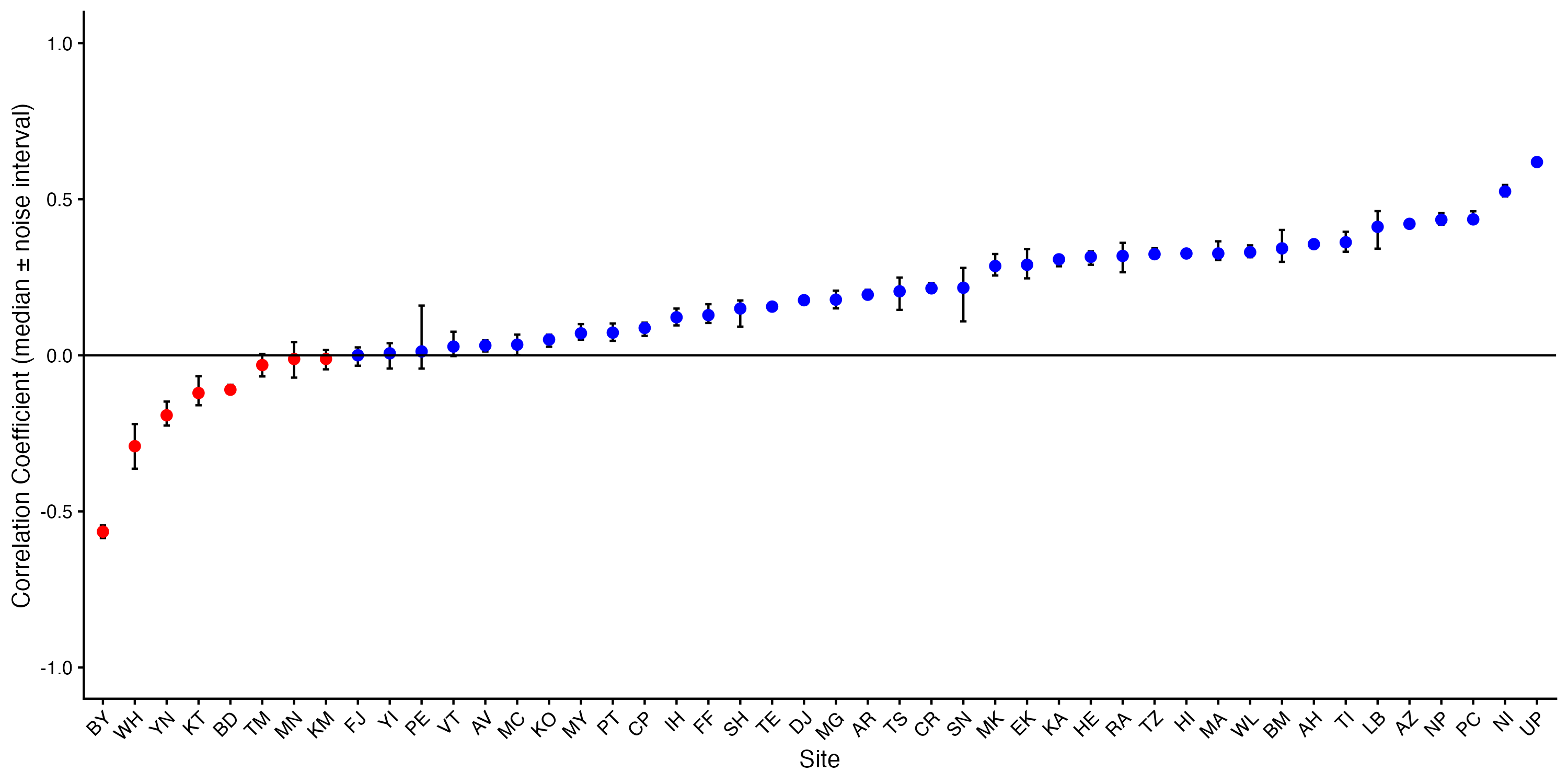}
    \caption{Support provisioning p.c. vs wealth p.c.}
\end{subfigure}
%\vspace{1em} % Increased space before the main figure caption
\caption{\textbf{Correlations between provisioning or accessing support and material wealth metrics, accounting for wealth uncertainty}. Access to support (i.e., out-degree) and provisioning of support (i.e., in-degree) are calculated using the ``composite'' network. Points show the median correlation across the 50 trials. Bars show the range from the second-lowest to second-highest correlation across the trials, effectively representing a 95\% confidence interval.}
\label{fig:within-site-multipanel-noised}
\end{figure}

\begin{figure}
    \centering
    \includegraphics[width=0.6\linewidth]{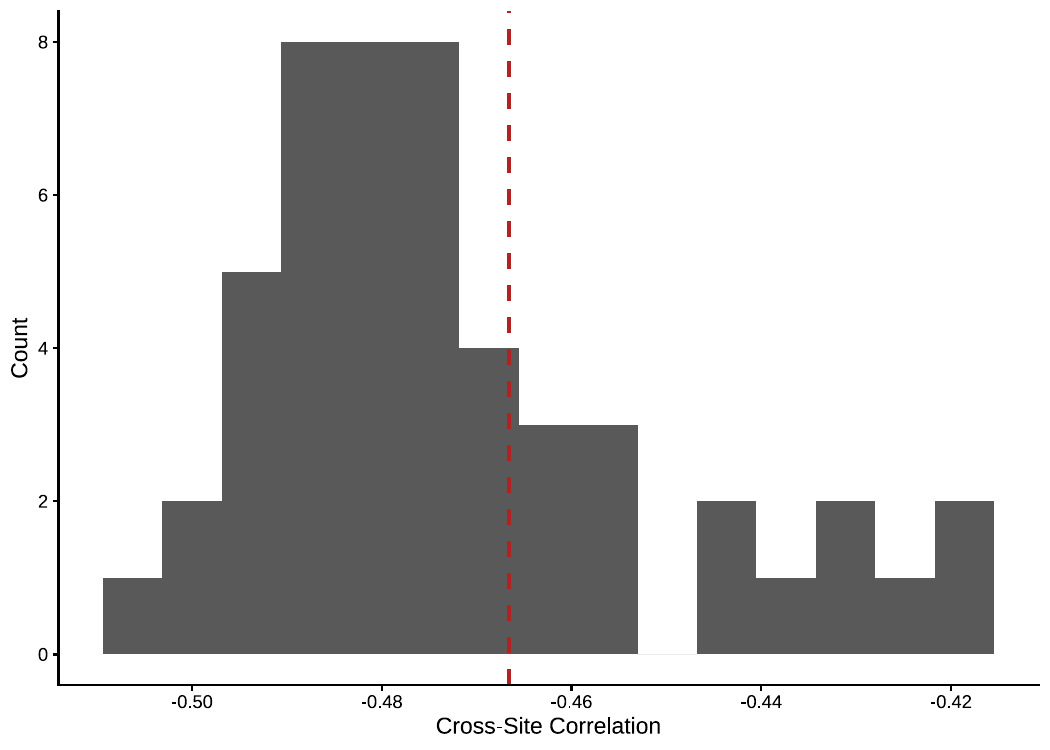}
    \caption{\textbf{Uncertainty due to wealth measurement for the cross-site correlation between the wealth per capita Gini and Relative Average Alter Wealth.} This shows the distribution of correlation coefficients, for each of the fifty trials with noisy perturbations. The dotted line shows the correlation in the observed data. }
    \label{fig:uncertainty-cross-site}
\end{figure}

For each of the 50 trials, we randomly multiply every item value $p_k$ by a noise factor of 0.8 with probability one half, and 1.2 with probability one half. We then recompute all of our downstream statistics from these noisy datasets. Figure~\ref{fig:ginis-with-noise} shows that our downstream wealth Gini coefficients are fairly invariant to this form of noise. In Figure~\ref{fig:within-site-multipanel-noised} we see that our within-site results in the cross-sharing-unit relationship between material and social wealth are very robust to this form of measurement error. Similarly, we see in Figure~\ref{fig:uncertainty-cross-site} that this form of measurement error does not seem to substantially adversely affect the strength of the correlation between the wealth per capita Gini and Relative Average Alter Wealth.

\clearpage
\subsection{Multiple Imputation of Missing Item Data}\label{sec:wealth-sensitivity-imputation}

\begin{figure}[h]
    \centering
    \includegraphics[width=0.75\linewidth]{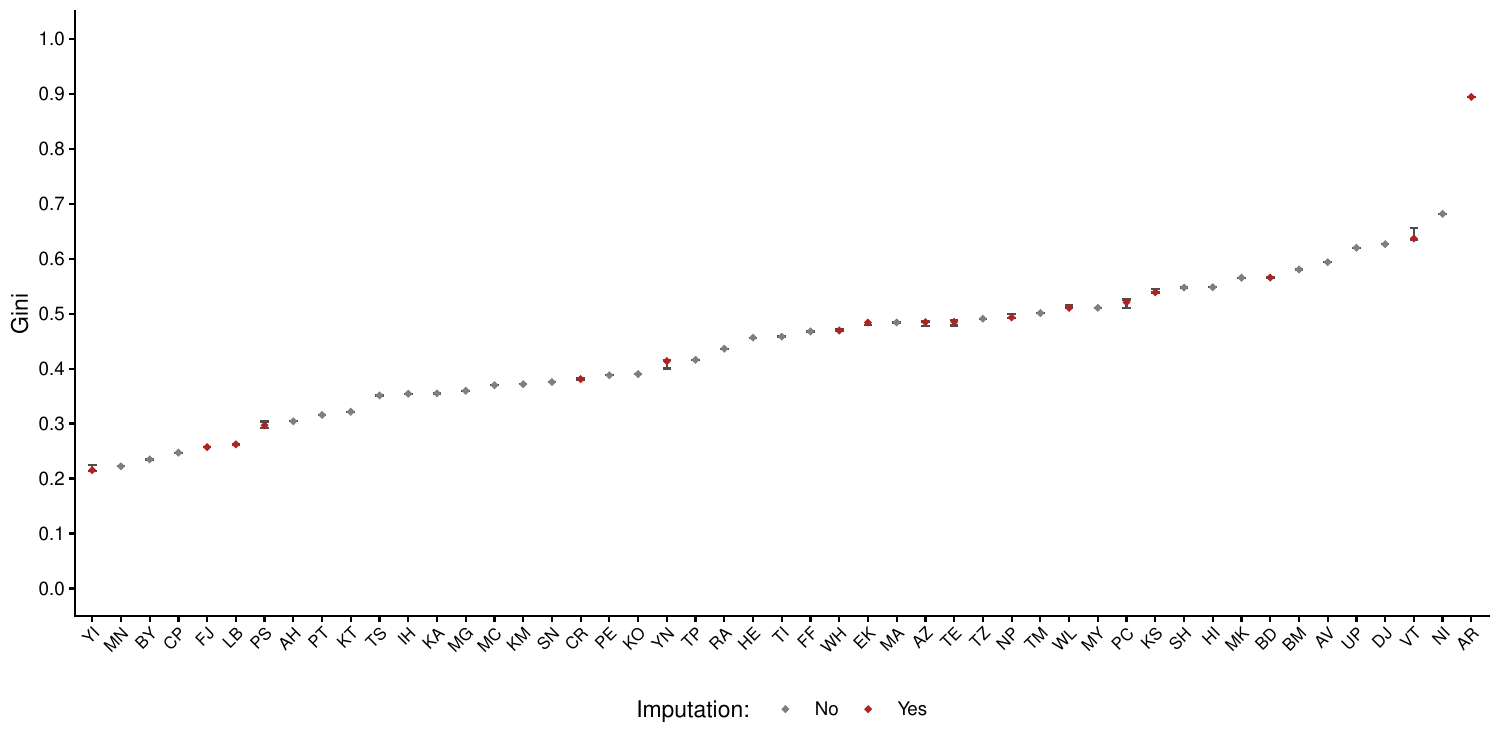}
    \caption{\textbf{Wealth Ginis, with Imputation.} For the handful of communities where some item inventories were incomplete (highlighted in red), we show the range of possible wealth Ginis across 50 imputed datasets. Points show the median wealth Gini across the 50 trials. Bars show the range from the second-lowest to second-highest wealth Gini across the trials, effectively representing a 95\% confidence interval. Fully observed cases show (in gray) the observed wealth Gini.}
    \label{fig:ginis-with-imputations}
\end{figure}

Some communities have incomplete item inventories for a subset of sharing units. In this case, we use the \texttt{mice} package in \textsf{R} \citep{mice} to impute missing values, constructing 50 imputed datasets per site. Figure~\ref{fig:ginis-with-imputations} shows that the sites where we impute item inventories for some sharing units do not systematically differ from sites in which imputations of wealth are not necessary, and that additionally the downstream Gini coefficients are fairly invariant to the (stochastic) realizations of the imputation algorithm.

\clearpage
\section{Kinship}\label{supp:kinship}

\begin{figure}
    \centering
    \begin{subfigure}{0.49\textwidth}
        \centering
        \includegraphics[width=\textwidth]{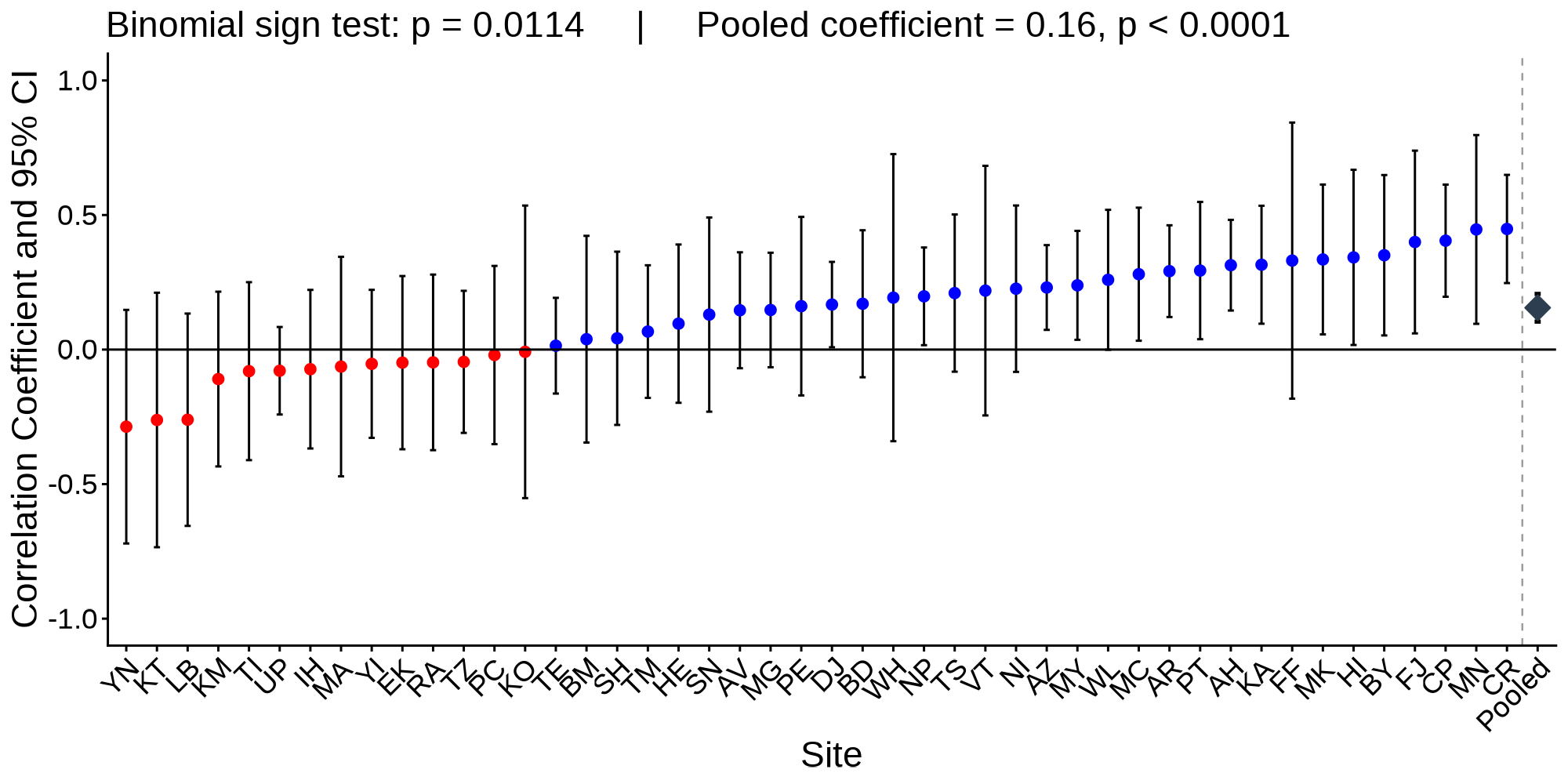}
        \caption{Kin-based support access p.c. vs wealth p.c.}
    \end{subfigure}%
    \vspace{1em}
    %\hspace{2em}
    \begin{subfigure}{0.49\textwidth}
        \centering
        \includegraphics[width=\textwidth]{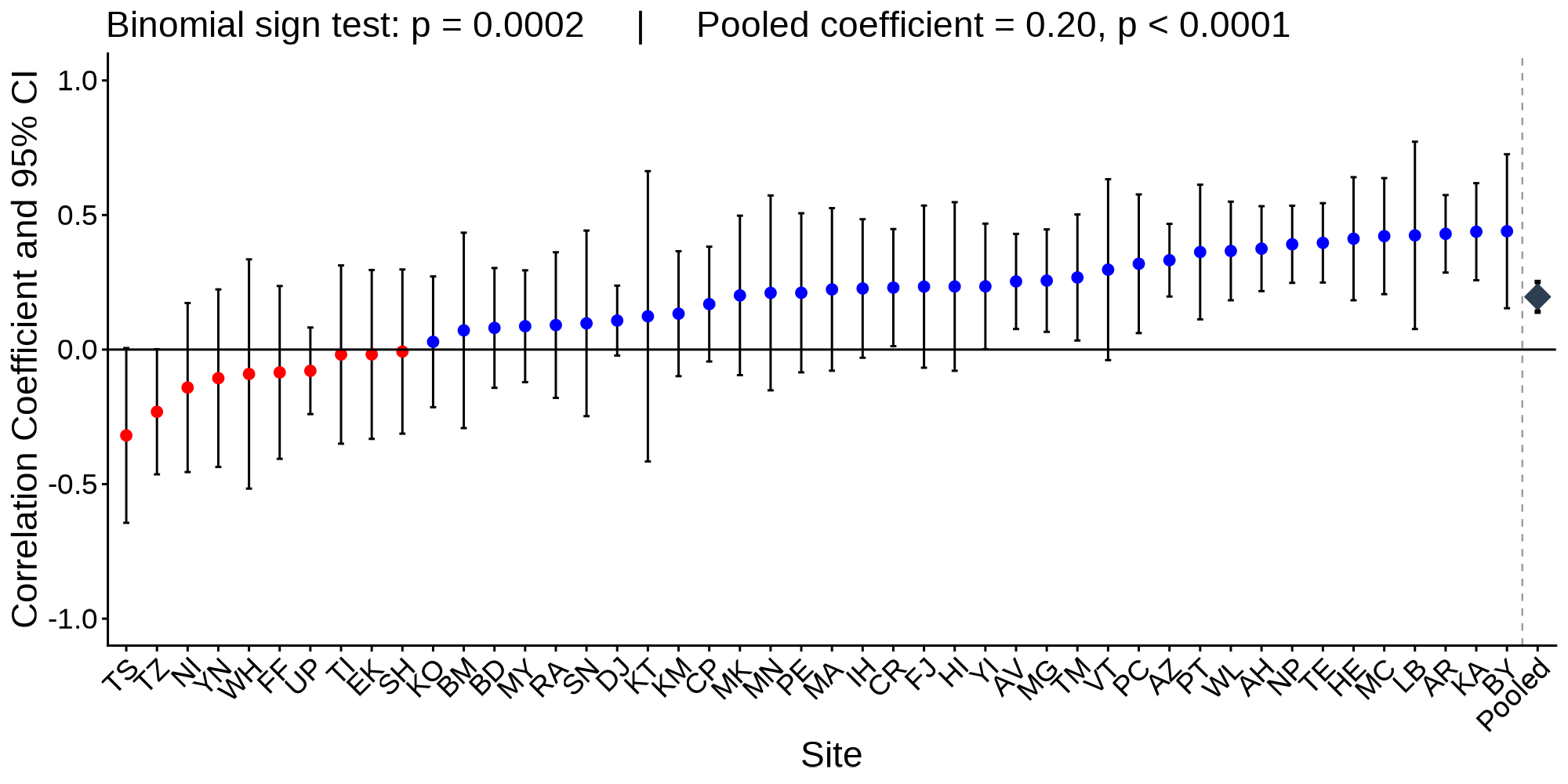}
        \caption{Non-kin-based support access p.c. vs wealth p.c.}
    \end{subfigure}
        \begin{subfigure}{0.49\textwidth}
        \centering
        \includegraphics[width=\textwidth]{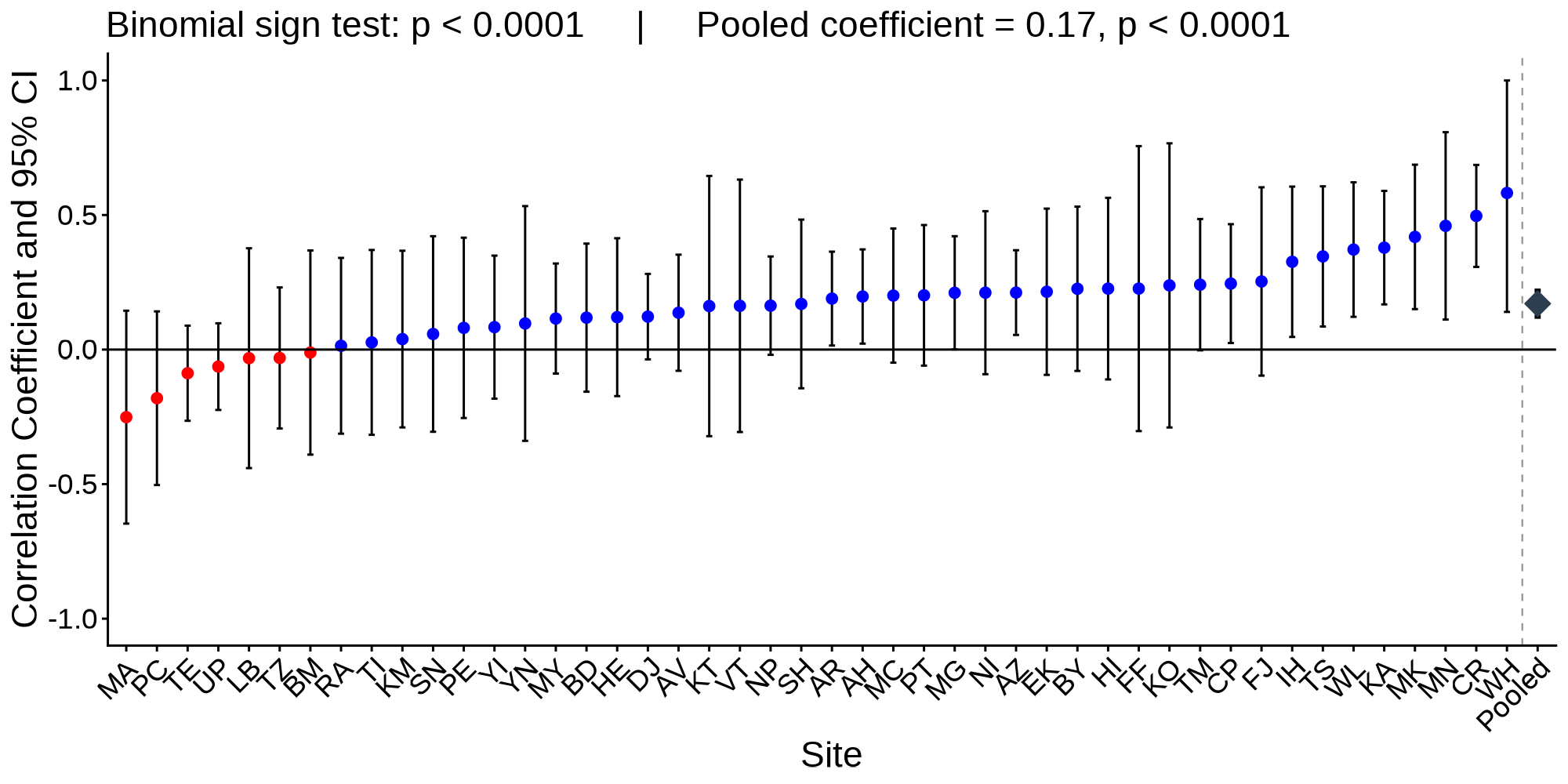}
        \caption{Kin-based support provision p.c. vs wealth p.c.}
    \end{subfigure}%
    \vspace{1em}
    %\hspace{2em}
    \begin{subfigure}{0.49\textwidth}
        \centering
        \includegraphics[width=\textwidth]{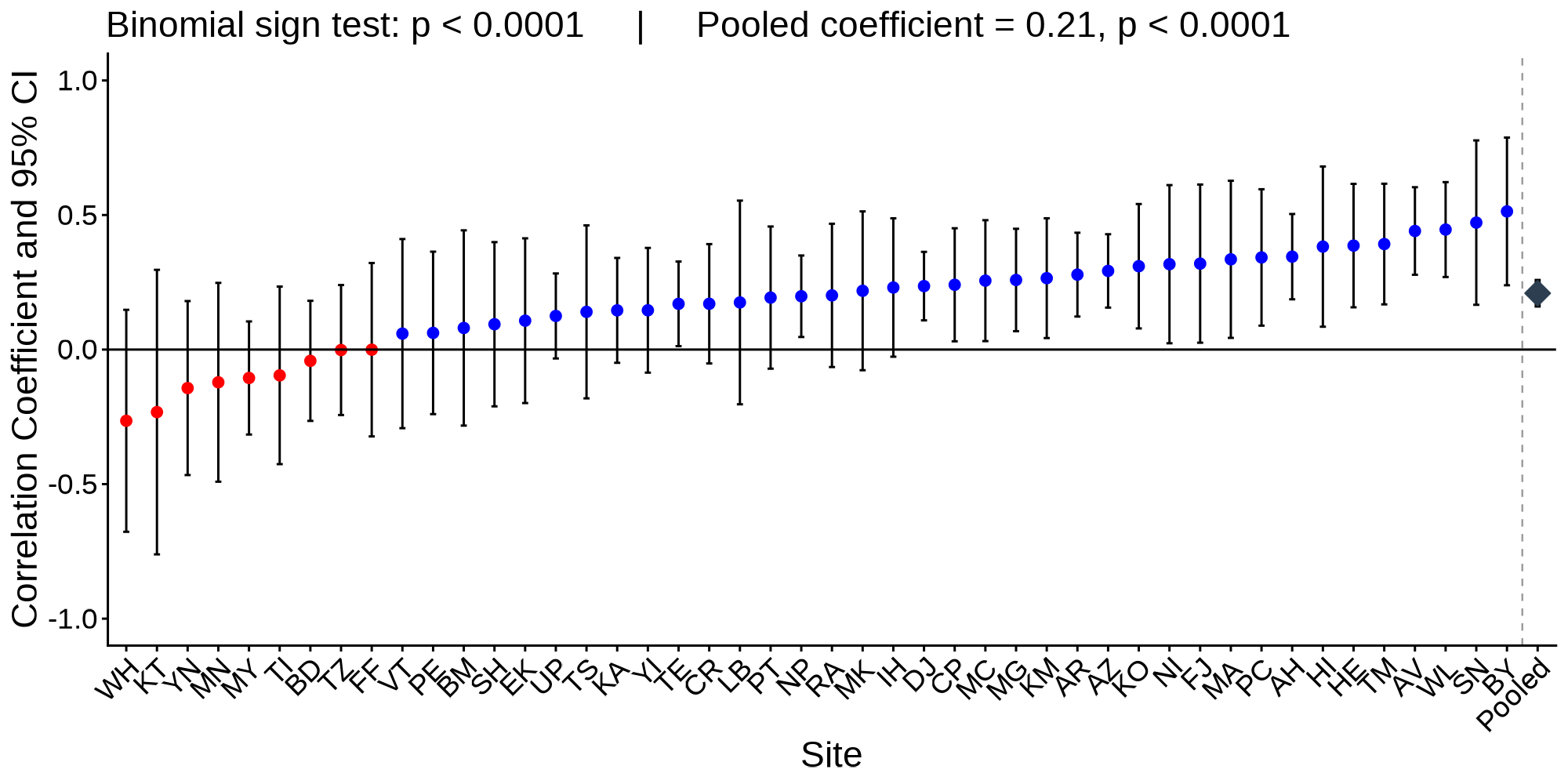}
        \caption{Non-kin-based support provision p.c. vs wealth p.c.}
    \end{subfigure}
    \begin{subfigure}{0.49\textwidth}
        \centering
        \includegraphics[width=\textwidth]{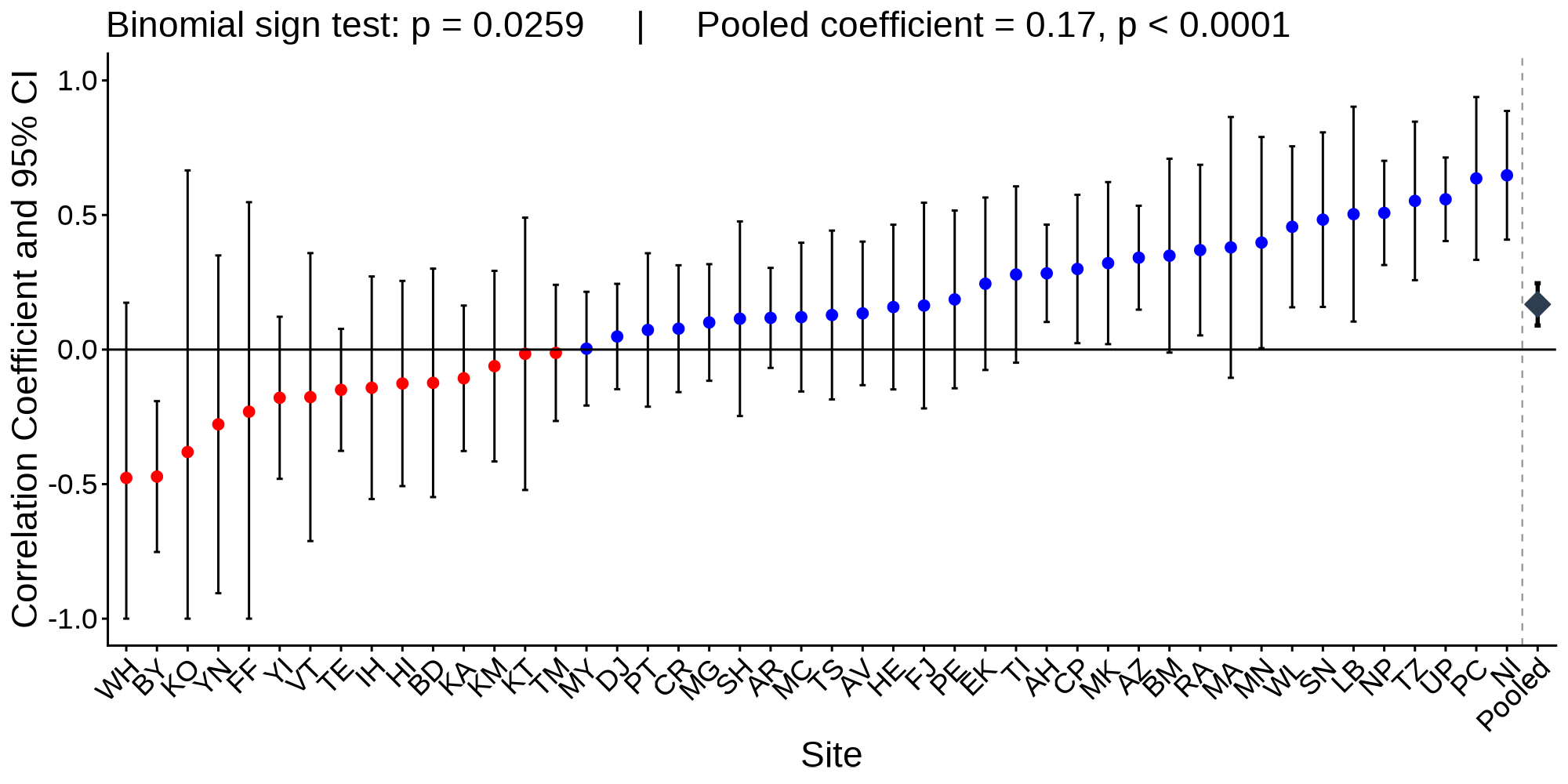}
        \caption{Kin-based $\text{AAW}_i$ (of supporters) vs wealth p.c.}
    \end{subfigure}%
    %\hspace{2em}
    \begin{subfigure}{0.49\textwidth}
        \centering
        \includegraphics[width=\textwidth]{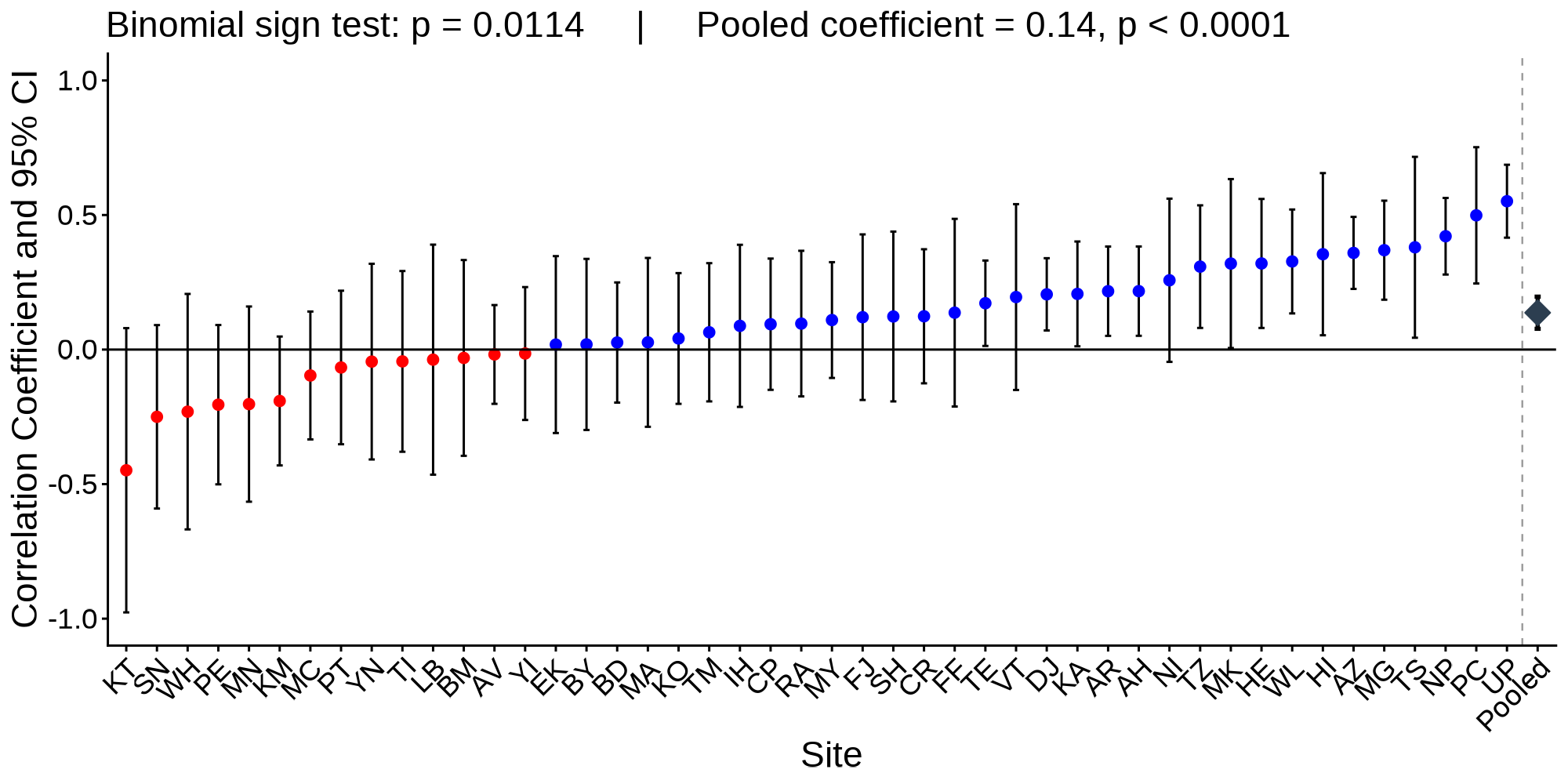}
        \caption{Non-kin-based $\text{AAW}_i$ (of supportees) vs wealth p.c.}
    \end{subfigure}
    \caption{\textbf{Correlations between network connections and material wealth, based on kinship-based or non-kinship-based connections}. Support access, support provisioning, and Average Alter Wealth (using outgoing nominations) are all calculated from the composite network, restricting it to include either connections only to primary kin (left) or only to non-primary kin (right). }
    \label{fig:kinship-site-multipanel}
\end{figure}

Kinship ties are an important source of support in many of the communities we study, so we examined how our results appear when focusing on kin versus non-kin ties. Figure \ref{fig:kinship-site-multipanel} shows the correlations between our network-based measures of social capital and material wealth (per capita) when restricting the composite network to retain only those connections that are to primary kin (left) or to retain only those connection that are to non-primary kin (right). The overall cross-site patterns are quite comparable and broadly align with those using the overall composite network (shown in the main text in Figure~\ref{fig:within-site-multipanel}).

\begin{figure}
    \centering
    \includegraphics[width=0.5\linewidth]{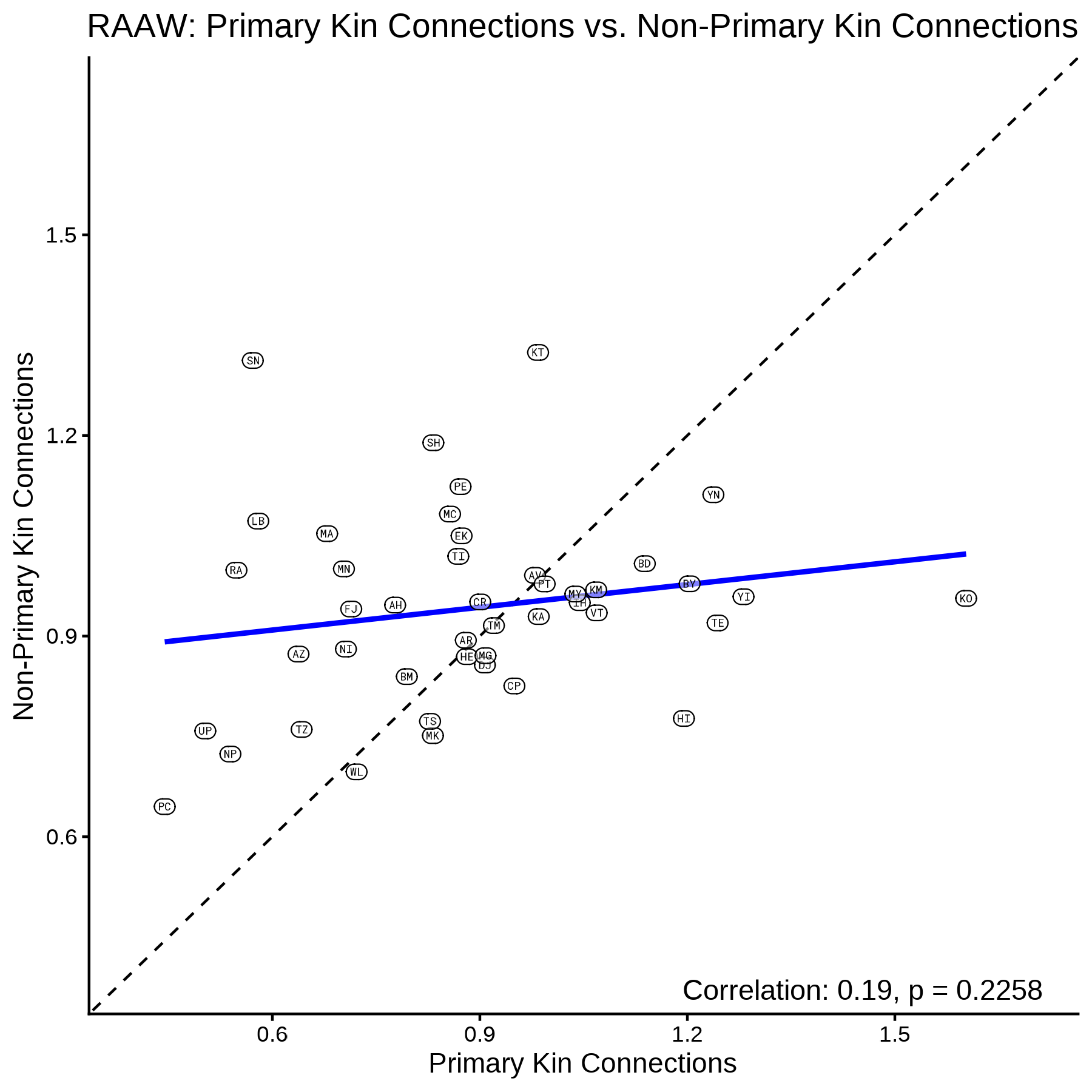}
    \caption{\textbf{Cross-site relationship between RAAW derived from kin and non-kin connections.} This plot shows the bivariate correlation between Relative Average Alter Wealth based on primary kin connections versus Relative Average Alter Wealth based on non-primary kin connections.}
    \label{fig:RAAW-kin-scatter}
\end{figure}

\begin{figure}
    \centering
    \includegraphics[width=0.75\linewidth]{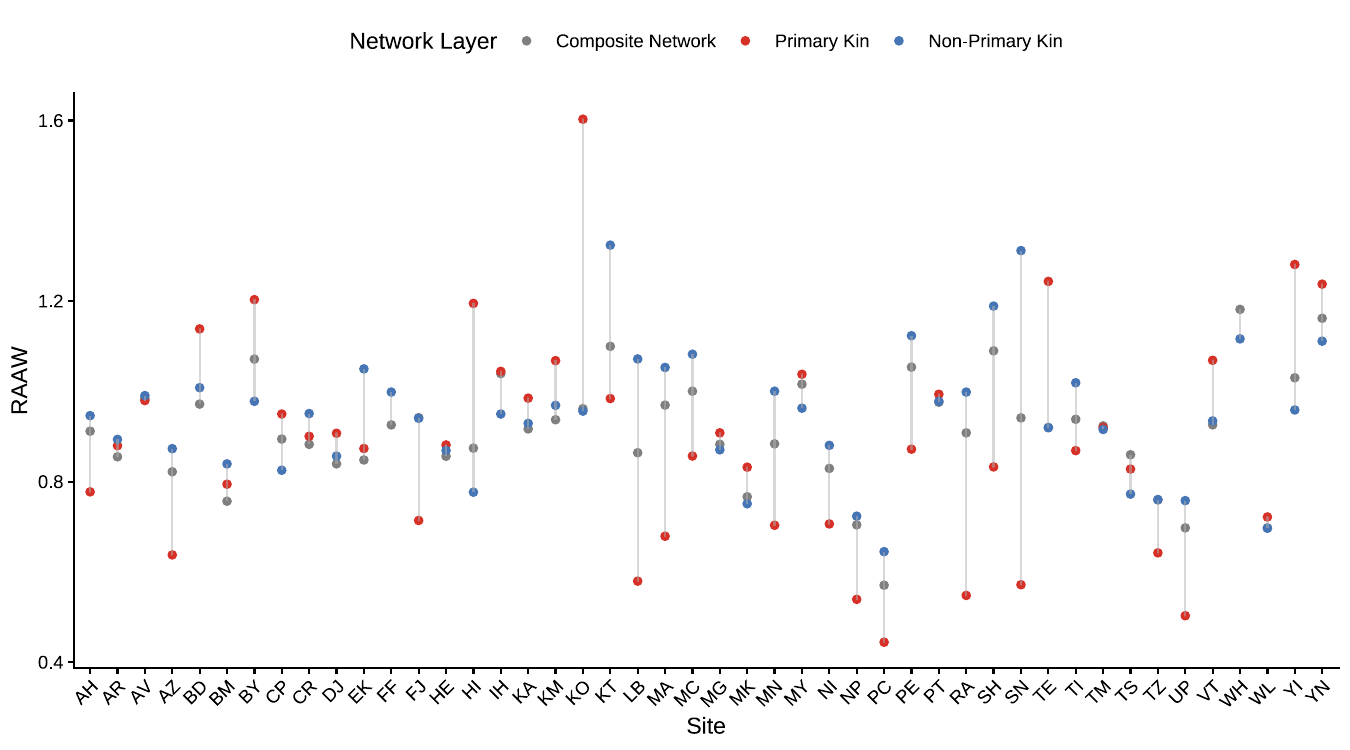}
    \caption{\textbf{Relative Average Alter Wealth overall, with kin connections, and non-kin connections.} This plot shows the Relative Average Alter Wealth for communities in the composite network and restricting the composite network to primary kin connections and non-primary kin connections.}
    \label{fig:RAAW-kin-dotplot}
\end{figure}

Across communities, we examine what kinds of connections contribute to Relative Average Alter Wealth by again separating kin and non-kin connections in the composite network. We find that Relative Average Alter Wealth derived from kin connections and from non-kin connections are uncorrelated across communities (Figure~\ref{fig:RAAW-kin-scatter}). Moreover, there is no consistent pattern for whether kin-based RAAW is higher or lower than that derived from non-kin connections (Figure~\ref{fig:RAAW-kin-dotplot}), suggesting that the relationship between RAAW and the wealth Gini is not being driven primarily by kinship connections.

Taken together, these results indicate that while kinship provides a baseline of support in all communities, it is not the main explanation of our results.

\clearpage
\section{The Relationship of Other Community Characteristics with Wealth Inequality and Economic Connectedness}
\label{supp:othervariables}

Here we explore how the other community characteristics discussed in Section \ref{across} relate to both wealth inequality and economic connectedness. Specifically, we consider the variables presented in Table~\ref{tab:variables}, which gives a basic description of each variable, and outlines the source of the information and the scale to which it refers. We show the exact values of these site-level variables in Table~\ref{tab:site_vars_other} (alongside the other site-level variables in Section~\ref{supp:site-vars}).

\input{tables/Varbs}

\clearpage
\subsection{Correlations of Community Characteristics with Wealth Inequality}\label{supp:predicing_ec}

Table~\ref{tab:corr_gini_wealth_per_capita} shows the bivariate correlations of these variables with the Gini of wealth per capita, corresponding to what is shown in the main text in Figure~\ref{fig:corrs-with-wpc-gini}.

\input{tables/CorrelationsWithgini_wealth_per_capita_only_raaw}

Given the large number of community-level explanatory variables that we have available, we perform a Lasso with wealth inequality as the predicted variable, and then we further analyze how these predict wealth inequality when combined.

% kinship: look to angelucci_consumption_2018, among others (obviously rosensweig, etc)

\begin{figure}[ht]
    \centering
    \includegraphics[width=0.8\linewidth]{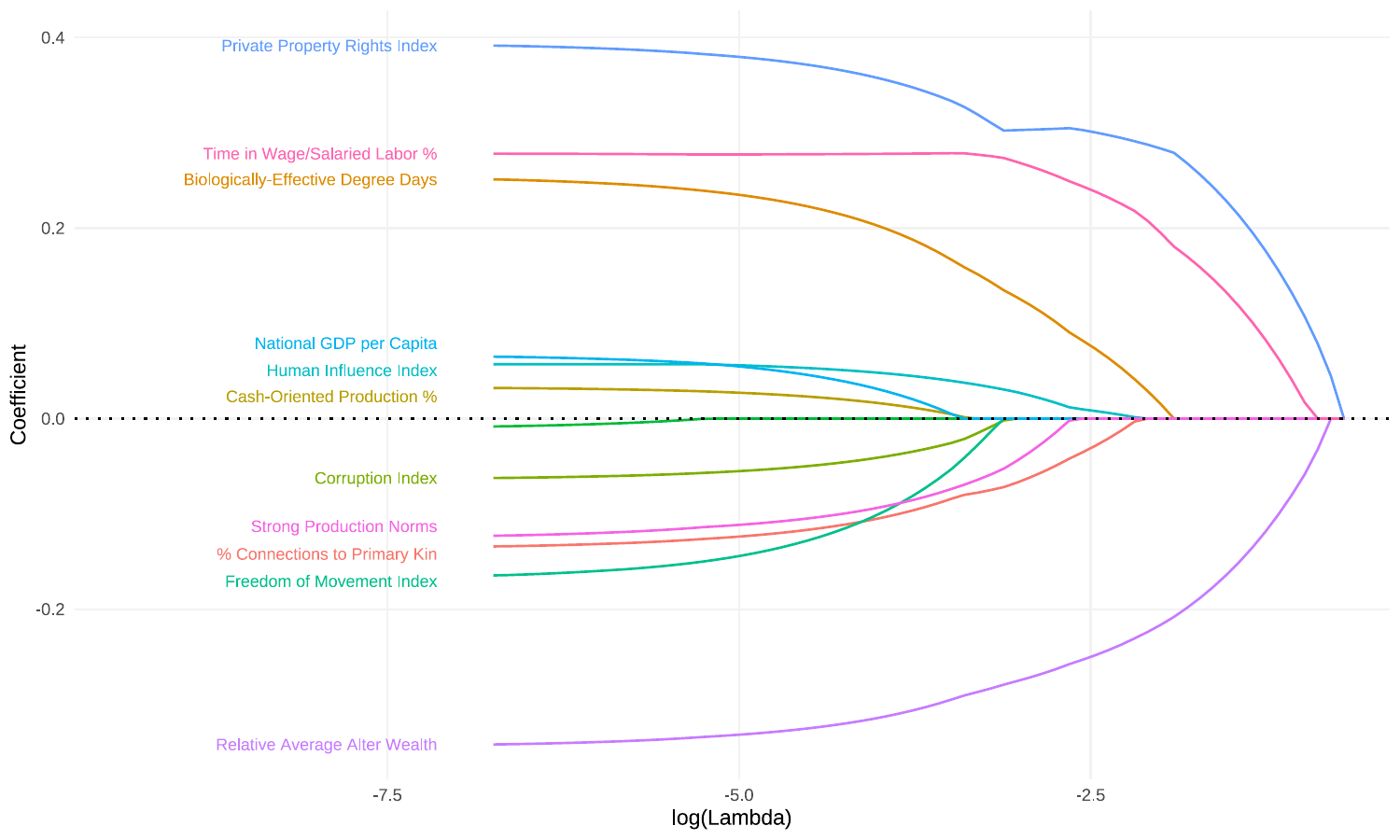}
    \caption{LASSO, with the Gini of Wealth per Capita as the dependent variable. Lambda denotes the penalty term in the LASSO.}
    \label{fig:lasso}
\end{figure}

\begin{figure}
    \centering
    \includegraphics[width=0.5\linewidth]{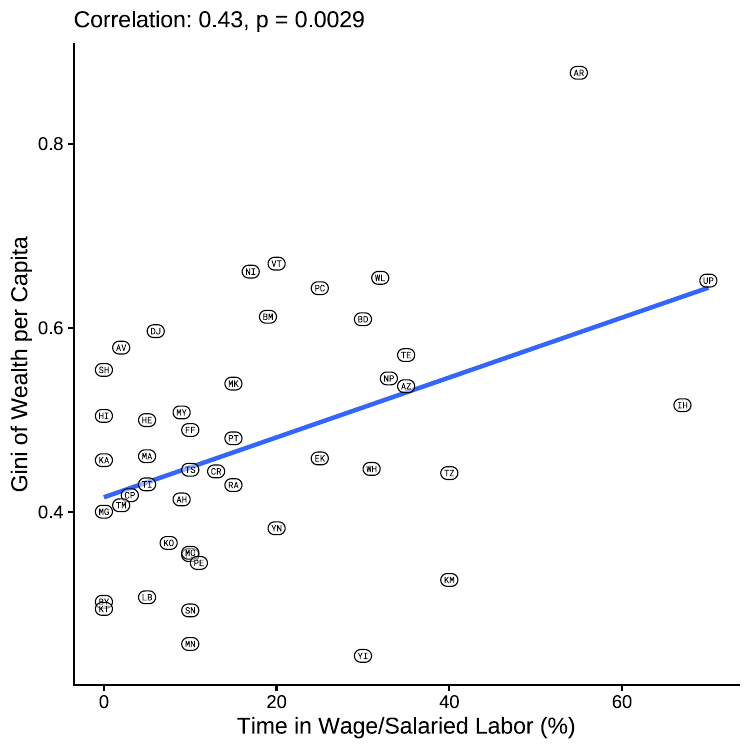}
    \caption{Time spent in Wage/Salaried Labor \% vs Gini wealth per capita}
    \label{fig:wage-labor}
\end{figure}

\begin{figure}
    \centering
    \includegraphics[width=0.5\linewidth]{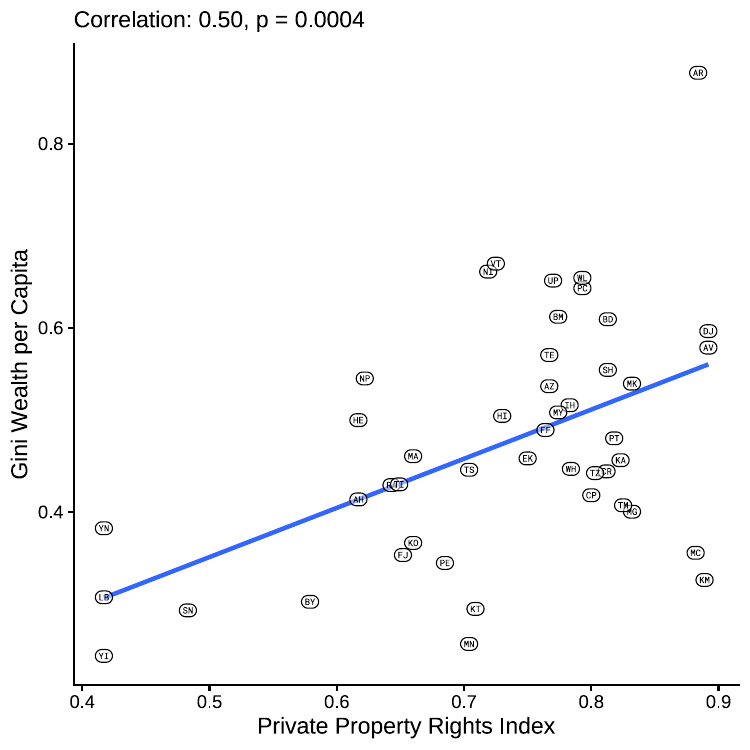}
    \caption{National Private Property Rights Index vs Gini wealth per capita}
    \label{fig:property-rights}
\end{figure}

\begin{figure}
    \centering
    \includegraphics[width=0.8\linewidth]{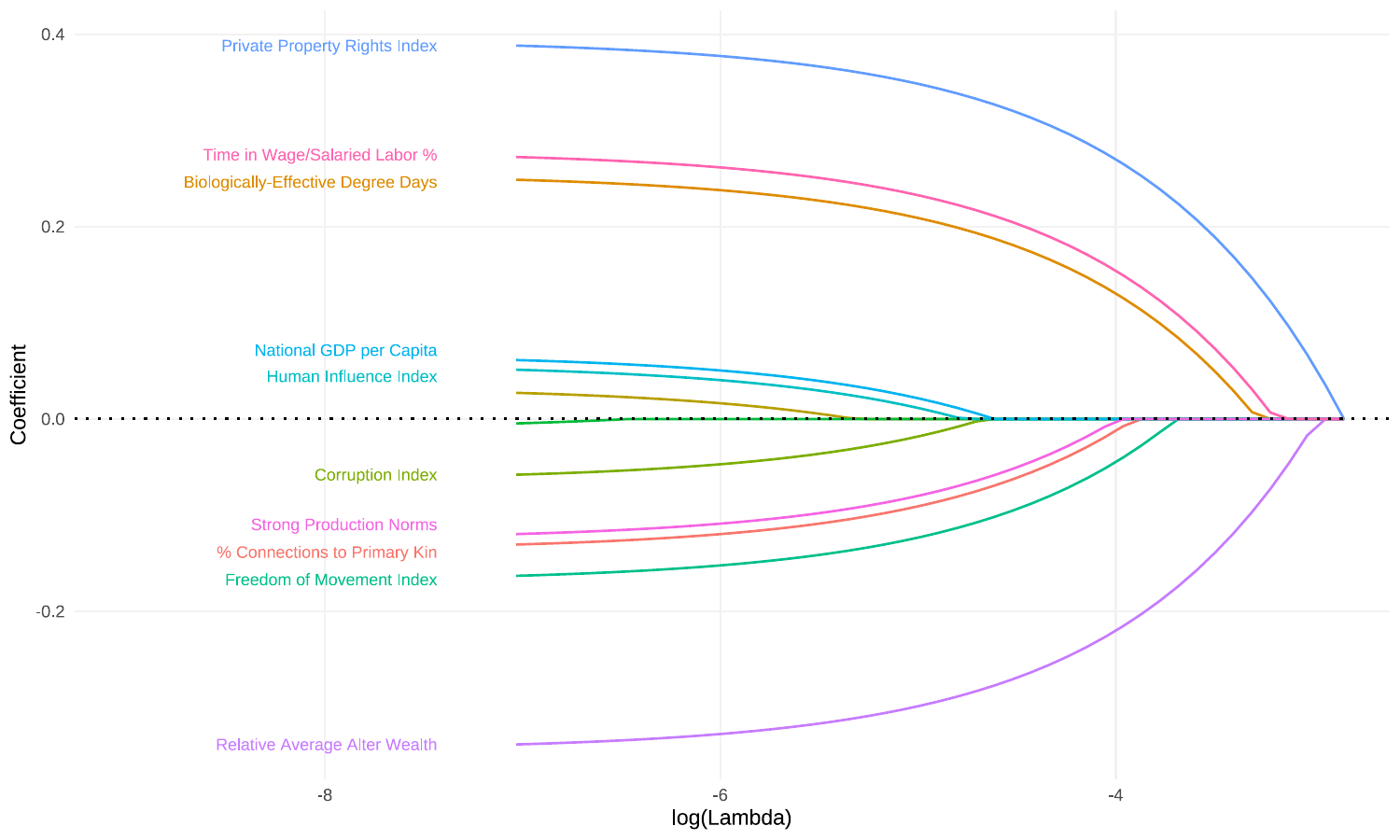}
    \caption{\textbf{LASSO after a Puffer transform, with Gini of Wealth per Capita as the dependent variable.}}
    \label{fig:LASSO-puffer}
\end{figure}

As we see from the results of the Lasso in Figure \ref{fig:lasso}, the variables can be divided into three main groupings.  The economic connectedness variable of Relative Average Alter Wealth, %\footnote{We work with this version of economic connectedness for consistency with earlier sections, but one can use any of the three.}
the private property rights index and the percent of time spent in wage/salaried labor come in first. As shown in the main text in Figure~\ref{fig:corrs-with-wpc-gini}, these variables also have significant bivariate associations with the Gini of wealth per capita (Figures~\ref{fig:wage-labor} and \ref{fig:property-rights} show their bivariate correlation).
Next are a measure of environmental productivity (Biologically-Effective Degree Days),
the Human Influence Index,
the percent of connections that are to close kin,
and a binary measure of whether the community has strong normative restrictions on who can undertake different types of productive work.
Variables that enter in later have relatively small resulting coefficients.
We obtain similar results when running the LASSO after applying a Puffer transform \citep{jia2015preconditioning}, in Figure~\ref{fig:LASSO-puffer}. Relative Average Alter Wealth, the private property rights index, and time in wage/salaried labor continue to be the strongest predictors of the Gini of wealth per capita (alongside the Biologically-Effective Degree Days measure).

We perform multiple regressions first with the variables that have significant bivariate correlations (three of which are the first selected in the Lasso), then add other variables highlighted by the Lasso, before including the full set of variables, as shown in Table~\ref{tab:stepwise_regression_lasso}.

\input{tables/Stepwise_Regression_from_Lasso}

As we see in Table \ref{tab:stepwise_regression_lasso}, the economic connectedness variable, Relative Average Alter Wealth, is significantly and strongly negatively correlated with inequality in all specifications. The better connected the poor are to the wealthy, the lower is inequality.
Time spent in wage/salaried labor shows a consistent positive effect. (%though there is noteworthy uncertainty about the coefficient in the full model).
Communities where residents spend more of their time engaged in wage or salaried work have greater inequality.
We also see that communities that are located in countries where private property rights are more widely shared have greater inequality.
None of the other variables are consistently informative predictors of inequality when grouped with the other covariates.

\clearpage
\subsection{Correlations of Community Characteristics with Economic Connectedness}\label{predicting-ec}

Given that economic connectedness is significantly predictive of inequality, and explains the greatest amount of the variation (among the variables we have in our data), one might wonder what characteristics of a community predict or correlate with its economic connectedness, here assessed with Relative Average Alter Wealth. This is comparable to recent efforts by \citet{otero_differences_2024} to look at country-level differences in social capital, including a measure of social capital that measures the socioeconomic status of social contacts, somewhat akin to our measures of economic connectedness.

\begin{figure}[h!]
    \centering
    \includegraphics[width=0.8\linewidth]{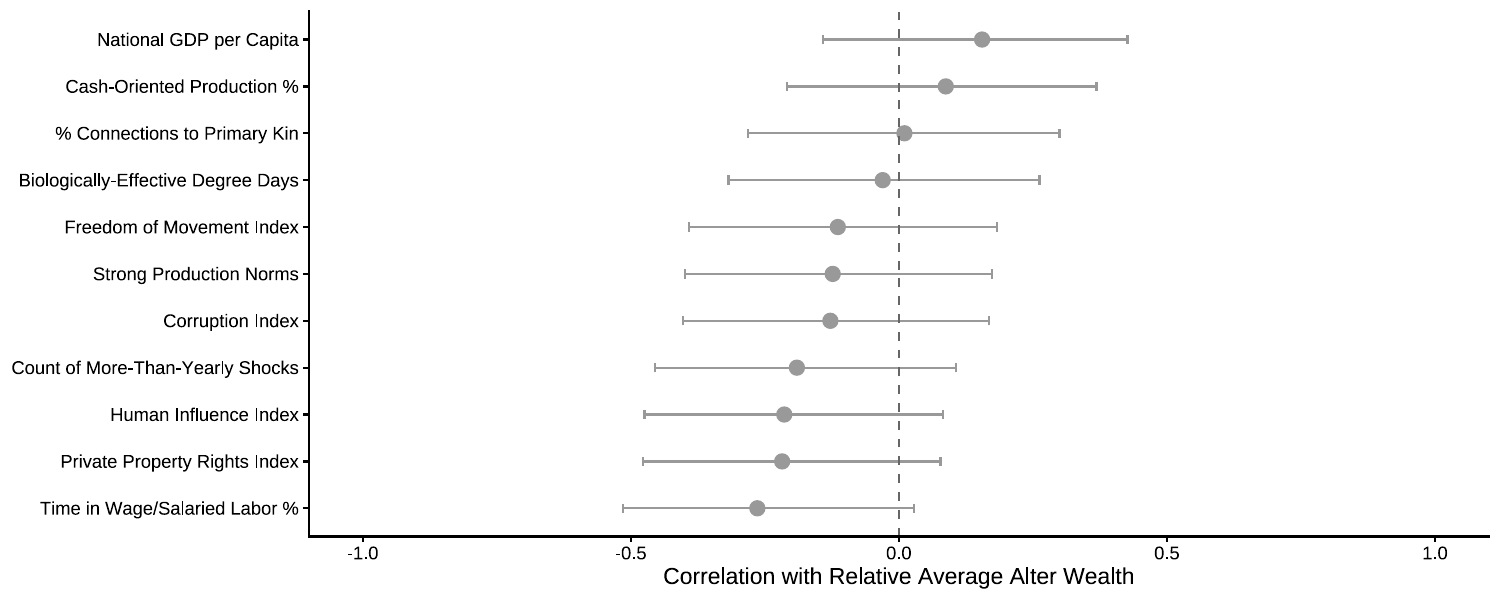}
    \caption{Bivariate Correlations with Relative Average Alter Wealth.}
    \label{fig:corrs-with-raaw}
\end{figure}

We have explored this, but have found no robust predictors of Relative Average Alter Wealth among the variables under consideration.
Figure~\ref{fig:corrs-with-raaw} and Table~\ref{tab:corr_rel_frank_capita} present bivariate correlations of the other variables with Relative Average Alter Wealth (using wealth per capita).
There, we see that Time Spent in Wage/Salaried Labor is at the margin of significance, with a negative association.
Communities where residents are involved in wage and salaried labor %and communities that are located in countries where more people have protected property rights
generally have fewer connections between poor and rich.
As we see in Table \ref{tab:stepwise_regression_lasso_ec}, however, this variable is not a consistent predictor of RAAW when considered in tandem with other variables, nor are any others.

\input{tables/CorrelationsWithrel_frank_capita}

\input{tables/Stepwise_Regression_from_Lasso_EC}

\clearpage
\section{Site-Level Variables}\label{supp:site-vars}

Here, we present the various site-level variables used across our analyses. Where appropriate, we show the measures derived from different accountings of sharing unit composition (absolute, per capita, per adult, and size-adjusted). For the network-derived social capital measures, we show the measures considering both supporters (i.e., those providing support) and supportees (i.e., those to whom support is provided).
Table~\ref{tab:site_vars_wealth} shows the material wealth inequality measures.
Table~\ref{tab:site_vars_deg_gini} shows the access and provisioning of support inequality measures.
Table~\ref{tab:site_vars_raaw} shows the Relative Average Alter Wealth measures.
Table~\ref{tab:site_vars_ec} shows the Economic Connectedness measures.
Table~\ref{tab:site_vars_rc} shows the Relative Connectedness measures.
Table~\ref{tab:site_vars_rc} shows the Normed Wealth Modularity measure.
table~\ref{tab:site_vars_other} shows the other site-level variables considered in our bivariate and Lasso regressions.

\clearpage
\input{tables/wealth_site_vars_table}

\clearpage
\input{tables/deg_gini_site_vars_table}

\clearpage
\input{tables/raaw_site_vars_table}

\clearpage
\input{tables/ec_site_vars_table}

\clearpage
\input{tables/rc_site_vars_table}

\clearpage
\input{tables/wealth_mod_site_vars_table}

\begin{landscape}
\setlength{\tabcolsep}{3pt}
\input{tables/other_site_vars_table}
\end{landscape}

\end{document}

%% file: tables/su_summary.tex
\begingroup\fontsize{8}{10}\selectfont

\begin{longtable}[t]{l>{\raggedright\arraybackslash}p{3cm}>{\raggedright\arraybackslash}p{2cm}r>{\raggedright\arraybackslash}p{3cm}rrrrrrrrrr}
\caption{\label{tab:su_summary}Summary of each community and its sharing unit characteristics, sampling, and wealth. The values for the material wealth of sharing units are reported in U.S. dollars, circa the time of data collection at the respective field sites. The Members columns show the average, standard deviation, and maximum number of sharing unit members in each community. The Sampling columns show the average number of people surveyed in each sharing unit for the network questions, and subsequently the percent of sharing units with network and wealth data. The Wealth columns show the mean, median, and standard deviation of the total wealth of sharing units in each community.}\\
\toprule
\multicolumn{6}{c}{ } & \multicolumn{3}{c}{Members} & \multicolumn{3}{c}{Sampling} & \multicolumn{3}{c}{Wealth} \\
\cmidrule(l{3pt}r{3pt}){7-9} \cmidrule(l{3pt}r{3pt}){10-12} \cmidrule(l{3pt}r{3pt}){13-15}
Code & Participants & Country & Year & Lead(s) & \# SU & Mean & SD & Max & \# Surv & \% w/ Net & \% w/ Wealth & Mean & Median & SD\\
\midrule
\endfirsthead
\caption[]{Summary of each community and its sharing unit characteristics, sampling, and wealth. The values for the material wealth of sharing units are reported in U.S. dollars, circa the time of data collection at the respective field sites. The Members columns show the average, standard deviation, and maximum number of sharing unit members in each community. The Sampling columns show the average number of people surveyed in each sharing unit for the network questions, and subsequently the percent of sharing units with network and wealth data. The Wealth columns show the mean, median, and standard deviation of the total wealth of sharing units in each community. \textit{(continued)}}\\
\toprule
Code & Participants & Country & Year & Lead(s) & \# SU & Mean & SD & Max & \# Surv & \% w/ Net & \% w/ Wealth & Mean & Median & SD\\
\midrule
\endhead

\endfoot
\bottomrule
\endlastfoot
AH & Ahus islanders & Papua New Guinea & 2018 & Michele Barnes and Joshua Cinner & 142 & 4.84 & 2.65 & 19 & 1.01 & 0.96 & 0.96 & 7488 & 5408 & 4412\\
AR & Argentinean & Argentina & 2024 & Madalena Monteban, Juan Pablo Ferreiro and Federico Fernandez & 176 & 3.10 & 1.81 & 10 & 1.00 & 0.90 & 0.90 & 100949 & 12602 & 839626\\
AV & Xwla & Benin & 2023 & Augusto Dalla Ragione & 122 & 4.65 & 3.37 & 17 & 1.75 & 0.95 & 1.00 & 417 & 220 & 567\\
AZ & Tamil & India & 2017 & Eleanor Power & 191 & 3.23 & 1.54 & 9 & 2.32 & 0.99 & 1.00 & 11929 & 8900 & 11032\\
BD & Bengalis & Bangladesh & 2017 & Mary Shenk and Nurul Alami & 79 & 3.78 & 1.58 & 8 & 1.00 & 1.00 & 1.00 & 9845 & 6277 & 20425\\
BM & Rancheros or Choyeros & Mexico & 2017 & Shane Macfarlan & 32 & 3.53 & 1.52 & 8 & 1.90 & 0.97 & 0.97 & 73895 & 49594 & 96031\\
BY & BaYaka & Congo-Brazzaville & 2018 & Sheina Lew-Levy & 43 & 6.58 & 3.39 & 14 & 2.73 & 0.95 & 0.98 & 108 & 104 & 45\\
CP & Chugurpampans & Peru & 2019 & Kathryn Oths & 90 & 3.22 & 1.81 & 8 & 1.11 & 1.00 & 1.00 & 18614 & 16212 & 9176\\
CR & Ecuadorians & Ecuador & 2019 & Christine Beitl and Wendy Ch\'{a}vez-P\'{a}ez & 92 & 4.33 & 2.66 & 16 & 1.07 & 0.95 & 0.89 & 24767 & 21192 & 18025\\
DJ & Xwla and Fon & Benin & 2023 & Augusto Dalla Ragione & 254 & 3.80 & 2.89 & 21 & 1.51 & 0.91 & 1.00 & 411 & 201 & 661\\
EK & Khasi Khynriam & India & 2020 & Alexandra Alvergne and Banrida Langstieh & 46 & 5.83 & 3.36 & 17 & 1.17 & 0.91 & 0.91 & 4213 & 2842 & 4092\\
FF & Damara, Nama, Ovaherero, and Ovambo & Namibia & 2018 & Michael Schnegg & 44 & 3.68 & 2.66 & 12 & 1.00 & 1.00 & 1.00 & 6158 & 3737 & 5786\\
FJ & Yasawans & Fiji & 2017 & Matthew Gervais & 44 & 3.98 & 2.20 & 9 & 1.00 & 0.95 & 0.95 & 15438 & 14021 & 7278\\
HE & Solos & Papua New Guinea & 2019 & Gianluca Grimalda & 64 & 6.36 & 2.46 & 12 & 1.00 & 0.98 & 0.98 & 2866 & 1923 & 3199\\
HI & Himba & Namibia & 2017 & Brooke Scelza and Sean Prall & 39 & 16.49 & 9.05 & 38 & 1.00 & 1.00 & 1.00 & 17766 & 11218 & 18133\\
IH & Mestizo & Colombia & 2018 & Cody Ross & 78 & 3.41 & 1.92 & 10 & 1.99 & 0.99 & 1.00 & 2458 & 2199 & 1589\\
KA & Karo & Indonesia & 2018 & Geoff Kushnick & 101 & 3.96 & 1.41 & 8 & 1.01 & 0.98 & 1.00 & 39799 & 38727 & 25091\\
KM & Koryak & Russia & 2019 & Drew Gerkey & 95 & 2.66 & 1.83 & 12 & NaN & 0.00 & 0.76 & 13735 & 11268 & 9264\\
KO & Kore & Ethiopia & 2017 & Mark Caudell & 67 & 7.28 & 3.76 & 20 & 1.84 & 1.00 & 1.00 & 5099 & 4341 & 3566\\
KT & Semai & Malaysia & 2023 & Vivek Venkataraman & 18 & 5.00 & 1.81 & 10 & 1.00 & 1.00 & 1.00 & 16657 & 15094 & 9394\\
LB & Mosuo & China & 2017 & Chun-Yi Sum and Siobh\'{a}n Cully & 30 & 3.83 & 1.42 & 7 & 1.00 & 1.00 & 1.00 & 16463 & 14492 & 7548\\
MA & Maasai & Tanzania & 2017 & Mark Caudell & 42 & 6.98 & 3.76 & 19 & 1.98 & 1.00 & 1.00 & 11430 & 8460 & 10769\\
MC & Macushi & Guyana & 2017 & Curtis Atkisson & 116 & 4.42 & 2.39 & 11 & 1.64 & 0.99 & 0.66 & 3503 & 2733 & 2618\\
MG & Betsimisaraka & Madagascar & 2019 & Bapu Vaitla and Christopher Golden & 101 & 3.25 & 1.61 & 9 & 1.00 & 1.00 & 1.00 & 4724 & 4300 & 3098\\
MK & Mikea & Madagascar & 2017 & Bram Tucker & 48 & 5.42 & 2.60 & 10 & 1.81 & 1.00 & 1.00 & 277 & 127 & 439\\
MN & Moseten & Bolivia & 2018 & Edmond Seabright & 31 & 4.58 & 1.88 & 9 & 1.27 & 0.97 & 0.97 & 9787 & 9496 & 4173\\
MY & Yucatac Maya & Mexico & 2017 & Karen Kramer & 93 & 5.84 & 2.93 & 15 & 1.00 & 0.96 & 0.97 & 1524 & 825 & 3535\\
NI & Mayangna & Nicaragua & 2018 & Jeremy Koster & 42 & 8.60 & 4.42 & 20 & 2.39 & 0.98 & 1.00 & 2932 & 1021 & 6708\\
NP & Nepali & Nepal & 2024 & Ivan Deschenaux & 163 & 2.71 & 1.37 & 8 & 2.04 & 0.99 & 1.00 & 27481 & 15527 & 28975\\
PC & Afrocolombian, Embera, and Mestizo & Colombia & 2017 & Cody Ross & 61 & 4.49 & 2.21 & 10 & 2.12 & 0.97 & 0.97 & 1613 & 1080 & 1762\\
PE & Pembans & Tanzania & 2018 & Emmanuel Maliti and Monique Borgerhoff Mulder & 44 & 6.95 & 2.88 & 13 & 1.93 & 1.00 & 1.00 & 9988 & 7634 & 8486\\
PT & Pantaneiros & Brazil & 2019 & Rafael Morais Chiaravalloti & 60 & 4.70 & 2.37 & 14 & 1.15 & 0.98 & 1.00 & 9424 & 7633 & 6175\\
RA & Haitians & Haiti & 2018 & Angelina Demarco & 55 & 3.63 & 2.37 & 12 & 1.00 & 0.98 & 0.98 & 622 & 480 & 527\\
SH & Shodagor & Bangladesh & 2017 & Katherine Starkweather & 47 & 4.74 & 2.37 & 17 & 1.05 & 0.94 & 0.96 & 6037 & 4248 & 6536\\
SN & Sena & Mozambique & 2019 & John Ziker and Karl Mertens & 34 & 9.12 & 5.53 & 22 & 1.85 & 1.00 & 1.00 & 1032 & 833 & 703\\
TE & Tamil & India & 2017 & Eleanor Power & 154 & 3.19 & 1.48 & 7 & 2.28 & 0.99 & 0.99 & 10280 & 6690 & 10870\\
TI & Tsimane' & Bolivia & 2019 & Bret Beheim & 43 & 5.23 & 2.39 & 10 & 1.95 & 0.88 & 0.91 & 3016 & 2141 & 2741\\
TM & Tzotzil & Mexico & 2022 & Werner Hertzog & 75 & 5.45 & 2.59 & 16 & 1.79 & 0.96 & 0.96 & 14397 & 8335 & 21310\\
TP & Tiriyó, Wayana, and Akuriyó & NA & NA & NA & 56 & 6.38 & 4.77 & 26 & 1.74 & 0.96 & 1.00 & 4379 & 3836 & 3389\\
TS & Tsimane & Bolivia & 2018 & Edmond Seabright & 65 & 5.18 & 2.34 & 10 & 1.06 & 0.72 & 0.77 & 4739 & 4214 & 3085\\
TZ & High Atlas Amazighs & Morocco & 2023 & Sarah Alami & 77 & 5.47 & 2.84 & 14 & 2.74 & 0.91 & 0.92 & 15592 & 10970 & 15193\\
UP & North Indian & India & 2025 & Komal Chauhan & 156 & 4.72 & 2.18 & 16 & 3.11 & 0.97 & 0.97 & 34969 & 19459 & 48983\\
VT & Ni-Vanuatu & Vanuatu & 2018 & Siobh\'{a}n Cully & 34 & 4.76 & 2.58 & 9 & 1.03 & 0.97 & 1.00 & 26745 & 6589 & 34433\\
WH & Mestizo & Colombia & 2019 & Cody Ross & 25 & 4.32 & 2.66 & 13 & 2.64 & 1.00 & 1.00 & 1624 & 960 & 1560\\
WL & Afrocolombian, Embera, and Mestizo & Colombia & 2017 & Cody Ross & 104 & 3.82 & 2.32 & 13 & 1.71 & 1.00 & 1.00 & 1303 & 872 & 1753\\
YI & Yi & China & 2018 & Madeline Brown & 74 & 5.38 & 1.42 & 9 & 1.00 & 0.93 & 0.99 & 62900 & 51987 & 28236\\
YN & Mosuo & China & 2017 & Siobh\'{a}n Cully & 40 & 4.97 & 1.83 & 10 & 1.00 & 0.98 & 1.00 & 16074 & 14245 & 12382\\*
\end{longtable}
\endgroup{}

%% file: tables/within_site_multilevel_tables.tex
\begin{table}[ht!]
\centering
\caption{\textbf{Multilevel estimates corresponding to Figure~\ref{fig:within-site-multipanel}, panels A and B.}}
\label{tab:within-site-multilevel-degree}
\begin{tabular}{lcc}
\toprule
 & \multicolumn{2}{c}{Dependent variable: wealth per capita rank} \\
\cmidrule(lr){2-3}
 & \shortstack{Panel A\\Support Access p.c. rank} & \shortstack{Panel B\\Support Provision p.c. rank} \\
\midrule
Pooled slope ($\beta$) & 0.220 & 0.234 \\
Standard error & (0.029) & (0.022) \\
95\% CI & [0.163, 0.277] & [0.191, 0.277] \\
\midrule
SD of site slopes ($\tau$) & 0.154 & 0.091 \\
Residual SD ($\sigma$) & 0.952 & 0.960 \\
Sharing units & 3,428 & 3,483 \\
Sites & 46 & 46 \\
\bottomrule
\end{tabular}

\vspace{0.5em}
\begin{minipage}{0.92\textwidth}
\footnotesize
\emph{Notes}: This table reports estimates from varying-slope linear mixed models: $Y_{is} = \beta_s X_{is} + \varepsilon_{is}$, with $\beta_s \sim N(\beta,\tau^2)$ and $\varepsilon_{is} \sim N(0,\sigma^2)$. Both the dependent variable and predictor are percentile ranks standardized within site. We omit intercepts because within-site standardization fixes each site's mean at zero. Support access (i.e., out-degree) and support provisioning (i.e., in-degree) are calculated on the composite network.
\end{minipage}
\end{table}

\begin{table}[ht!]
\centering
\caption{\textbf{Multilevel estimates corresponding to Figure~\ref{fig:within-site-multipanel}, panels C and D.}}
\label{tab:within-site-multilevel-aaw}
\begin{tabular}{lcc}
\toprule
 & \multicolumn{2}{c}{Dependent variable: wealth per capita rank} \\
\cmidrule(lr){2-3}
 & \shortstack{Panel C\\$AAW_i$ of supporters rank} & \shortstack{Panel D\\$AAW_i$ of supportees rank} \\
\midrule
Pooled slope ($\beta$) & 0.191 & 0.179 \\
Standard error & (0.031) & (0.031) \\
95\% CI & [0.131, 0.251] & [0.117, 0.240] \\
\midrule
SD of site slopes ($\tau$) & 0.165 & 0.170 \\
Residual SD ($\sigma$) & 0.956 & 0.956 \\
Sharing units & 3,330 & 3,257 \\
Sites & 46 & 46 \\
\bottomrule
\end{tabular}

\vspace{0.5em}
\begin{minipage}{0.92\textwidth}
\footnotesize
\emph{Notes}: This table reports estimates from varying-slope linear mixed models: $Y_{is} = \beta_s X_{is} + \varepsilon_{is}$, with $\beta_s \sim N(\beta,\tau^2)$ and $\varepsilon_{is} \sim N(0,\sigma^2)$. The predictor is the percentile rank of average alter wealth per capita, using alters who are supporters or supportees. Both the dependent variable and predictor are percentile ranks standardized within site. We omit intercepts because within-site standardization fixes each site's mean at zero. We use the composite network.
\end{minipage}
\end{table}

%% file: tables/Varbs.tex
\begin{table}[ht] \centering 
  \caption{Additional variables characterizing each community.} 
  \label{tab:variables} 
%\resizebox{\textwidth}{!}
\footnotesize
{\begin{tabularx}{\textwidth}{l X c c}
\\[-1.8ex]\hline 
\hline \\[-1.8ex] 
\\[-1.8ex] & Description & Scale & Source \\ 
\hline \\[-1.8ex] 
 %RAAW & Relative Average Alter Wealth, see Section~\ref{sec:definitions-measures} for full definition & Community & survey data \\ 
 %EC & Economic Connectedness, see Section~\ref{sec:definitions-measures} for full definition & Community & survey data \\ 
 %RC & Relative Connectedness, see Section~\ref{sec:definitions-measures} for full definition & Community & survey data \\ 
 \% Connections to Primary Kin & The percent of support ties that are to close kin & Community & survey data \\ 
 Strong Production Norms & Whether there are strong normative restrictions on who can undertake different types of productive work  & Community & anthropologist \\ 
 Time in Wage/Salaried Labor \% & Percent of productive time spent in the community doing wage or salaried labor & Community & anthropologist \\ 
 Cash-Oriented Production \% & Percent of produced goods in the community that are sold for cash & Community & anthropologist \\
 Count of More-Than-Yearly Shocks & Of climate, health, economic, and violence-related shocks, how many occur at least more-than-yearly  & Immediate Area & anthropologist \\
 %Population Density & & Immediate Area & \citep{ciesin_gridded_2017} \\
 Biologically-Effective Degree Days & Measure of thermal suitability/favorability for growth. The sum of daily mean temperatures above 10°C and less than 30°C, over 10 days. (higher = greater growth potential) & Immediate Area & \citep{nobakht_agroclimatic_2019} \\
 Human Influence Index & Index combining population density, infrastructure, and accessibility (higher = greater influence) & Immediate Area & \citep{sanderson_march_2022}  \\
 Property Rights Index & Index of property rights for men and women, ``$v2xcl\_prpty$'',  (higher = more widely-shared rights) & Nation & \citep{vdem2025} \\ 
 Freedom of Movement Index & Index of freedom of movement within the country for men and women, ``$v2xcl\_dmove$'',  (higher = more freedom) & Nation & \citep{vdem2025}  \\ 
 Corruption Index & Pervasiveness of public sector, executive, legislative, and judicial corruption, ``$v2x\_corr$'', (higher = more corrupt) & Nation & \citep{vdem2025} \\ 
 National GDP per Capita & Gross Domestic Product per capita (constant 2015 US\$) & Nation & \citep{worldbank2025wdi} \\ 
\hline 
\hline \\[-1.8ex] 
\end{tabularx}} 
\end{table}

%% file: tables/CorrelationsWithgini_wealth_per_capita_only_raaw.tex
\begin{table}[!h]
\centering
\caption{\label{tab:corr_gini_wealth_per_capita}Pairwise correlations with Wealth per Capita Gini. }
\centering
\fontsize{9}{11}\selectfont
\begin{threeparttable}
\begin{tabular}[t]{lrr}
\toprule
Variable & Correlation & 95\% CI\\
\midrule
Private Property Rights Index & \textbf{0.50} & {}[0.25, 0.69]\\
Time in Wage/Salaried Labor % & \textbf{0.43} & {}[0.16, 0.64]\\
Human Influence Index & \textbf{0.32} & {}[0.03, 0.56]\\
Freedom of Movement Index & 0.22 & {}[-0.07, 0.48]\\
Biologically-Effective Degree Days & 0.20 & {}[-0.09, 0.47]\\
Count of More-Than-Yearly Shocks & 0.16 & {}[-0.14, 0.43]\\
Cash-Oriented Production % & 0.10 & {}[-0.20, 0.38]\\
Strong Production Norms & 0.00 & {}[-0.29, 0.29]\\
Corruption Index & -0.02 & {}[-0.31, 0.27]\\
National GDP per Capita & -0.03 & {}[-0.32, 0.26]\\
% Connections to Primary Kin & -0.16 & {}[-0.43, 0.14]\\
Relative Average Alter Wealth & \textbf{-0.47} & {}[-0.67, -0.20]\\
\bottomrule
\end{tabular}
\begin{tablenotes}
\item \textit{Note:} 
\item Point estimates are Pearson correlations between each listed variable and Wealth per Capita Gini; intervals are 95 percent confidence intervals based on Fisher's z-transformation. Bolded rows denote intervals that exclude zero.
\end{tablenotes}
\end{threeparttable}
\end{table}

%% file: tables/Stepwise_Regression_from_Lasso.tex
% Table created by stargazer v.5.2.3 by Marek Hlavac, Social Policy Institute. E-mail: marek.hlavac at gmail.com
% Date and time: Mon, Aug 10, 2026 - 13:07:14
\begin{table}[!htbp] \centering 
  \caption{Stepwise regression results predicting the Gini of wealth per capita.} 
  \label{tab:stepwise_regression_lasso} 
\small 
\begin{threeparttable}
\begin{tabular}{@{\extracolsep{5pt}}lccc} 
\\[-1.8ex]\hline 
\hline \\[-1.8ex] 
 & \multicolumn{3}{c}{\textit{Dependent variable:}} \\ 
\cline{2-4} 
\\[-1.8ex] & \multicolumn{3}{c}{Wealth per Capita Gini} \\ 
\\[-1.8ex] & (1) & (2) & (3)\\ 
\hline \\[-1.8ex] 
 Relative Average Alter Wealth & $-$0.300$^{**}$ (0.119) & $-$0.315$^{**}$ (0.119) & $-$0.344$^{**}$ (0.133) \\ 
  Time in Wage/Salaried Labor \% & 0.265$^{**}$ (0.125) & 0.314$^{**}$ (0.133) & 0.278$^{*}$ (0.149) \\ 
  Private Property Rights Index & 0.381$^{***}$ (0.118) & 0.298$^{**}$ (0.128) & 0.394$^{**}$ (0.180) \\ 
  Human Influence Index & 0.063 (0.126) & 0.063 (0.130) & 0.057 (0.146) \\ 
  Biologically-Effective Degree Days &  & 0.209 (0.129) & 0.255 (0.152) \\ 
  \% Connections to Primary Kin &  & $-$0.123 (0.118) & $-$0.136 (0.158) \\ 
  Strong Production Norms &  & $-$0.138 (0.123) & $-$0.126 (0.137) \\ 
  National GDP per Capita &  &  & 0.067 (0.166) \\ 
  Count of More-Than-Yearly Shocks &  &  & $-$0.011 (0.141) \\ 
  Freedom of Movement Index &  &  & $-$0.169 (0.180) \\ 
  Corruption Index &  &  & $-$0.064 (0.139) \\ 
  Cash-Oriented Production \% &  &  & 0.033 (0.128) \\ 
 \hline \\[-1.8ex] 
Observations & 46 & 46 & 46 \\ 
R$^{2}$ & 0.466 & 0.514 & 0.533 \\ 
Adjusted R$^{2}$ & 0.415 & 0.427 & 0.368 \\ 
Residual Std. Error & 0.756 (df = 42) & 0.749 (df = 39) & 0.786 (df = 34) \\ 
F Statistic & 9.163$^{***}$ (df = 4; 42) & 5.888$^{***}$ (df = 7; 39) & 3.237$^{***}$ (df = 12; 34) \\ 
\hline 
\hline \\[-1.8ex] 
\end{tabular} 
\begin{tablenotes}
\small
\item \textit{Note:} Variables first included based on significant bivariate correlation, then those highlighted by the LASSO, and finally the full set of variables under consideration. $^{*}p<0.1$; $^{**}p<0.05$; $^{***}p<0.01$. Standard errors in parentheses.
\end{tablenotes}
\end{threeparttable}
\end{table}

%% file: tables/CorrelationsWithrel_frank_capita.tex
\begin{table}[!h]
\centering
\caption{\label{tab:corr_rel_frank_capita}Pairwise correlations with Relative Average Alter Wealth. }
\centering
\fontsize{9}{11}\selectfont
\begin{threeparttable}
\begin{tabular}[t]{lrr}
\toprule
Variable & Correlation & 95\% CI\\
\midrule
National GDP per Capita & 0.16 & {}[-0.14, 0.43]\\
Cash-Oriented Production \% & 0.09 & {}[-0.21, 0.37]\\
\% Connections to Primary Kin & 0.01 & {}[-0.28, 0.30]\\
Biologically-Effective Degree Days & -0.03 & {}[-0.32, 0.26]\\
Freedom of Movement Index & -0.11 & {}[-0.39, 0.18]\\
Strong Production Norms & -0.12 & {}[-0.40, 0.17]\\
Corruption Index & -0.13 & {}[-0.40, 0.17]\\
Count of More-Than-Yearly Shocks & -0.19 & {}[-0.46, 0.11]\\
Human Influence Index & -0.21 & {}[-0.47, 0.08]\\
Private Property Rights Index & -0.22 & {}[-0.48, 0.08]\\
Time in Wage/Salaried Labor \% & -0.26 & {}[-0.51, 0.03]\\
\bottomrule
\end{tabular}
\begin{tablenotes}
\item \textit{Note:} 
\item Point estimates are Pearson correlations between each listed variable and Relative Average Alter Wealth; intervals are 95 percent confidence intervals based on Fisher's z-transformation. Bolded rows denote intervals that exclude zero.
\end{tablenotes}
\end{threeparttable}
\end{table}

%% file: tables/Stepwise_Regression_from_Lasso_EC.tex
% Table created by stargazer v.5.2.3 by Marek Hlavac, Social Policy Institute. E-mail: marek.hlavac at gmail.com
% Date and time: Mon, Aug 10, 2026 - 13:07:14
\begin{table}[!htbp] \centering 
  \caption{Stepwise regression results predicting Relative Average Alter Wealth.} 
  \label{tab:stepwise_regression_lasso_ec} 
\small 
\begin{threeparttable}
\begin{tabular}{@{\extracolsep{5pt}}lccc} 
\\[-1.8ex]\hline 
\hline \\[-1.8ex] 
 & \multicolumn{3}{c}{\textit{Dependent variable:}} \\ 
\cline{2-4} 
\\[-1.8ex] & \multicolumn{3}{c}{Relative Average Alter Wealth} \\ 
\\[-1.8ex] & (1) & (2) & (3)\\ 
\hline \\[-1.8ex] 
 Time in Wage/Salaried Labor \% & $-$0.200 (0.157) & $-$0.208 (0.174) & $-$0.201 (0.187) \\ 
  Private Property Rights Index & $-$0.164 (0.148) & $-$0.170 (0.168) & $-$0.161 (0.227) \\ 
  Human Influence Index & $-$0.097 (0.160) & $-$0.081 (0.173) & $-$0.160 (0.183) \\ 
  Biologically-Effective Degree Days &  & 0.004 (0.172) & 0.106 (0.192) \\ 
  \% Connections to Primary Kin &  & $-$0.070 (0.157) & $-$0.184 (0.198) \\ 
  Strong Production Norms &  & $-$0.090 (0.162) & $-$0.147 (0.173) \\ 
  National GDP per Capita &  &  & 0.217 (0.207) \\ 
  Count of More-Than-Yearly Shocks &  &  & $-$0.220 (0.175) \\ 
  Freedom of Movement Index &  &  & 0.006 (0.229) \\ 
  Corruption Index &  &  & $-$0.061 (0.177) \\ 
  Cash-Oriented Production \% &  &  & 0.080 (0.162) \\ 
 \hline \\[-1.8ex] 
Observations & 46 & 46 & 46 \\ 
R$^{2}$ & 0.109 & 0.120 & 0.225 \\ 
Adjusted R$^{2}$ & 0.047 & $-$0.013 & $-$0.019 \\ 
Residual Std. Error & 0.966 (df = 43) & 0.995 (df = 40) & 0.998 (df = 35) \\ 
F Statistic & 1.753 (df = 3; 43) & 0.905 (df = 6; 40) & 0.923 (df = 11; 35) \\ 
\hline 
\hline \\[-1.8ex] 
\end{tabular} 
\begin{tablenotes}
\small
\item \textit{Note:} Variables included in the same order as done when predicting the Gini of wealth per capita. $^{*}p<0.1$; $^{**}p<0.05$; $^{***}p<0.01$. Standard errors in parentheses.
\end{tablenotes}
\end{threeparttable}
\end{table}

%% file: tables/wealth_site_vars_table.tex
\begingroup\fontsize{9}{11}\selectfont

\begin{longtable}[t]{ccccccc}
\caption{\label{tab:site_vars_wealth}Site-Level Wealth Inequality Variables}\\
\toprule
Site & \makecell{Absolute\\Gini} & \makecell{Per Capita\\Gini} & \makecell{Per Adult\\Gini} & \makecell{Size Adj.\\Gini} & \makecell{Per Capita\\90-10 ratio} & \makecell{Per Capita\\80-20 ratio}\\
\midrule
\endfirsthead
\caption[]{Site-Level Wealth Inequality Variables \textit{(continued)}}\\
\toprule
Site & \makecell{Absolute\\Gini} & \makecell{Per Capita\\Gini} & \makecell{Per Adult\\Gini} & \makecell{Size Adj.\\Gini} & \makecell{Per Capita\\90-10 ratio} & \makecell{Per Capita\\80-20 ratio}\\
\midrule
\endhead

\endfoot
\bottomrule
\endlastfoot
AH & 0.30 & 0.41 & 0.36 & 0.30 & 5.96 & 3.37\\
AR & 0.89 & 0.88 & 0.85 & 0.89 & 22.28 & 7.35\\
AV & 0.59 & 0.58 & 0.58 & 0.56 & 18.63 & 7.68\\
AZ & 0.48 & 0.54 & 0.50 & 0.49 & 24.09 & 8.62\\
BD & 0.57 & 0.61 & 0.60 & 0.57 & 17.55 & 6.69\\
BM & 0.58 & 0.61 & 0.61 & 0.60 & 17.01 & 9.82\\
BY & 0.23 & 0.30 & 0.27 & 0.22 & 3.32 & 2.57\\
CP & 0.25 & 0.42 & 0.32 & 0.25 & 7.49 & 4.44\\
CR & 0.38 & 0.44 & 0.40 & 0.38 & 7.78 & 3.36\\
DJ & 0.63 & 0.60 & 0.59 & 0.59 & 15.59 & 6.79\\
EK & 0.48 & 0.46 & 0.39 & 0.45 & 6.86 & 3.65\\
FF & 0.47 & 0.49 & 0.49 & 0.44 & 8.06 & 4.58\\
FJ & 0.26 & 0.35 & 0.31 & 0.25 & 5.88 & 3.26\\
HE & 0.46 & 0.50 & 0.48 & 0.46 & 5.23 & 2.97\\
HI & 0.55 & 0.50 & 0.53 & 0.50 & 22.42 & 9.42\\
IH & 0.35 & 0.52 & 0.45 & 0.35 & 10.60 & 5.09\\
KA & 0.35 & 0.46 & 0.39 & 0.36 & 26.22 & 4.94\\
KM & 0.37 & 0.33 & 0.31 & 0.31 & 4.03 & 2.74\\
KO & 0.39 & 0.37 & 0.39 & 0.37 & 7.39 & 3.88\\
KT & 0.32 & 0.29 & 0.31 & 0.29 & 3.30 & 2.18\\
LB & 0.26 & 0.31 & 0.28 & 0.26 & 3.21 & 2.64\\
MA & 0.48 & 0.46 & 0.43 & 0.45 & 7.67 & 3.58\\
MC & 0.37 & 0.36 & 0.37 & 0.34 & 5.34 & 3.09\\
MG & 0.36 & 0.40 & 0.35 & 0.35 & 6.41 & 3.00\\
MK & 0.57 & 0.54 & 0.56 & 0.53 & 7.11 & 3.34\\
MN & 0.22 & 0.26 & 0.32 & 0.21 & 3.34 & 2.89\\
MY & 0.51 & 0.51 & 0.48 & 0.49 & 7.83 & 3.31\\
NI & 0.68 & 0.66 & 0.66 & 0.66 & 19.77 & 5.47\\
NP & 0.49 & 0.55 & 0.51 & 0.49 & 15.13 & 5.11\\
PC & 0.52 & 0.64 & 0.58 & 0.52 & 45.87 & 11.68\\
PE & 0.39 & 0.34 & 0.35 & 0.34 & 5.00 & 2.63\\
PT & 0.32 & 0.48 & 0.42 & 0.33 & 4.66 & 2.67\\
RA & 0.44 & 0.43 & 0.49 & 0.40 & 5.78 & 3.27\\
SH & 0.55 & 0.55 & 0.59 & 0.56 & 20.61 & 7.87\\
SN & 0.38 & 0.29 & 0.31 & 0.25 & 3.20 & 1.88\\
TE & 0.49 & 0.57 & 0.56 & 0.50 & 14.12 & 5.02\\
TI & 0.46 & 0.43 & 0.42 & 0.42 & 7.91 & 5.24\\
TM & 0.50 & 0.41 & 0.38 & 0.41 & 5.90 & 2.88\\
TS & 0.35 & 0.45 & 0.35 & 0.37 & 5.68 & 3.50\\
TZ & 0.49 & 0.44 & 0.38 & 0.44 & 16.61 & 5.05\\
UP & 0.62 & 0.65 & 0.63 & 0.65 & 54.67 & 18.80\\
VT & 0.64 & 0.67 & 0.66 & 0.64 & 48.87 & 11.09\\
WH & 0.47 & 0.45 & 0.37 & 0.42 & 6.62 & 4.56\\
WL & 0.51 & 0.65 & 0.55 & 0.51 & 31.10 & 7.28\\
YI & 0.22 & 0.24 & 0.24 & 0.21 & 2.76 & 1.84\\
YN & 0.41 & 0.38 & 0.35 & 0.37 & 6.05 & 3.54\\*
\end{longtable}
\endgroup{}

%% file: tables/deg_gini_site_vars_table.tex
\begingroup\fontsize{9}{11}\selectfont

\begin{longtable}[t]{ccccccccc}
\caption{\label{tab:site_vars_deg_gini}Site-Level Access and Provisioning Support Gini Variables}\\
\toprule
\multicolumn{1}{c}{ } & \multicolumn{4}{c}{Support Access Gini} & \multicolumn{4}{c}{Support Provisioning Gini} \\
\cmidrule(l{3pt}r{3pt}){2-5} \cmidrule(l{3pt}r{3pt}){6-9}
Site & \makecell{Per Capita} & \makecell{Absolute} & \makecell{Per Adult} & \makecell{Size-Adjusted} & \makecell{Per Capita} & \makecell{Absolute} & \makecell{Per Adult} & \makecell{Size-Adjusted}\\
\midrule
\endfirsthead
\caption[]{Site-Level Access and Provisioning Support Gini Variables \textit{(continued)}}\\
\toprule
\multicolumn{1}{c}{ } & \multicolumn{4}{c}{Support Access Gini} & \multicolumn{4}{c}{Support Provisioning Gini} \\
\cmidrule(l{3pt}r{3pt}){2-5} \cmidrule(l{3pt}r{3pt}){6-9}
Site & \makecell{Per Capita} & \makecell{Absolute} & \makecell{Per Adult} & \makecell{Size-Adjusted} & \makecell{Per Capita} & \makecell{Absolute} & \makecell{Per Adult} & \makecell{Size-Adjusted}\\
\midrule
\endhead

\endfoot
\bottomrule
\endlastfoot
AH & 0.27 & 0.37 & 0.31 & 0.26 & 0.32 & 0.40 & 0.35 & 0.31\\
AR & 0.33 & 0.46 & 0.41 & 0.31 & 0.40 & 0.44 & 0.41 & 0.34\\
AV & 0.37 & 0.47 & 0.35 & 0.36 & 0.44 & 0.46 & 0.41 & 0.38\\
AZ & 0.29 & 0.31 & 0.27 & 0.27 & 0.34 & 0.33 & 0.35 & 0.30\\
BD & 0.26 & 0.39 & 0.37 & 0.26 & 0.38 & 0.44 & 0.47 & 0.35\\
BM & 0.21 & 0.35 & 0.30 & 0.22 & 0.37 & 0.42 & 0.38 & 0.36\\
BY & 0.32 & 0.34 & 0.30 & 0.29 & 0.22 & 0.36 & 0.29 & 0.22\\
CP & 0.41 & 0.49 & 0.43 & 0.35 & 0.48 & 0.52 & 0.46 & 0.37\\
CR & 0.28 & 0.38 & 0.29 & 0.27 & 0.36 & 0.40 & 0.33 & 0.33\\
DJ & 0.38 & 0.49 & 0.37 & 0.35 & 0.51 & 0.59 & 0.48 & 0.38\\
EK & 0.28 & 0.34 & 0.34 & 0.27 & 0.35 & 0.38 & 0.31 & 0.32\\
FF & 0.45 & 0.53 & 0.51 & 0.29 & 0.57 & 0.59 & 0.56 & 0.36\\
FJ & 0.25 & 0.32 & 0.33 & 0.23 & 0.30 & 0.33 & 0.30 & 0.26\\
HE & 0.27 & 0.35 & 0.32 & 0.27 & 0.36 & 0.40 & 0.37 & 0.35\\
HI & 0.29 & 0.49 & 0.48 & 0.29 & 0.45 & 0.43 & 0.43 & 0.38\\
IH & 0.52 & 0.62 & 0.57 & 0.31 & 0.56 & 0.64 & 0.57 & 0.32\\
KA & 0.17 & 0.32 & 0.25 & 0.17 & 0.35 & 0.40 & 0.36 & 0.34\\
KM & 0.29 & 0.39 & 0.34 & 0.25 & 0.44 & 0.50 & 0.45 & 0.36\\
KO & 0.21 & 0.34 & 0.32 & 0.22 & 0.34 & 0.42 & 0.38 & 0.31\\
KT & 0.29 & 0.27 & 0.32 & 0.26 & 0.31 & 0.24 & 0.23 & 0.24\\
LB & 0.29 & 0.39 & 0.36 & 0.24 & 0.40 & 0.45 & 0.40 & 0.33\\
MA & 0.21 & 0.38 & 0.29 & 0.20 & 0.32 & 0.30 & 0.30 & 0.28\\
MC & 0.26 & 0.36 & 0.23 & 0.24 & 0.47 & 0.51 & 0.49 & 0.42\\
MG & 0.19 & 0.36 & 0.27 & 0.19 & 0.61 & 0.64 & 0.61 & 0.55\\
MK & 0.33 & 0.44 & 0.37 & 0.31 & 0.39 & 0.43 & 0.38 & 0.38\\
MN & 0.26 & 0.33 & 0.36 & 0.26 & 0.26 & 0.32 & 0.33 & 0.26\\
MY & 0.22 & 0.34 & 0.31 & 0.22 & 0.29 & 0.30 & 0.28 & 0.26\\
NI & 0.20 & 0.22 & 0.29 & 0.15 & 0.33 & 0.30 & 0.33 & 0.29\\
NP & 0.30 & 0.32 & 0.29 & 0.28 & 0.34 & 0.37 & 0.35 & 0.33\\
PC & 0.38 & 0.48 & 0.42 & 0.28 & 0.43 & 0.44 & 0.44 & 0.32\\
PE & 0.25 & 0.32 & 0.31 & 0.25 & 0.33 & 0.36 & 0.40 & 0.33\\
PT & 0.29 & 0.39 & 0.36 & 0.28 & 0.44 & 0.47 & 0.45 & 0.37\\
RA & 0.29 & 0.38 & 0.36 & 0.29 & 0.46 & 0.46 & 0.45 & 0.36\\
SH & 0.30 & 0.36 & 0.39 & 0.30 & 0.35 & 0.34 & 0.37 & 0.31\\
SN & 0.27 & 0.37 & 0.33 & 0.25 & 0.38 & 0.28 & 0.29 & 0.25\\
TE & 0.27 & 0.31 & 0.28 & 0.24 & 0.35 & 0.33 & 0.34 & 0.30\\
TI & 0.29 & 0.27 & 0.30 & 0.26 & 0.37 & 0.32 & 0.34 & 0.28\\
TM & 0.38 & 0.41 & 0.41 & 0.36 & 0.36 & 0.41 & 0.37 & 0.31\\
TS & 0.28 & 0.35 & 0.31 & 0.25 & 0.43 & 0.50 & 0.43 & 0.39\\
TZ & 0.31 & 0.37 & 0.35 & 0.29 & 0.40 & 0.42 & 0.40 & 0.37\\
UP & 0.28 & 0.29 & 0.29 & 0.26 & 0.31 & 0.33 & 0.30 & 0.29\\
VT & 0.26 & 0.41 & 0.31 & 0.24 & 0.38 & 0.37 & 0.37 & 0.26\\
WH & 0.46 & 0.48 & 0.49 & 0.36 & 0.41 & 0.46 & 0.41 & 0.30\\
WL & 0.40 & 0.53 & 0.45 & 0.35 & 0.43 & 0.56 & 0.47 & 0.37\\
YI & 0.32 & 0.35 & 0.35 & 0.31 & 0.32 & 0.34 & 0.37 & 0.27\\
YN & 0.37 & 0.45 & 0.46 & 0.26 & 0.49 & 0.54 & 0.52 & 0.36\\*
\end{longtable}
\endgroup{}

%% file: tables/raaw_site_vars_table.tex
\begingroup\fontsize{9}{11}\selectfont

\begin{longtable}[t]{ccccccccc}
\caption{\label{tab:site_vars_raaw}Site-Level Relative Average Alter Wealth Variables}\\
\toprule
\multicolumn{1}{c}{ } & \multicolumn{4}{c}{RAAW (of supporters)} & \multicolumn{4}{c}{RAAW (of supportees)} \\
\cmidrule(l{3pt}r{3pt}){2-5} \cmidrule(l{3pt}r{3pt}){6-9}
Site & \makecell{Per Capita} & \makecell{Absolute} & \makecell{Per Adult} & \makecell{Size-Adjusted} & \makecell{Per Capita} & \makecell{Absolute} & \makecell{Per Adult} & \makecell{Size-Adjusted}\\
\midrule
\endfirsthead
\caption[]{Site-Level Relative Average Alter Wealth Variables \textit{(continued)}}\\
\toprule
\multicolumn{1}{c}{ } & \multicolumn{4}{c}{RAAW (of supporters)} & \multicolumn{4}{c}{RAAW (of supportees)} \\
\cmidrule(l{3pt}r{3pt}){2-5} \cmidrule(l{3pt}r{3pt}){6-9}
Site & \makecell{Per Capita} & \makecell{Absolute} & \makecell{Per Adult} & \makecell{Size-Adjusted} & \makecell{Per Capita} & \makecell{Absolute} & \makecell{Per Adult} & \makecell{Size-Adjusted}\\
\midrule
\endhead

\endfoot
\bottomrule
\endlastfoot
AH & 0.96 & 0.91 & 0.93 & 0.94 & 0.98 & 0.93 & 0.92 & 0.96\\
AR & 0.92 & 0.86 & 0.88 & 0.92 & 0.91 & 0.86 & 0.89 & 0.90\\
AV & 0.98 & 0.98 & 1.00 & 0.97 & 0.94 & 1.00 & 1.00 & 0.96\\
AZ & 0.78 & 0.82 & 0.82 & 0.78 & 0.78 & 0.79 & 0.78 & 0.76\\
BD & 0.90 & 0.97 & 0.95 & 0.92 & 0.90 & 0.96 & 0.89 & 0.95\\
BM & 0.75 & 0.76 & 0.70 & 0.71 & 0.75 & 0.91 & 0.77 & 0.79\\
BY & 0.88 & 1.07 & 1.14 & 1.03 & 0.89 & 1.12 & 1.16 & 1.10\\
CP & 0.89 & 0.89 & 0.94 & 0.90 & 0.82 & 1.01 & 0.88 & 0.85\\
CR & 0.86 & 0.88 & 0.99 & 0.92 & 0.85 & 0.90 & 0.90 & 0.86\\
DJ & 0.88 & 0.84 & 0.90 & 0.90 & 0.85 & 0.90 & 0.84 & 0.93\\
EK & 0.97 & 0.85 & 1.01 & 0.84 & 1.07 & 0.92 & 1.02 & 0.86\\
FF & 0.96 & 0.93 & 1.11 & 1.07 & 0.91 & 0.83 & 1.02 & 1.06\\
FJ & 0.93 & 0.94 & 0.95 & 0.95 & 0.97 & 0.89 & 0.87 & 0.96\\
HE & 1.00 & 0.86 & 0.97 & 0.96 & 0.87 & 0.81 & 0.91 & 0.84\\
HI & 0.84 & 0.87 & 0.82 & 0.88 & 0.69 & 0.75 & 0.76 & 0.77\\
IH & 0.93 & 1.04 & 0.88 & 0.94 & 0.72 & 0.93 & 0.88 & 0.72\\
KA & 0.94 & 0.92 & 0.91 & 0.91 & 0.86 & 0.86 & 0.84 & 0.85\\
KM & 1.06 & 0.94 & 1.08 & 1.07 & 0.98 & 1.00 & 0.99 & 1.08\\
KO & 0.89 & 0.96 & 0.95 & 0.99 & 0.93 & 0.94 & 0.93 & 1.02\\
KT & 0.95 & 1.10 & 0.98 & 1.01 & 0.95 & 1.04 & 0.88 & 1.03\\
LB & 0.93 & 0.86 & 1.00 & 0.84 & 0.95 & 0.90 & 1.07 & 0.86\\
MA & 0.96 & 0.97 & 0.97 & 0.89 & 0.87 & 0.90 & 0.95 & 0.91\\
MC & 0.99 & 1.00 & 1.02 & 0.95 & 0.98 & 0.94 & 0.95 & 0.93\\
MG & 0.90 & 0.88 & 0.91 & 0.88 & 1.00 & 0.98 & 0.92 & 0.96\\
MK & 0.92 & 0.77 & 0.75 & 0.82 & 0.89 & 0.90 & 0.84 & 0.90\\
MN & 1.01 & 0.88 & 1.17 & 1.01 & 1.04 & 0.95 & 1.00 & 1.01\\
MY & 0.97 & 1.02 & 0.97 & 1.01 & 0.94 & 0.95 & 1.00 & 0.97\\
NI & 0.84 & 0.83 & 0.79 & 0.82 & 0.80 & 0.77 & 0.76 & 0.78\\
NP & 0.75 & 0.70 & 0.79 & 0.74 & 0.73 & 0.73 & 0.79 & 0.73\\
PC & 0.66 & 0.57 & 0.76 & 0.73 & 0.75 & 0.68 & 0.82 & 0.80\\
PE & 1.01 & 1.05 & 1.07 & 1.01 & 0.96 & 1.03 & 1.05 & 0.99\\
PT & 1.00 & 0.98 & 0.97 & 1.03 & 0.93 & 0.97 & 0.96 & 0.96\\
RA & 0.96 & 0.91 & 1.02 & 0.96 & 0.96 & 0.90 & 0.91 & 0.86\\
SH & 0.98 & 1.09 & 0.97 & 0.94 & 0.97 & 1.11 & 0.97 & 0.96\\
SN & 0.93 & 0.94 & 1.01 & 0.96 & 0.98 & 0.77 & 0.95 & 0.94\\
TE & 0.95 & 0.92 & 0.96 & 0.94 & 0.93 & 0.90 & 0.94 & 0.92\\
TI & 1.03 & 0.94 & 0.96 & 0.95 & 0.97 & 0.90 & 0.94 & 0.87\\
TM & 0.90 & 0.92 & 0.90 & 0.96 & 0.90 & 1.01 & 1.00 & 1.01\\
TS & 0.90 & 0.86 & 0.82 & 0.85 & 0.86 & 0.73 & 0.77 & 0.82\\
TZ & 0.83 & 0.76 & 0.81 & 0.79 & 0.88 & 0.81 & 0.88 & 0.83\\
UP & 0.68 & 0.70 & 0.68 & 0.69 & 0.64 & 0.66 & 0.65 & 0.66\\
VT & 0.99 & 0.93 & 0.96 & 0.98 & 1.06 & 0.92 & 1.00 & 1.04\\
WH & 1.18 & 1.18 & 1.23 & 1.31 & 1.08 & 0.85 & 1.18 & 1.20\\
WL & 0.79 & 0.70 & 0.77 & 0.78 & 0.74 & 0.68 & 0.81 & 0.79\\
YI & 1.04 & 1.03 & 0.97 & 1.07 & 1.04 & 1.01 & 0.92 & 1.04\\
YN & 0.99 & 1.16 & 1.02 & 1.13 & 1.15 & 1.21 & 1.09 & 1.26\\*
\end{longtable}
\endgroup{}

%% file: tables/ec_site_vars_table.tex
\begingroup\fontsize{9}{11}\selectfont

\begin{longtable}[t]{ccccccccc}
\caption{\label{tab:site_vars_ec}Site-Level Economic Connectedness Variables}\\
\toprule
\multicolumn{1}{c}{ } & \multicolumn{4}{c}{EC (to supporters)} & \multicolumn{4}{c}{EC (to supportees)} \\
\cmidrule(l{3pt}r{3pt}){2-5} \cmidrule(l{3pt}r{3pt}){6-9}
Site & \makecell{Per Capita} & \makecell{Absolute} & \makecell{Per Adult} & \makecell{Size-Adjusted} & \makecell{Per Capita} & \makecell{Absolute} & \makecell{Per Adult} & \makecell{Size-Adjusted}\\
\midrule
\endfirsthead
\caption[]{Site-Level Economic Connectedness Variables \textit{(continued)}}\\
\toprule
\multicolumn{1}{c}{ } & \multicolumn{4}{c}{EC (to supporters)} & \multicolumn{4}{c}{EC (to supportees)} \\
\cmidrule(l{3pt}r{3pt}){2-5} \cmidrule(l{3pt}r{3pt}){6-9}
Site & \makecell{Per Capita} & \makecell{Absolute} & \makecell{Per Adult} & \makecell{Size-Adjusted} & \makecell{Per Capita} & \makecell{Absolute} & \makecell{Per Adult} & \makecell{Size-Adjusted}\\
\midrule
\endhead

\endfoot
\bottomrule
\endlastfoot
AH & 1.12 & 1.13 & 1.01 & 1.08 & 1.05 & 1.11 & 1.05 & 0.99\\
AR & 1.09 & 1.05 & 0.96 & 1.09 & 1.08 & 1.11 & 1.08 & 1.08\\
AV & 1.38 & 1.27 & 1.27 & 1.24 & 1.11 & 1.27 & 1.22 & 1.16\\
AZ & 0.93 & 0.92 & 0.97 & 0.87 & 0.86 & 0.93 & 0.88 & 0.81\\
BD & 0.77 & 0.86 & 0.85 & 0.77 & 0.91 & 1.16 & 1.00 & 0.95\\
BM & 0.81 & 0.83 & 0.69 & 0.61 & 0.64 & 1.01 & 0.75 & 0.65\\
BY & 1.03 & 1.47 & 1.26 & 1.09 & 1.03 & 1.48 & 1.33 & 1.19\\
CP & 1.19 & 1.22 & 1.14 & 1.18 & 0.74 & 1.33 & 1.02 & 0.76\\
CR & 0.88 & 1.13 & 1.31 & 1.14 & 0.84 & 1.19 & 1.14 & 1.01\\
DJ & 1.23 & 1.13 & 1.13 & 1.24 & 1.02 & 1.09 & 0.92 & 1.08\\
EK & 1.24 & 1.07 & 1.17 & 1.01 & 1.20 & 1.09 & 1.07 & 0.94\\
FF & 0.99 & 1.05 & 1.34 & 1.24 & 0.97 & 0.64 & 1.42 & 1.18\\
FJ & 0.84 & 1.03 & 1.04 & 0.88 & 0.97 & 0.98 & 0.99 & 0.92\\
HE & 1.23 & 1.05 & 1.17 & 1.14 & 1.04 & 1.04 & 1.16 & 1.02\\
HI & 1.21 & 1.07 & 0.99 & 1.02 & 0.84 & 0.84 & 0.74 & 0.80\\
IH & 0.94 & 1.28 & 1.17 & 0.94 & 0.68 & 1.08 & 1.07 & 0.68\\
KA & 1.00 & 1.04 & 0.97 & 1.07 & 0.94 & 1.00 & 1.00 & 0.90\\
KM & 1.24 & 1.15 & 1.20 & 1.30 & 1.00 & 1.05 & 1.03 & 1.14\\
KO & 1.03 & 1.33 & 1.26 & 1.20 & 0.82 & 1.06 & 1.07 & 1.05\\
KT & 1.04 & 1.00 & 0.97 & 0.95 & 0.95 & 0.99 & 0.98 & 0.92\\
LB & 1.14 & 1.02 & 1.16 & 0.99 & 1.11 & 1.01 & 1.25 & 0.95\\
MA & 1.22 & 1.08 & 1.19 & 1.11 & 0.88 & 0.99 & 0.95 & 0.95\\
MC & 1.17 & 1.27 & 1.29 & 1.10 & 1.01 & 1.22 & 1.06 & 1.00\\
MG & 1.36 & 1.21 & 1.28 & 1.28 & 1.19 & 1.12 & 1.11 & 1.11\\
MK & 1.19 & 1.01 & 0.80 & 1.04 & 0.92 & 1.07 & 0.90 & 0.87\\
MN & 1.09 & 1.06 & 1.32 & 1.03 & 1.20 & 1.22 & 1.52 & 1.20\\
MY & 1.02 & 1.00 & 1.02 & 1.00 & 0.94 & 1.08 & 1.14 & 0.98\\
NI & 1.17 & 1.12 & 0.97 & 1.02 & 0.81 & 0.87 & 0.87 & 0.80\\
NP & 0.82 & 0.79 & 0.89 & 0.82 & 0.78 & 0.73 & 0.87 & 0.78\\
PC & 0.82 & 0.61 & 1.04 & 0.98 & 0.71 & 0.75 & 1.17 & 0.90\\
PE & 0.96 & 1.16 & 1.26 & 0.99 & 1.10 & 1.24 & 1.38 & 1.09\\
PT & 1.17 & 1.14 & 1.19 & 1.14 & 0.95 & 1.17 & 1.16 & 0.91\\
RA & 1.32 & 1.28 & 1.48 & 1.32 & 1.00 & 1.05 & 1.17 & 0.98\\
SH & 1.21 & 1.29 & 1.22 & 1.07 & 0.96 & 1.31 & 1.20 & 1.19\\
SN & 1.42 & 1.05 & 1.21 & 1.23 & 1.05 & 0.78 & 1.02 & 0.73\\
TE & 1.07 & 1.08 & 1.05 & 1.01 & 0.91 & 1.03 & 1.07 & 0.99\\
TI & 1.27 & 1.07 & 1.11 & 1.16 & 1.13 & 0.97 & 1.15 & 0.99\\
TM & 1.22 & 1.32 & 1.23 & 1.17 & 1.13 & 1.24 & 1.25 & 1.06\\
TS & 1.18 & 1.08 & 1.15 & 1.00 & 0.85 & 0.92 & 0.80 & 0.82\\
TZ & 0.95 & 0.93 & 0.80 & 0.94 & 1.01 & 0.79 & 0.81 & 0.76\\
UP & 0.66 & 0.74 & 0.68 & 0.70 & 0.54 & 0.63 & 0.55 & 0.58\\
VT & 1.31 & 1.06 & 1.14 & 1.31 & 1.24 & 0.99 & 1.07 & 1.24\\
WH & 0.93 & 1.19 & 1.06 & 1.05 & 0.95 & 0.88 & 1.16 & 1.19\\
WL & 0.91 & 1.06 & 1.05 & 0.90 & 0.78 & 0.99 & 0.96 & 0.77\\
YI & 1.12 & 1.14 & 1.20 & 1.10 & 0.96 & 1.09 & 1.08 & 1.05\\
YN & 1.04 & 1.07 & 0.82 & 1.08 & 0.98 & 1.07 & 0.98 & 1.15\\*
\end{longtable}
\endgroup{}

%% file: tables/rc_site_vars_table.tex
\begingroup\fontsize{9}{11}\selectfont

\begin{longtable}[t]{ccccccccc}
\caption{\label{tab:site_vars_rc}Site-Level Relative Connectedness Variables}\\
\toprule
\multicolumn{1}{c}{ } & \multicolumn{4}{c}{RC (to supporters)} & \multicolumn{4}{c}{RC (to supportees)} \\
\cmidrule(l{3pt}r{3pt}){2-5} \cmidrule(l{3pt}r{3pt}){6-9}
Site & \makecell{Per Capita} & \makecell{Absolute} & \makecell{Per Adult} & \makecell{Size-Adjusted} & \makecell{Per Capita} & \makecell{Absolute} & \makecell{Per Adult} & \makecell{Size-Adjusted}\\
\midrule
\endfirsthead
\caption[]{Site-Level Relative Connectedness Variables \textit{(continued)}}\\
\toprule
\multicolumn{1}{c}{ } & \multicolumn{4}{c}{RC (to supporters)} & \multicolumn{4}{c}{RC (to supportees)} \\
\cmidrule(l{3pt}r{3pt}){2-5} \cmidrule(l{3pt}r{3pt}){6-9}
Site & \makecell{Per Capita} & \makecell{Absolute} & \makecell{Per Adult} & \makecell{Size-Adjusted} & \makecell{Per Capita} & \makecell{Absolute} & \makecell{Per Adult} & \makecell{Size-Adjusted}\\
\midrule
\endhead

\endfoot
\bottomrule
\endlastfoot
AH & 0.97 & 0.91 & 0.91 & 0.96 & 1.00 & 0.93 & 0.89 & 0.97\\
AR & 0.91 & 0.85 & 0.77 & 0.91 & 0.93 & 0.87 & 0.85 & 0.93\\
AV & 0.97 & 0.96 & 1.03 & 0.95 & 0.90 & 1.00 & 1.06 & 0.94\\
AZ & 0.71 & 0.75 & 0.75 & 0.66 & 0.69 & 0.73 & 0.67 & 0.63\\
BD & 0.79 & 0.90 & 0.82 & 0.78 & 0.90 & 0.91 & 0.82 & 0.95\\
BM & 0.56 & 0.69 & 0.50 & 0.45 & 0.55 & 0.82 & 0.63 & 0.55\\
BY & 0.83 & 1.08 & 1.25 & 1.07 & 0.88 & 1.11 & 1.26 & 1.21\\
CP & 0.83 & 0.89 & 0.93 & 0.87 & 0.71 & 1.03 & 0.85 & 0.75\\
CR & 0.72 & 0.81 & 0.96 & 0.86 & 0.70 & 0.83 & 0.87 & 0.79\\
DJ & 0.91 & 0.78 & 0.85 & 0.93 & 0.84 & 0.86 & 0.75 & 0.93\\
EK & 1.05 & 0.77 & 1.01 & 0.80 & 1.23 & 0.88 & 1.00 & 0.86\\
FF & 0.93 & 0.97 & 1.21 & 1.28 & 0.87 & 0.82 & 1.24 & 1.28\\
FJ & 0.95 & 0.82 & 0.81 & 1.00 & 1.02 & 0.76 & 0.75 & 0.96\\
HE & 1.00 & 0.79 & 0.98 & 0.92 & 0.87 & 0.74 & 0.90 & 0.84\\
HI & 0.85 & 0.75 & 0.70 & 0.75 & 0.66 & 0.61 & 0.61 & 0.59\\
IH & 0.78 & 1.14 & 1.00 & 0.78 & 0.57 & 0.89 & 0.82 & 0.57\\
KA & 0.93 & 0.85 & 0.83 & 0.85 & 0.88 & 0.79 & 0.78 & 0.79\\
KM & 1.11 & 1.05 & 1.09 & 1.15 & 0.95 & 1.02 & 1.04 & 1.08\\
KO & 0.88 & 1.01 & 0.96 & 1.06 & 0.88 & 0.96 & 0.91 & 1.13\\
KT & 1.13 & 1.15 & 0.88 & 0.99 & 1.06 & 1.11 & 0.76 & 1.04\\
LB & 1.12 & 0.97 & 1.29 & 0.85 & 1.17 & 1.00 & 1.31 & 0.85\\
MA & 0.94 & 0.99 & 1.01 & 0.88 & 0.79 & 0.89 & 0.93 & 0.84\\
MC & 0.97 & 1.02 & 1.03 & 0.95 & 0.94 & 0.94 & 0.91 & 0.90\\
MG & 0.92 & 0.87 & 0.95 & 0.88 & 1.12 & 1.04 & 0.99 & 1.04\\
MK & 0.92 & 0.66 & 0.55 & 0.74 & 0.92 & 0.76 & 0.68 & 0.82\\
MN & 1.03 & 0.85 & 1.21 & 1.06 & 1.00 & 0.92 & 1.09 & 1.11\\
MY & 0.98 & 0.99 & 0.96 & 1.04 & 0.90 & 0.95 & 0.98 & 0.96\\
NI & 0.80 & 0.85 & 0.70 & 0.77 & 0.68 & 0.78 & 0.66 & 0.76\\
NP & 0.63 & 0.58 & 0.72 & 0.62 & 0.59 & 0.53 & 0.71 & 0.59\\
PC & 0.51 & 0.37 & 0.75 & 0.62 & 0.51 & 0.46 & 0.81 & 0.67\\
PE & 1.01 & 1.21 & 1.14 & 1.00 & 0.96 & 1.11 & 1.08 & 0.97\\
PT & 0.95 & 0.94 & 1.09 & 1.02 & 0.88 & 0.97 & 0.99 & 0.94\\
RA & 1.03 & 0.91 & 1.09 & 1.01 & 0.99 & 0.84 & 0.94 & 0.86\\
SH & 1.07 & 1.30 & 0.93 & 0.90 & 0.94 & 1.23 & 0.91 & 0.97\\
SN & 0.98 & 0.87 & 1.04 & 1.02 & 0.98 & 0.68 & 1.02 & 1.03\\
TE & 0.89 & 0.88 & 0.93 & 0.88 & 0.84 & 0.85 & 0.92 & 0.91\\
TI & 1.09 & 0.92 & 0.96 & 1.00 & 1.00 & 0.85 & 0.92 & 0.85\\
TM & 0.95 & 1.01 & 0.94 & 0.95 & 0.98 & 1.09 & 1.00 & 1.00\\
TS & 0.85 & 0.80 & 0.77 & 0.79 & 0.70 & 0.71 & 0.61 & 0.72\\
TZ & 0.73 & 0.72 & 0.72 & 0.69 & 0.87 & 0.72 & 0.77 & 0.68\\
UP & 0.45 & 0.53 & 0.49 & 0.50 & 0.40 & 0.48 & 0.42 & 0.46\\
VT & 1.10 & 0.89 & 0.89 & 1.10 & 1.17 & 0.86 & 0.90 & 1.17\\
WH & 1.17 & 1.17 & 1.53 & 1.44 & 0.93 & 0.82 & 1.24 & 1.29\\
WL & 0.68 & 0.67 & 0.79 & 0.68 & 0.63 & 0.64 & 0.83 & 0.70\\
YI & 1.10 & 1.10 & 1.09 & 1.16 & 1.11 & 0.98 & 0.90 & 1.14\\
YN & 0.94 & 1.34 & 1.10 & 1.35 & 0.97 & 1.58 & 1.11 & 1.87\\*
\end{longtable}
\endgroup{}

%% file: tables/wealth_mod_site_vars_table.tex
\begingroup\fontsize{9}{11}\selectfont

\begin{longtable}[t]{ccc}
\caption{\label{tab:site_vars_wealth_mod}Site-Level Normed Wealth Modularity}\\
\toprule
\multicolumn{1}{c}{ } & \multicolumn{2}{c}{Normed Wealth Modularity} \\
\cmidrule(l{3pt}r{3pt}){2-3}
Site & Per Capita & Absolute\\
\midrule
\endfirsthead
\caption[]{Site-Level Normed Wealth Modularity \textit{(continued)}}\\
\toprule
\multicolumn{1}{c}{ } & \multicolumn{2}{c}{Normed Wealth Modularity} \\
\cmidrule(l{3pt}r{3pt}){2-3}
Site & Per Capita & Absolute\\
\midrule
\endhead

\endfoot
\bottomrule
\endlastfoot
AH & 0.006 & -0.001\\
AR & 0.006 & 0.007\\
AV & -0.011 & -0.011\\
AZ & 0.035 & 0.035\\
BD & 0.002 & 0.004\\
BM & -0.009 & 0.018\\
BY & -0.038 & 0.020\\
CP & -0.009 & 0.005\\
CR & 0.009 & 0.015\\
DJ & -0.001 & 0.026\\
EK & 0.007 & -0.013\\
FF & -0.027 & -0.015\\
FJ & -0.016 & -0.003\\
HE & 0.009 & -0.009\\
HI & -0.012 & 0.030\\
IH & -0.033 & -0.027\\
KA & 0.007 & 0.001\\
KM & -0.025 & -0.039\\
KO & -0.019 & 0.008\\
KT & -0.046 & 0.007\\
LB & -0.012 & -0.043\\
MA & -0.002 & -0.015\\
MC & -0.029 & -0.017\\
MG & 0.007 & 0.006\\
MK & 0.002 & 0.001\\
MN & -0.028 & -0.037\\
MY & -0.018 & -0.018\\
NI & 0.010 & 0.000\\
NP & 0.046 & 0.047\\
PC & 0.074 & 0.056\\
PE & -0.017 & -0.006\\
PT & -0.016 & -0.023\\
RA & -0.010 & -0.016\\
SH & -0.045 & -0.033\\
SN & -0.013 & -0.006\\
TE & 0.003 & 0.002\\
TI & -0.012 & -0.029\\
TM & -0.006 & 0.002\\
TS & -0.032 & -0.020\\
TZ & 0.023 & 0.020\\
UP & 0.060 & 0.073\\
VT & -0.008 & -0.029\\
WH & -0.018 & -0.033\\
WL & 0.053 & 0.046\\
YI & -0.043 & -0.033\\
YN & -0.052 & -0.058\\*
\end{longtable}
\endgroup{}

%% file: tables/other_site_vars_table.tex
\begingroup\fontsize{9}{11}\selectfont

\begin{longtable}[t]{cccccccccccc}
\caption{\label{tab:site_vars_other}Other Site-Level Variables}\\
\toprule
Site & \makecell{Primary Kin\\Connect. \%} & \makecell{Strong Prod.\\Norms} & \makecell{Wage/Sal.\\Labor \%} & \makecell{Cash-Orient.\\\%} & \makecell{Yearly+\\Shocks} & \makecell{Biol. Degree\\Days} & \makecell{Human\\Influence Index} & \makecell{Prop. Rights\\Index} & \makecell{Freedom of\\Movement Index} & \makecell{Corruption\\Index} & \makecell{Natl. GDP\\per Capita}\\
\midrule
\endfirsthead
\caption[]{Other Site-Level Variables \textit{(continued)}}\\
\toprule
Site & \makecell{Primary Kin\\Connect. \%} & \makecell{Strong Prod.\\Norms} & \makecell{Wage/Sal.\\Labor \%} & \makecell{Cash-Orient.\\\%} & \makecell{Yearly+\\Shocks} & \makecell{Biol. Degree\\Days} & \makecell{Human\\Influence Index} & \makecell{Prop. Rights\\Index} & \makecell{Freedom of\\Movement Index} & \makecell{Corruption\\Index} & \makecell{Natl. GDP\\per Capita}\\
\midrule
\endhead

\endfoot
\bottomrule
\endlastfoot
AH & 0.17 & 0 & 9.0 & 75.0 & 2 & 0.00 & 1233.80 & 0.62 & 0.78 & 0.70 & 2518.46\\
AR & 0.24 & 0 & 55.0 & 19.0 & 3 & 3323.52 & 2719.39 & 0.88 & 0.82 & 0.39 & 12774.35\\
AV & 0.05 & 0 & 2.0 & 60.0 & 2 & 6538.71 & 2620.33 & 0.89 & 0.92 & 0.18 & 1263.62\\
AZ & 0.08 & 1 & 35.0 & 90.0 & 0 & 6745.78 & 3138.02 & 0.77 & 0.75 & 0.62 & 1788.70\\
BD & 0.08 & 0 & 30.0 & 75.0 & 2 & 5808.28 & 2432.04 & 0.81 & 0.74 & 0.90 & 1373.75\\
BM & 0.35 & 0 & 19.0 & 79.0 & 3 & 4275.40 & 362.36 & 0.77 & 0.82 & 0.68 & 10193.77\\
BY & 0.29 & 1 & 0.0 & 2.0 & 0 & 5693.93 & 1285.46 & 0.58 & 0.35 & 0.83 & 1909.96\\
CP & 0.31 & 0 & 3.0 & 45.0 & 2 & 1809.13 & 1412.07 & 0.80 & 0.83 & 0.53 & 6626.26\\
CR & 0.55 & 1 & 13.0 & 90.0 & 2 & 5683.02 & 1359.89 & 0.81 & 0.90 & 0.63 & 5971.39\\
DJ & 0.21 & 0 & 6.0 & 72.5 & 2 & 6542.34 & 3109.74 & 0.89 & 0.92 & 0.18 & 1263.62\\
EK & 0.49 & 0 & 25.0 & 10.0 & 3 & 3806.70 & 1582.30 & 0.75 & 0.62 & 0.66 & 1806.50\\
FF & 0.10 & 0 & 10.0 & 30.0 & 2 & 3367.58 & 646.48 & 0.76 & 0.80 & 0.19 & 4400.48\\
FJ & 0.10 & 1 & 10.0 & 10.0 & 2 & 5540.83 & 3122.99 & 0.65 & 0.82 & 0.25 & 5500.68\\
HE & 0.15 & 0 & 5.0 & 28.0 & 1 & 6483.60 & 1643.03 & 0.62 & 0.78 & 0.69 & 2572.56\\
HI & 0.22 & 0 & 0.0 & 10.0 & 3 & 2667.04 & 1296.62 & 0.73 & 0.74 & 0.19 & 4476.73\\
IH & 0.27 & 0 & 67.0 & 0.0 & 0 & 1110.07 & 3799.76 & 0.78 & 0.70 & 0.38 & 6353.55\\
KA & 0.11 & 0 & 0.0 & 70.0 & 1 & 5471.26 & 1757.13 & 0.82 & 0.76 & 0.73 & 3701.32\\
KM & 0.17 & 0 & 40.0 & 25.0 & 3 & 163.72 & 2259.20 & 0.89 & 0.82 & 0.72 & 9882.02\\
KO & 0.01 & 0 & 7.5 & 19.5 & 4 & 3488.15 & 921.95 & 0.66 & 0.60 & 0.55 & 678.99\\
KT & 0.54 & 0 & 0.0 & 30.0 & 0 & 5601.05 & 1602.93 & 0.71 & 0.79 & 0.32 & 11445.39\\
LB & 0.36 & 0 & 5.0 & 10.0 & 0 & 381.73 & 1677.00 & 0.42 & 0.28 & 0.48 & 9221.35\\
MA & 0.07 & 1 & 5.0 & 20.0 & 3 & 4343.69 & 887.52 & 0.66 & 0.78 & 0.32 & 1001.56\\
MC & 0.25 & 0 & 10.0 & 10.0 & 2 & 6356.44 & 1058.26 & 0.88 & 0.90 & 0.39 & 5997.18\\
MG & 0.20 & 0 & 0.0 & 10.0 & 2 & 5266.06 & 2017.00 & 0.83 & 0.72 & 0.86 & 467.28\\
MK & 0.41 & 0 & 15.0 & 25.0 & 3 & 5225.66 & 1362.92 & 0.83 & 0.72 & 0.82 & 456.85\\
MN & 0.29 & 0 & 10.0 & 80.0 & 2 & 4791.04 & 1313.11 & 0.70 & 0.85 & 0.65 & 3204.54\\
MY & 0.52 & 0 & 9.0 & 16.0 & 0 & 6030.77 & 2314.65 & 0.77 & 0.82 & 0.68 & 10193.77\\
NI & 0.32 & 0 & 17.0 & 20.0 & 2 & 5431.16 & 373.28 & 0.72 & 0.33 & 0.90 & 2107.34\\
NP & 0.08 & 1 & 33.0 & 10.0 & 0 & 1971.59 & 2667.57 & 0.62 & 0.65 & 0.62 & 1179.95\\
PC & 0.29 & 0 & 25.0 & 42.0 & 4 & 5609.30 & 4697.96 & 0.79 & 0.65 & 0.40 & 6309.68\\
PE & 0.18 & 0 & 11.0 & 50.0 & 3 & 5978.02 & 1850.37 & 0.69 & 0.78 & 0.30 & 1024.02\\
PT & 0.47 & 0 & 15.0 & 40.0 & 2 & 6032.93 & 179.41 & 0.82 & 0.83 & 0.55 & 8771.44\\
RA & 0.18 & 0 & 15.0 & 10.0 & 4 & 5021.81 & 1462.06 & 0.64 & 0.75 & 0.89 & 1437.35\\
SH & 0.23 & 0 & 0.0 & 70.0 & 3 & 5808.28 & 2652.20 & 0.81 & 0.74 & 0.90 & 1373.75\\
SN & 0.29 & 0 & 10.0 & 25.0 & 1 & 5417.51 & 1320.22 & 0.48 & 0.67 & 0.54 & 617.14\\
TE & 0.07 & 1 & 35.0 & 90.0 & 0 & 6745.78 & 3181.62 & 0.77 & 0.75 & 0.62 & 1788.70\\
TI & 0.38 & 0 & 5.0 & 23.0 & 4 & 6183.49 & 149.98 & 0.65 & 0.64 & 0.67 & 3207.39\\
TM & 0.36 & 1 & 2.0 & 40.0 & 0 & 3701.16 & 2615.85 & 0.82 & 0.75 & 0.55 & 10013.25\\
TS & 0.64 & 0 & 10.0 & 20.0 & 2 & 6078.47 & 1381.06 & 0.70 & 0.85 & 0.65 & 3204.54\\
TZ & 0.10 & 0 & 40.0 & 20.0 & 2 & 2382.59 & 1613.67 & 0.80 & 0.55 & 0.67 & 3421.22\\
UP & 0.10 & 1 & 70.0 & 20.0 & 3 & 5532.83 & 3231.14 & 0.77 & 0.78 & 0.68 & 2523.44\\
VT & 0.14 & 0 & 20.0 & 30.0 & 1 & 4362.49 & 2217.76 & 0.72 & 0.72 & 0.40 & 3319.76\\
WH & 0.11 & 0 & 31.0 & 85.0 & 3 & 4662.23 & 715.34 & 0.78 & 0.64 & 0.38 & 6439.96\\
WL & 0.12 & 0 & 32.0 & 40.0 & 2 & 4662.23 & 902.68 & 0.79 & 0.65 & 0.40 & 6309.68\\
YI & 0.21 & 0 & 30.0 & 60.0 & 2 & 2525.08 & 1094.77 & 0.42 & 0.28 & 0.48 & 9798.58\\
YN & 0.15 & 0 & 20.0 & 45.0 & 1 & 521.69 & 1974.60 & 0.42 & 0.28 & 0.48 & 9221.35\\*
\end{longtable}
\endgroup{}